KTH Informations- och
kommunikationsteknik

# A Verification Framework for Component Based Modeling and Simulation

## "Putting the pieces together"

Imran Mahmood

KTH School of Information and
Communication Technology
SE-164 40 Kista
Sweden

Akademisk avhandling som med tillstand av Kungliga Tekniska högskolan framlägges till offentlig granskning för avläggande av teknologie doktorsexamen tisdagen den 26 feb 2013 kl 14.00 i Sal E, Forum, Isafjordsgatan 39, Kista.

// Acknowledgement

I would like to dedicate this manuscript to my loved ones, including some dignitaries, my parents who recently passed away, my guardians Mr. & Mrs. Sajid Latif who raised me well to make me see this day and most important of all: my beloved wife and my little daughter. Their sacrifice for being apart and for my long absence cannot be compensated for anything.

I offer my deepest gratitude to my supervisor Professor Rassul Ayani for this devotion and support. Instead of just giving me the directions he actually grabbed my hand and took me to the destination like a true guide. I am honored to work under his supervision. I am thankful to Assoc. Professor Vladimir Vlassov who gave sound advice and provided valuable contributions in my research. I offer my affectionate tribute to the esteemed palace of knowledge, the Royal Institute of Technology, and specially the school of Information and Communication Technology.

I am thankful for continuous support and encouragement from Dr. Farshad Moradi from Swedish Defense Research agency (FOI). I am grateful for the constructive critics I received from my opponent Dr. Gary Tan and the member of the evaluation committee Dr. Oliver Dale.

I am very grateful for the Higher Education Commission of Pakistan to provide entire financial support for my studies. I thank Mrs. Mumtaz Begum for her support. I would like to offer special thanks to Mr. Awais Ali Sohrawardi and Dr. B. Muhammad for their moral support during my study period. I thank all my colleagues, friends and especially the cricket team for wonderful time.

Finally I thank Sweden for its hospitality, care and warm memories.

Imran Mahmood

January 2013, Stockholm

# Abstract

The discipline of component-based modeling and simulation offers promising gains including reduction in development cost, time, and system complexity. This paradigm is very profitable as it promotes the use and reuse of modular components and is auspicious for effective development of complex simulations. It however is confronted by a series of research challenges when it comes to actually practise this methodology. One of such important issue is *Composability verification*. In modeling and simulation (M&S), *composability* is the capability to select and assemble components in various combinations to satisfy specific user requirements. Therefore to ensure the correctness of a composed model, it is verified with respect to its requirements specifications.

There are different approaches and existing component modeling frameworks that support composability. Though in our observation most of the component modeling frameworks possess none or weak built-in support for the composability verification. One such framework is Base Object Model (BOM) which fundamentally poses a satisfactory potential for effective model composability and reuse. However it falls short of required semantics, necessary modeling characteristics and built-in evaluation techniques, which are essential for modeling complex system behavior and reasoning about the validity of the composability at different levels.

In this thesis a comprehensive verification framework is proposed to contend with some important issues in composability verification and a verification process is suggested to verify composability of different kinds of systems models, such as reactive, real-time and probabilistic systems. With an assumption that all these systems are concurrent in nature in which different composed components interact with each other simultaneously, the requirements for the extensive techniques for the structural and behavioral analysis becomes increasingly challenging. The proposed verification framework provides methods, techniques and tool support for verifying composability at its different levels. These levels are defined as foundations of consistent model composability. Each level is discussed in detail and an approach is presented to verify composability at that level. In particular we focus on the Dynamic-Semantic Composability level due to its significance in the overall composability correctness and also due to the level of difficulty it poses in the process. In order to verify composability at this level we investigate the application of three different approaches namely (i) Petri Nets based Algebraic Analysis (ii) Colored Petri Nets (CPN) based State-space Analysis and (iii) Communicating Sequential Processes based Model Checking. All the three approaches attack the problem of verifying dynamic-semantic composability in different ways however they all share the same aim i.e., to confirm the correctness of a composed model with respect to its requirement specifications. Beside the operative integration of these approaches in our framework, we also contributed in the improvement of each approach for effective applicability in the composability verification. Such as applying algorithms for automating Petri Net algebraic computations, introducing a state-space reduction technique in CPN based state-space analysis, and introducing function libraries to perform verification tasks and help the modeler with ease of use during the composability verification. We also provide detailed examples of using each approach with different models to explain the verification process and their functionality. Lastly we provide a comparison of these approaches and suggest guidelines for choosing the right one based on the nature of the model and the available information. With a right choice of an approach and following the guidelines of our component-based M&S life-cycle a modeler can easily construct BOM based composed models and can verify them with respect to the requirement specifications.

# Table of Contents

# List of Figures

# List of Tables

# List of Acronyms

| | |
|---|---|
| **ABV** | Assertion-based Verification |
| **Ac** | Communicating Arcs |
| **ACP** | Algebra of Communicating Processes |
| **ALSP** | Aggregate Level Simulation Protocol |
| **AOI** | Area-of-Interest |
| **AP** | Atomic propositions |
| **API** | Application programming interface |
| **ARC** | Adelaide Refinement Checker |
| $A_{SV}$ | State-variable arc |
| $A_T$ | Transiting arc |
| **BB** | Basic BOM |
| **BDD** | Binary Decision Diagram |
| **BHQ** | Battalion Headquarters |
| **BHQSM** | Battalion Headquarters State-machine |
| **BID** | Battery ID |
| **BL** | Behavioral Layer |
| **BOM** | Base Object Model |
| **CB** | Composed BOM |
| **CBM&S** | Component Based Modeling and Simulation |
| **CBSE** | Component Based Software Engineering |
| **CBT** | Composable Behavioral Technologies |
| **CCA** | Common Component Architecture |
| $C_{CP}$ | Color set of communicating port |
| **CCS** | Milner's Calculus of Communicating Systems |
| **CL** | Communication Layer |
| **CM** | Conceptual Model |
| **CODES** | Composable Discrete-Event scalable Simulation |
| **COST** | Component Oriented Simulation Toolkit |
| **CP** | Communicating Port |
| **CPN** | Colored Petri Nets |
| **CPN-CM** | Colored Petri Nets Component Model |
| **CPN-ML** | ML scripting language for Colored Petri Nets |
| **CSP** | Hoare's Communicating Sequential Processes |
| $C_{SV}$ | Color set of State variable |
| **CTL** | Computation Tree Logic |
| **DEDS** | Discrete Event Dynamic Systems |
| **DES** | Discrete Event Systems |
| **DEVS** | Discrete Event System Specification |
| **DIS** | Distributed Interactive Simulation |
| **DMC** | Discovery, Matching & Composition |
| **DOT** | DOT file format |
| **EC** | Event Controller |
| **EFSM** | Extended Finite State-machine |
| **EIC** | DEVS input port couplings |
| **EOC** | DEVS output port couplings |
| **EXPR** | Expression |
| **FA** | Field Artillery |
| **FD** | Field Data |
| **FDC** | Fire Direction Center |
| **FSM** | Finite State-Machine |
| **HLA** | High Level Architecture |
| **HPC** | High Performance Computing |
| **IC** | DEVS Internal Coupling |
| **IDE** | Integrated Development Environment |
| **INT** | Integer |
| **ISV** | Initialization function of State-variable |
| **JCSP** | Java based Communicating Sequential Process |
| **JSIMS** | Joint Simulation System |
| **JUNG** | Java Universal Network Graph library |
| **LCIM** | Levels of Conceptual Interoperability |
| **LTL** | Linear Temporal Logic |
| **LVC** | Live, virtual, or constructive Simulation |
| **MBSC** | Model based simulation composition |
| **MCT** | Model Coupling Toolkit |
| **MDF** | Matrix Definitional Form |
| **MDP** | Markov Decision Processes |
| **MOCCA** | Component based Grid Environment |
| **MPD** | Markov decision processes |
| **MUSCLE** | A Multi-scale Coupling Library and Environment |
| **NET** | Network |
| **OMT** | High Level Architecture Object Model Template |
| **OSA** | Open Simulation Architecture |
| **OWL** | Web Ontology Language |
| **PAT** | Process Analysis Toolkit |
| **PIPE** | Platform Independent Petri Net Editor |
| **PLTL** | Probabilistic Linear Temporal Logic |
| **PN** | Petri Net |
| **PNML** | Petri Net Markup Language |
| **POI** | BOM Pattern Of Interplay |
| **RS** | Requirement Specifications |
| **SAT** | Boolean Satisfiability |
| **SCT** | Semantic Composability Theory |
| **SCXML** | State Chart extensible markup language |
| **SE** | Software Engineering |
| **SIMNET** | Simulation Networking |
| **SISO** | Simulation Interoperability Standards Organization |
| **SL** | Structural Layer |
| **SM** | Syntactic Matching |
| **SML** | Scripting language |
| **SMM** | State-Machine Matching |
| **SSM** | Static-Semantic Matching |
| **SV** | State Variable |
| **$SV_{IN}$** | Input State Variable |
| **$SV_{OUT}$** | Output State Variable |
| **TCSP** | Timed Communicating Sequential Processes |
| **TENA** | Test and Training Enabling Architecture |
| **TOT** | Time On Target |
| **UML** | Unified Modeling Language |
| **$V_{CP}$** | Communication Port Variable |
| **V&V** | Verification and Validation |
| **VVA** | Verification, Validation and Accreditation |
| **VVT** | Verification, Validation and Testing |
| **XML** | Extensible Marking Language |
| **XMSF** | Extensible Modeling and Simulation Framework |
| **$X_T$** | Firing Vector |
| | |
| | |

# Part I

# Episteme

***Epistêmê*** *in Greek means "to know". It is the theoretical knowledge; a principled system of understanding; fundamental body of ideas and collective presuppositions that determine the knowledge which is intellectually certain at any particular period of time; Pure-Science; episteme deals with "what" and "why" of the subject.*

Part-I covers the epistemology of the research under discussion where the theory, concepts, principles, paradigms, philosophy and rationale of the problem domain and the solution domain are sketched. In essence Part-I contains theoretical knowledge and the background information required to understand the problem and proposed solution discussed in the second part.

*"If you can't explain it simply, you don't understand it well enough".*

*- Albert Einstein*

# Chapter 1
# Introduction

*This chapter provides the opening statement and general information about the research presented in this thesis. It outlines background, history, the formal definition and the basic philosophy of the problem under question and covers the motivation, goals and scope of the research and the contributions of the thesis. In the end, a section on the thesis organization is rendered.*

## 1.1 Background and the opening perspective

Over the last fifty years, there has been a revolutionary development influencing almost all of the sciences. This progress is mainly instigated by the astonishing growth of the use of the digital computers and the subsequent rise of the age of computer simulations [**1**]. It is the emergence and widespread availability of computing power and resources that have made possible the new dimension of experimentation with complex models and their simulations [**2**]. Computer simulations are now widely used in various scientific disciplines and application domains. They are used for studying complex systems and gaining insight into the operation of an existing system without disturbing the actual system. Furthermore they are used for testing new concepts of the systems before implementation, visualizing and predicting behavior of a future system. Besides, they are used for analyzing and solving problems, drawing conclusions and aiding the process of crucial decision making [**3**]. Therefore computer simulation is regarded as third branch of science [**4**] and stands alongside of the first two branches namely *theory* and *experimentation*.

Modeling and Simulation (M&S) is a discipline with its own body of knowledge, theory, and research methodology [**4**]. The goals of M&S are aligned with the systems theory, and include modeling & analysis, design & synthesis, control, performance evaluation and optimization of a real system. The M&S community has demonstrated a longstanding focus on providing support for these goals. With the advent of the net-centric era of methods and technologies in designing complex simulation systems, the focus of M&S industry has been driven by the most recognized potential benefits of reduced development cost, time and system complexity [**5**]. This is because M&S development process is costly, time consuming, resource intensive. Models can be large, complex and require a great deal of time, resource and domain specific expertise to develop. Beside this, an enormous effort is required to evaluate that the model is correct and meets its requirements. Therefore M&S community has taken a deep interest in the quality design principles and their underlying supportive theories to alleviate these challenges. It has been realized that constructing a model from scratch each time it is needed is inefficient. Instead, the practice of model reuse has been increasingly appreciated and is inspired from the vision of software reuse, which was originally introduced in 1968 [**6**]. Apparently this approach looks very appealing however it poses many obstacles in implementing, such as lack of flexibility and adaptability in design, difficulty of integration, mismatched interface, incomplete specification etc. [**7**]. These obstacles are considered elusive research challenges and are now the primary research interests of the software engineering and M&S communities [**8**]

### 1.1.1 Component based Software Engineering

Component-based software engineering (CBSE) has been identified as a key enabler in the construction of complex systems by combining software components that are already developed and prepared for integration [**8**].

**Software Component**

*A software component is defined as a unit of composition which is independently developed and can be combined with other components to build larger units. It must have clearly specified interfaces to communicate with its environment while the implementation must be encapsulated in the component and is not directly reachable from the environment* [**9**]*, and therefore can be easily used by the third party without having to know implementation details* [**8**], [**10**].

Building software from components contributes to a major paradigm shift in software engineering. The basic philosophy behind the idea of component-based development is to carry out the software development process by (quickly) producing software applications through assembling prefabricated software components and to archive these interoperable software components in form of an increasingly large repository for further reuse [**11**]. CBSE promotes the principle of modularity. That essentially helps to master the complexity of the reality by decomposing it into parts [**12**] and enables the designer to use and reuse appropriate parts for different purposes. These parts are the sub-systems built in a component-based fashion. These subsystem components may have been separately developed by different teams. They may also have been developed for different purposes unrelated to the current context of the usage. CBSE has many advantages, such as effective management of complexity, logical separation, reduced time and cost, increased productivity, improved quality, a greater degree of consistency, increased dependability, and a wider range of usability. In addition, the growing connectivity of real world problems is reflected in the requirement to compose cross domain solutions [**13**], and therefore support knowledge sharing to a wider user community. CBSE is therefore a discipline of software engineering that deals with the composition of components to construct software systems which are capable of performing functions according to the user's requirements [**14**].

In CBSE, component integration and component composition are two distinguished terms. Component integration is merely the task of connecting components together whereas composition also includes reasoning about the semantic behavior of the resulting assembly [**14**]. With the advent of component technology the integration problems are becoming a difficulty of the past. Instead more crucial problems of predicting the emergent behavior of assemblies and the problem of reasoning about how well components will play together are now in debate. Component composition supports this type of reasoning and provides a foundation for fundamental reasoning to justifying validity of the resulting assemblies, their run-time compatibility and emergent behavior. The main reason for the difference between integration and composition is due to the fact that component interfaces do not provide enough information to determine how well the composed components will play together [**14**]. An interface can only help to determine if the component can be connected to some other component but cannot supports reasoning about emergent properties of the assemblies [**14**], [**15**]. Component composition promises such rationale; however is still a subject of open research.

### 1.1.2 Component based Modeling & Simulation

Inspired by the discipline of component based software engineering, M&S community has also started to develop simulation models by reusing previously developed and validated "simulation components", and composing them in a new simulation model according to the desired user objectives [**16**], [**17**], [**18**], [**19**], [**20**]. The basic and effective strategy for tackling any large and complex simulation problem is "divide and conquer." One major idea in component-based simulation development is to create models that are themselves self-contained and independently deployable. Thus different simulationist will be able to work on different components independently, without needing much communication among each other, (and particularly without the need to share the classified domain knowledge) and yet the components will work together seamlessly. In addition, during the maintenance phase, it is possible to modify some of the components without affecting all of the others [**21**].

In simulation community the research on component based development falls under the rubric of composability [**22**], where simulation models are considered to be the building blocks and are referred as "model-components".

**Model Component**[1]

> *A model component is an independent element of a simulation model that conforms to certain component standard, has well-defined functionalities (inputs/outputs) and behaviors, presented through its interface describing its communication with other components and a formalized description of its internal behavior. A model component is not a stand-alone component, but can be independently deployed, and it is subject to third-party composition with or without modification* [**19**].

In component based development, some basic reusable model components are composed together to create complex and sophisticated simulations, without building them from scratch. The model components can be composed if their inputs and outputs physically match each other however it is difficult to say whether this combination is meaningful. Besides it cannot be said for sure if it will perform according to the desired requirements unless the correctness of the composability is checked.

Composability is the property of the models, as it essentially contends with the alignment of issues on the modeling level [**13**], where it is viewed as creation of complex models by selection and integration of basic reusable model-components. A set of components can be integrated if their inputs and outputs are compatible, but in order to guarantee that their combination is valid in the required executable scenarios, we study the degree of composability.

With a slightly greater number of components, which are somewhat complex in nature, the composability becomes an increasingly challenging problem. In the presence of functional and non-functional application requirements it poses severe implications on the effort involved in verifying the requirements, and increasing dynamism. Even though, the individual components are pre-verified; their verification is usually done in a limited context, with assumptions that may not hold after composition. As a result, the complexity of system verification grows exponentially with the number of applications [**23**][2].

[1]The term Model component should be differentiated from the term Component Model, which in text refers to the underlying technology being used by the component based software engineering platforms such as CORBA, EJB etc.

### 1.1.3 Modeling and Analysis using Petri Nets

Petri nets (PN) is a mechanism of modeling complex systems, in which states and events can be manipulated according to certain rules and explicit conditions. PN formalism was introduced by Carl Adam Petri in 1962. It provides an elegant and useful graphical and mathematical formalism for modeling concurrent systems and their behaviors [**24**].

PN graphs are quite suitable for representing Discrete Event Systems (DES) in which operations depend on potentially complex control schemes. PN graphs are intuitive and capture a lot of structural and behavioral information about the system. Another motivation for considering PN for the modeling of DES is the body of analysis techniques that have been developed for over three decades and are used for reasoning about structural and behavioral properties of PN models. These techniques include reachability analysis, state-space analysis, and model-checking as well as linear-algebraic techniques [**25**].

The PN research has been developed in two directions for the past three decades: (i) PN theory that focused on the development of basic tools, techniques and concepts needed for the PN application; (ii) Applied PN theory which is mainly concerned with the PN application for the modeling of systems and their analysis. Successful work in this direction requires good knowledge of the application area in which PN are applied and PN theories and techniques [**26**].

### 1.1.4 Modeling and Analysis using Process Algebra

Process Algebra is an algebraic approach for the modeling and analytical study of concurrent processes. It has a diverse family of algebraic formalisms for modeling concurrent systems. These formalisms comprise of algebraic language for the specification of processes and provide calculi in form of algebraic laws that allow process descriptions to be manipulated and analyzed, and permit formal reasoning about their correctness and equivalence [**27**]. The main Process algebraic formalisms are:

- CCS, Milner's Calculus of Communicating Systems
- CSP, Hoare's Communicating Sequential Processes
- ACP, Algebra of Communicating Processes
- LOTOS, Language Of Temporal Ordering Specification

### 1.1.5 Model Verification

In M&S, verification is concerned with building the model right. It is typically defined as a process of determining whether the model has been implemented correctly [**28**] and whether it is consistent with its specifications [**29**]. In principle, verification is concerned with the accuracy of transforming the model's requirements into a conceptual model and the conceptual model into an executable model [**29**]. The distinction of a conceptual model and executable model is of great importance and is a fundamental principle in M&S. A conceptual model is abstract description of a real system [**30**], captured based on given requirements and modeling objectives. This is later refined and implemented into a more concrete executable model. In these terms, conceptual modeling is a subset of model design [**31**]. Conceptual modeling is about moving from a problem situation, through model requirements to a definition of what is going to be modeled, and is independent of its implementation details [**30**], which are later addressed in form of an executable model.

---

[2] Even though the referred text corresponds to the electronic components which are physically composable, however the problem of composability complexity is the same and is mutually understood by different communities.

## 1.2 Summary of the opening perspective

In essence, component-based approach is highly favored in M&S community for building large and complex models. But to ensure that the model is correct and meets its requirement specifications, a substantial effort is required to evaluate its degree of composability. In M&S community, the discipline of Model Verification provides basic concepts and fundamental principles for the compressive study of the degree of composability and reasoning its correctness with respect to the given specifications. However the existing component-based simulation frameworks offer limited built-in extensive verification techniques or none at all. Therefore third party approaches such as PN analysis techniques and process algebra are considered for the thorough examination of composed models at various levels of depth.

The sub-topics: (i) Component-Based Modeling & Simulation, (ii) PN /CSP Analysis and (iii) Model-Verification are the elementary pillars and theoretical foundations of this thesis and are expanded in details in chapter 2, 3 & 4 respectively.

## 1.3 Preliminaries

Based on the previous discussion, the formal definition of the problem of this thesis is furnished in this section. In order to define the problem statement, following definitions are used:

### 1.3.1 Definition 1: Set of Components

Let $\mathbf{C = \{c_1, c_2, c_3 \ldots, c_n\}}$ be a given set of components discovered and selected from a component repository R, as per the abstraction of the real-system.

### 1.3.2 Definition 2: Requirement Specification

The Requirement specification of the system model is defined as a tuple:

$$\mathbf{RS = \langle O, S \rangle} \text{ where:}$$

$\mathbf{O = \{o_1, o_2, o_3 \ldots, o_n\}}$ is a set of objectives (or goals) and

$\mathbf{S = \{s_1, s_2, s_3 \ldots, s_n\}}$ is a set of system constraints (or system properties).

**Objective:**

An objective $o_i \in O$ can be defined as a reachable "final-state" of the composed model or an aggregated desirable output (a data value or event) produced by the composed model which cannot be produced by individual components.

**System Constraint:**

> In modeling terms, a system constraint $s_i \in S$ is defined as a system property that must be satisfied; for instance a good state; which must be reached or a bad state; which must be avoided (never be reached) during the execution.

The notions of constraints are different from Objectives, because they can be necessary requirements but not the ultimate goals. E.g., a manufacturing system should not only produce the desired products (objective) but also fulfill safety requirements (constraints).

### 1.3.3 Definition 3: Composition & Composability Pattern

Let $CM\langle c_1, c_2, c_3 \ldots, c_n\rangle$ be a composition of a set of given components C, composed using a particular composability pattern P. A pattern describes how the components are attached to each other, i.e., the topology of the components. And provide important information for composability verification. A pattern of composability can be sequential, parallel, fork, join, iterative, or composite.

### 1.3.4 Definition 4: Satisfiability Operator

For each element in the requirement specification RS, a Satisfiability operator $\models$ maps a given composed model **CM** to a Boolean (True or False) formally described as follows:

- $CM\langle c_1, c_2, c_3 \ldots, c_n\rangle \models_i o_i \in O \rightarrow true \mid false$
- $CM\langle c_1, c_2, c_3 \ldots, c_n\rangle \models_j s_j \in S \rightarrow true \mid false$

For each relation $\models_i$ we define a verification function (algorithm or theorem) based on which the satisfiability operator maps the resultant value. This verification function determines whether a given composed model satisfies a required property.

## 1.4 Problem Statement

Based on the above definitions the problem statement is defined as follows:

> "Given a composed model CM, composed from a set of components C using a pattern P, and the requirement specification RS, can we verify that CM fulfills all the objectives and satisfy all the constraints given in the requirement specification".

Formally:

$$CM = Compose([c_1, c_2, c_3 \ldots, c_n], P) \wedge RS = \langle O, S\rangle \rightarrow \{CM \models_i \forall o_i \in O \wedge CM \models_j \forall s_j \in S\} \quad (1.1)$$

This problem statement is considered as an initial point and basis of the research presented in this thesis proposal. In this work it will be shown how a modeler can correctly compose component models and verify the composition at different levels through utilization of our proposed verification framework.

## 1.5 Approach

In this section an overview of the approach and methodology is presented. Based on the software engineering principle, this section is divided into two main parts (i) Problem Domain and (ii) Solution Domain.

### 1.5.1 Problem Domain

Problem domain (or problem space) is an engineering term referring to all information that defines the problem and its constraints. It includes the goals that the problem-owner [3] wishes to achieve, the context within which the problem exists, and all rules that define required essential functions or other aspects of any solution product. It represents the environment in which a solution will have to operate [Wikipedia].

All the information provided in this thesis related to modeling & Simulation, component-based model development, conceptual modeling, model components, composability, model-verification and the problem of composability correctness correspond to the problem-domain. In particular, Chapter 2 covers the main aspect of the problem domain where the component based modeling and simulation is discussed in detail. Following sub-sections briefly describe the selected method of specification of the problem domain.

**Base Object Model (BOM) as Composability Framework**

BOM is selected in this thesis as a component specification standard which can be used as a foundation for developing model components at conceptual level. They are composed and are subjected to the composability verification process to evaluate that they satisfy given requirements, hence represent component framework of the problem domain.

**Requirement Specification Template**

A "Requirement Specification Template" is defined which is used to formulate requirement specifications. It essentially contains a set of objectives and constraints (of standard or scenario-specific properties), which are required to be satisfied for the proof of correctness of the composed model.

### 1.5.2 Solution Domain

Solution domain (or solution space) is a term referring to all information that defines the proposed solution of the problem. It includes the concepts, principles, methods, techniques, algorithms, programs, software architects, frameworks, processes and recommended practices, which help in solving the problem under study.

Following sub-section gives a brief overview of the approach used in this thesis:

---

[3] A problem owner can be the customer, solution buyer, organization or a prospective target community. A problem owner sees the problem as an opportunity, whereas the solution engineer sees the problem for which he/she has to provide a solution.

**Multi-tier Composability Verification**

The composed model undergoes multiple iterations for composability verification at different levels. Each level corresponds to a tier in the verification process. When the composability at a particular level is successfully verified, next level is iterated. When all the levels are completed, the components are said to be fully composable. These levels are discussed in detail in chapter 2. The verification of these levels is discussed in chapter 5.

**PN Formalism**

PN formalism (and specially the Colored Petri Nets extension) is chosen for creating executable models of the BOM based conceptual models. The proposed framework automatically transforms BOM components in form of an executable PNML[4] or CPN-based component which can be executed or undergo a verification process using the corresponding PN execution environment.

**CSP Formalism**

CSP[5] formalism with an extension of Timed-CSP is picked as another executable modeling language for BOM based conceptual models. The proposed framework transforms BOM components into executable CSP process components and composed for execution and verification.

**Automatic Transformation Tools**

An automatic transformation tool is proposed, which transforms a BOM component model into the selected executable modeling formalism such as PNML, CPN based or CSP based executable model. It may also be required to provide additional details, which cannot be modeled or represented by BOM.

**Dynamic Analysis Approach**

Three main dynamic analysis approaches are selected for composability verification of BOM base composed models at dynamic-semantic composability level:

**Algebraic analysis approach**

This approach is used to transform a BOM composition into a classical PN model using PNML format and verifies the properties using PN algebra.

**State-space analysis approach**

This approach is based on using Colored Petri Nets and State-Space analysis. CPN tool is a strong simulation and verification tool. State-space analysis is a very accurate correctness reasoning technique; however it is costly in terms of computational power and memory. Therefore a reduction technique is also proposed to reduce a state-space graph of a composed model, in order to avoid state-space explosion.

**Model Checking approach**

CSP based model checking is used for the formal verification of BOM based composed model. In formal logic, model checking designates the problem of determining whether a formula or a correctness property $\phi$ defined using LTL, CTL[6] or similar property specification formalism, evaluates to true or false in an interpretation of a system K, written as $K \models \phi$. Efficient algorithms are selected to determine whether $K \models \phi$ holds [**32**].

[4] Petri Net Markup Language
[5] Communicating Sequential Process
[6] Linear Temporal Logic, Computational Tree Logic

In summary, these three approaches are extensively being used in formal verification for over a couple of decades and therefore equipped with rich theoretical foundations and practical tools and techniques. We however believe that they are being considered in this thesis for the composability verification of BOM based models (or for that matter any M&S composition framework) for the first time and will prove to be very promising and effective. A basic foundation is built using these approaches in this thesis, and their usage are shown though appropriate examples. Also necessary guidelines are provided for developing new verification methods using these approaches and tools, in order to address various verification issues.

This research aims to propose a multi-tier verification life cycle for defining, development, archiving, discovering, matching, selection and composing, transforming, executing, verifying and finally reasoning about the correctness of the composed models. This life-cycle extensively relies on the integrated component development, composition and verification framework that is being proposed in this research. This life-cycle follows our proposed process to perform verification of a composed model at different levels. This life-cycle can be adapted by M&S practitioners for rapid model construction, analysis, refinements and reuse and thus it will boost the process of modeling and simulation of complex dynamic systems.

### 1.5.3 Solution Statement

Based on the proposed approach the solution statement is described as follows:

> "A verified composed model guarantees that the selected components are composable at all composability levels, and they meet the requirement specification by satisfying given objectives and fulfilling the required constraints".

A correctly composed model, promotes reuse of base components thus support rapid model development and can be reused as yet another component later on.

## 1.6 Scope of the Thesis

In this section, the scope and the boundaries of the thesis are outlined.

### 1.6.1 Correctness

In this thesis, "Correctness" is the main focus of the research. The approach, methods, process and framework mainly deal with the correctness issue of the composability verification. The other issues such as performance, efficiency and cost estimation of the solution are currently beyond the scope of this thesis and considered as future work.

### 1.6.2 Validation

Validation is a vital part of model evaluation and always goes hand in hand with verification. However it is beyond the scope of this thesis. Although we believe that, our framework is flexible and open-ended. Therefore it can accommodate necessary extensions to support validation with a minor effort.

### 1.6.3 Emergence

Emergent behavior due to composition of sub-systems is an important and open research topic in the composability domain. We however do not address this issue in this thesis and consider it a future work.

### 1.6.4 Generalization

Currently, the proposed approach is based on Base Object Model, only as a demonstration of how our approach can be applied on an existing component standard. However the framework presented in this thesis is open-ended and can be generalized to accommodate any other component standard. Furthermore, heterogenic composability can also be supported. We however do not address generalization issues in this thesis.

## 1.7 Summary of the Contributions

The existing work in the area of component based modeling and simulation is fragmentary in nature, especially when the verification of component composability of model at a conceptual level is concerned. Furthermore, even though different composability verification approaches exist, but they have not been studied in depth at different granular levels.

In this research, composability of BOM based model is studied in depth, focusing mainly on the different levels. A multi-tier component based verification life-cycle is proposed that tackles key issues of such as model development, discovery, selection, matching, composition, requirement specification, transformation, implementation, execution, analysis and most importantly verification.

In terms of verification, the major contributions of this thesis include development of a composability verification framework, which integrates different methods, and techniques to support different tasks in the composability verification process of a composed model. These different tasks are categorized in different phase of a proposed component based modeling and simulation (CBM&S) life-cycle. We propose methods for evaluating structural and behavioral consistency of the composed BOMs. For structural evaluation we propose a set of static analysis techniques to verify that the components can be correctly connected and their communication is semantically consistent, meaningful and is understood as intended.

For behavioral consistency of the composition we suggest a state-machine matching technique. It verifies that the components can correctly interact with each other in a right causal order to reach final states. For the further evaluation of the behavioral composability our framework incorporates three main approaches: (a) PN Algebraic technique (b) CPN-based State-space analysis technique and (c) CSP based model checking. For each approach we develop automatic transformation tool that transforms a BOM based composed model into the executable model of the corresponding approach. We present three different case studies for the proof of concept and for the evaluation of our verification framework.

We also suggest various extensions in each approach to suit the needs of composability verification. For instance we propose algorithms for automation of the PN algebraic approach. Also a CPN based component model is proposed for the State-space algebraic approach in order to describe a BOM component (or any other simulation component) in form of an executable model that can be executed using

CPN execution environment. We also introduce a State-space reduction technique for the CPN based state-space analysis approach to avoid the risk of state-space explosion. For the CSP based model checking approach we propose an external function library for methods to support various modeling tasks such as definition of probability distribution functions for probabilistic system models.

## 1.8 Structure of the Thesis

This thesis is divided into two main parts:

**Part I Episteme:** This part mainly covers the theoretical concepts, principles and discussions. It comprises of chapters 1, 2, 3 & 4.

**Chapter 1: Introduction:**

Chapter 1 gives a bird's eye view of the research presented in this thesis. It addresses the concept, historical background and the basic philosophy of composability. The problem is defined and the approach is briefly introduced. A section on the scope of the thesis and main contributions are also presented.

**Chapter 2: Component Based Modeling and Simulation:**

Chapter 2 introduces and discusses component based modeling and simulation in details, as it is the foundation of the problem domain. This chapter mainly covers the theory, issues, different levels, framework and the formalism of model composition. It also introduces Base Object Model (BOM) in details as a choice of Model composition standard of this thesis.

**Chapter 3: Executable Modeling Formalism**

This chapter provides introduction, theory, basic definition and classification of PN and CSP as executable modeling formalisms and regarded as solution domain. It also describes basic concepts of the analysis techniques that are used later in this thesis.

**Chapter 4: Verification and Analysis**

Chapter 4 discusses theory and principles of verification. It also categorizes some of the important verification techniques that are used in this thesis.

**Part II Techne:** This part contains practical aspects including approaches, methods, tools, development frameworks and lifecycle. It also contains examples related to our proposed solutions for the proof of concept. It comprises of chapters 5, 6, 7, 8 & 9.

**Chapter 5: Proposed Approach and Framework**

Chapter 5 is the center of the thesis as it provides the most important details of our contributions. It describes the proposed verification framework and verification life-cycle. It covers our proposed methods, techniques, algorithms, procedures as our contributions at different phases of composability verification process. These phases and their concerning activities are outlined as composability verification life-cycle.

**Chapter 6: Composability Verification Process**

This chapter presents the proposed composability verification process. It provides essential guidelines of how to use our proposed composability verification framework (discussed in chapter 5). It uses work flow diagrams to describe the overall process and gives necessary guidelines to the modeler at each step.

**Chapter 7: Fairness verification using PN Algebraic Technique**

Chapter 7 describes a case study of a manufacturing system as an example to explain how the proposed framework helps to verify fairness property in a composed system. The purpose of this chapter is to practically demonstrate algebraic verification method.

**Chapter 8: Model verification using State-space analysis technique**

Chapter 8 covers an example of the verification of a Field Artillery Model. It practically demonstrates how state-space analysis is used to verify a composed system. The field artillery model is introduced in detail along with requirement specifications and it is shown how the proposed approach can help to verify its composability.

**Chapter 9: Model Checking**

This chapter demonstrates an example of verification using CSP based Model Checking. The field artillery model discussed in chapter 8 is modified into a real-time probabilistic system and is verified using CSP based model checking.

**Chapter 10: Conclusion and Future work**

This chapter provides summary and conclusion, discussion and future work of the thesis.

# Chapter 2
# Component Based Modeling and Simulation

*Composability is an important quality characteristic and an effective means to achieve several benefits in M&S discipline, but in reality, it is a challenging and daunting problem. The community has conducted active research on its theoretical and practical intricacies. In theory, composability is studied under various facets and views primarily distinguished, by its different "layers" or "levels" as identified by different research groups. Whereas in practice, various practical challenges associated with it are investigated. Most important of these issues are component specification, development, integration, composability verification and validation, collectively referred to as phases of a Component based life-cycle. In this chapter both theoretical and practical aspects of composability are discussed in detail.*

## 2.1 Composability in M&S

In M&S applications, composability has been defined in different ways. Much of these definitions have been collected by A. Tolk in his article [**13**]. Harkrider and Lunceford defined composability as:

> *The ability to create, configure, initialize, test, and validate an exercise by logically assembling a unique simulation execution from a pool of reusable system components in order to meet a specific set of objectives* [**33**].

Kasputis and Ng defined composability as:

> *The ability to compose models across a variety of application domains, levels of resolution, and time scales* [**16**]

Petty and Weisel recommended the following definition in their article on theory of composability, which later was appended by P. K. Davis:

> *Composability is the capability to select and assemble simulation components in various combinations into valid simulation systems to satisfy specific user requirements, meaningfully* [**17**] [**34**].

It has been realized that composing models is more difficult than composing general software components. This argument is predicated on the assumptions that models are more complex; they are developed for particular purposes, and they depend on context-sensitive assumptions [**8**] [**17**]. Model development is a hard design task, mainly due to the complexity involved in the process. Nowadays this complexity is increasing to levels in which the utilization of pre-defined models is considered very useful to cut short the development time. Thus model composition is a paradigm, where existing components are the building blocks for the construction of new larger and more sophisticated models. When a model is composed, it must be evaluated in terms of correctness with respect to its requirements. In short the predictability of guaranteeing the correctness of model composition is called Composability.

## 2.2 A Brief History of Composability and related work

### 2.2.1 Initiation

Composability in M&S has primarily been investigated by the defense research sector. The earliest uses of the term composability within the context of defense simulation dates back to the Composable Behavioral Technologies (CBT) project during the mid-1990s [**35**]. Later on the Joint Simulation System (JSIMS) project investigated composability as a system objective [**36**]. In 1998, a project on model based simulation composition (MBSC) was started in which a prototype composition environment for JSIMS was developed. In 1999 Page and Opper investigated the composability problem from a computability and complexity theoretic perspective [**35**]. Composability became a key system objective for OneSAF project in 1999 [**22**].

### 2.2.2 Theoretical evolution

Later on a series of numerous articles were published which addressed various issues of and methodologies of composability and became the theoretical foundations for further research. Important works in this series include: Kasputis and Ng [**16**]; Davis et al. [**37**]; Petty & Weisel [**38**]. Petty and Weisel extended the work of Page and Opper, provided a broad survey of the uses of the term composability, and examined the composite validation problem within the context of automata theory and computable functions. Later a comprehensive report was published by Davis and Anderson in 2003 [**17**] that provides a broad survey of the composability and suggests its applications for the DoD[7] in this area.

### 2.2.3 Standards & Frameworks

Later on, the research on composability remained focused on the development of standard composition frameworks and its practical application in various domains of modeling and Simulation. In 2005 the Extensible Modeling and Simulation Framework (XMSF) was initiated by the Naval Postgraduate School to develop a web-based simulation environment [**39**]. Advances in M&S technologies, gave rise to different distributed simulation standards and protocols such as Simulation Networking (SIMNET), Distributed Interactive Simulation (DIS), Aggregate Level Simulation Protocol (ALSP) and the High Level Architecture (HLA). The details of these standards are well documented by Moradi [**19**]. Due to the complex nature of the standards, and distributed simulation itself, different composability frameworks were introduced to co-op with these requirements. More general-purpose frameworks such as the Discrete Event System Specification (DEVS) [**40**], the Open Simulation Architecture (OSA) [**41**], the Base Object Model (BOM) [**42**], and the Component Oriented Simulation Toolkit (COST) emerged and contributed to various issues of composability in different ways.

### 2.2.4 Technological Advances

Due to the technological advances in computer engineering, many approaches emerged with the aim to address issues and high end requirements of modeling and simulation such as representation of Complex, Dynamic and Adaptive Systems; integration of large interdependent Systems; multi-resolution and multi-scale modeling [**43**], and much more. In this period, many tools and techniques were developed using composability paradigm. Model Coupling Toolkit (MCT) was developed to support and simplify the construction of parallel coupled models [**44**]. MUSE is another composable simulation environment for astrophysical applications in which different simulation models of star systems are incorporated into a single framework [**45**]. Some frameworks such as Common Component Architecture (CCA) [**46**] and Component based Grid Environment (MOCCA) [**47**], were proposed to be used in high-performance computing, where scientific components are directly connected by their Users and Providers ports. A Multi-scale Coupling Library and Environment (MUSCLE) provided a software framework for building composable simulations according to the complex automata theory [**48**]. Compo-HLA is an environment proposed for supporting HLA component [**49**].

---

[7] United State Department of Defense

### 2.2.5 Composability verification and Validation

Most of these frameworks lack strong built-in composability evaluation support. Therefore some third-party composition, verification and validation frameworks were developed by individual research teams such as Composable Discrete-Event scalable Simulation (CODES) [**20**] and Semantic Web-based BOM composition framework [**19**], where verification and validation of composability are strongly focused.

## 2.3 Theory of Composability

The formal theory of composability was pioneered by Petty and Weisel [**34**], [**38**], [**50**] in an initiative developed at the Virginia Modeling, Analysis & Simulation Center (VMASC). It was also called "semantic composability theory" (SCT). The aim of the SCT is to check and prove the semantic composability of components using formal descriptions and reasoning. A *model* is defined as a computable function: $y = f(x)$, where function is calculable by a finite procedure and relates each input to a unique output, as shown in **Figure 1**

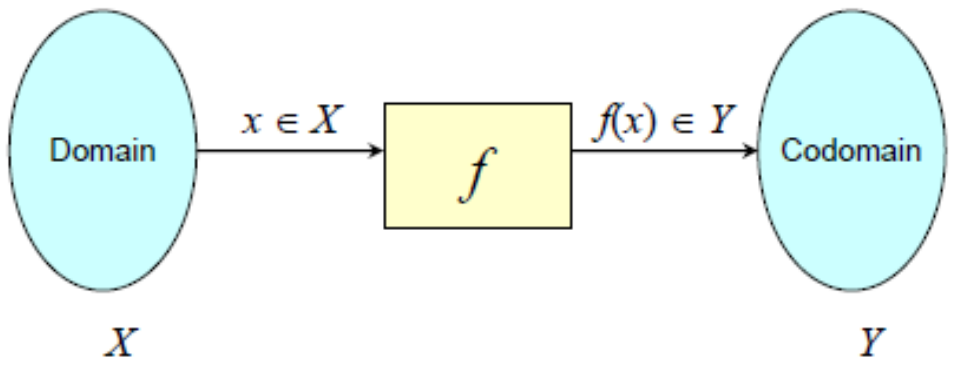

**Figure 1: A model as computable function** (acquired from [**34**])

A simulation is a sequence of executions of a model $f(x)$, where the output from each execution step is the input to the next step of the execution:

$$(m_n, o_n) = f(m_{n-1}, i_{n-1}) \tag{2.1}$$

Where i = input value; m=memory value; o=output value and n=current iteration, as shown in **Figure 2**

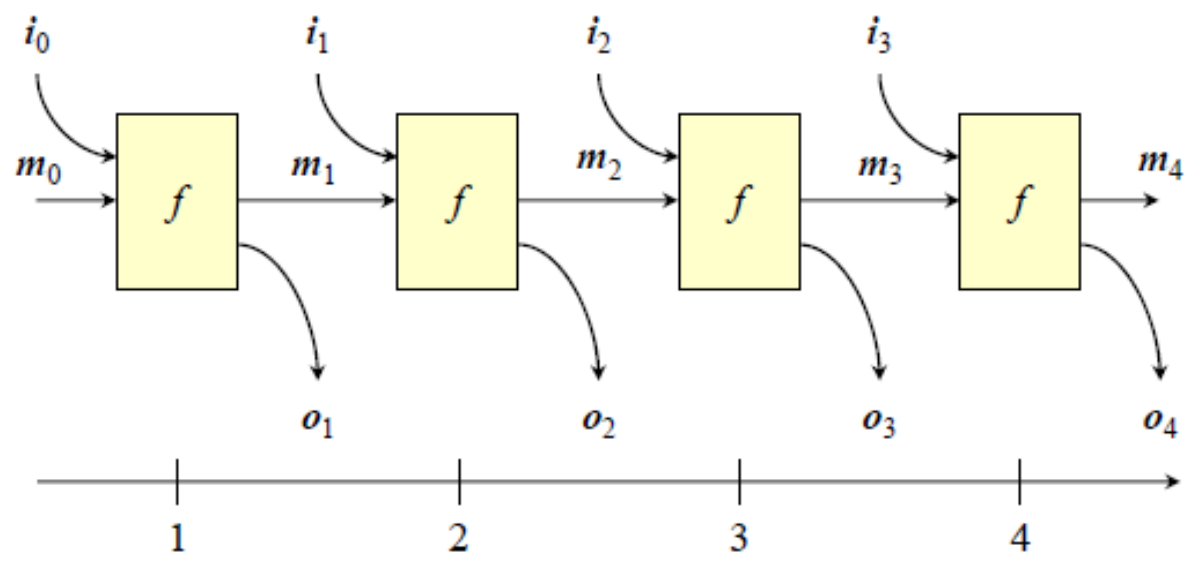

**Figure 2: Sequence of executions** (acquired from [**50**])

The composition is defined as output of one function to be the input of another:

$$h(x) = f(g(x)) \tag{2.2}$$

**Figure 3** shows the representation of a composed model, which is developed through composing other models (f1, f2 & f3). A composed model as a whole has also a set of inputs, outputs, current states and next-states as shown in **Figure 3**.

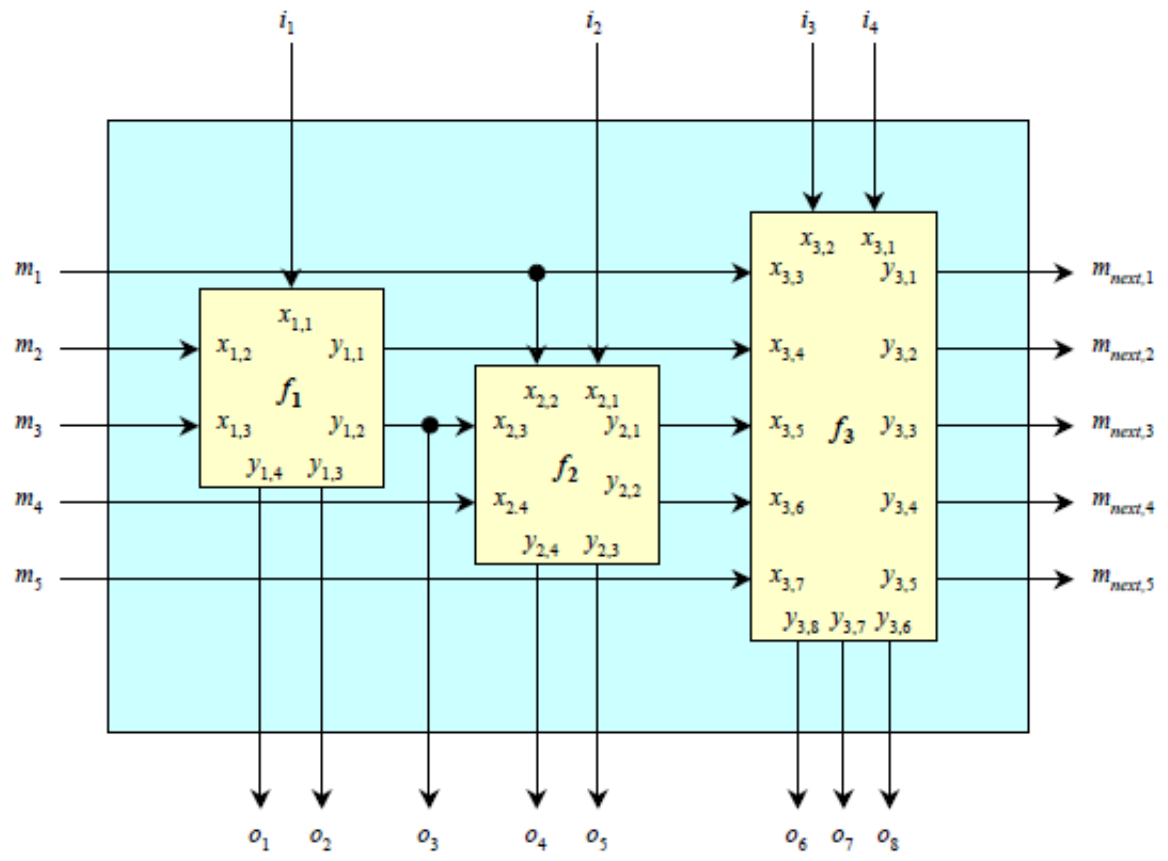

**Figure 3: Composed Model** (acquired from [**50**])

The composition of models in SCT is in fact the composition of functions. Since a set of computable functions is closed under composition any set of models can be composed if the composition exists, but there is no guarantee that the resultant will be a useful model. Thus focus of SCT is semantic composability, the question of whether the model composition is meaningfully valid or not.

### Validity

*A model is defined as valid, if it is an accurate representation of the real-world with respect to the intended use. For formal validation, the simulation of a composition is represented as Labeled Transition System where nodes are model states, edges are function executions, and labels are model inputs. A composition is valid if and only if its simulation is close to the simulation of a perfect model.*

### Perfect Model

*A model is perfect with respect to a natural system N* [8] *if and only if it represents a system of perfect observations of the natural system* [**50**].

[8] A natural system N is a real or imagined system.

For details of different classes of models, their equivalence relations, formal theorems and proofs of equivalence, interested readers should refer to [**50**]. The basic concepts of a formal theory of semantic composability include formal definitions for model, simulation, validity, and composition. A theory of composability can facilitate the convenient reuse of simulation components, which holds the potential to the time and cost of simulation development [**34**] [**38**] [**50**].

# 2.4 Concepts related to Composability

In this sub-section, some of the concepts and idea related to composability are discussed.

## 2.4.1 Composability vs. Reusability

Composability is differentiated from reusability in many aspects. Balci et al. define *Reusability* as the degree to which an artifact, method, or strategy is capable of being used again or repeatedly [**5**]. Robinson et al. on the other hand suggest that the term simulation model reuse can be taken to mean various things from the reuse of small portions of code, through component reuse, to the reuse of complete models [**51**]. Composability offers means to achieve reusability, but reusability might not always be the ultimate objective of model composition. For instance, in a particular situation, a set of modular components are purpose-fully built and composed to construct a large model, but they cannot be reused in a different project, due to their highly specific design. To be widely reusable, a component must be sufficiently general, scalable, and adaptable. A requirement for reusability may lead to another development approach, for example, a design on a more abstract level [**9**]. The comparison between usability and reusability of composable components poses a tradeoff between them being very specific in function and behavior so that they can be used in a particular case to satisfy specific user's requirements or them being very generic and abstract so that they can be reused in different situations again and again.

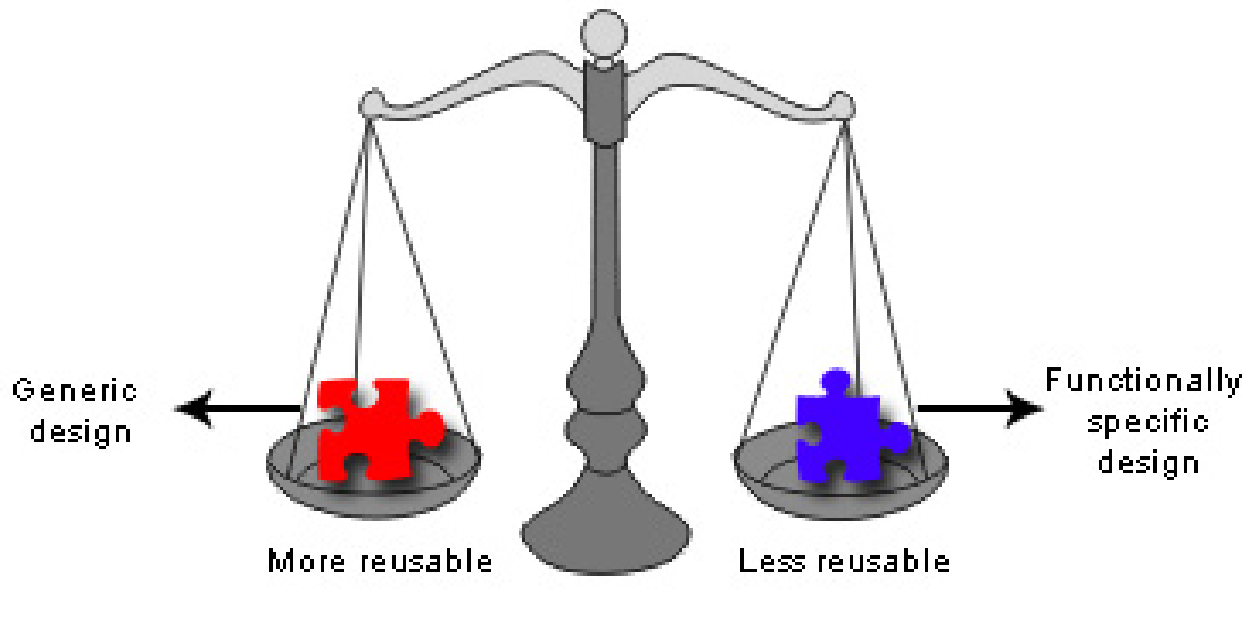

**Figure 4: Generic vs. Specific component design**

**Figure 4** illustrates a component is often more reusable if it has a generic design and less reusable if it has functionally specific design.

Both use and reuse of composable components share three levels of transparency. [**7**]. A component can be seen as a box, which contains the interfaces and internal implementation. Three levels of composability transparency are defined:

**Black Box Composition**

In black box composition, the user sees the interface, but not the implementation of the component. The user documentation is provided that contains the details of the inputs and outputs, requirements and restrictions of the component. All the implementation details are hidden. The clients will get what the contract promises. The changes are not feasible at the deployment end. The advantage of black-box composition is that the testing done at the development side is persevered and there is no need of further testing at the deployment side.

**Glass Box Composition**

In glass box composition the inside structure of a component can be viewed, but it is not possible to modify. This solution has an advantage when compared to black box reuse, as the modeler can understand the box and its use better. However it is not possible to make any changes in the implementation. The advantage of this level remains the same as that of black-box composition however an additional benefit is that the user can gain knowledge of the internal implementation and can understand the mechanics of the component.

**White Box Composition**

In white box composition it is possible to see and change the inside of the box as well as its interface. A white box can share its internal structure and implementation with another box through inheritance or delegation. The advantage of this level is greater flexibility due to the provision of modifications. However this level incurs an extra burden of testing at the deployment end.

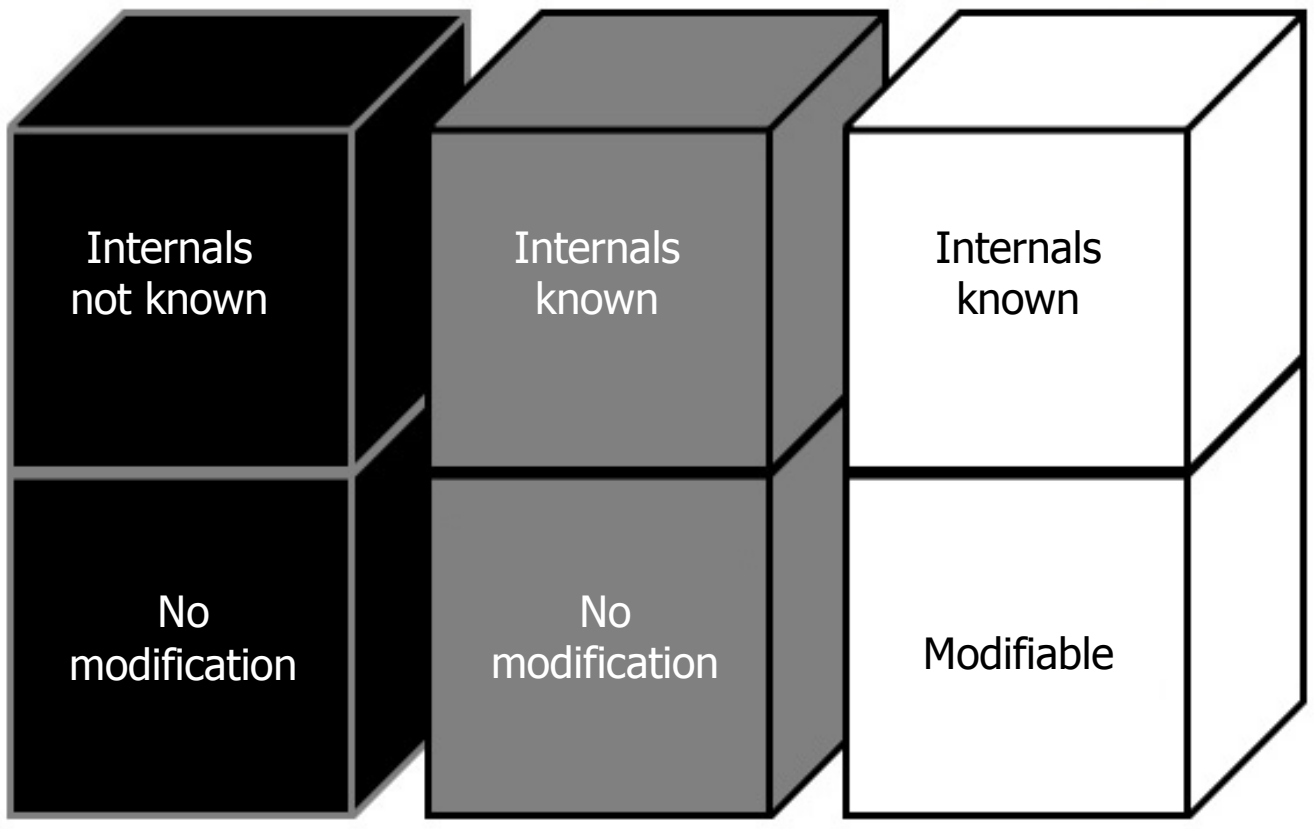

**Figure 5: Black Box, Glass Box, White Box**

**Figure 5** illustrates difference between black box, glass box and white box composition.

## 2.4.2 Composability vs. Interoperability

Bearing in mind the definition of composability mentioned previously, the IEEE definition of interoperability is:

> *The ability of two or more systems or components to exchange information and to use the information that has been exchanged*

The concept of interoperability is mainly about inter-connecting systems of various types developed for different purposes; for different platforms, and about their syntactically and semantically agreed upon communication [**13**]. In the context of modeling and simulation, interoperability is the ability of different simulations connected in a distributed system to collaboratively simulate a common scenario [**19**]. Page et al. [**52**] distinguishes composability and Interoperability as follows:

> *Composability contends with the alignment of issues on the modeling level. The underlying models are purposeful abstractions of reality used for the conceptualization being implemented by the resulting systems; whereas Interoperability contends with the software and implementation details of interoperations; this includes exchange of data elements via interfaces, the use of middleware, mapping to common information exchange models.*

## 2.5 Composability Levels

Petty and Weisel emphasized on two basic types of composability: syntactic and semantic in their theory of composability [**38**] [**50**]. According to which the syntactic composability requires that the composable components should be constructed with compatible implementation details such as parameter passing mechanisms, external data accesses, and timing assumptions. The question of syntactic composability is whether the components can be connected. In contrast, semantic composability is a question of whether the models can be meaningfully composed to form a composed simulation system and whether the combined computation is semantically valid. It is possible that two components may be syntactically linked, so that one can pass data to the other, but they can be semantically invalid. **Figure 6** represents the difference between syntactic and semantic composability metaphorically.

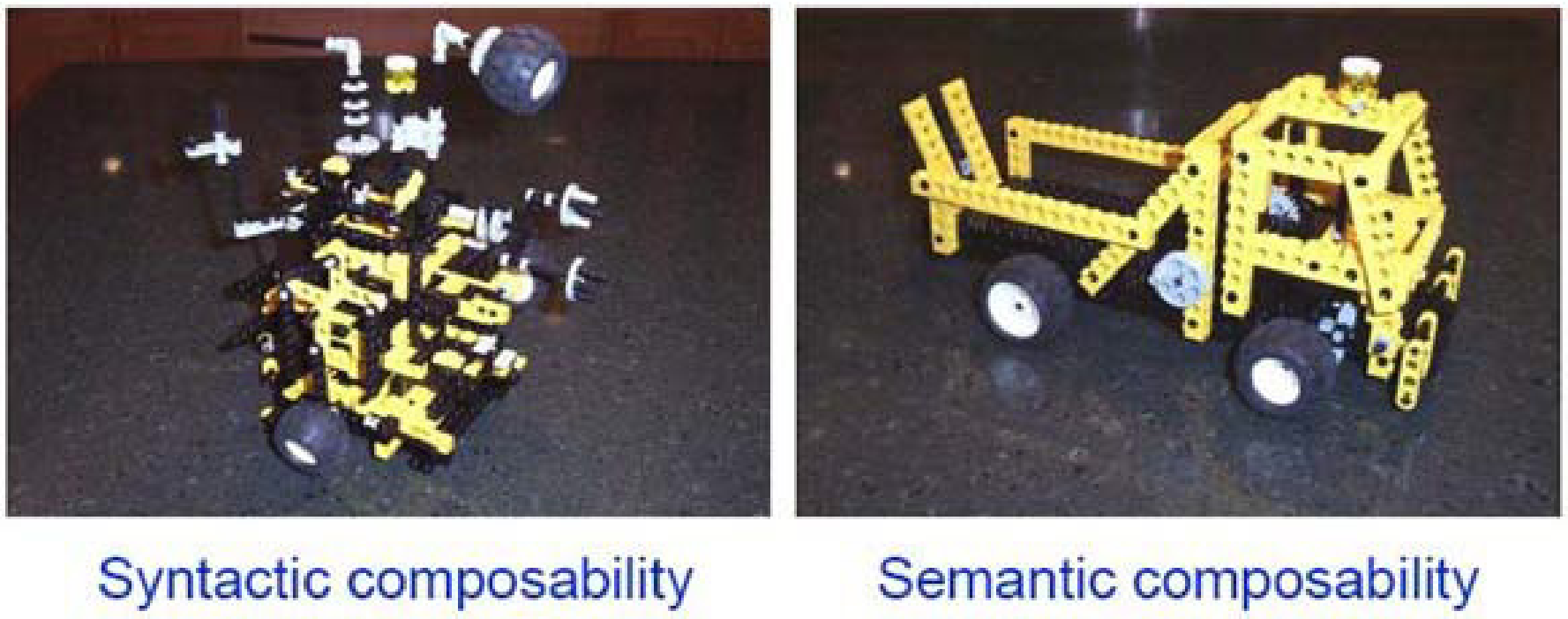

**Figure 6 Syntactic vs. Semantic Composability** (acquired from [**38**])

Composability is studied in more depth under different levels, as identified by different research groups. Several levels of understanding and agreement are required between the models in order for them to be meaningfully composed—that is, for their composition to produce meaningful results [**17**].

Davis recommended five distinctions of levels namely: *syntax, semantics, pragmatics, assumptions,* and *validity* to study composability [**43**]. He describes these levels as different consistencies of composability, which all together are examined for the correctness of model composability. Petty & Weisel have suggested nine levels of composability in terms of composition units. These levels are: *Application, Federate, Package, Parameter, Module, Model, Data, Entity* and *Behavior* [**38**]. Tolk described a six layered model called Levels of Conceptual Interoperability (LCIM) to study composability and interoperability. This model includes: technical layer, syntactic layer, semantic layer, pragmatic layer, dynamic layer, and the conceptual layer [**13**]. Similarly Medjahed & Bouguettaya introduced a composability stack in which the composability of semantic web services is checked at four levels: Syntactic, Static Semantic, Dynamic Semantic and Qualitative level [**53**]. First three levels of Medjahed & Bouguettaya's composability stack were adopted by Moradi, et al. to study the degree of composability of Base object Model (BOM) components [**54**]. In this thesis, these levels are considered as fundamental benchmarks for the evaluation of model composability. The notion of model composability and its correctness strongly depend on the consistency of these levels as explained in the following subsections.

### 2.5.1 Syntactic level:

At this level, the structure of the components is studied to know if they can fit together i.e., the output of one can be read as an input to the other and that the syntactic information of the connected components, such as message name, mode of action and number of parameters match each other e.g., A "passenger airplane" component will be a syntactic misfit in a military training simulation, where a "fighter jet" component is required whose input will be a signal from "ground station" component to engage a target and output will be an airstrike on the "target" component. A passenger plane can neither take a target designation as input, not it can fire on a ground target. So this component is not composable at syntactic level.

### 2.5.2 Static-Semantic level:

It is concerned with the meaningful interaction of the composed components. Static-Semantic level of composability involves in having a concise and mutual understanding of the data exchanged by the components participating in the composition. At this level, it is ensured that all the components possess the same understanding of the terms, parameters, data types and units, so basically this level deals with the interpretation of same meaning of concepts for the information exchanged between the composed components. For instance, if two components being composed interpret units of quantities in a different way, they will incorrectly process data values during the information exchange and thus result in a situation not intended by the user e.g., if a integer data value is intended to be the bearing of a target (in degrees) but interpreted as target distance (in Km) by the other component then it is a semantic mismatch.

The term "static" is prefixed, because all the information that is required to evaluate this level is static and does not change during the entire component interaction.

### 2.5.3 Dynamic-Semantic level:

Dynamic Semantic Composability implies that the components are dynamically consistent, i.e., they have suitable state-full behavior, necessary to reach the desired goals and subsequently satisfy user requirements. The dynamic level of composability ensures in having a behavioral consistency and coherency among the participating components in achieving the common goals. The dynamic semantic composability can only be achieved if the components are at the right states during their interaction. Also they should possess required behavior to make a collective progress. E.g., in a composed model of a restaurant, a waiter component may have two different behaviors (i) Classical restaurant where a waiter takes order from customer, serves food and then collects payment or (ii) Fast food restaurant where waiter takes order, collects payment and then serves food. The selection of the correct behavior and the correct customer component (the one who can correctly interact with the classical restaurant waiter or fast food waiter) will affect the overall composability of the model. This example presents how the components should be at right states to make progress. A customer (expecting classical treatment) will wait forever for the (fast food waiter) to serve food and vice versa.

Even if the components are at the right states, but their behavior is not correct, the composition may not reach its goals. E.g., in a manufacturing system two machine components produce two different parts that are later combined to make a finished good, and they share a single robot component for input of raw material, it is required that the robot component should be fair so that both machines get more or less equal chance to proceed. If the robot is not fair the proportion of good produced will be unbalanced and therefore the system will fail to meet its objectives even though the components are at right states and continue to progress.

The term dynamic is prefixed, because the information such as current state of components changes dynamically during component interaction.

### 2.5.4 Pragmatic level:

Consistency of meaning is not always straightforward because the same word means very different things depending on context [**43**]. Pragmatic consistency refers to a context based meaningful composition of the components. In linguistics *the study of the relations between linguistic phenomena and aspects of the context of language use is called pragmatics* whereas Context is defined as *something that consists of the ideas, situations, events, or information that relates to it and makes it possible to fully understand it* [**55**].

The pragmatic level of composability evaluates the difference of actual effect of the messages with the intended effect of messages during communication [**43**]. The research of pragmatic level of composability involves in-depth study of computational linguistics, cognitive technologies and contextual computing [**55**]. An important issue at this level is pragmatic ambiguity. Pragmatic ambiguity arises when the message is not specific, and the context does not provide the information needed to clarify the statement, and due to which the components do not interact according to the desired objectives. An example of pragmatic ambiguity is the story of King Croesus and the Oracle of Delphi (derived from [**56**]):

*"King Croesus consulted the Oracle of Delphi before warring with Cyrus of Persia. The Oracle replied that, "If Croesus went to war with Cyrus; he would destroy a mighty kingdom". Delighted, Croesus attacked Persia, and Croesus' army and kingdom were crushed. Croesus complained bitterly to the Oracle's priests, who replied that the Oracle had been entirely right. By going to war with Persia, Croesus had destroyed a mighty kingdom – his own."*

In essence, a set of components can possibly fit together (syntactically), and their communication is meaningful and understood (semantically), but unless all components preserve essential behavior (dynamically) in order to reach the desired composition goals, and they share the correct contextual knowledge (pragmatically), the composability cannot be qualified as correct with respect to given requirement specifications.

## 2.6 Composability frameworks

Composability essentially relies on a suitable composition framework that can provide accurate reasoning of its correctness and support means to be able to leverage certain component standard. Various component standards and their respective frameworks have been developed for M&S to support composability. Some of these frameworks contribute to conceptual modeling by providing the needed formalism and influence the ability to develop and compose model components at conceptual level, while others support model composition at executable level. These frameworks practically support composability, as they usually offer features such as model specification, development, and execution. A brief description of some of the composability frameworks is provided below:

### 2.6.1 Discrete Event System Specification (DEVS)

DEVS [**57**] is a component based formalism based on dynamic systems theory. It was developed for the purpose of describing the structure and behavior of systems. It supports the concept of hierarchical and modular model construction through coupling of components [**19**]. DEVS is basically a model specification formalism however it incorporates different implementation frameworks such as DEVS-Java, DEVS-C++ and DEVS-Sharp which are used to implement DEVS models into executable form.

Two types of DEVS models exist, namely, atomic and coupled [**20**].

An atomic DEVS is a tuple $\mathbf{M = \langle X, S, Y, \delta_{int}, \delta_{ext}, \lambda, \tau \rangle}$ where:

$X = \{(p, v) \mid p \in InPorts, v \in Xp\}$ is the set of input ports and values

$Y = \{(p, v) \mid p \in OutPorts, v \in Yp\}$ is the set of output ports and values

S is the set of states

$\delta_{int} : S \rightarrow S$ is the internal transition function

$\delta_{ext}: Q \times X \rightarrow S$ is the external transition function, where

$Q = \{(s, e) \mid s \in S, 0 \leqslant e \leqslant \tau(s)\}$ is the total state set

e is the time elapsed since last transition

$\lambda : S \rightarrow Y$ is the output function

$\tau: S \rightarrow R^{+}_{0,\infty}$ 0, ∞ is the time advance function

A DEVS atomic component has inputs X, outputs Y, and a set of S states. At a given moment, a DEVS model is in a state $s \in S$. In the absence of external events, it remains in that state for a lifetime defined by $\tau(s)$. When $\tau(s)$ expires, the model outputs the value $\lambda$ (s) through a port $y \in Y$, and it then changes to a new state given by $\delta_{int}(s)$. A transition that occurs due to the consumption of time indicated by $\tau(s)$ is called an internal transition. On the other hand, an external transition occurs due to the occurrence of an external event. In this case, the external transition function determines the new state, given by $\delta_{ext}$ (s, e, x), where s is the current state, e is the time elapsed since the last transition, and $x \in X$ is the external event that has been received. The time advance function can take any real value between 0 and $\infty$. A state for which $\tau(s)=0$ is called a transient state (which will trigger an instantaneous internal transition). In contrast, if $\tau(s)=\infty$, then s is said to be a passive state, in which the system will remain perpetually unless an external event is received.

A coupled DEVS is a tuple: $M = (X, Y, D, \{M_d \mid d \in D\}, EIC, EOC, IC, Select)$ where:

$X = \{(p, v) \mid p \in InPorts, v \in Xp\}$ is the set of input ports and values
$Y = \{(p, v) \mid p \in OutPorts, v \in Yp\}$ is the set of output ports and values
D is the set of component names
$M_d$ is a DEVS model with
$X_d = \{(p, v) \mid p \in InPorts_d, v \in Xp\}$
$Y_d = \{(p, v) \mid p \in OutPorts_d, v \in Yp\}$
EIC is the set of input port couplings
$EIC \subseteq \{((N, ip_N), (d, ip_d)) \mid ip_N \in InPorts, d \in D, ip_d \in InPorts_d\}$
EOC is the set of output port couplings
$EOC \subseteq \{((d, op_d), (N, op_N)) \mid op_N \in OutPorts, d \in D, op_d \in OutPorts_d\}$
IC is the set of internal couplings
$IC \subseteq \{((a, op_a), (b, ip_b)) \mid a, b \in D, op_a \in OutPorts_a, ip_b \in InPorts_b\}$
Select is the tie-break function

A system modeled using DEVS can be described as a composition of atomic and coupled components. A coupled model comprises a set of input and output ports, a set of component names D, a set of DEVS components $M_d$, input port EIC and output port EOC couplings, and, a set of internal couplings IC connecting internal components with each other. The tie-break function decides which component to proceed when two or more components have internal transitions scheduled at the same time.

Figure 7 describes a DEVS example. In this example two atomic component A & B are coupled together. Both components have two states Send $\tau(s)=0.1$ and Wait $\tau(s)=\infty$. Input port: *?receive* and Output port: *!send* are defined and connected to each other in coupled DEVS.

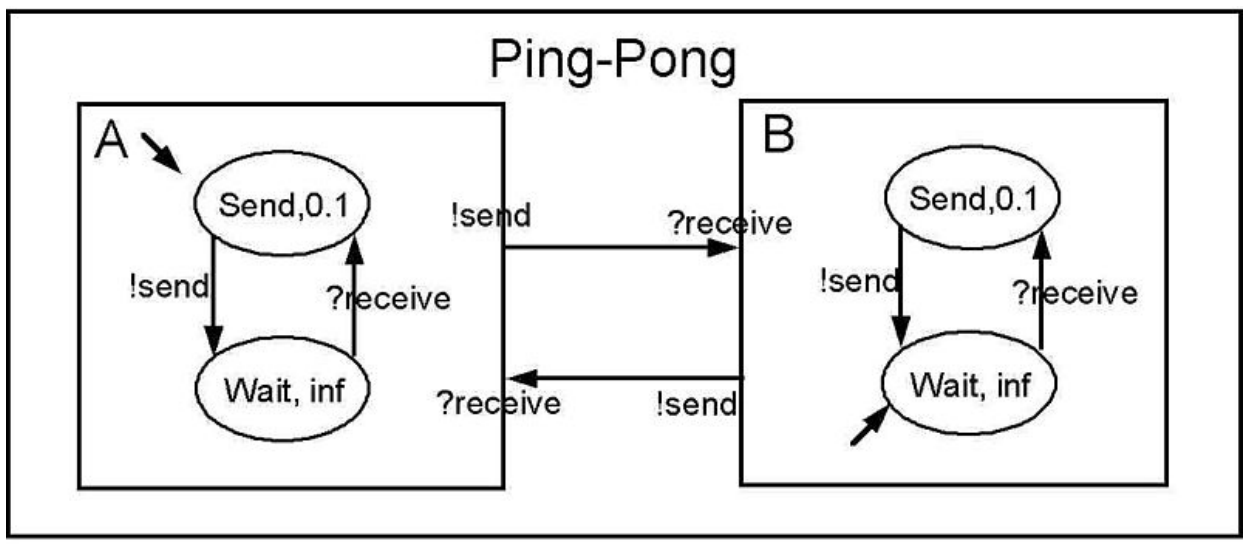

Figure 7: Ping-Pong DEVS [Wikipedia]

## 2.7 Base Object Model (BOM) framework

The SISO[9] standard BOM is defined as, "a piece part of a conceptual model composed of a group of interrelated elements, which can be used as a building block in the development and extension of simulations and simulation environments" [**58**].

BOM provides a simulation standard that allows model developers and simulation engineers to create modular conceptual models in form of composable objects, which can be used as the basis for a simulation or simulation environment [59], [60]. The concept of BOM is based on the assumption that components of models, simulations, and federations can be reused as building blocks in the development of a new simulation or a federation [54].

[9] Simulation Interoperability Standards Organization

BOMs are unique because they provide a means to represent aspects of a conceptual model that captures structural and behavioral descriptions of items abstracted from the real system (simuland). Then they allow these conceptual models to be mapped to one or more class definitions, which may be used by a software design, variety of programming languages, or distributed simulation architectures such as HLA or TENA[10] [61], [62].

BOM standard also offers a general purpose modeling architecture for defining components to be represented within a live, virtual, or constructive (LVC) simulation environment. It is well suited for characterizing models including the structural and anticipated behavior of interacting systems, individuals, and other entities. Primarily BOMs framework poses a satisfactory potential for effective composability of conceptual models at syntactic and semantic levels, resulting in a framework for the assembly of a system (i.e. simulation) or system of systems (i.e. distributed simulation environment) [62].

In spite of these reasonable qualities, BOM framework still falls short of required behavioral semantics and necessary built-in evaluation techniques, which are essential for modeling complex system behavior and reasoning about the correctness of the composability at each of its different level. Therefore it becomes a most suitable candidate and a preferred choice of a composition framework (in this thesis) for studying model composability in depth and applying proposed methods on BOM based compositions to explain the approach.

### 2.7.1 Structure of BOM

A BOM is constituted of elements specifying metadata information, conceptual model and the class structure information defined using HLA OMT constructs [59]. Figure 8 presents different parts of BOM, explained as follows:

#### Model Identification

Model Identification associates the metadata information with the BOM. Its purpose is to document certain key identifying information within the BOM description. It provides a minimum but sufficient degree of descriptive information about a BOM

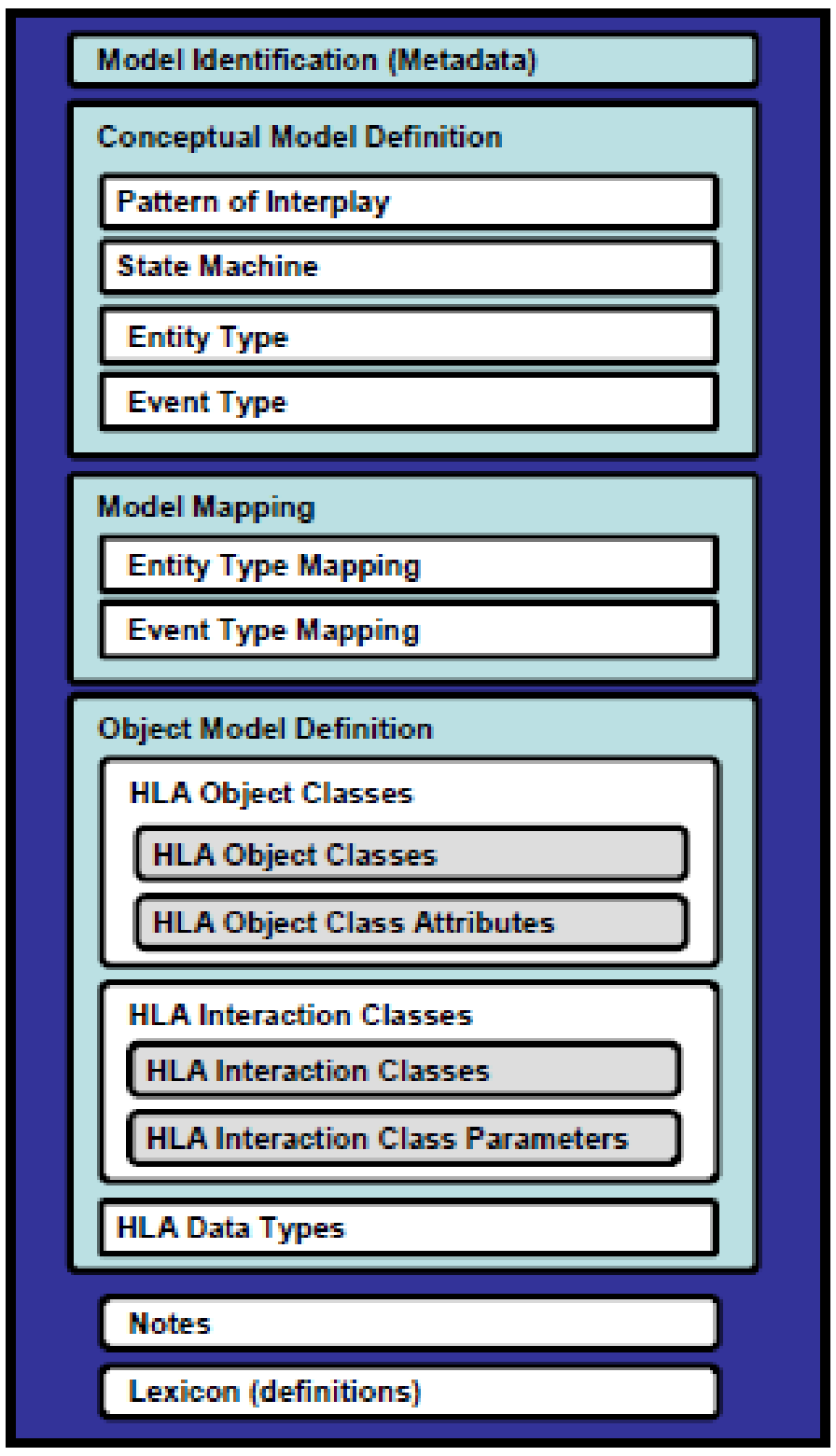

**Figure 8: BOM structure**
(acquired from [59])

[10] Test and Training Enabling Architecture

### Conceptual Model Definition

From the composability point of view, this is the most important part of BOM and therefore the main focus of this thesis. To understand this part, the definition of a conceptual model should first be considered:

**Conceptual Model**

*A Conceptual model is an abstract description or an appropriate simplification of a real (or proposed) system, which is later, refined and implemented in to a more concrete executable model (or simulation model). In these terms, conceptual modeling is a subset of model design which is formed through an iterative process according to the objectives of system modeling* [**63**], [**64**].

The term conceptual model is used in different ways in the literature. A conceptual model could be a specific diagram like UML class diagram or it could be documentation of a particular aspect of the simuland[11] [**29**]. To better understand the concept of BOMs, consider the home construction analogy. When a new house is to be built the conceptual understanding of features of the building is captured in architectural drawings, which is analogous to a conceptual model (BOM) [**60**]. BOM Conceptual Model definition consists of following parts:

### Pattern of Interplay (POI)

POI models a specific purpose or capability and is represented by one or more pattern actions. For each pattern action, one or more senders and receivers are specified to provide a means for understanding and the behavioral relationship among conceptual entities. POI is represented by UML sequence diagram [**60**].

### State Machine

The state machine is used to model the behavior of a BOM's conceptual entity. The state machine is specified by a set of states where each state may transit to a subsequent state called next state, upon an exit action, which is identified in a pattern of interplay. UML state-machine diagram is used to represent BOM's state-machine [**60**].

### Entity Type

A conceptual entity is an abstraction of a real world entity. It defines a relationship with other entities within a pattern of interplay and acts as a sender or receiver of the events [**60**].

### Event Type

Conceptual events include information about the source, target, and content (parameters) of a message or trigger. The difference between a trigger and a message is that a trigger is used to broadcast information whereas the messages are directed exchanges of information where the sender knows about the intended receiver of the message [**60**].

Entities and Events represent data about the real world objects and their interaction (physical description), whereas the pattern of interplay and state-machine collectively represents the dynamic behavior of the component.

[11] A simuland is the real world system of interest. It is the object, process, or phenomenon to be simulated [**29**].

## 2.7.2 BOM Assembly

The BOM concept provides a mechanism for combining BOMs and creating High-Level BOMs, called BOM Assemblies, as shown in **Figure 9**. A BOM Assembly representing a composition of BOMs, is built in a hierarchical manner and includes information about composed BOMs, which in turn is used to identify a composite interface, and represent a federate, federation within the simulation space[12]. Typically a developer of a simulation would search a BOM repository for suitable BOM candidates for use in a simulation and combine those into a BOM Assembly (i.e. a simulation model), which is then used to create the actual simulation [**19**].

A BOM assembly contains Model Identification, and pattern of the interplay among conceptual entities being represented, which is provided through the association of BOMs to the various Pattern Description actions that the BOM Assembly identifies, within the Conceptual Model view [**60**].

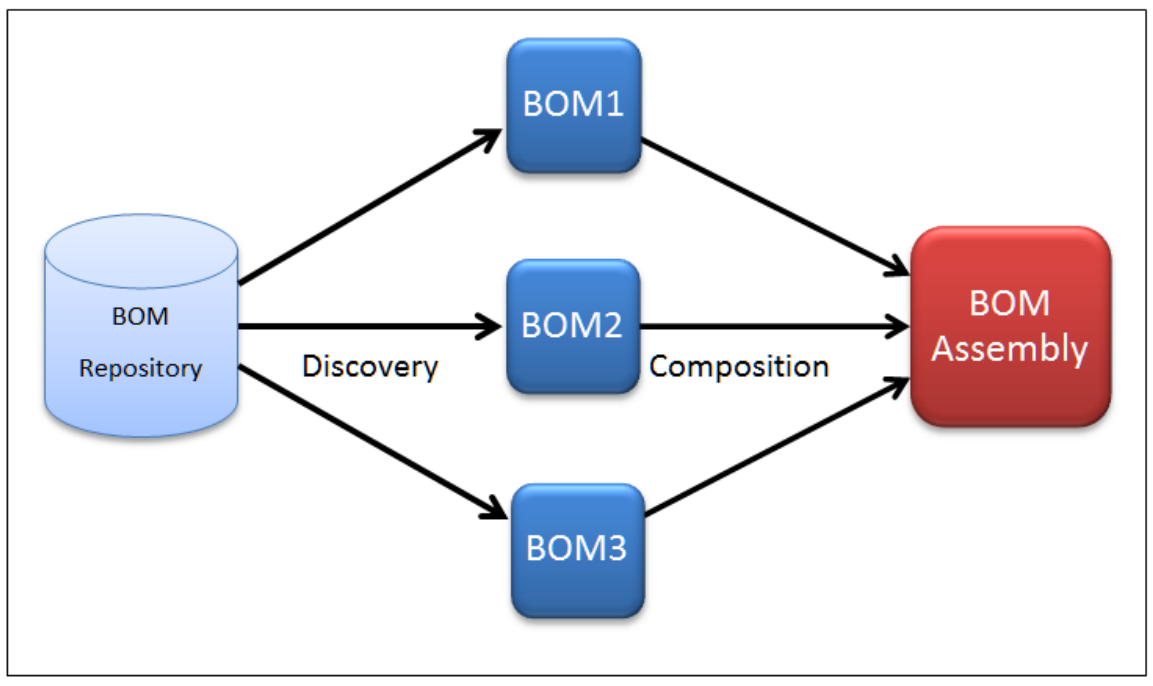

**Figure 9: BOM Assembly**

BOM models can be created using XML script. But for constructing BOM models graphically, a free IDE tool called BOM Works [**65**] is available. **Figure** 10 represents an example developed using BOM Works. It is similar to the DEVS example shown in Figure 7, to compare the difference.

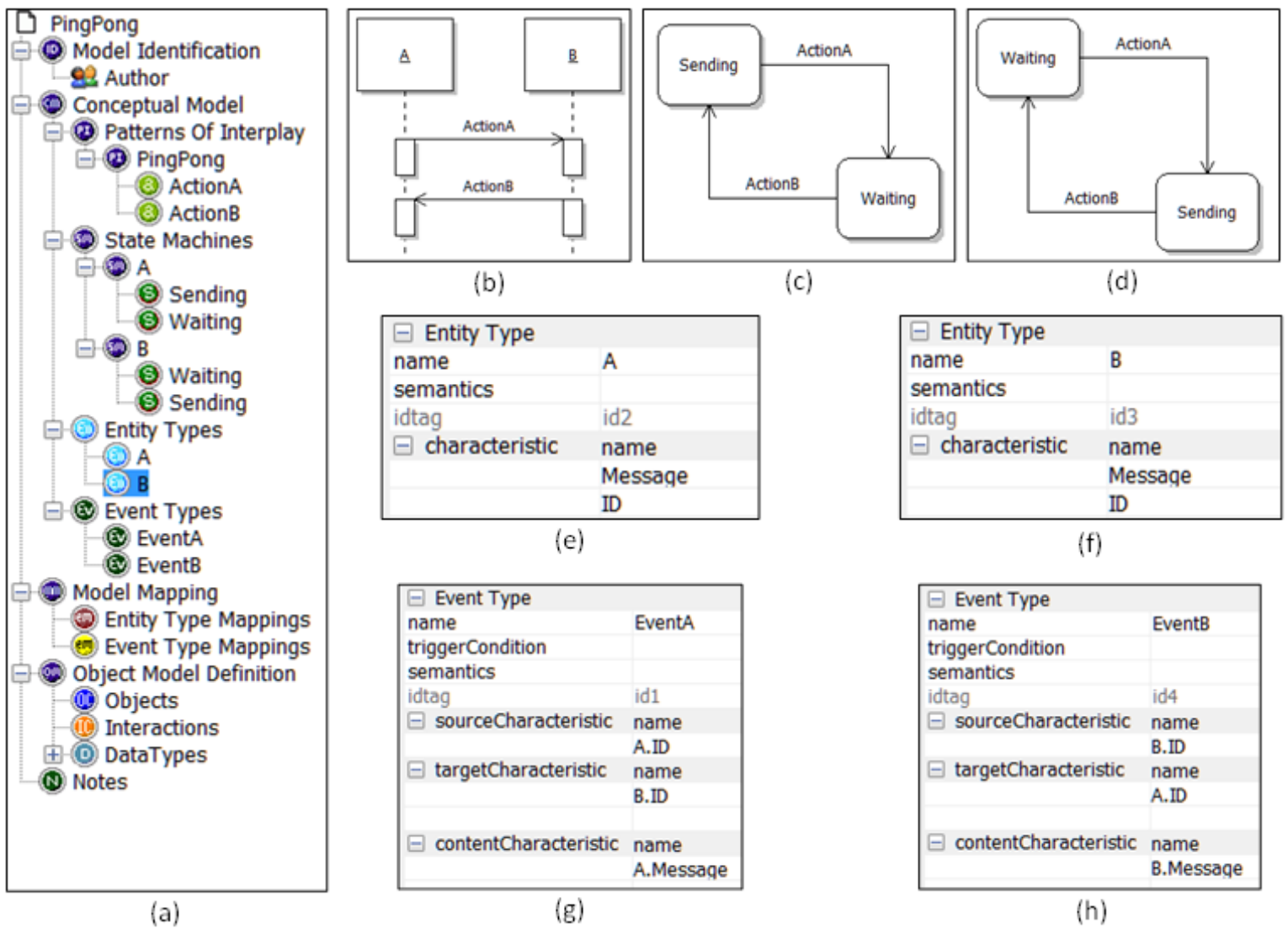

**Figure 10: (a) PingPong BOM in BOM Works (b) POI (c) State-machineA (d) State-machineB (e) EntityA (f) EntityB (g) EventA (h) EventB**

12 Although use of HLA is not a mandatory subsequent step, it is likely that BOM assemblies are intended to support an HLA based federation [**59**].

### 2.7.3 Model Mapping and Object Model Definition

The model mapping provides a mechanism for mapping between the elements of the conceptual model and the class structure elements of the Object Model Definition that are described using HLA OMT[13] specification constructs. The object model definition defines the structure of an object and interaction class, and their associated attributes and parameters. HLA Object classes include HLA attributes and HLA interaction classes include HLA parameters. These parts of BOM are not used in this thesis, however interested readers can find more details in [**58**], [**59**], [**60**], [**61**], [**62**].

### 2.7.4 Formal specification for the Compositon of BOM

Unlike DEVS, BOM does not have a graphical and mathematical formalism for specifying how components are composed (even though parts of BOM such as state-machine and POI can be represented in UML and BOM documents can be described using XML). This initiates a need for a graphical and formal representation of BOM composition.

In this section, we introduce a formal and graphical specification of BOM[14]. We define two types of BOM: (i) Basic BOM and (ii) Composed BOM. A basic BOM is an undividable atomic BOM component, with an assumption that it represents only one conceptual entity at the most. A composed BOM is a hierarchical combination of basic and other composed BOM.

**Basic BOM**

We propose that a basic BOM (BB) can formally be defined as:

$$BB = \langle\, EnT, EvT, AcT,\ S \,\rangle \tag{2.3}$$

Where:

- **EnT** is an entity type. We assume that a basic BOM has only one entity. EnT is defined as:

  EnT = Name {Characteristic: Type}

Where Name is the name of an entity uniquely defined by an identifier[15] and characteristic is a set of attributes of an entity. Each characteristic is uniquely defined by an identifier and has a type[16]

- **EvT** is a set of event types, each with sender, receiver and content

  Evt = {(Name, Sender, Receiver, {Content: Type}) | Name ∈ Identifier, Sender & Receiver ∈ EnT, Content∈ Identifier: Type ∈ type}

- **S** is a set of states, each has an exit-condition and a next state:

  S = {(Name, ExitCondition{Action, NextState})} | Name ∈ Identifier, Action ∈ Act, NextState ∈ S

[13] High Level Architecture Object Model Template
[14] These concepts are not new and exist in literature for other component-based approaches [**21**]. In this thesis, their application in BOM is intended for facilitating specification and ease of understanding
[15] An identifier is a unique sequence of letters & digits, starting with a letter.
[16] Type := Integer | String | Double | Complex

- **AcT** is a set of actions, each has name, sender, receiver and an associated event:

  AcT= {(Name, Sender, Receiver, Event) | Name ∈ Identifier, Sender & Receiver ∈ EnT, Event ∈ EvT}

**Composed BOM**

A composed BOM (CB) can formally be defined as:

$$CB = \langle AcT_{IN}, AcT_{OUT}, POI \rangle \quad (2.3)$$

Where:

- **$AcT_{IN}$** is a set of input actions that are received from other BOM. This set can be empty if the Composed BOM is closed.

$AcT_{IN}$ = {(Name, Sender, Receiver, BOM) | Name ∈ Identifier, Sender & Receiver ∈ EnT, BOM ∈ File}

- **$AcT_{OUT}$** is a set of input actions that are sent to other BOM. This set can also be empty if the Composed BOM is closed.

$AcT_{OUT}$ = {(Name, Sender, Receiver, BOM) | Name ∈ Identifier, Sender & Receiver ∈ EnT, BOM ∈ File}

- **POI** is the pattern of interplay that defines how basic or composed BOMs are connected to each other (through actions). It maps a list of send actions to a list of receive actions. ‘ ! ’ symbol means send and ‘ ? ’ symbol means receive.

POI = {({!$AcT_{SEND}$} , {?$AcT_{RECV}$})} | $AcT_{SEND}$ & $AcT_{RECV}$ ∈AcT

**Example**

As an example, BOMs from Figure 10 can formally be represented as:

**$BB_0 = \langle$ EnT, EvT, AcT, S $\rangle$ where:**

EnT = EntityA {$C_0$(Message:String)}

EvT = {$E_0$(EventA, $BB_0$, $BB_1$, $BB_0.C_0$), { $E_1$(EventB, $BB_1$, $BB_0$, $BB_1.C_0$)}

Act = { $A_0$(ActionA, $BB_0$, $BB_1$, $E_0$), $A_1$(ActionB, $BB_1$, $BB_0$, $E_1$)}

S = { $S_0$(Sending, $A_0$, $S_1$), $S_1$(Waiting, $A_1$, $S_0$)}

**Table 1: Entity A**

$$BB_1 = \langle EnT, EvT, S, AcT \rangle \text{ where:}$$

$EnT$ = EntityB {$C_1$(Message:String)}

$EvT$ = {$E_2$(EventA, $BB_0$, $BB_1$, $BB_0.C_0$), { $E_3$(EventB, $BB_1$, $BB_0$, $BB_1.C_1$)}

$Act$ = { $A_2$(ActionA, $BB_0$, $BB_1$, $E_2$), $A_3$(ActionB, $BB_1$, $BB_0$, $E_3$)}

$S$ = { $S_2$(Waiting, $A_2$, $S_3$), $S_3$(Sending, $A_3$, $S_2$)}

**Table 2: Entity B**

Similarly a composed BOM $CB_0$ can be formally described as:

$$CB_0 = \langle AcT_{IN}, AcT_{OUT}, POI \rangle \text{where:}$$

$AcT_{IN} = \varnothing$ (since there is no incoming actions from any other BOM)

$AcT_{OUT} = \varnothing$

$POI$ = {I/$O_0$(!$A_0$ , ?$A_2$), I/$O_1$(!$A_3$, ?$A_1$)}

**Table 3: Composed BOM**

We propose a graphical notation for representing basic BOM and their composition shown in **Figure 11**. In this figure two basic BOM EntityA and EntityB are composed.

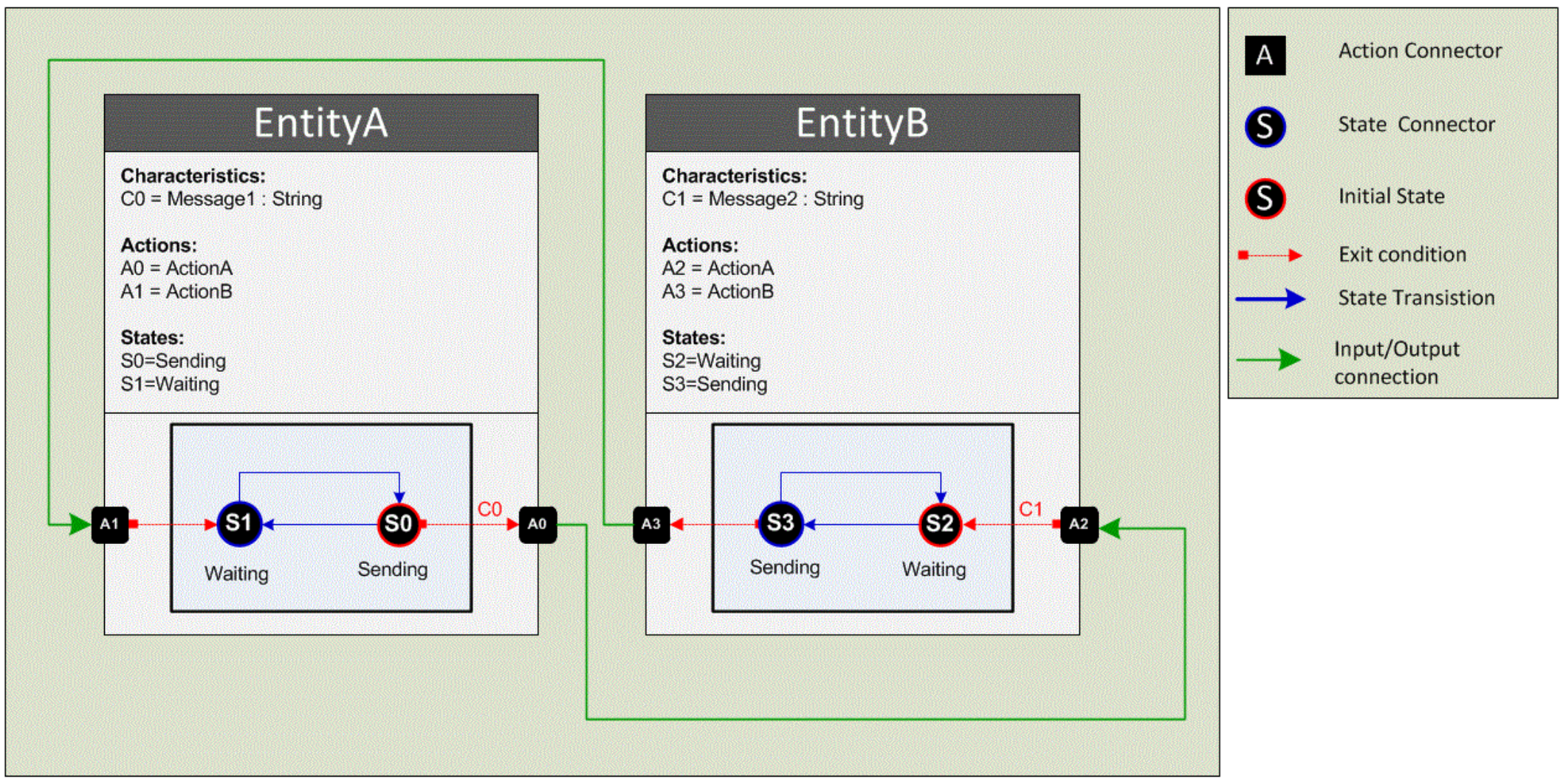

**Figure 11: Composed BOM**

The general information of a component such as entity name, characteristics, actions and states are defined in the main block. In the lower block the states and their transitions (with blue arrow) are shown. Each transition is mapped with actions (in red arrow) with parameter labels (the IDs of characteristics). The direction of the arrow shows the type of the associated action (send or receive). The composition of BOMs is shown through connectors (in green color).

### 2.7.5 Summary

In this thesis, we harness the capability of BOM as a conceptual modeling framework, because it provides a component standard using an XML specification; gives guidelines for the further development of the executable model and helps determine the appropriateness of the model or its parts for model reuse; and most importantly due to its strong support for syntactic and semantic composability. It will be shown, how BOM with its existing potential can be facilitated by composability evaluation for accurate and rapid construction and modification of its corresponding federates in HLA based simulations and hence brings forth an improvement in the distributed simulation community.

# Chapter 3
# Executable Modeling Formalisms

*In this chapter two popular model description formalisms are discussed namely Petri Nets and Communicating Sequential Processes (CSP)[17], which are normally used for modeling, execution (or simulation) and verification of concurrent systems. This chapter provides an introduction, theory, properties, classification, modeling methods and analysis techniques of PN and CSP. PN and CSP are both considered as a part of solution domain in this thesis, because of their impressive accumulation of knowledge in concurrency modeling and analysis techniques. These aspects are imported in this thesis and used for composability verification.*

PN and CSP formalisms are relatives since they are used to model same class of systems called concurrent systems. Unlike other systems such as transitions systems or automata, the formalisms of concurrent systems are strongly based on concurrency theory. One of the major contributors of concurrency theory are: Carl Adam Petri who initiated concept of interacting sequential processes and introduced Petri Nets; C. A. R. Hoare who focused on developing programming language (CSP) for concurrent systems; and Robin Milner who introduced Calculus of Communicating System (CCS) and $\pi$-Calculus. These are variants of approaches for formally modeling concurrent systems and are the member of the family of mathematical theories of concurrency known as process algebras, or process calculi. CSP is also a member of process algebra. The main difference between PN and CSP is that the former are based on graphs, while the latter are based on a textual description. However both offer strong formal semantics for modeling executable systems and share a broad pool of knowledge of theoretical principles and practical techniques for the analysis and verification of models of complex behavior. In this thesis, we propose using these two formalisms to model executable form of components and study their composability.

## 3.1 Petri Nets

PN were introduced by Carl Adam Petri (and named after him) in 1962. They provide an elegant and useful graphical and mathematical formalism [**24**]. With PN the main idea is to represent states of subsystems separately. In this way, the distributed activities of a system can be represented very effectively. PN are widely used for modeling and control in a variety of the sorts of systems. Particularly, in Discrete Event Dynamic Systems (DEDS)[18] in which many properties such as synchronization, sequentiality (producer-consumer problem), concurrency and conflict (mutual exclusion) concurrency, and choices can be well presented and analyzed using PN [**66**]. Their structural and behavioral properties have been successfully exploited for solving various problems of complex and dynamic systems. Significant progress in these directions was made over three decades. Most essential features of PN are the principles of locality, concurrency, graphical and algebraic representation. They can be used not only for the specification and analysis of the structural system design but also for design of the system behavior. [**66**], [**67**].

---

[17] The "Sequential" word of the CSP name is now something of a misnomer, since modern CSP allows component processes to be defined both as sequential processes, and parallel [Wikipedia].

[18] Examples of DEDS are air traffic control systems; automated manufacturing systems; computer and communication networks; embedded and networked systems; and software systems etc. The activity in these systems is governed by operational rules designed by humans and their dynamics is often driven by asynchronous occurrences of discrete events [**67**].

PN present two interesting characteristics. Firstly, they make it possible to model and visualize systems with complex behaviors including parallelism, concurrency, synchronization and resource sharing. Secondly the properties of these nets, their analysis and theorems have been extensively studied [**68**].

### 3.1.1 PN Definitions and Concept

In PN, two basic elements of modeling are places and transitions. **Events** are associated with ***transitions*** which occur when some conditions are satisfied. Information related to these conditions is contained in ***places***. There are two types of places namely: Input places and Output places. Input places are associated with the conditions required for this transition to occur. Output places are associated with conditions that are affected by the occurrence of this transition [**25**]. Transitions, places, and certain relationships between them define the basic components of a ***Petri net graph***. A PN graph has two types of nodes, places and transitions, and arcs connecting these. It is a bipartite graph in the sense that arcs cannot directly connect nodes of the same type; rather, arcs connect place nodes to transition nodes and transition nodes to place nodes [**25**].

### 3.1.2 Petri net graph

Mathematically a PN is a 5 tuple: $PN = \langle P, T, F, W, M_0 \rangle$ where:

- P is a finite set of places $P = \{p_1, p_2 ... p_m\}$ represented as oval shaped node in the PN graph
- T is a finite set of transitions $T = \{t_1, t_2 ... t_n\}$ represented as a line or a rectangular shaped node in the graph
- F is a flow function such that $F \subseteq (P \times T) \cup (T \times P) \rightarrow N$ [19]
- $W: F \rightarrow N^+$ where $N \in \{1, 2, 3 ...\}$ is arc weight function.
- $M_0$: $P \rightarrow N$ is a function called the initial marking, where each element $M_0(p)$ has N number of **tokens**[20] initially in place p where N is a set of non-negative integers.
- For each transition $t \in T$ a set of input places denoted as $\bullet t$ are those places which are connected to t through incoming arcs:

$$\bullet t = \{p_i \mid (p_i, t) \in F\} \tag{3.1}$$

- Similarly, for each transition $t \in T$ a set of output places denoted as $t\bullet$ are those places to which t is connected through outgoing arcs:

$$t\bullet = \{p_i \mid (t, p_i) \in F\} \tag{3.2}$$

---

[19]Such that $P \cap T = \emptyset$ (i.e. P&T are disjunctive sets) and $P \cup T \neq \emptyset$ (i.e. neither P nor T are isolated). Also an arc can be connected from place to transition (input arc) or from transition to place (output arc) but not to the node of same type.

[20] In classical PN, tokens are represented as black dots. They are assigned to, and can be thought to reside in, the places of a Petri net.

**Definition: Marking**

A marking is an assignment of tokens to the places of a PN. The number and position of tokens defines a system state, and it may change when the tokens move. This movement of tokens due to the firing of transitions causes the execution of a PN [**26**]. The marking M can be defined as an n-vector, $M = (m_0, m_1, m_2 \dots m_n)$, where $n = |P|$ (no. of places), and each $m_n \in N$, $i = 1...n$. The vector M gives for each place $p_i$ in a PN the number of tokens in that place.

**Definition: PN State-space**

The state of a PN model is defined by its marking. The firing of a transition represents a change in the marking of the net. The state space of a PN with n places is the set of all markings. State-space will be discussed in detail later in this chapter.

**Definition: Enabling of a Transition**

A transition $t$ in a given PN is called enabled or fire-able by a marking $M_i$ **iff** for each input place $p \in \bullet t$ its marking is equal or greater than the weight of the arc from it to $t$, (or $t$ has no input place). Mathematically, a transition $t$ is fire-able **iff**

$$\forall p \in \bullet t \mid M(p) \geq W(p, t) \vee \bullet t = \emptyset \quad (3.3)$$

**Definition: Firing of a Transition**

If a transition t is enabled, it may fire by removing $W(p, t)$ number of tokens from each input place p and putting $W(t, p')$ tokens in each output place p', due to which a new marking $M_{n+1}$ is generated.

$$M_{n+1} \xrightarrow{t} M_n \mid M(p') = M(p) - W(p, t) + W(t, p') \ \forall p \in P \quad (3.4)$$

$M_{n+1}$ is immediately reachable from $M_n$. $M_n$ is reachable from $M_0$ if firing a sequence $\sigma = t_1, t_2 \dots t_k$ of enabled transitions leads $M_0$ to $M_n$, written as $M_0 \xrightarrow{\sigma} M_n$

**Example**[21]

Consider the PN model $PN = \langle P, T, F, W, M_0 \rangle$ as shown in **Figure 12** where:

$P = \{p_1, p_2, p_3, p_4, p_5\}$ and $T = \{t_1, t_2, t_3, t_4\}$,

Let W = 1 for all arcs

Initial marking $M_0 = [1\ 0\ 0\ 0\ 0]$

[21] This example is inspired from [**68**]

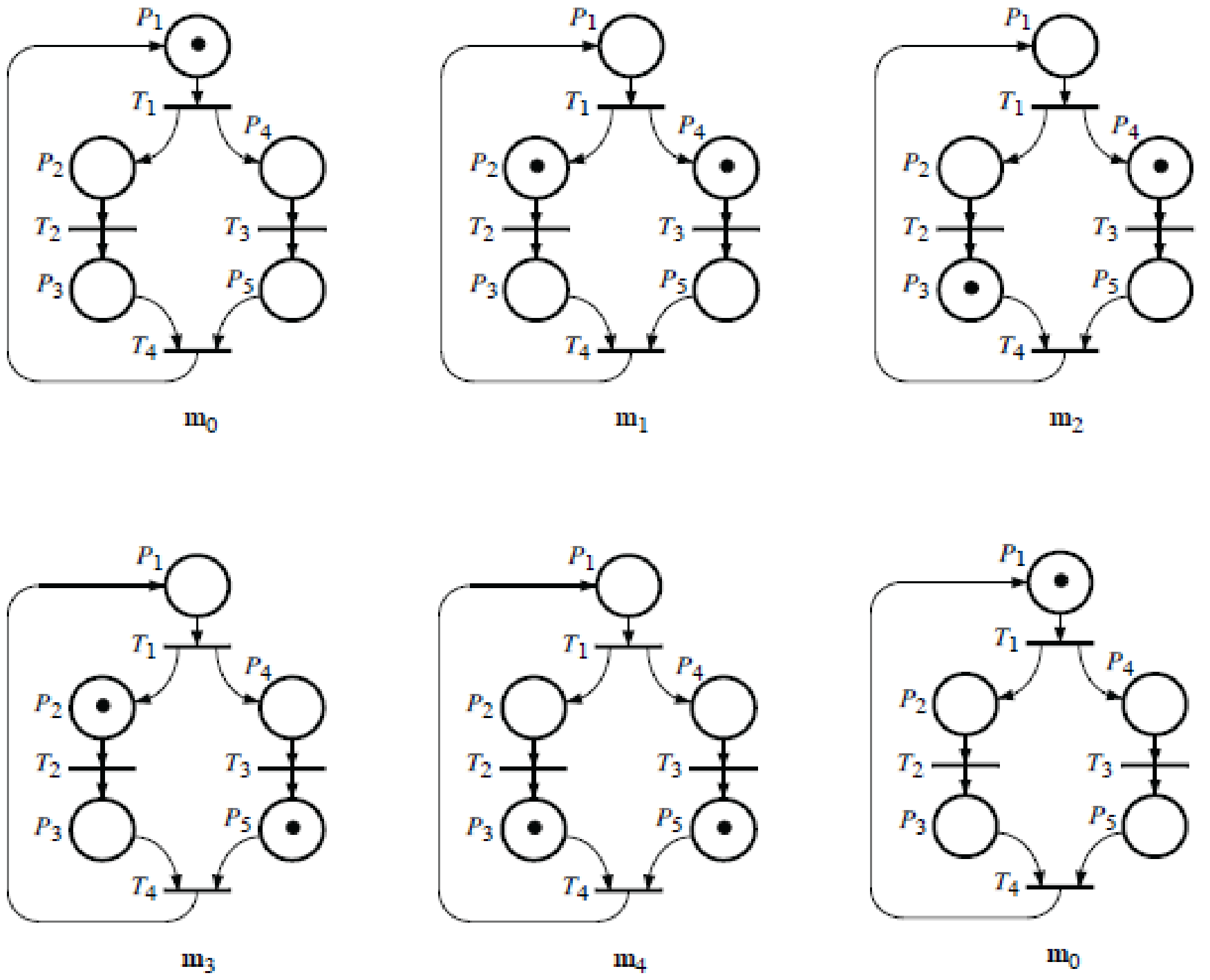

**Figure 12: Transition firing sequence** (acquired from [**68**])

σ1: $M_0 = [1\ 0\ 0\ 0\ 0] \xrightarrow{T1} M_1 = [0\ 1\ 0\ 1\ 0] \xrightarrow{T2} M_2\ [0\ 0\ 1\ 1\ 0] \xrightarrow{T3} M_4\ [0\ 0\ 1\ 0\ 1] \xrightarrow{T4} M_4\ [1\ 0\ 0\ 0\ 0]$

σ2: $M_0 = [1\ 0\ 0\ 0\ 0] \xrightarrow{T1} M_1 = [0\ 1\ 0\ 1\ 0] \xrightarrow{T3} M_3\ [0\ 1\ 0\ 0\ 0] \xrightarrow{T2} M_4\ [0\ 0\ 1\ 0\ 1] \xrightarrow{T4} M_4\ [1\ 0\ 0\ 0\ 0]$

In this example there are two possible transition firing sequences σ1= T1, T2, T3, T4 and σ2 = T1, T3, T2, T4

### 3.1.3 Properties of PN

Just like other models, PN are constructed from informal requirement specifications, which is not a trivial task, and requires a great deal of modeling experience. If a system being modeled is very complex, a PN model may differ considerably from its original specification. A model can only be useful if it is logically correct with respect to its specifications [**69**]. Different concepts of correctness exist. A system is said to be correct when two aspects, namely the specification and the implementation, are equivalent, or when the system satisfies a set of desirable properties. These desirable properties allow the system designer to identify the correctness of the system [**69**].

In PN literature a "basic kit of PN properties" is referred to a set of properties that are related to frequently occurring problems or the key issues related to the logical structure and behavior of complex systems, therefore they are classified into two main categories namely (i) Structural Properties and (ii) Behavioral Properties. It is important to note that fulfillment of these properties answer many questions of system correctness, therefore they contribute in the analysis of PN models. Some of the selected behavioral PN properties are listed and briefly discussed informally[22] below.

**Reachability**

Reachability is a fundamental property for studying the dynamic behavior of a system. In PN, reachability property is studied to analyze if a particular system state (in terms of markings) can be reached or not. A marking Mn is said to be reachable from an initial marking $M_0$ if there exists a sequence of firings that transforms $M_0$ to Mn. In reachability analysis, a set of all possible firing sequences from $M_0$ are populated in a reachability graph $R(N, M_0)$ and the reachability problem for PN is the problem of finding if a given marking $Mn \in R(N, M_0)$ [**70**].

**Boundedness**

In classical systems theory, a state variable that is allowed to grow to infinity is generally an indicator of instability in the system [**25**]. Therefore it is desirable that a system holds boundedness. A PN is said to be bounded (or k-bounded) if the number of tokens in each place does not exceed a finite number k for any marking reachable from initial marking, i.e., $M(p) \leq k$ for every place p and every marking Mn $\in R(N, M_0)$ [**70**].

**Deadlock-free and Liveness**

A PN is said to be deadlock-free if from any reachable marking at least one transition can always occur. A stronger condition than deadlock-freeness is liveness. A transition is live if it is potentially fire-able in all reachable markings. In other words, a transition is live if it never loses the possibility of firing. A net is live if all transitions are live [**69**].

**Reversibility**

A PN is said to be reversible if, from each marking Mn, the initial marking $M_0$ is reachable. Thus, in a reversible net one can always get back to the initial marking or state [**70**].

**Fairness**

Fairness has different meanings and understanding in literature. In specific terms, fairness means to give some contenders an equal number of chances, such that no one proceeds for more than "k-times" without letting the others to take their turn. In PN s, two transitions $t_1$ and $t_2$ are said to be in a bounded-fair (or B-fair) relation if the maximum number of times that either one can fire while the other is not firing is bounded. A PN is said to be a B-fair net if every pair of transitions in the net are in a B-fair relation [**70**].

**Mutual Exclusion**

This property captures constraints such as the impossibility of a simultaneous access of a critical section (resource) by two or more processes. In PN, mutual exclusion can be defined in terms of places or transitions. Two places p and q are mutually exclusive in a PN if their token counts cannot be both positive in the same marking, i.e., $\forall m \in RS \; m(p) \cdot m(q) = 0$. Similarly, two transitions in a PN are mutually exclusive if they cannot be both enabled in any marking [**71**].

[22] In literature these properties are discussed in detail with mathematical definitions and proofs [**70**]. In this chapter they are only discussed for background concept. Some of these properties are used later in this thesis.

Some of the important structural properties of PN are defined below:

**Controllability:**

A PN is said to be completely controllable if any marking is reachable from any other marking [**70**].

**Conservativeness:**

A PN N is said to be (partially) conservative if there exists a positive integer y(p) for every place p such that the weighted sum of tokens, is a constant, for every marking. Given a PN model, we are often required to ensure conservation with respect to certain weights representing the fact that resources are not lost or gained.

**Persistence**

A PN is said to be persistent if, for any two enabled transitions, the firing of one transition will not disable the other. A transition in a persistent net, once it is enabled, will stay enabled until it fires [**25**].

### 3.1.4 PN Analysis

The major strength of PN is the modeling of systems that exhibit concurrency. However modeling by itself is of little use. It is necessary to be able to analyze the modeled system. The analysis leads to important insights into the structure and behavior of the modeled system [**26**]. There are many techniques available for the analysis of PN models and can be employed for verification depending upon the nature of the model. Each technique may also have different variants. In this section two of the most commonly used techniques for the analysis of a PN model are discussed:

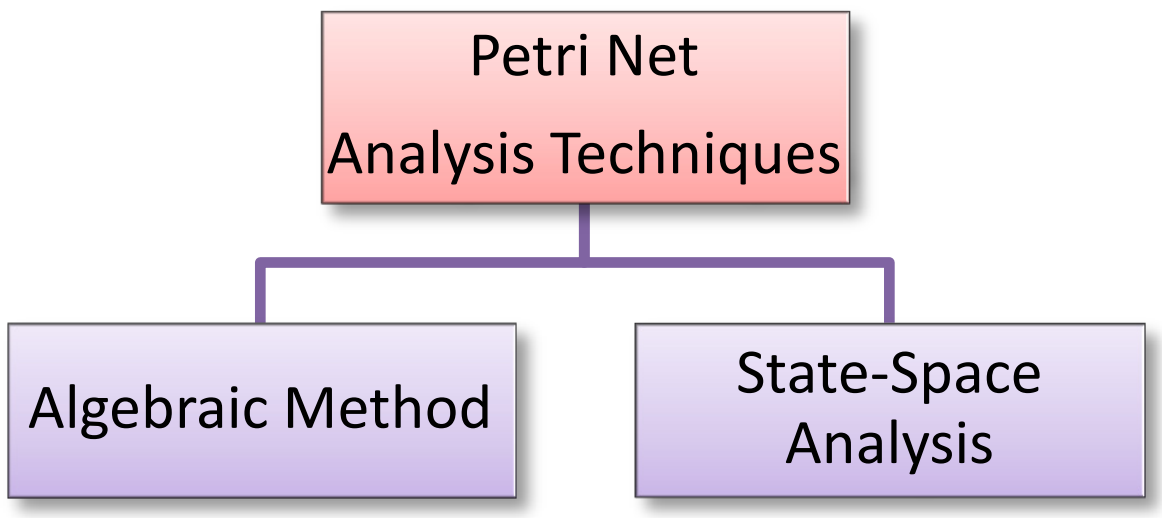

**Figure 13: Petri Net Analysis Techniques**

These techniques provide solutions and mechanism for verifying the properties mentioned in the previous section. In this thesis, these techniques are selected for composability verification and their application is shown in Part II with suitable examples. In this chapter, they are briefly explained and discussed, with their advantages and limitations.

### Algebraic Method

This technique is also called Linear-Algebraic Technique (or Linear Invariant due to its abundant use of invariants). In the framework of using algebraic techniques for reasoning about PN, solving a PN problem is reduced to finding a solution for an algebraic equation associated with the PN [24]. Due to the nature of this technique, the method is in general efficient (and in most cases, polynomial in the size of the PN). The dynamic behavior of PN models can be described by algebraic equations. In order to work with Algebraic method, the following basic concepts are applied:

### Matrix Definitional Form (MDF)

A PN model has a Matrix Definitional Form (MDF) that consists of three n×m[23] matrices:

**(i) Output matrix A+**

$$A^{+} = [a_{ij}^{+}]^{n \times m}, \text{ where } a_{ij}^{+} = w(t_i, p_j); \text{ if } p_j \in t_i\bullet, \text{ and } i \in n; j \in m \tag{3.5}$$

i.e., if $p_j$ is connected to the output of $t_i$ then $a_{ij}^{+}$ is equal to the weight of output arc; 0 otherwise [70].

**(ii) Input matrix $A^-$**

$$A^{-} = [a_{ij}^{-}]^{n \times m}, \text{ where } a_{ij}^{-} = w(p_j, t_i); \text{ if } p_j \in \bullet t_i \tag{3.6}$$

i.e., if $p_j$ is connected to the input of $t_i$ then $a_{ij}^{-}$ is equal to the weight of output arc; 0 otherwise [70].

**(iii) Incidence matrix A**

$$A = A^{+} - A^{-}, \text{ where } [a_{ij}] = [a_{ij}^{+} - a_{ij}^{-}] \tag{3.7}$$

In the incidence matrix A, each entry $a_{ij}$ represents the change of tokens in place j when transition i fires once [70].

### Firing Count Vector

A marking $M_k$ is an m × 1 column vector. The j^th entry of $M_k$ denotes the number of tokens in place j after the k^th firing in some firing sequence. An **n×1** column vector **X** of nonnegative integers is called firing count vector, where the i^th entry of **X** denotes the number of times transition **t** must be fired to transform $\mathbf{M_{k-1}}$ to $\mathbf{M_k}$ [70].

### State Equation

State equation for a PN is written as:

$$M_k = M_{k-1} + A.X \tag{3.7}$$

Where:

$\mathbf{M_{k-1}}$ is the current marking

[23] n×m refers n transitions and m places.

**$M_k$** is the new marking

**A** is incidence matrix

**X** is the firing count vector

### Example

An example of a Producer-Consumer PN model is shown in Figure 14.

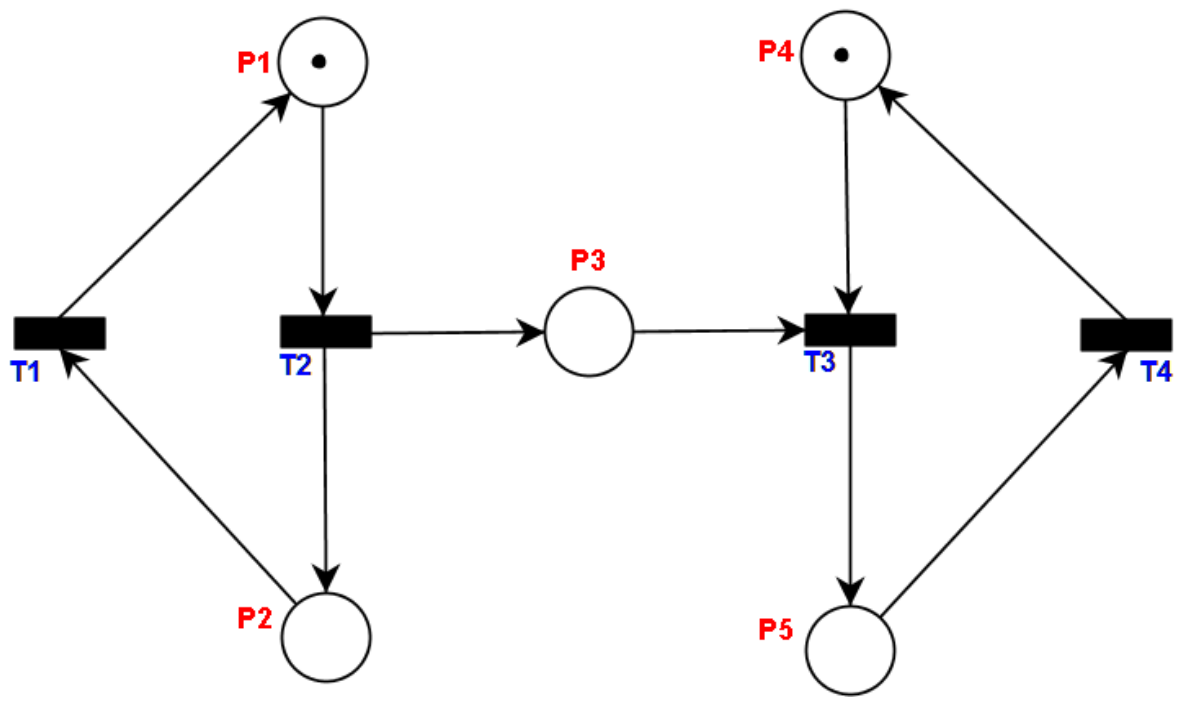

**Figure 14: Producer Consumer Example**

Using equation 3.7 the incidence matrix A of this model is calculated as follows:

| A+ | T1 | T2 | T3 | T4 |
|---|---|---|---|---|
| P1 | 1 | 0 | 0 | 0 |
| P2 | 0 | 1 | 0 | 0 |
| P3 | 0 | 1 | 0 | 0 |
| P4 | 0 | 0 | 0 | 1 |
| P5 | 0 | 0 | 1 | 0 |

-

| A- | T1 | T2 | T3 | T4 |
|---|---|---|---|---|
| P1 | 0 | 1 | 0 | 0 |
| P2 | 1 | 0 | 0 | 0 |
| P3 | 0 | 0 | 1 | 0 |
| P4 | 0 | 0 | 1 | 0 |
| P5 | 0 | 0 | 0 | 1 |

=

| A | T1 | T2 | T3 | T4 |
|---|---|---|---|---|
| P1 | 1 | -1 | 0 | 0 |
| P2 | -1 | 1 | 0 | 0 |
| P3 | 0 | 1 | -1 | 0 |
| P4 | 0 | 0 | -1 | 1 |
| P5 | 0 | 0 | 1 | -1 |

**Table 4: Incidence Martic A**

In this model, the initial marking is [1 0 0 1 0]. With a firing sequence $\sigma$ = t2, t1, t2 the firing count vector will be [1 2 0 0]. Using the state equation, the marking $M_x$ can be generated as follows:

| M0 | |
|---|---|
| P1 | 1 |
| P2 | 0 |
| P3 | 0 |
| P4 | 1 |
| P5 | 0 |

+

| A | T1 | T2 | T3 | T4 |
|---|---|---|---|---|
| P1 | 1 | -1 | 0 | 0 |
| P2 | -1 | 1 | 0 | 0 |
| P3 | 0 | 1 | -1 | 0 |
| P4 | 0 | 0 | -1 | 1 |
| P5 | 0 | 0 | 1 | -1 |

.

| X | |
|---|---|
| T1 | 1 |
| T2 | 2 |
| T3 | 0 |
| T4 | 0 |

=

| Mx | |
|---|---|
| P1 | 0 |
| P2 | 1 |
| P3 | 2 |
| P4 | 1 |
| P5 | 0 |

**Table 5: State equation**

**Figure 15** graphically illustrates, how a firing sequence of $\sigma$ = t2, t1, t2 can lead $M_0$ to $M_3$. Green color highlights the firing of a transition. It can be noted that the marking $M_3$ in the lower right corner matches the marking generated by matrix state-equation in **Table 5**.

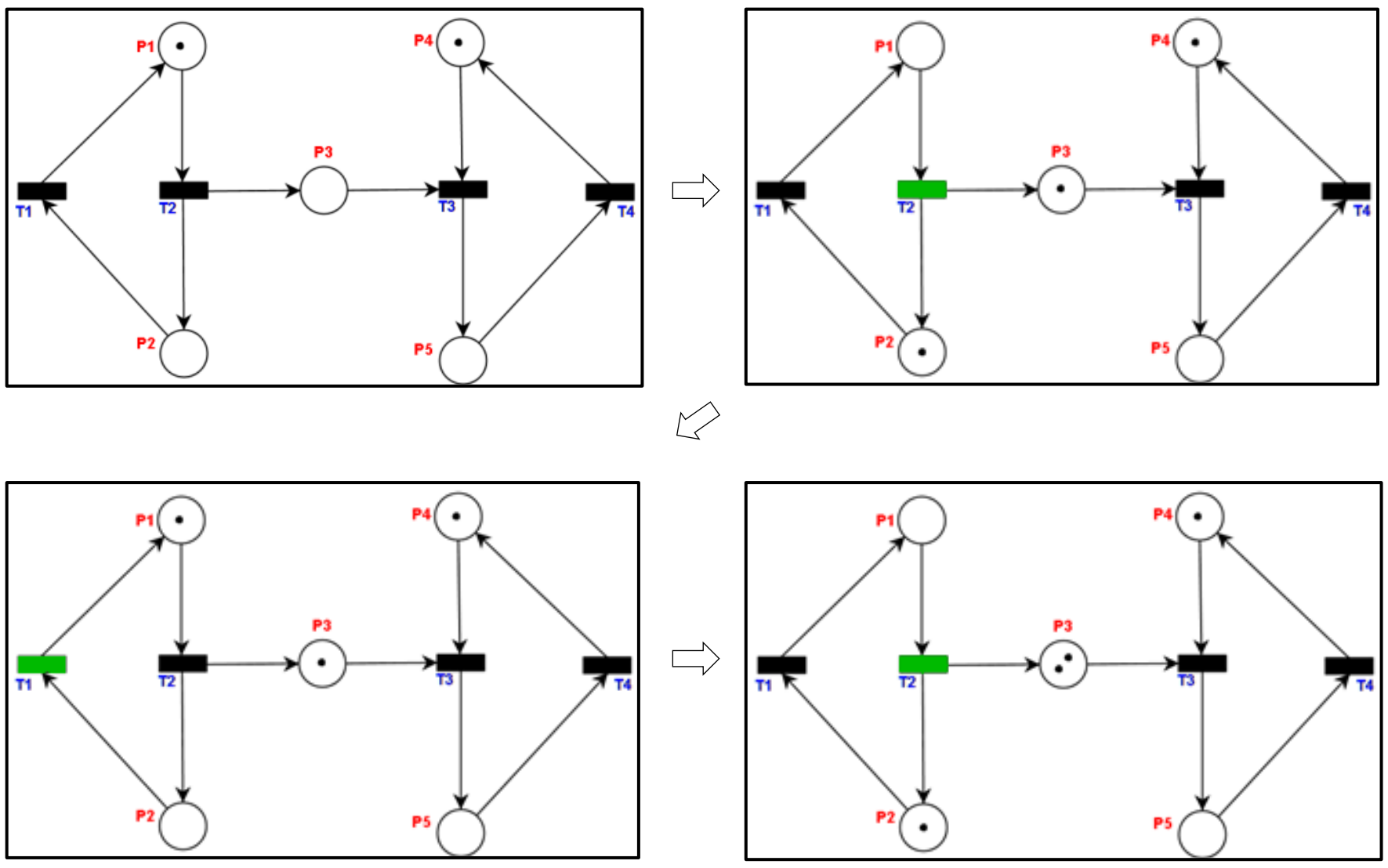

**Figure 15: $M_0$ to $M_3$ throguh firing sequece $\sigma$ = t2, t1, t2**

State equation alone can only help to algebraically compute a future marking. In order to analyze the model algebraically, some more concepts are used, such as PN Invariants.

## PN Invariants

Occurrences of transitions transform the token distribution of a net, but they often respect some global properties of markings, regarded as Linear Invariant Laws. Invariants are very useful for analyzing structural and behavioral properties of PN. From an initial marking, the marking of a PN can evolve by the firing of transitions (and if there is no deadlock) the number of firings is unlimited. However, not just any marking can be reached, all the reachable markings have some properties in common; a property which does not vary when the transitions are fired is said to be *invariant*. Similarly, not just any transition sequence can be fired; some invariant properties are common to the possible firing sequences. Hence, invariants enable certain properties of the reachable markings and firable transitions to be characterized, irrespective of the evolution.

**Figure 16** illustrates a PN model of different seasons in a year. It can be seen that, regardless of the change of seasons, there will always be one and only one token for all 4 places. Thus at all times, $M(p_1) + M(p_2) + M(p_3) + M(p_4) = 1$. This invariant property has an obvious meaning that at all time there is one and only one season [**68**]. It also means that the net is structurally bounded.

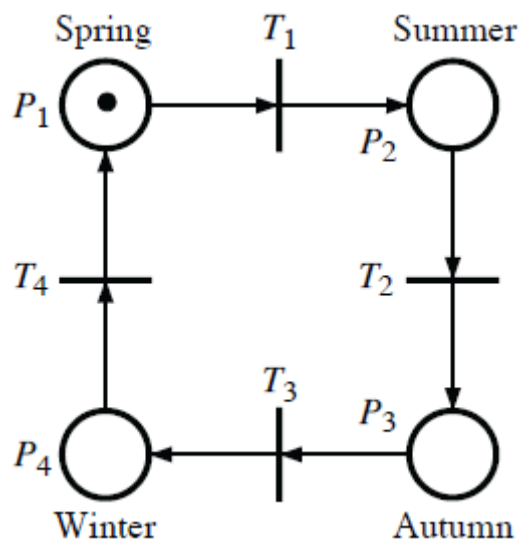

**Figure 16: Seasons in a year** (acquired from [**68**])

There are two important types of invariants of PN:

**P-Invariant**

Place Invariants formalize invariant properties regarding places in PN, e.g., if in a set of places the sum of tokens remains unchanged independently of any firing, then this set can define a place invariant. They are useful to evaluate structural properties of PN. In simple words, a place belonging to a P-invariant is bounded [**24**], [**70**].

A P-invariant exists in a PN if

$$\sum_{p=1}^{n} m \, . \, y_p = \sum_{p=1}^{n} m_0 . \, y_p \quad \forall \mathrm{m} \in \mathrm{R(N, m_0)} \tag{3.8}$$

Where y is an m × 1 column vector of integers such that $\exists \, \mathbf{y} = (y_1, y_2 \ldots y_n) > 0$ i.e., has at least one positive non-zero entry [**71**]. It means the firing of any transition does not change the weighted sum of tokens in the PN. More generally, a vector y is called P-Invariant if

$$\mathrm{A} \, . \, \mathrm{y} = 0$$

It is easy to see that if there is a P-invariant, for all p ∈ P, then the PN is guaranteed to be structurally bounded. Hence, place invariants can be used for reasoning about structural boundedness [**24**]. P-invariant is a P-semi-flow if every element of it is non-negative [**67**].

**T-Invariants**

Transition Invariants on the other hand formalize properties regarding transition firing sequences applicable to a PN. They are useful to evaluate behavioral properties such as liveliness and fairness [**24**], [**70**].

A n × 1 firing count vector X, is called a T-Invariant if

$$\mathrm{A} \, . \, \mathrm{X} = 0$$

i.e., firing each transition the number of times specified in X, brings the PN back to its initial marking $M_0$ [**24**]. T-invariant is a T-semi-flow if every element of J is non-negative [**67**].

A T-Invariant X is a minimal T-invariant, if there is no other T-invariant X′ such that $x'_i \leq x_i$ for all i∈T. There can be multiple T-invariants for a PN. A minimal T-Invariant is called the *Reproduction vector* of the net.

The intrinsic difference between P- and T-invariants are the facts that all places in a PN if covered by P-invariants is a sufficient condition for boundedness, whereas the existence of T- invariants is only a necessary condition for a PN model to be able to return to a starting state, because there is no guarantee that a transition sequence with transition count vector equal to the T- invariants can actually be fired [**71**].

**Advantages and Disadvantages**

The advantage of algebraic analysis is that the net structure is much less than the number of reachable markings and therefore there is no risk of state-space explosion. Various properties of PN consequently can be proven using linear algebraic techniques. However the weakness of this method is that it only entertains limited set of properties and provides only sufficient or necessary conditions. Also this method involves complex underlying mathematical theorems, each one different for different property verification and thus cannot be generalized for automated reasoning.

### State-Space Analysis

State space analysis is one of the most prominent approaches for conducting formal analysis and verification. In contrast to algebraic techniques, it is relatively simpler approach for analyzing the behavior of a model. The basic idea in this approach is to calculate all possible system states and the events which cause the change of states and represent them in a directed graph. When the graph is completely constructed, different search techniques can be applied to analyze the model.

In PN terms, this method is also commonly known as Reachability graph analysis. The state-space analysis of a PN model is performed by exhaustively generating all the reachable markings from a given initial marking, and then reasoning about the PN properties of the model by examining the structure of the reachability graph.

The reachability graph consists of vertices which correspond to reachable markings and of arcs corresponding to firing of transitions resulting in the passing from one marking to another. A simple example of reachability graph is shown in **Figure 17**.

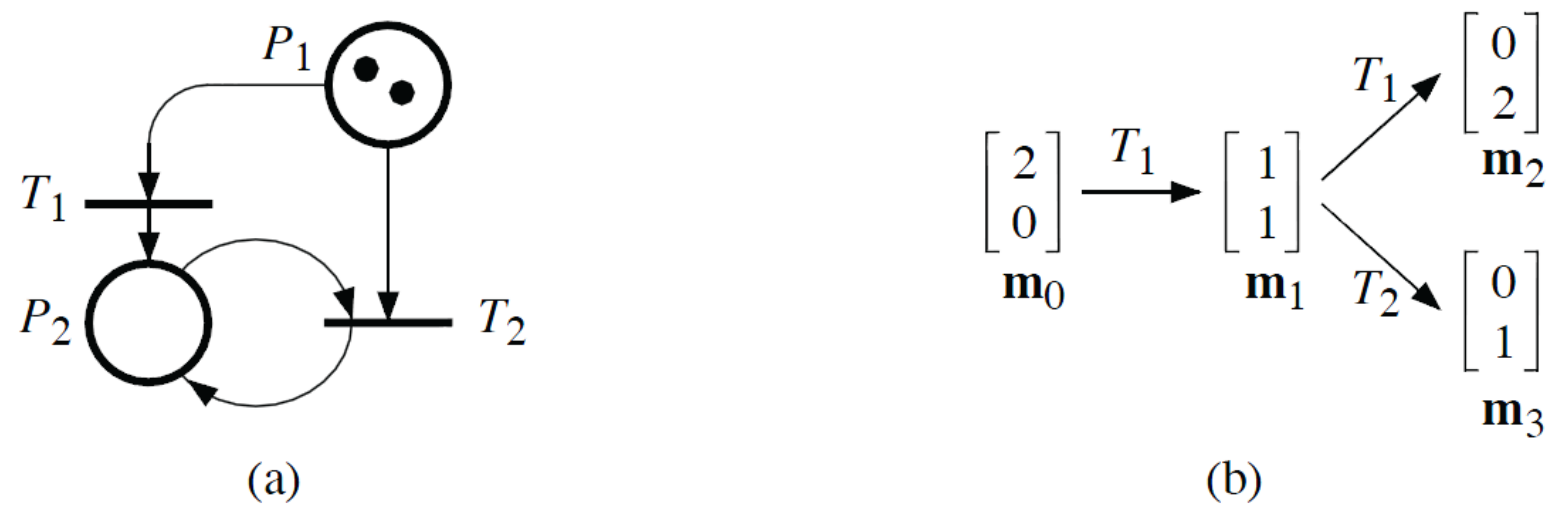

**Figure 17: (a) PN Model (b) Reachability Graph** (acquired from [68])

In some cases, the construction of reachability graphs becomes infinite if the PN or some of its parts are repetitive and the net is unbounded, or in other words the PN has infinite number of reachable markings. Therefore instead of keep on constructing nodes of the graph infinitely, an alternative technique is used, in which a finite graph is constructed by abstracting out certain details and inserting the symbol ω (the symbol of "infinity") to representing the marking of an unbounded place. This is called cover-ability graph. The coverability graph of the Producer-Consumer PN model is shown in **Figure 18**

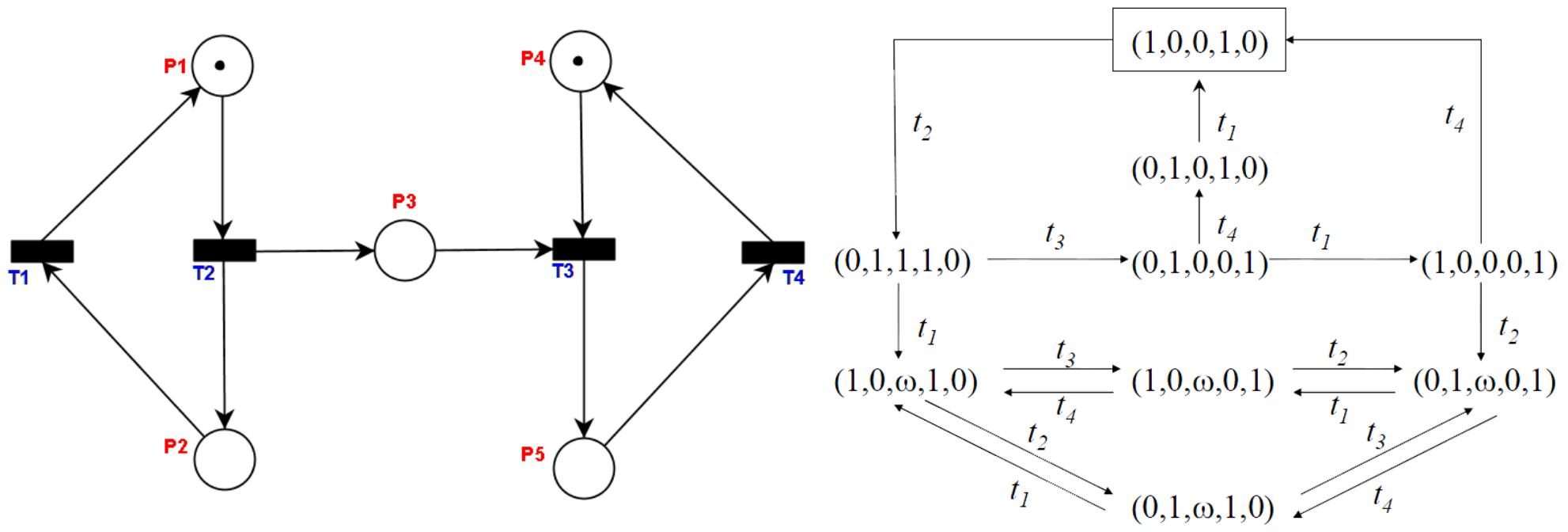

**Figure 18: Producer Consumer PN Model and its Coverability Graph**

It can be seen that the markings in which place P3 is unbounded contain ω symbols.

A constructed state space can help in answering a large set of analytical questions concerning the structure and behavior of the model such as verifying deadlock-freedom, absence of live-locks; presence of liveness, the possibility of being able to reach good states, and impossibility of reaching bad states and the guarantee of fulfilling the objectives. Following are some examples of how state space analysis help in model verification:

**Boundedness**

The problem of boundedness is easily solved using a coverability tree with an assumption that a PN is bounded if the symbol ω never appears in its coverability tree. Since ω represents an infinite number of tokens in some place, therefore its absence can guarantee that the PN is structurally bounded [**25**].

**Deadlock freedom**

A deadlock freedom problem is solved, if there is no node in the graph (which is not a final node), and yet it does not have an outgoing arc meaning there is no further enabled transition. Existence or one or more such nodes shows that the model has possibility of deadlock and can also help to find out the exact cause of it.

**Live lock freedom**

Similarly, a live-lock can be detected using state space analysis. For concurrent systems, a process is tasked to perform some particular actions [**72**]. These actions are normally intended to make progress and are called *progress actions*. A live lock is detected, if there exists a cycle within the reachability graph, in which no progress action is being executed.

**State Reachability**

Reachability of good states (or bad states) can be guaranteed using state space analysis. A state is reachable if there is a valid firing sequence that leads to that state from the initial marking. (In graph, there exists a path from the initial node to the corresponding node of the desired state). There could be multiple paths in a graph that reach the desired state. A shortest path analysis can be useful to analyze the minimum number of steps required to reach that state.

For details on how state space analysis are conduced, interested readers are recommended to refer to a very informative step by step tutorial on PN state space analysis [**73**].

**Advantages and Disadvantages**

The main advantage of state space method is that it is a way to explore all the possible states of the system. Also it provides counter examples as to why an expected property does not hold. Furthermore, the automatic calculation and generation of state-space provides an ease of use, due to the fact that the computer tool hides a large portion of the underlying complex mathematics from the user, who is only required to formulate the property which is to be investigated and a suitable query function to evaluate it [**74**].

The main disadvantage of using state spaces is the state explosion problem. The construction of the reachability graph is very expensive and intensive from a computational point of view. This is because the size of the state space may grow exponentially with respect to the size of the PN model (measured, for example, by the number of places). Even relatively small systems may have an astronomical or infinite number of reachable states. This problem escalates severely, when the models includes time. A lot of effort has been invested in the development of reduction methods to alleviate this problem. Reduction methods represent the state space in a compact form. The reduction should not affect the properties of the system and they should be preserved and can still be derived from the reduced state space. However, due to the complexity and diversity in verification, there is no single reduction method which works well in all situations. Therefore the choice of a reduction method completely depends on the nature of the system being verified [**74**]. Some of the important reduction methods are Sweep line method [**75**], Hash Compaction Method [**76**], Symmetry Method [**77**] and Equivalence Method [**78**].

In this thesis, we propose another reduction method which suits our need (Composability verification) and can help to alleviate the state explosion problem, if the model under consideration becomes large and resource intensive.

### 3.1.5 PN Classes

The computational power of basic or classical PN is weak as it has been shown that PN are not as expressive as Turing machines, making them inadequate for modeling certain real-world systems. To overcome this shortcoming, a number of extended PN have been introduced to enhance the expressive capabilities of PN. There are different ways to classify PN. In structural sense, they can be classified into three main categories [**79**]:

**Level-1 PN:** are characterized by '*Boolean tokens*', i.e. places are marked with at most one unstructured token.

**Level-2 PN:** are characterized by '*Integer tokens*', i.e. places are marked with several unstructured tokens - they represent counters.

**Level-3 PN:** are characterized by high-level tokens, i.e. places are marked with structured tokens where information is attached to them.

There are many extensions of PN formalism. In this section we only discuss some of the extensions of PN, which are used in this thesis.

**Colored Petri Nets (CPN)**

CPN is a level-3 extension of PN, in which places are marked with structures token representing data. CPN is a graphical language for constructing models of concurrent systems and analyzing their properties. CPN is a general purpose discrete event language which combines the capabilities of PN, as a foundation of the graphical notation and a programming language (CPN ML), which is based on Standard ML [**80**] functional programming language, that provides the primitives for the definition of data types and for specifying data manipulation routines [**78**].

CPN is formally defined by the tuple [**81**]:

**CPN = (P, T, A, Σ, V, C, G, E, I)** where:

**P** is a finite set of places

**T** is a finite set of transitions such that: $P \cap T = \varnothing$

$A \subseteq P \times T \cup T \times P$ is a set of directed arcs.

**Σ** is a finite set of non-empty color sets.

**V** is a finite set of typed variables such that: $Type[v] \in \Sigma$ for all variables $v \in V$

**C: P→Σ** is a color set function that assigns a color set to each place.

**G: T → Expression** is a guard function that assigns a guard to each transition t

**E: A→ Expression** is an arc expression function that assigns an arc expression to each arc a

**I: P → Expression** is an initialization function that assigns an initialization expression to each place p.

Tokens of an ordinary PN have no types. With CPN it is possible to define token using data types and complex data manipulation i.e., each token has attached a data value called the token color. The token colors can be investigated and modified by the occurring transitions [**81**].

"CPN Tools" is a software package for the editing, simulation, state space analysis, and performance analysis of CPN models [**82**]. The tool acts as an integrated development environment (IDE) for the construction of CPN models. It provides a canvas for creating PN graphs, offers features for writing CPN ML code with a facility of incremental syntax checking. It also comes along with a bundled simulator that efficiently handles the execution of untimed and timed nets. The most important feature of CPN tool from our point of view is the generation and analysis of state spaces. The analysis of state space includes various built-in state-space querying functions, and support for creating analysis report which altogether greatly contributes to the verification process. For further details of CPN formalism and its application [**78**], [**81**] are referred.

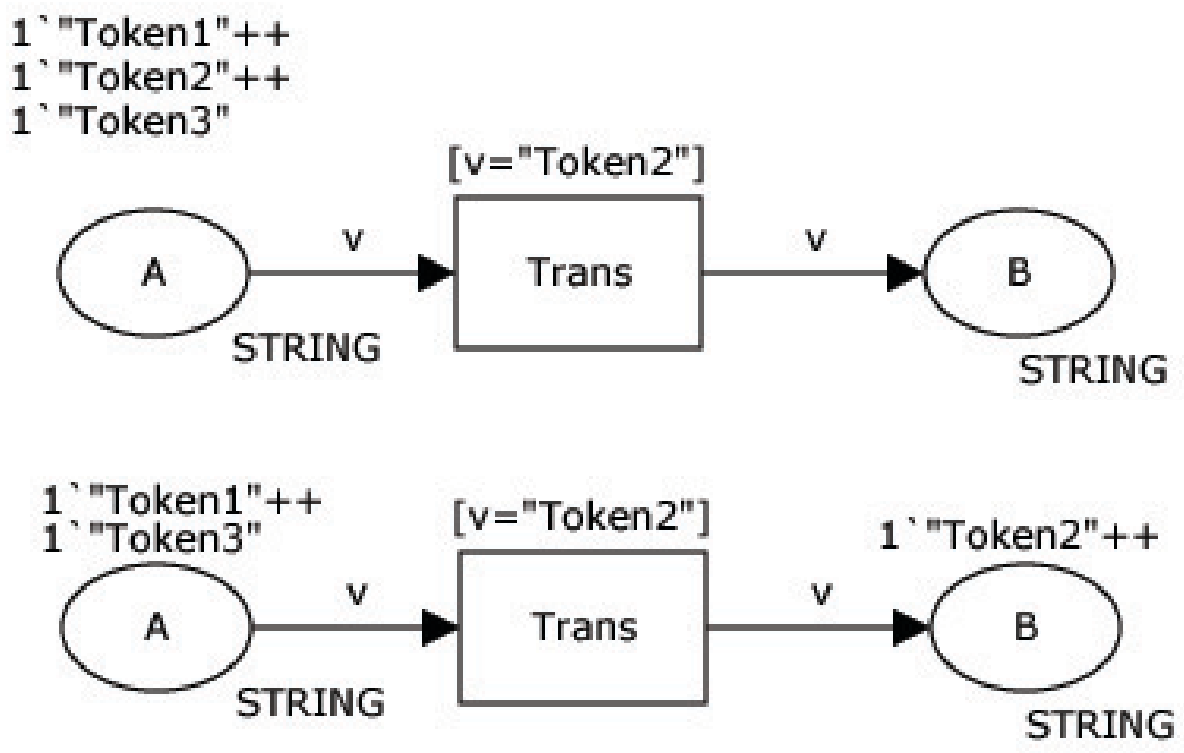

**Figure 19: A CPN Model**

Figure 19 shows a basic example of a CPN model. The nodes A and B in oval shape represent places. The place is initialized with three tokens of String type. The rectangular shaped node represents transition. An input arc connects Place A with the transition with an arc variable v of type String (to carry tokens of the same type).

Similarly an output arc connects transition to place B. The transition has a guard expression that checks the token value. If the expression is true only then the transition can be fired. The second part of the Figure 19 shows the result of the firing of transition, i.e., the token "Token2" being deposited to place B.

### Hierarchical CPN

CPN model can be organized as a set of modules; where modules can be seen as black boxes which make it possible to work at different abstraction levels, concentrating on one at a time.

### Substitute Transitions

CPN tools offer facility to construct hierarchical CPN models. In hierarchical nets a transition can represent an entire piece of net structure. Such a transition is called substitution transition [**82**].

### Sub-page /Super-page

A page that contains a substitution transition is called a super-page. When a CP-net uses a substitution transition the logic that the transition represents is kept on a page called a subpage [**82**].

### Ports and sockets

Super-pages and sub-pages are connected by ports and sockets. A socket is a place in the super-page that has at least one arc between a substitution transition and a socket. A port on the other hand is a place in a subpage, marked with one of the port-type tags: (i) In-Port (ii) Out-Port or (iii) In/Out-Port. It is bound with a socket in the main page using *Port & socket assignment.* This relationship is used to define how a subpage should be connected with the surroundings of its super-page. Some of the assignment rules are as follows:

- A port with an In-tag must be assigned to a socket which is an input arc of the substitution transition.
- An Out-tag indicates that the port must be related to a socket which is an output arc,
- I/O-tag indicates that the socket must be both an input and output arc [**82**].

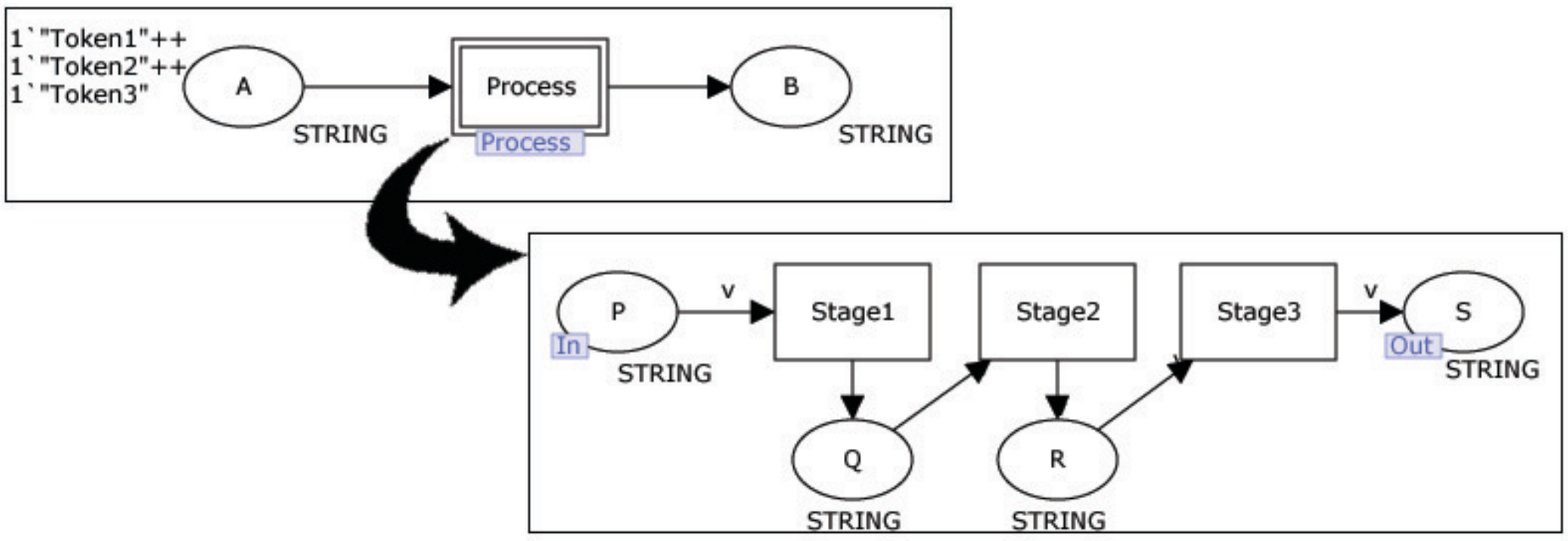

**Figure 20: Hierarchical Colored Petri Net**

Figure 20 presents an example of hierarchical CP-net. In the super-page (above), a substitute transition Process is shown which represents a sub-module (below). A process has three stages, and input and an output marked with In and Out ports which are connected with A and B socket places in the super-page.

### Timed Petri Nets

PN with timing dependencies can be classified according to the way of specifying timing constraints. These constraints can be timing intervals or single numbers, or elements of the net these constraints are associated with i.e., places, transitions or arcs [**83**]. The next criterion is an interpretation of the timing constraints. When associated with a transition, the constraint can be viewed as

#### (i) Firing time

A transition consumes the input tokens when it becomes enabled, but does not create the output tokens until the delay time associated with it has elapsed [**83**].

#### (ii) Holding time

When the transition fires, the actions of removing and creating tokens are performed instantaneously, but the tokens created are not available to enable new transitions until they have been in their output places for the time specified as the duration time of the transition which created them [**83**].

#### (iii) Enabling time

A transition is forced to be enabled for a specified period of time before it can fire, and tokens are removed and created in the same interval [**83**].

Timed extensions are known also for high-level PN. One of them is timed Colored Petri nets [**78**], in which the time concept is based on introducing a global clock used to represent the model time. Tokens are equipped with time stamps, which describe the earliest model times at which they can be used to fire a transition. Stamps are modified according to expressions associated either with transitions, or with their output arcs. Timing intervals can be interpreted as periods of non-activity of tokens, and the transitions are fired according to the strong earliest firing rule [**78**].

Formally a time PN is a tuple:

$$N = (P, T, F, m_0, Eft, Lft)$$

Where:

$(P, T, F, m_0)$ is a PN,

Eft = Earliest firing time for each $t \in T$

Lft = Latest firing time for each $t \in T$

## 3.2 Communicating Sequential Processes

CSP is the second formalism that is selected in this thesis for the modeling of executable components. CSP is a language developed by Sir Charles Antony Richard Hoare [**84**]. It aimed to be used for specification and reason about the concurrent interaction of the system processes. The idea of CSP was conceived for the study of concurrent processes using formal notation with required expressive power and algebraic laws. The formal notation and the associated algebraic laws allow the process models to be controlled and analyzed. They also enable formal reasoning about their correctness and prove equivalences between the processes. They also provide sufficient theoretical foundations for the development of the necessary tools for these purposes.

### 3.2.1 Basic Concepts and Definitions

The main primitives of CSP formalism are (i) Processes (ii) Events and (iii) Algebraic Operators.

**Process**

In CSP terms, a process is an independent, self-contained, modular description of an entity and a basic unit to capture behavior. A process has particular interface, captured by events that are used to interact with the environment which itself is a process, called the universe of the system ($\Sigma$). The environment can be viewed as a system of concurrently evolving processes. In any run a process performs a sequence of events. A process has a name, list of parameters and expression which determines its computational logic:

$$\text{Process (parameters)} = \text{Expression}$$

Expression is behavior of a process which can be described as an occurrence of an event or the sequence of some events, known as a trace. A process can only perform a finite number of events in any finite time, and thus all traces have finite length [**85**].

**Events**

The ultimate unit in the behavior of a process is an event [**85**]. Events characterize communications or interactions. Events are abstraction of observations. Each event forms an interaction between the process and its environment. If the interaction does not occur then the process is blocked. Event can be defined with no data or data with typed values. A set of all events of a Process P are called Alphabet of P ($\alpha$P). The following line describes a simple vending machine which takes in a coin and dispatches a coffee every time [**84**].

$$\text{VM()} = \text{insert-coin} \rightarrow \text{coffee} \rightarrow \text{VM();}$$

Where VM() is a process (with no parameters) and its expression contains a sequence of atomic events: insert-coin and coffee and then the process is self-referenced (recursion). Events can be written in compound form, i.e., with parameters as shown in the following line:

$$\text{VM()} = \text{insert-coin.10} \rightarrow \text{coffee.1} \rightarrow \text{VM();}$$

Also there could be data operations using statement blocks inside the event body:

$$\text{VM()} = \text{insert-coin.10\{Balance= Balance +10\}} \rightarrow \text{coffee.1\{coffee--\}} \rightarrow \text{VM();}$$

A statement block could be a complete sequential program contains assignment statements, if-then-else clauses, for or while loops and math functions etc.

**Input/output Channels**

Processes may also communicate through channels. Channels are special type of events, called communication events. Usually a communication on a channel results from an input and output occurring in parallel. The input channel is represented by '?' symbol whereas the output channel is represented by '!' symbol. The channel parameters can be send or received using the form: c ! x or c ? y

**Algebraic operators**

There are many different useful operators that are used to represent different notions of process behavior and their compositions [**85**]. Some are described as follows:

- **Prefix** $a \rightarrow P$

  The prefix operator combines an event and a process to produce a new process.

- **Sequential composition** P ; Q

  It composes two processes P and Q in a sequential order i.e. the latter only starts when the former terminates

- **Deterministic Choice P □ Q**

  The deterministic (or external) choice operator allows choosing between two component processes, and allows the environment to resolve the choice externally.

- **Non-deterministic Choice P ⊓ Q**

  The nondeterministic (or internal) choice operator allows a choice between two component processes, but does not allow the environment any control over which one of the component processes will be selected.

- **Conditional Choice if *cond* P *else* Q**

  The choice depends on the evaluation of a condition to choose between P or Q.

- **Interleaving P ||| Q**

  The interleaving operator represents completely independent concurrent activity between the processes P and Q i.e., without barrier synchronization.

- **Parallel Composition P || Q**

  The parallel composition operator represents concurrent activity between P and Q that requires barrier synchronization between the component processes. If an event is in the alphabet of both P and Q, then it can only occur when both processes are able to engage in that event.

### 3.2.2 CSP Analysis Techniques

Many techniques have been developed for the analysis of CSP models however Model Checking has surpassed them all in many aspects and is commonly favored by most of the CSP based modeling environments. In this section Model Checking technique is briefly described.

**Model Checking**

> *"Model checking is an automated technique that, given a finite-state model of a system and a formal property, systematically checks whether this property holds for that model* [**86**]*"*

The instigation and rapid advancements of model checking methods is one of the towering achievements in the area of model based software verification, especially with the advent of difficulties faced by the computing communities when the struggle of sequential program verification was followed by even more daunting exertion of verifying concurrent programs [**87**]. The growing difficulty in error tracing of such programs is due to the increase of complexity of the system behavior and the arbitrariness of large portion caused by emergent system states which cannot be easily tacked by ordinary testing and debugging methods.

Starting from late 70's Model checking and other similar algorithmic and automata theoretic approaches are the result of efforts of notable researchers who pioneered different standards that can be marked as a collective foundation of principles that shaped the modern model checking techniques [**87**].

Model checking became successful in different communities due to following reasons:

- Unlike traditional testing methods it is an exhaustive approach that provides an in-depth analysis of a system model to certify absence of bugs (instead of just finding few of them through debugging).
- Model checking returns answers — either successful outcomes or counterexamples showing the exact trace of errors and their causes
- Improvements in model checking techniques have effectively alleviated the risk of state-space explosion problem [**87**].
- Model Checking has a sound and mathematical underpinning and is based on theory of graph algorithms, data-structures, and logic [**86**].
- Model checking support formalism both for the specification of the input models (such as FSM, PN, CSP or others) and the specification of system properties being verified (which are mostly in the form of LTL or CTL or their extensions). Therefore any 3rd party community can use a model checker as a black box without knowing the insights and complexity of the process.

Beside its various strengths some of the weaknesses include:

- Most model checkers require the models to have reduced details using compact and less expressive states and without specifying enumerations due to the risk of state-explosion. Therefore the reduction in the system expressiveness may cost extra effort and possibly lead to overlooking important features and getting inadequate verification results.
- Despite the development of several very effective methods and improved data-structures to combat the state-explosion problem, models of realistic systems may still be too large to verify.

### Types of Model Checking

Model checking approaches are classified into two types: (i) Explicit and (ii) Symbolic based on how they enumerate states [**88**].

**Explicit model checking techniques** store the explored states in a hash table, where each entry corresponds to a single system state. For just a few hundred states the nodes in the state space graph becomes as large as ~$10^{11}$ [**88**]. On the other hand explicit model checkers support state-enumeration that gives detailed expressiveness of the system states.

**Symbolic model checking techniques** store sets of explored states symbolically by using efficient data structures represented by canonical structures such as Binary Decision Diagrams (BDDs) [**89**], and traverse the state-space symbolically by exploring a set of states in a single step. The use of these BDD-based methods has greatly improved scalability in comparison to explicit state enumeration techniques, yet they have performance degradation because BDDs constructed in the course of symbolic traversal grow extremely large, and BDD size is critically dependent on variable ordering. This causes a newer trend of research towards separating Boolean reasoning and representation. Hence Boolean Satisfiability (SAT) [**90**] has been studied and explored for Boolean reasoning and efficient semi-canonical representations which results in the development of SAT-solvers which are efficient and have compact representation compared to BDDs. SAT, together with efficient representation, have become a viable alternative to BDDs for model checking applications [**88**].

**Bounded model Checking** is a model checking approach where the number of steps in forward traversal of the state space are bounded and checks whether a property violation can occur in k or fewer steps [**88**]. The approach reports either "violation found" or "no violation possible within the bounded depth (i.e., k steps), which can be incremented to look ahead for possible violation of the property. This method is promising because it does not cause state-space explosion or at least let the user control its possibility.

In this thesis all three model checking approaches are accompanied by the tools selected for composability verification of CSP based models.

## 3.2.3 Temporal Logics

Logic provides formal languages containing formulas for the representation of the statements and their logical reasoning within some area of application [**91**]. Generally, a logical language is given by an alphabet of different symbols and the definition of the set of formulas which are strings over the alphabet [**91**]. In logic, the term temporal logic is used for representing and reasoning about propositions qualified in terms of time. Temporal logic has found an important application in formal verification, where it is used to specify system requirements. Linear Temporal Logic (LTL) and Computational Tree Logic (CTL) are its two main variants. LTL formulas are interpreted on computation paths. Let A and B be atomic predicates and $\neg$ , $\wedge$ , $\vee$ , $\leftrightarrow$ and True be the operators of classical logic, whereas $\bigcirc$, $\square$, $\diamond$ and $\mathcal{U}$ are the operators of linear temporal logic called *Next*, *Always* and *Eventually* and *Until.*

The intuitive meanings of some LTL statements are:

- $\neg$ A : A does not hold
- A $\wedge$ B : Both A & B hold
- $\bigcirc$A: A holds at the next state
- $\Box$A: A holds in all states
- $\Diamond$A: A will eventually hold
- A U B: A will hold until B holds.

In CTL there are additional path quantifiers '∃' and '∀' denoting 'there exists a path' and 'for all paths', respectively. CTL formulas are interpreted on computation trees. With respect to a tree the intuitive meanings of the formulas mentioned above are:

- $\exists\bigcirc$A: There exists a path in which A holds at the next state
- $\forall\Box$A: For all paths A always holds in all states

### 3.2.4 Time CSP

CSP has been in evolution for decades. One of the major extensions of CSP is devised with timing primitives, denoted as TCSP, to support time sensitive process modeling [**92**]. In TCSP, each of the untimed CSP operators is interpreted in a timed context, and two primitive timing operators are added: (i) timeout and (ii) interrupt, with a Newtonian Time assumption (i.e., that all the processes have a single global clock with same progress rate).

- **Timeout $P \rhd^{d} Q$**

  Timeout operator can be used to introduce delay in the processes.
- **Timed Interrupt $P \triangle_{e} Q$**

  Interrupt is used if the process is permitted to run for no more than a particular length of time.

The concept of TCSP is used later in this thesis to model and perform verification of real-time systems.

### 3.2.5 Probabilistic Systems

Systems that exhibit probabilistic aspects essential for designing randomized algorithms, modeling unreliable or unpredictable behavior or specifying model-based performance evaluation are called probabilistic systems [**86**]. In order to model random phenomena in such systems, transition systems are enriched with probabilities. Probabilistic systems can be specified in different ways. Two very popular ways are: (i) Markov chains (MC) and (ii) Markov decision processes (MPD). In this thesis, we considered MPDs as specification formalism for probabilistic systems because they support both nondeterministic and probabilistic choices and unlike MC they can model the interleaving behavior of the concurrent processes in an adequate manner [**86**].

A Markov Decision Process is a tuple ⟨S, Act, P, $l_{init}$, AP, L ⟩ [86].
Where:
S = Set of states
Act = set of actions
P: S × Act × S → [0, 1] is the transition probability function such that for all states s∈S and actions α∈ Act:

$$\sum_{s' \in S} P(s, \alpha, s') \in \{0,1\}$$

$l_{init}$: S → [0, 1] is the initial distribution such that:

$$\sum_{s' \in S} l_{init}(s) = 1$$

AP is a set of atomic propositions
L: S → $2^{AP}$ is a labeling function

The concept of MDP is used later in this thesis to model and perform verification of probabilistic systems.

### 3.2.6 CSP Implementation Tools

There are a variety of implementation support tools and languages for developing CSP models such as CTJ (Java), CSP++ (C++), CSP.NET, PyCSP (Python), JCSP (Java) and CSP# (C-Sharp) [93].

Similarly various techniques exist for CSP analysis such as:

- FDR2 model checker is developed by Formal Systems Europe Ltd [94].
- ARC, the Adelaide Refinement Checker, is a CSP verification tool [95].
- ProB is an animator and model-checker and support refinement checking and LTL model-checking of CSP [96].
- PAT is a model checker, simulator and refinement checker for CSP [97].

In this thesis we selected PAT model checker because of its user friendly environment for modeling CSP models, fast simulator and model checker and above all its support for CSP extensions such as Real-Time CSP, Probabilistic CSP and Real-time Probabilistic CSP.

### 3.2.7 Process Analysis Toolkit (PAT)

PAT is an established tool developed by National University of Singapore in concurrent system verification and has been used in real-world industrial projects. PAT is designed to develop, compose, simulate and analyze event-based system models using an extension of CSP formalism called CSP-Sharp (or CSP#[24]). This extension comprises of some additions such as shared variables and asynchronous message passing. Moreover it supports using complex data types (such as Set, Queue, and Stacks) and functions from external libraries written in C# therefore allow to model complex process behaviors. PAT also supports automated refinement checking and model checking of LTL extended with events [**98**].

[24] It uses C# like syntax for the specification of CSP processes

PAT is an appropriate modeling, composition, simulation, verification and reasoning framework of CSP based process models. These models can be of different nature such as concurrent, real-time and probabilistic systems. The main strength of this framework is that it implements various model checking techniques and provide verification support for different properties. That includes general system properties such as deadlock-freeness, divergence-freeness or reachability and user specific properties defined in terms of LTL assertions. It also includes refinement checking, model checking of real-time and probabilistic systems. To achieve good performance, advanced optimization techniques are also implemented in PAT, such as partial order reduction using BDD, symmetry reduction and parallel model checking [**97**].

## 3.3 Summary

In this chapter we have discussed two executable modeling formalisms namely: (i) Petri Nets and (ii) Communicating Sequential Processes and their associated concepts, tools and techniques. Both formalisms are used in this thesis for describing executable models. The conceptual background of both PN and CSP is required to understand the approach presented later in this thesis.

# Chapter 4
# Verification and Analysis

*Verification and Validation are important aspects of any software engineering expedition. They are independent procedures with different characteristics that are used to check that a program, service, model or a system is correct, meets requirements specifications and that it fulfills its intended purpose. They are critical constituents for achieving the necessary levels of quality assurance, and are essential prerequisites for a credible and reliable use of the delivered product. The main focus of this chapter is on Verification and its different analysis techniques. The aim of this chapter is to outline basic concepts, principles, issues and different approaches of software verification. This chapter can be viewed as a manual to understand the verification process being proposed later in this thesis.*

The correctness of a program is a relative concept, meaning that the program is doing no less than prescribed by its specification [**99**]. Verification, Validation and Testing (VVT) in combination is a broader and more complex discipline of system engineering. In M&S the combination of Verification, Validation and Accreditation (VVA) is generally referred where "Accreditation" is the formal certification that a model or simulation is acceptable to be used for a specific purpose [**100**]. Nevertheless the goal is to assure the quality of the product and the impetus behind this assurance is intensified when the systems are highly critical, either because they are very expensive to produce, such as land rovers investigating outer planets, or because human lives depend on them, such as computers controlling airplanes and cars, and life assisting real-time systems in hospitals [**101**]. These systems need to be correct, because their failure can lead to loss of human lives or enormous economic losses. Moreover correct systems can be used in a wrong manner which can also results in a failure. This is a general problem when systems are designed in a modular fashion, and are implemented with assumptions on a new environment. A similar case caused a drastic failure at the launch of Ariane-5 expendable rocket launch system, because a software module was reused from Ariane-3 with certain assumptions that did not hold for Ariane-5 which self-destructed just because one single variable of 64 bit floating point value was erroneously converted to a 16 bit integer causing the system to crash [**102**]. So for critical systems it is worth the effort to have a guarantee that they are correct and have no errors.

Verification and validation aim to increase the credibility of models and simulation results by providing evidence and indication of correctness and suitability. Verification in particular deals with the correctness of the model perceived from a real-system, whereas validation deals with the suitability or fitness of the model with respect to its real-system. Testing on the other hand aims to uncover incorrectness in the system. In the following section, definitions and concepts of these inter-related terms are discussed.

## 4.1 Some Basic Concepts in Modeling and Simulation

The first applied technical discipline that began to struggle with the methodology and terminology of V&V was the operations research (OR) community, also referred to as systems analysis or modeling and simulation (M&S) community [**103**].

### Verification

According to the Department of Defense (DoD) Defense Modeling and Simulation Office verification is defined: *as a process of determining that a model implementation accurately represents the developer's conceptual description and specification* [**104**].

In general verification refers to an evaluation process that determines whether a product is consistent with its specifications or compliant with applicable regulations. In M&S, verification is typically defined as the process of determining if a model is consistent with its specification [**29**]. Verification deals with the model correctness and is concerned with building the model right [**28**], i.e., a model which works correctly and has no bugs. In principle, verification is concerned with the accuracy of transforming the model's requirements into a conceptual model and the conceptual model into an executable model [**29**].

For the sake of clarity the notions of correctness are defined as follows:

**Correct:** *Free from error; accurate; in accordance with the fact, truth, or reason; Conforming to the acknowledged standards of a method, routine or behavior* [Oxford Dictionary]

**Correctness**

*The degree to which a program, model or a system as a whole is free from defects in its specification, design, and implementation* [**105**]

*The ability of a software product (or a simulation model) to perform the exact task, as defined by its specification* [**106**].

We define a composed model to be correct if its structure and behavior matches its specification. Correctness of a composed model is therefore relative to its specifications. A software entity can exist in three apparent states of correctness namely: (i) correct when it has been established correct against its specification; (ii) defective when it has been established incorrect against its specification and (iii) unknown when its correctness has not been established against a specification [**107**]. In SE a software entity's specification is the sum of all its passing unit-tests [**107**]. We define specification to be a set of goals (or objectives) and property constraints (see 1.3.2) that must be fulfilled by the composed model to be established as correct.

### Validation

According to the Department of Defense (DoD) Defense Modeling and Simulation Office validation is defined: *as a process of determining the degree to which a model is an accurate representation of the real world from the perspective of intended uses of the model* [**104**].

Model validation on the contrary, deals with building the right model, i.e., the model which is an accurate representation of the real system [**28**]. Model validation is usually defined to mean "substantiation that a computerized model within its domain of applicability possesses a satisfactory range of accuracy consistent with the intended application of the model [**108**].

### Testing

Model Testing on the other hand, ascertains whether inaccuracies or errors exist in the model. The objective of testing is to show that the model (or system) is incorrect (rather than proving that it is correct). Testing can only find errors but cannot guarantee the absence of errors; therefore it is more of an ad-hoc and inexpensive method of necessity, where the correctness is established merely on the fact that all tests have passed, which is insufficient and unreliable. When the test fails, it succeeds in revealing an error. When a test is passed, it fails to detect an error. If a number of tests fail to detect a bug, they increase a confidence level in the system even if the correctness cannot be guaranteed [**99**].

## 4.1.1 Verification and Validation in a Modeling Process

A Modeling Process has been defined by Sargent [**108**] as shown in **Figure 21**. In this process Verification is referred to as an activity which ensures that the computer programming and implementation of the conceptual model is correct.

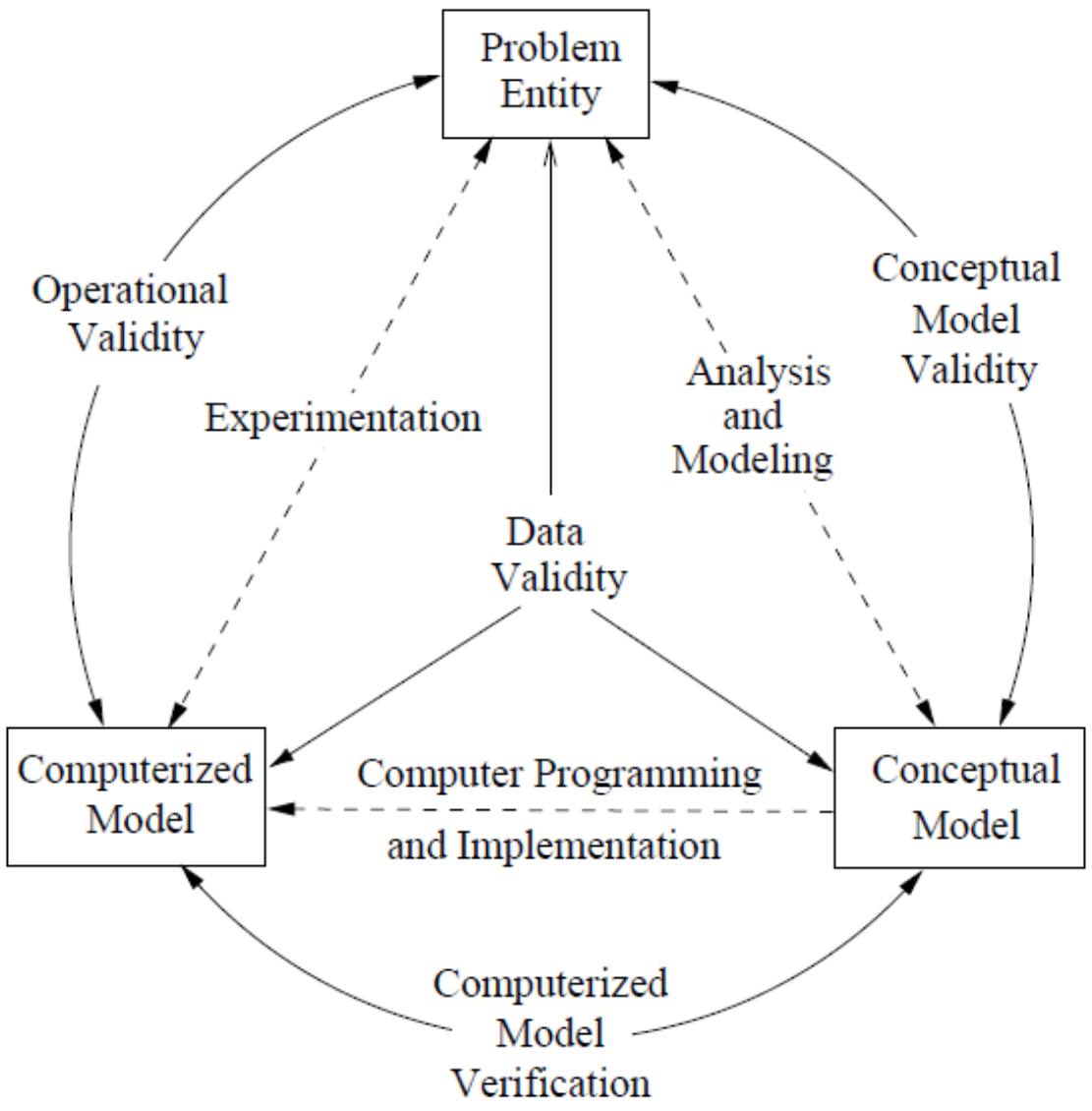

**Figure 21: Modeling Process** (acquired from [**108**])

Whereas validation is defined in three perspectives:

**Conceptual model validity** is defined as determining that the assumptions underlying the conceptual model are correct and that the model representation of the problem entity (simuland) is "reasonable" for its intended purpose.

**Operational validity** is defined as determining that the model's output behavior has sufficient accuracy for the model's intended purpose.

**Data validity** is defined as ensuring that the data necessary for the model execution and model experiments to solve the problem are adequate and correct [**108**].

Mike Petty in his article [**29**] also clarifies the difference between the two terms at different stages of model evaluation process as illustrated in **Figure 22**.

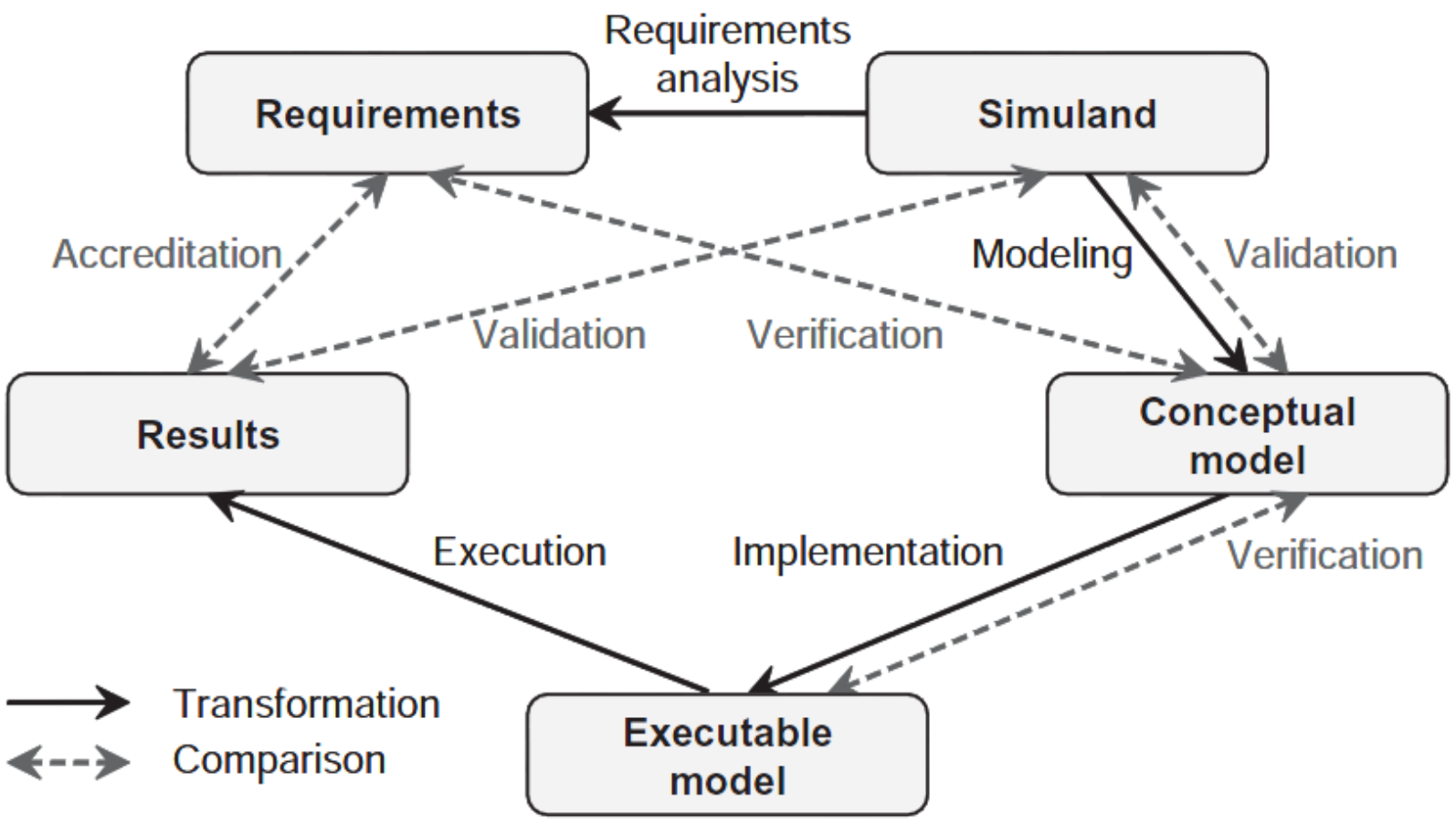

**Figure 22: Modeling Process** (acquired from [**29**])

A *simuland* is the real system that is to be simulated whereas a model is a representation of the simuland, developed with its intended application in mind and therefore captures only the necessary abstractions of the simuland and omit others. The requirements are driven by the intended application. Conceptual models document those aspects of the simuland including the structural and behavioral aspects such as objects, entities, events, functions, environmental phenomena etc. The executable model is the computer program that can be executed and is intended to simulate the simuland as detailed in the conceptual model. Therefore the conceptual model can be viewed as a design specification for the executable model. The results are the output produced by a model during a simulation.

**Figure 22** presents Verification and Validation as activities that compare one thing to another. Verification compares the requirements with the conceptual model. In this comparison, verification seeks to determine if the conceptual model satisfies the given requirements. The second comparison is between the conceptual model and the executable model, where the goal is to determine if the implemented executable model is consistent with respect to the conceptual model. Validation compares the simuland with the conceptual model to determine if the simuland has been accurately described in the conceptual model. The second comparison is between the simuland and the results which determine if the output of the simulation is sufficiently accurate with respect to the actual behavior of the simuland [**29**].

Another comprehensive VV&T model is presented by Balci [**28**] in the form of a simulation study life-cycle as shown in **Figure 23**. The phases are shown by oval symbols. The dashed arrows describe the processes which relate the phases to each other. The solid arrows refer to the credibility assessment stage. Every phase of the life-cycle has an associated VV&T activity. Problem Formulation (or problem definition) is the process of formulating a problem which is sufficiently well-defined to enable specific research action and the investigation of suitable solution techniques. The output of system investigation results in the System and objective definition which further aids in model formulation. Model formulation is the process of defining a conceptual model which abstracts or envisions the real system under study. The conceptual model is further represented inform of a Communicative Model which is a model representation and can be communicated to other designers and can be compared against the system and the study objectives. It is further transformed into an executable model through the process of programming. An Experimental Model is the programmed model incorporating an executable description of operations along with the design of experiments, for experimenting with the simulation model with a specific purpose. The process of experimentation produces the Simulation Results, which are presented for decision makers for their acceptance and implementation or undergo refinements if required.

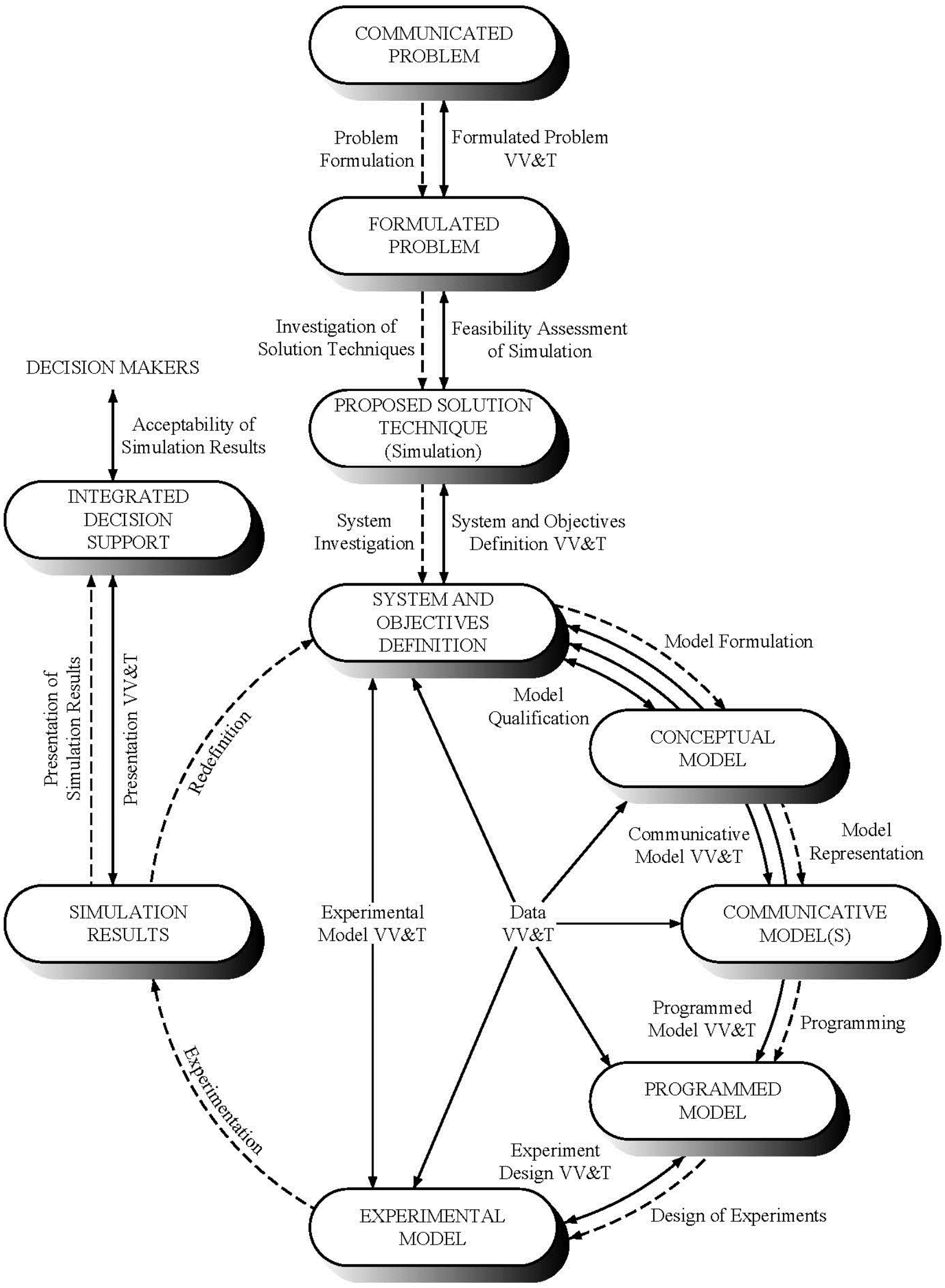

**Figure 23: Simulation study life-cycle** (acquired from [28])

The model-evaluation life-cycles shown in **Figure 21**, **Figure 22** and **Figure 23** have been considered as guidelines and they are used as inspiration for the verification life-cycle proposed and presented later in this thesis.

## 4.2 The Principles of Top-Down Refinement

The principle of top-down refinement has been appreciated in the area of model verification. Constructing a highly detailed model that satisfies all levels of correctness in one attempt is very difficult. Instead it is easy to construct a less detailed abstract model at first. Let $S_1$ be an initial model. To get from $S_1$ to the final shape of the model, the Top-Down Refinement paradigm advocates the derivation of an (ordered) sequence $S_1$, $S_2 \ldots S_f$ of models of S. For i = 1...f, model $S_{i+1}$ is a refinement of its immediate predecessor model $S_i$ if the following conditions are met:

(i) $S_{i+1}$ is more expressive than $S_i$

(ii) $S_{i+1}$ is less abstract than $S_i$

(iii) It is relatively easy to evaluate $S_{i+1}$ on the basis of verified $S_i$

Consequently, the last model in the refinement sequence should be correct by construction. The following are some consequences of the top-down refinement paradigm. First, $S_{i+1}$ is harder to understand than $S_i$ and therefore harder to prove on its own; it is precisely the refinement step that allows the verification of $S_{i+1}$ under the assumption that $S_i$ has already been proved correct [**99**].

In this thesis the proposed verification process is based on this fundamental principle where the verification is performed iteratively and on a relatively refined shape of the model.

## 4.3 Verification techniques

There exist a large variety of verification methods. The diversity is due to the range of different simulation project types, different subjects (simuland), and different types of data. Most of the verification methods are inspired from software engineering domain, because the executable models in simulation projects are almost always realized as software [**29**].

In literature, Verification techniques are generally classified into four main categories as show in **Figure 24**.

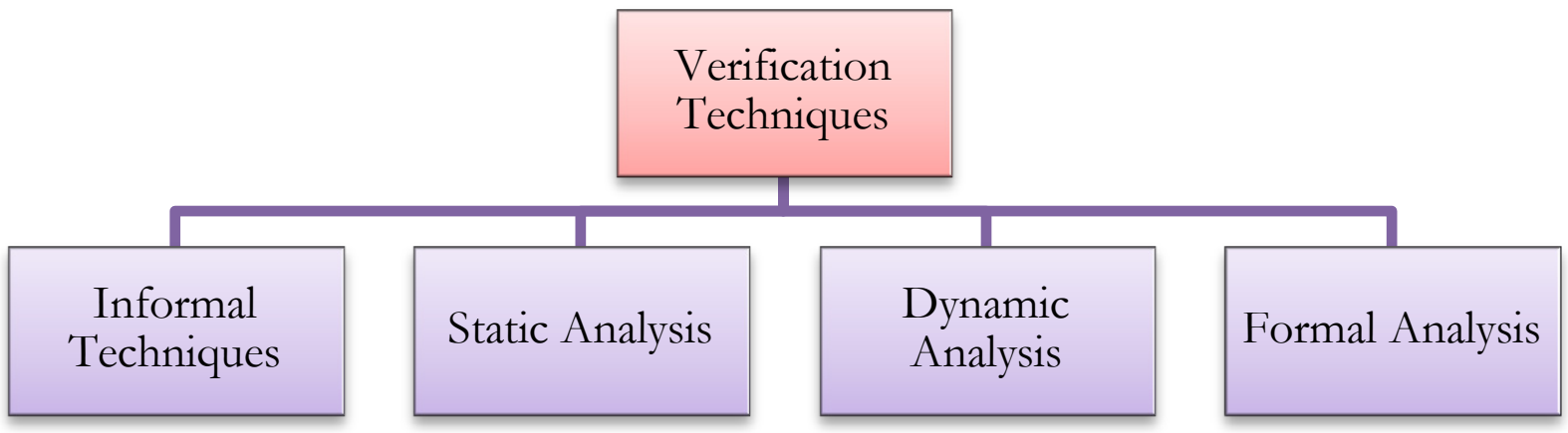

**Figure 24: Verification Techniques**

### 4.3.1 Informal Techniques

These techniques are most commonly used. They are called informal because the tools and methods used rely heavily on human reasoning and inspection without any underlying mathematical formalism [**28**]. These techniques are well structured and are conducted with proper guidelines by following standard policies and procedures, however these techniques are tedious and not very much effective [**109**].

Some of the commonly used informal methods are shown in Table 6.

| | |
|---|---|
| **Audit** | An audit is undertaken to assess how adequately the system study is conducted with respect to established plans, policies, procedures, standards and guidelines [28]. |
| **Desk checking** | Desk checking or self-inspection is a thorough examination performed by an individual as a first step. In this method syntax checking, specification comparison, code, control flow graph analysis are performed [28]. |
| **Inspections** | Inspections are conducted by a team and performed at different phases of developments such as problem definition, conceptual modeling, executions etc. Inspections are conducted to find and document faults [28]. |
| **Turing Tests** | Turing test is performed by domain experts (of the system under study). They are presented with two sets of output data obtained one from the model and one from the specification (without identifying which one is which) and are asked to differentiate both and based on their feedback model corrections are made [28]. |

**Table 6: Informal Verification Techniques**

### 4.3.2 Static Analysis:

These techniques are applied to assess the static model design and the implementation (source code), without executing the model. They aim at checking the structure of the model, the dataflow and control flow, the syntactical accuracy, and the consistency. Some of the commonly used static analysis methods are shown in Table 7.

| | |
|---|---|
| **Structure Analysis** | Structure Analysis is used to examine the model structure. It is conducted by constructing a control flow graph of the model structure [28]. |
| **Data Analysis** | It involves data dependency tests and data flow analysis to ensure that data used by the model is properly defined and proper operations are applied to data objects [28]. |
| **Cause-Effect Graphing** | Cause-Effect graphing assists model correctness evaluation by answering "what causes what" questions in the model representation. It is performed by identifying causes and effects in the model and checking if they are reflected accurately in the specification [28]. |
| **Syntactic Analysis** | Syntactic analysis is usually performed by the compiler of the simulation language being used. Syntactic analysis can also be performed using a set of rules applied on the model representation to verify if it satisfies given specification. |
| **Semantic Analysis** | This technique is used to determine the modeler's intent and verify that the true intent is accurately reflected in the model representation [28]. |

**Table 7: Static Analysis Techniques**

### 4.3.3 Dynamic Analysis:

Dynamic analysis techniques are based on the execution of the model in order to evaluate its behavior. They do not simply examine the output of an execution but also observe the model as it is being executed. The insertion of additional code into the model called *instrumentation* is needed to collect or monitor the behavior during its execution [**109**]. Table 8 presents some of the important dynamic analysis verification techniques.

| | |
|---|---|
| **Assertion Checking** | An assertion is a statement that should be true during the execution of a model. Assertions are placed in various parts of the model and monitored during execution [**28**]. |
| **Bottom up Checking** | This technique is used in conjunction with the bottom up model development strategy. The sub models are checked individually. Then the parents at the higher level are checked [**28**]. |
| **Fault/Failure insertion** | This approach is used to insert a fault or a failure in the model and observe whether the expected incorrect behavior is produced. This approach is effective to detect unexplained behavior and hence uncover errors [**28**]. |
| **Functional Testing** | This technique is used to assess the accuracy of model input-output transformation, to evaluate how accurately a model transforms a given input into a set of output data [**28**]. |
| **Sensitivity Analysis** | Sensitivity analysis is performed by changing the values of model input variables and parameters over some range of interest and observing the effect on model behavior. Unexpected effects may reveal errors [**28**]. |

**Table 8: Dynamic Analysis Techniques**

### 4.3.4 Formal Analysis

Formal analysis refers to mathematical analysis of proving or disproving the correctness of a system with respect to a certain unambiguous specification or property. The methods for analysis are known as formal verification methods, and unambiguous specifications are referred as formal specifications. Formal verification can provide complete coverage on an abstract model of the system, modeled using finite state machines, PN or any other specification formalism. However it should be noted that formal verification can ensure the correctness of a design only with respect to certain properties that it is able to prove [**88**]. There are many formal analysis techniques, which we classify in four main groups:

| | |
|---|---|
| **Equivalence Checking** | It is also called Reference Model Checking, which is widely used verification technique that allows two behavioral models to be compared with each other. In general, one of the two is taken as the reference model and represents the so-called golden model (or perfect model). It verifies that the behavior of two models is the same for the exercised scenarios. This technique has limitation that it does not actually verify that the design is bug free, and provides proof of relative correctness [**109**]. |
| **Theorem Proving** | This method involves verifying the truth of mathematical theorems that are postulated or inferred throughout the design using a formal specification language. The procedure involves two main components: (i) proof checker (which can be completely automated in most cases) and (ii) an inference engine (which may require occasional human guidance) [**109**]. |
| **Property Verification** | Formal properties specify the requirements of the correct system design. The objective of this method is to check whether an implementation satisfies these requirements. Static Assertion-based Verification (ABV) and dynamic [**110**]. |
| **Model Checking** | Model checking establishes a solid confidence in a reliable V&V process. Model checking is an automated and comprehensive verification technique that can be used to verify whether the properties specified (usually using Temporal Logic) for a given design or its components are satisfied for all legal design inputs. Model checking also faces a limitation, since it suffers from the well-known state explosion problem. In a worst-case scenario, the state space of the design may grow exponentially large with the number of state variables. Model checking can be fully automated for design verification and can yields results much more quickly than theorem proving [**109**]. |

**Table 9: Formal Analysis Techniques**

Some of these techniques have been adopted in our proposed verification framework.

## 4.4 Summary

In this chapter, different concepts of verification, validation and testing are discussed as they collectively contribute to proving the correctness and accuracy of a model. Some existing model development processes (devised mainly by M&S community) are also discussed, since they are the bases of the proposed verification life-cycle presented later in this thesis. The proposed framework essentially focuses on Verification (however its design is also open to adopt validation techniques). Different verification techniques are classified into four main groups and some of the selected techniques are briefly explained, as they will be used later in this thesis.

# Part II

# Techne

***Technê*** *in Greek is translated as craftsmanship or craft or art. In science it is the practice of knowledge; Techne resembles Epistēmē in the implication of knowledge of principles, although techne differs in that its intent is making or doing, as opposed to "in-depth understanding"; Applied-Science; It deals with "How" of the subject.*

Part-II covers the technology of the research under discussion, where the theoretical concepts provided in Part I are applied, and technically discussed under an integrated framework of methods, techniques, algorithms and processes and their practical implications are provided in the form of a proposed solution.

*"Without knowledge the practice is useless, and without practice the knowledge is useless"*

*– Ali bin Usman Hajvery*

*(Kashaf-Almahjoob)*

# Chapter 5
# Proposed Methodology and the Verification Framework

*This chapter renders the core of the solution framework proposed in this thesis. In this chapter, a collection of methods, techniques, algorithms, sub-processes, activities and approaches are presented, as proposed solution to various issues in the composability verification of BOM based model components. All these contributions are integrated into a unified framework which we refer to as: Composability Verification Framework.*

The proposed verification Framework consists of different methods, techniques, algorithms, sub-processes, activities and approaches which all together encompass the component based modeling & simulation (CBM&S) life-cycle.

## 5.1 Component-based Modeling & Simulation life-cycle

CBM&S life-cycle is inspired by different modeling architectures proposed by Sargent, Petty and Balci and discussed in section 4.1.1. It is extended with our proposed contributions at its different stages. The proposed CBM&S life-cycle is mainly divided into four main quadrants: (i) Inception (ii) Modeling (iii) Execution and (iv) Analysis. Each quadrant has different phases and in each phase there are multiple activities (or cycle of activities). Each activity consists of methods and techniques pertinent to its respective phase. These phases are revisited iteratively during the life-cycle; where each iteration represents a tier; hence the entire CBM&S life-cycle is a multi-tier process; whilst each tier results into a refinement of the solution of the problem under investigation; as it follows the principle of top down refinement, discussed in section 4.2. All the above mentioned features of the CBM&S life-cycle are shown in **Figure 25** divided into four quadrants:

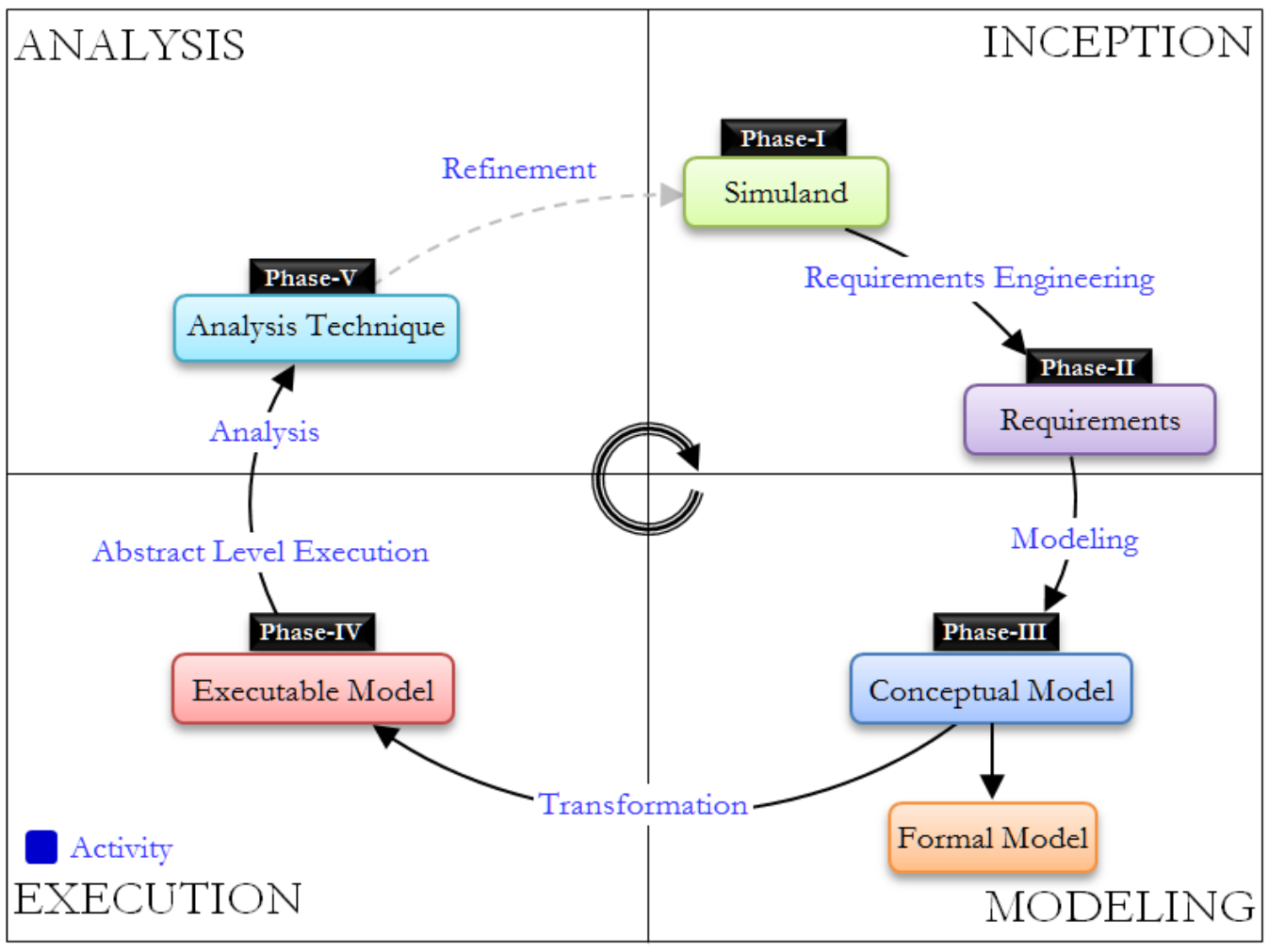

**Figure 25: CBM&S life-cycle**

The following sub-sections provide microscopic details of each quadrant along with their associated inside activities, methods and techniques.

## 5.2 Inception

The first quadrant of the CBM&S life-cycle called "Inception" initiates the process. At first the abstraction of a real-system is accumulated as simuland. A simuland can be ingested in the form of UML diagrams (**Figure 26**) or using any other formal or informal representation.

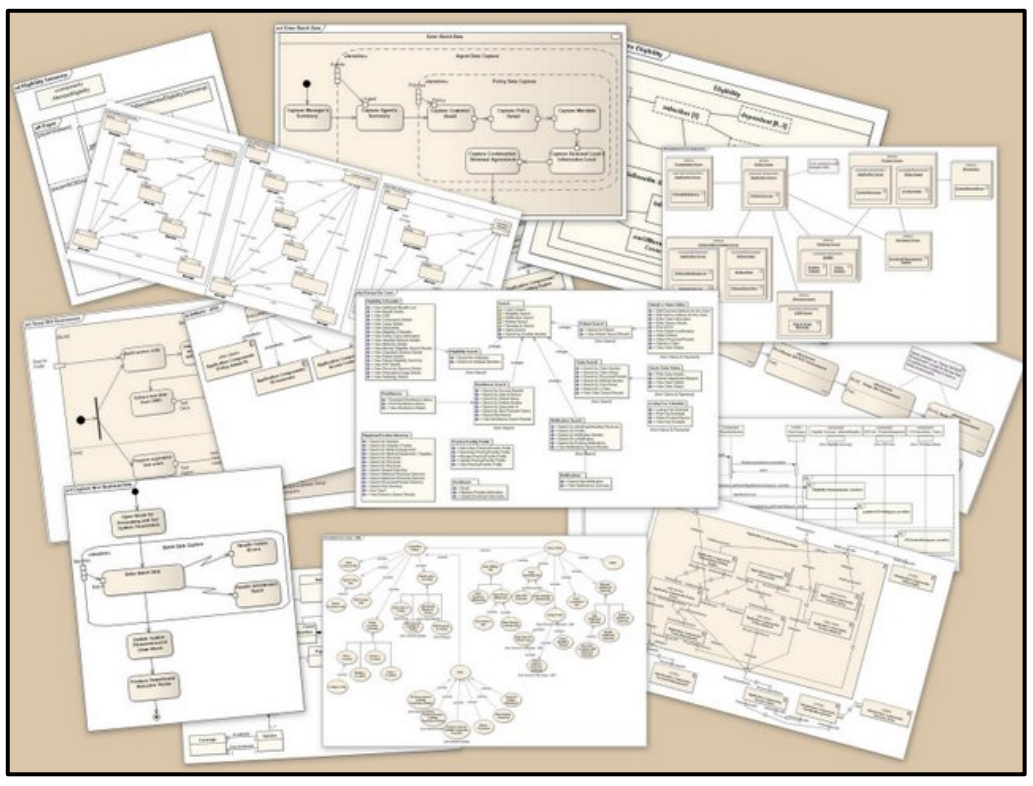

**Figure 26: Simuland using UML Diagrams**

The basic idea is to gather the body of knowledge so that the modelers can envision the real system under a certain frame of reference i.e., the context under which the system is being studied. When the simuland is ingested into the framework, it is used (i) to gather requirements, through the process of requirement engineering and (ii) to search and discover suitable components from a BOM repository for the construction of a composed model. If a required component does not exist in the repository then it is built from scratch and added in the repository. The outcome of the requirement engineering activity results in formulation of requirements specifications. The requirement specification formalism (as defined in section 1.3.2) is used to express formal requirements for this framework:

$$\mathbf{RS = \langle O, S \rangle}$$

Where

$\mathbf{O = \{o_1, o_2, o_3 \ldots, o_n\}}$ is a set of objectives or goals that must ultimately be fulfilled. These goals are usually defined in the context of the scenario of the modeling domain. Therefore the properties expressed as goals or objectives may be scenario-specific and not the standard system properties e.g. in a restaurant model the objectives could be that the customers are served food and payments are collected, and not that the model should be deadlock free (which however might be a necessary condition).

$\mathbf{S = \{s_1, s_2, s_3 \ldots, s_n\}}$ is a set of system constraints (system properties or scenario-specific safety/liveness properties). Deadlock freedom (or other similar system properties) could be the required constraints necessary to fulfill the above objectives and therefore must be satisfied. We propose to define the following mandatory (or default) constraints in the requirements specification of the composability verification framework:

| |
|---|
| ***S1*** = *All the interacting components*[25] *should be composable at Syntactic level* |
| ***S2*** = *All the interacting components should be composable at static-semantic level* |
| ***S3a*** = *State-machines of the interacting components should match each other such that they can continue to progress until they reach the final or goal states*[26]. |
| ***S3b*** = *If the conceptual model is transformed into an executable model, the latter should correctly represent the structure and behavior of the former.* |

**Table 10: Mandatory constraints in composability verification**

We assert that **[S1 ∧ S2 ∧ (S3a ∧ S3b)]** is a necessary condition for the overall composability verification. S1 and S2 ensure that the composed model is structurally consistent. Whereas S3a confirms that the behavior of the composed model is coherent for reaching given objectives. The satisfaction of S3b obeys the definition of Model verification (see section 4.1) in the sense that it confirms the second part of the definition that is: "*the accuracy of transforming the conceptual model into an executable model*" and therefore the overall success of the verification process depends on the satisfaction of S3b constraint. The conjunction of these default constraints impose the three C's of requirements namely (i) Consistency, (ii) Completeness, and (iii) Correctness [**111**]. Consistency is required for the evenness in the input and output connections of the composed components. Completeness is required for the totality of the information of the components being composed to check that the composition does not lack required inputs for making progress. Correctness is needed to confirm that the composed components interact in a correct way as they are supposed to.

If all the objectives are fulfilled and all the constraints are satisfied and then we say that the model is composable at all levels and is verified with respect to its specifications. The overall objective of our proposed framework is to provide environment and tool support to assess this postulation.

The outcome of discovery results in a set of candidate BOMs and their matching with the simuland and the requirements results in a selection of BOMs suitable for the composition. This selection is composed to form a conceptual model.

## 5.3 Modeling

In the Modeling quadrant, a BOM based composed model is taken as an input and the conceptual model is formed. Also a formal model and its graphical notation (as proposed in section 2.7.4) are produced for the purpose of documentation of the conceptual model[27]. Considering that BOM itself is a conceptual framework and is used to model passive components which cannot undergo any form of execution therefore the conceptual model is subjected to a series of extensions and refinements using external input and our proposed model transformation algorithms so that it can be implemented into executable forms and sent to the "Execution" quadrant (**Figure 25**) for abstract level execution. Our proposed extensions and refinements are listed as follows:

- BOM State-machines to State Chart XML (Transformation)
- Composed-BOM to Petri Net –PNML (Transformation)
- Basic-BOM to Extended-BOM (Extension)
- Extended-BOM (E-BOM) component to Colored Petri Net (CPN) Component Model (Transformation)
- Basic-BOM to Extended-BOM with Time (Extension)
- Basic-BOM to Extended-BOM with probabilistic factors (Extension)
- BOM to CSP based Process Model (Extension & Transformation)

[25] In a composed model it is not necessary that every component interacts with every other component for instance A, B and C are composed such that A interacts with B and B interacts with C but A does not interact with C.

[26] If there are no final-states defined in a model and the model is non-terminating then we assume that certain important states called goal-states are present in the model, reachability of which confirms that the goals are fulfilled.

[27] This step is optional but beneficial if different teams are working on different phases of the development life-cycle. This documentation makes it easy to understand the structure and behavior of basic components and their composition.

In the later section these extensions, refinements and transformations will be explained in detail. It is important to note that each time the conceptual model is extended or refined the Modeling quadrant is revisited in iteration.

## 5.4 Execution

As previously discussed this quadrant is mainly for the abstract-level execution activities. It takes following implemented and executable forms of the conceptual model from the Modeling quadrant as input:

- State Chart XML (SCXML)
- Petri Net –PNML
- Colored Petri Net (CPN) Composed Component Model
- Communicating Sequential Process (CSP) based Component Processes

In the later section these executable forms and their abstract level execution processes will be discussed in detail.

## 5.5 Analysis

The outcome of an execution process yields some results. These results are analyzed in the Analysis quadrant. Our verification framework supports different analysis techniques listed as follows:

- State-machine matching Analysis
- Petri Nets based Algebraic Analysis
- Colored Petri Net based State-Space Analysis
- Model Checking Analysis

These analysis techniques will be discussed in later section. When all the necessary steps in the composability verification are complete and the composed model under investigation is said to be verified with respect to the given requirement specification then the CBM&S life-cycle proceeds to the further steps for implementation and simulation as shown in **Figure 27**. The details of these steps are out of the scope of this thesis.

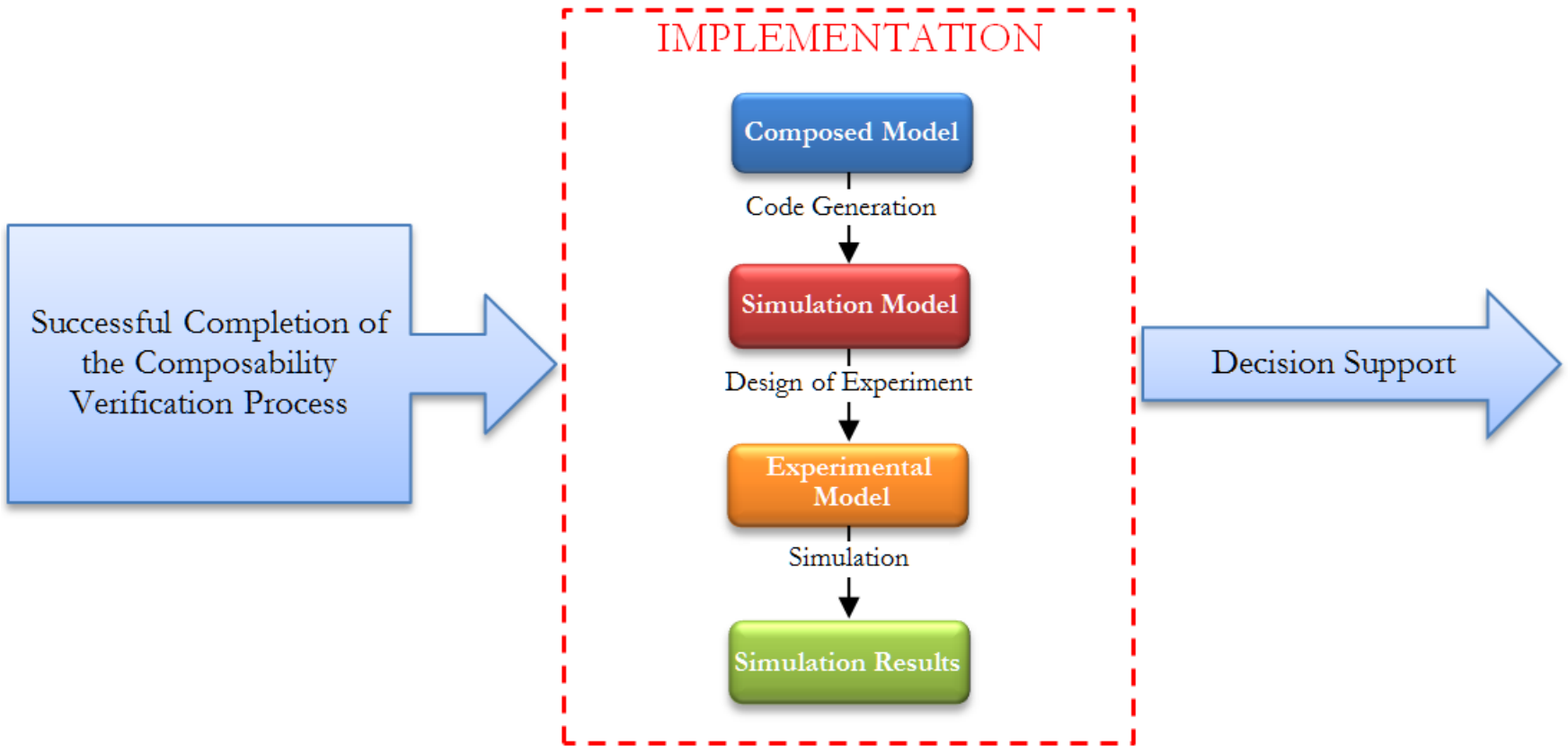

**Figure 27:** Implemenation and Simulation

## 5.6 Composability Verification Framework

In this section different method, techniques, procedures, algorithms and modules of our proposed composability verification framework are discussed in detail and considered as building blocks in the CBM&S life-cycle and will be connected to its different phases. These details are necessary to understand the composability verification process being presented in chapter 6.

### 5.6.1 Discovery Matching and Composition (DMC)

In component based development, it is a normal practice to construct reusable components and store them in a library or repository so that they can be reused later as required. To reuse an existing component, a Discovery, Matching, Composition (DMC) paradigm [**19**] is used. We assume that a library of BOM components is maintained in a repository. Using the information given in the simuland a modeler attempts to search and discover BOM components from the repository. If a collection of candidate components is retrieved, they are filtered through matching process. A matching process matches the candidate components from the simuland and requirement specifications and results in a selection of components suitable for the composition. The aspects of syntactic and semantic matching during the discovery and selection of BOM components are proposed and discussed in detail in [**54**]. In this article a set of discovery rules are presented which must be fulfilled while matching a candidate selection from the simuland. We apply these rules for the syntactic and semantic matching of the candidate selection with the simuland. We further suggest matching the candidate selection with given requirements, because a selection may match with its respective simuland but if it does not match with its requirements then the composability verification will fail. We implement the concept of DMC process in our framework as shown in **Figure 28**. It is also assumed that if a required component does not exist in the repository, then it is constructed from scratch and is added in the repository for reuse. The result of DMC process is a BOM-based composed model. This composed model is taken as input in the Modeling quadrant and considered as a conceptual model of the system. It is recommended that the modelers also use our proposed formal specification and graphical notation presented in section 2.7.4 to construct a formal model. This formal model can be used for documentation and shows how the components are composed. It is however an optional step and is not considered as a phase in our CBM&S life-cycle. In chapter 7 & 8 the formal models of the examples are also described for reader's understanding.

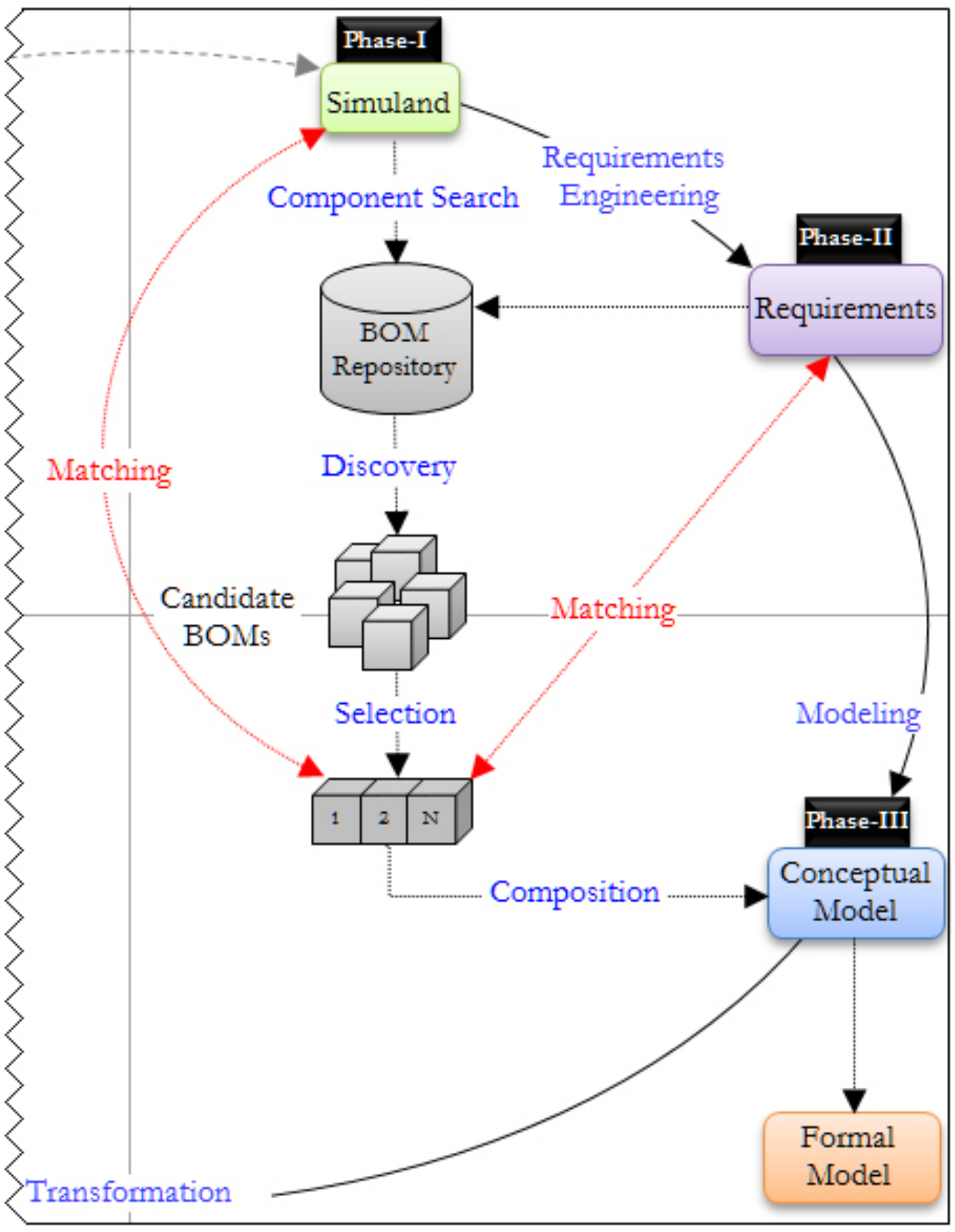

**Figure 28: Discovery, Matching, Composition (DMC)**

## 5.6.2 Structural and Behavioral Evaluation

The conceptual model ingested in the Modeling quadrant requires structural and behavioral evaluation so that we can confirm that the model is consistent, complete and correct. And it is suitable for thorough verification at different levels of composability. Checking the structure and behavior of the conceptual model before subjecting it to the deeper levels of composability verifications is useful. If the model is structurally and behaviorally consistent then the confidence level is increased based on which different useful assumptions can be made later during the in-depth verification.

If there are discrepancies in the structure or behavior of the model then we can skip further steps, save time and computational resources and perform necessary design refinements before the entire process is repeated. This setup obeys the principle of top-down refinement as discussed in section 4.2. The structure of the model is analyzed using static analysis techniques (see section 4.3.2), whereas the behavior of the model is evaluated using dynamic analysis techniques.

## 5.6.3 Static Analysis

We propose two types of Static analysis procedures (i) Syntactic Matching and (ii) Static-Semantic Matching. These procedures are used to evaluate the structure and verify composability at syntactic and static-semantic levels. They are called static analysis because they are evaluated based on pre-defined rules and do not require any form of execution and the information on which these rules are applied is static.

### Syntactic Matching (SM)

This module is responsible for evaluating BOM composability at syntactic level based on the following rules. The outcome of this module verifies that the components can be correctly connected to each other syntactically. These rules were introduced in a BOM matching technique presented in [**54**].

> **SM-Rule 1:**
> *The name of each event[28] exchanged between the two components should be same i.e., the send-event should have the same name as the receive-event.*

A send-event is defined in the BOM's event types where the sender is the BOM itself and the receiver is some other BOM (in the composition) whereas a receive-event is the definition of an event in the BOM event types, where the sender is some other BOM (in the composition) and the receiver is the BOM itself.

> **SM-Rule 2:**
> *Each send-event should have at least one corresponding receive-event and vice-versa i.e., the send/receive pair should be complete.*

> **SM-Rule 3:**
> *The number of parameters (content characteristics of event types) of the send-events should be the same as the number of parameters of the receive-events.*

The satisfaction of Syntactic Matching rule1, rule2 and rule3 fulfills the default constraint S1 (see Table 10) which is a necessary condition for the overall composability verification. **Figure 29** shows different steps in the syntactic matching activity.

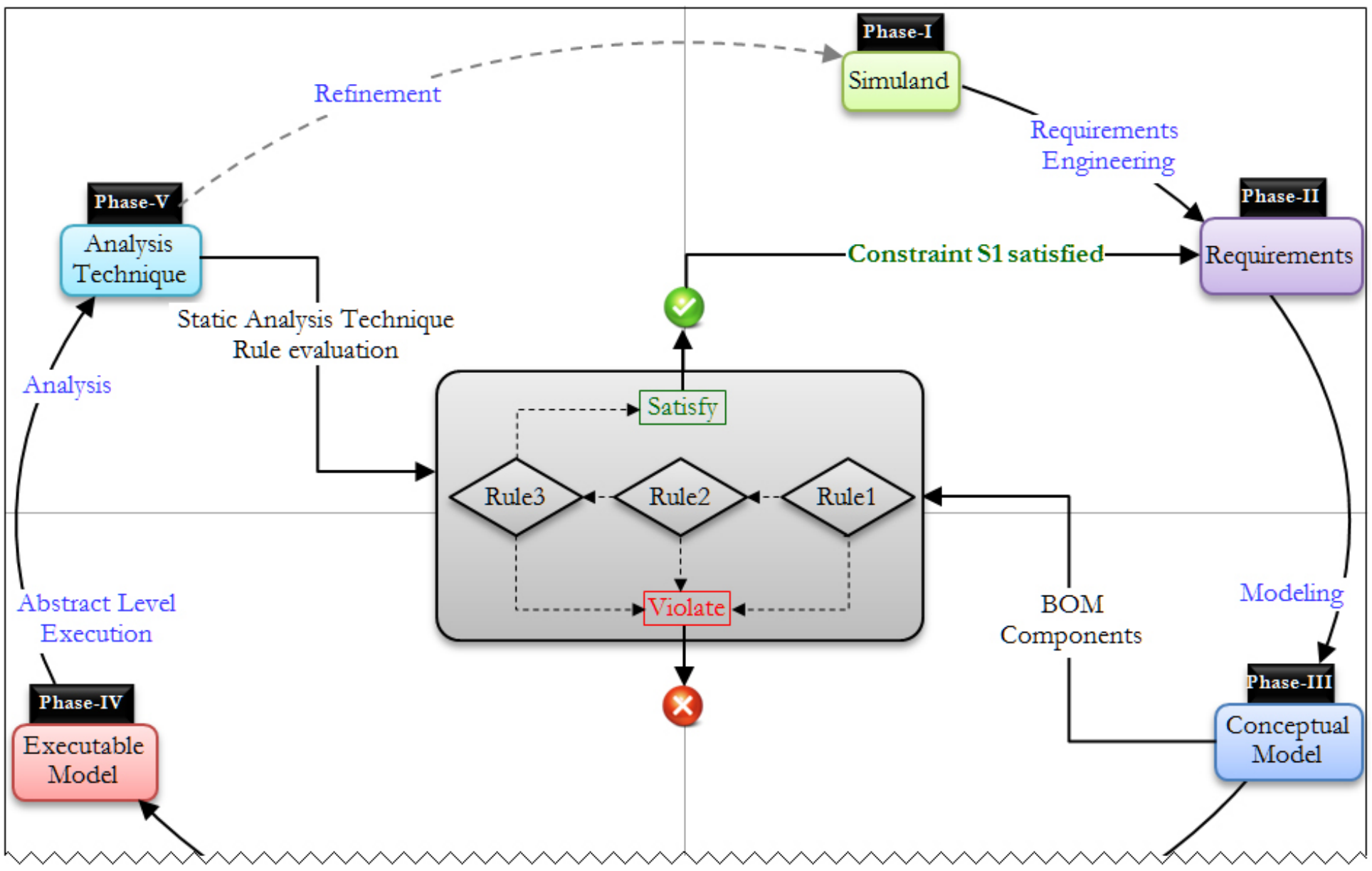

**Figure 29: Syntactic Matching**

---

[28] It is assumed that in the BOM construction the events and their corresponding actions are given the same name

### Static-Semantic Matching (SSM)

This module is responsible for evaluating BOM composability at static-semantic level based on certain rules. The outcome of this module verifies that the composition of the components is meaningful and the communication between the components is understood as intended. In order to certify these facts we propose static-semantic matching at two levels: (i) Operational Level matching and (ii) Message level [**53**]:

#### (i) Operational Level matching

In BOM-based composed models Operations are described by Pattern-of-Interplay (POI). POI is formed by a collection of actions from the basic BOMs being composed. In operational-level semantic matching, it is ensured that the composed components share the same "domain of interest" and they are composed for the same purpose (or aim) so that we can guarantee that the composition is (static) semantically meaningful and without any pragmatic ambiguity. Even with the same domain of interest, the component may serve for varied purposes e.g., in Military domain a Battalion Head Quarter (BHQ) component may have many purposes and can take part in many different operations. Therefore it is also important that the purpose of the selected components should be clear for a meaningful outcome.

In order to ensure semantic consistency at operational-level we propose to specify following semantic-attributes [29] in the definition of actions at the time of the construction of Basic BOMs and in the POI when the basic BOMs are being composed. In the static-semantic matching these attributes are used to compare that the correct actions are involved in the BOM composition.

- **Area-of-Interest:** It describes the area or the domain of interest of the system that is being modeled using the components and the operation. We propose to define "Area-of-Interest" as a semantic-attribute in each action of Basic BOM and also in the POI. This attribute will confirm that all the components share the same domain knowledge. If of some general purpose components that may belong to multiple-domains (e.g., Queues etc.) we propose to construct a specialization of the component and make it a member of the selected area-of-interest. E.g., In a restaurant composed model a generic queue component can be specialized into a restaurant-queue with actions *JoinRestaurantQueue()* and *ServeCustomer()* instead of Put() and Get() actions.

- **Purpose:** Purpose describes the aim or goal of the entire operation. In BOM composition, POI represents a single operation being performed by the composed components. However it is also possible that one or more composed components may be designed to serve multiple purposes; and in a given scenario only some part of the multi-purpose components is involved in the composition. e.g., a Customer component could be generic and can have multiple purposes whereas a Restaurant waiter component is specific to a restaurant scenario, so it is important that if a Customer component is selected in a Restaurant scenario then its purpose should be aligned with the other components in this scenario. Hence we define "purpose" as a semantic-attribute of actions in the basic BOM (with multiplicity $\geq 1$).

[29] In BOM the conceptual modeling elements (Entities, Events States and Actions) support semantic fields [**65**]

**(ii) Message Level matching**

BOM represent event driven components and function by sending or receiving events (messages). At the message level it is required that the communication between composed components is meaningful and semantically understood by the receivers as intended by the senders. At this level we propose to match Data-Types and Units of measurements of the parameters of send-events and receive-events [**53**] [**54**].

It is assumed that the BOM components have corresponding OWL attachments as proposed in [**54**]. The BOM-OWL attachments are used to define semantic classes of the domain ontology, their properties, data-types and the individuals and stored in the BOM repository. In order to evaluate static-semantic matching at both Operational and Message levels, we apply following rules:

**SSM-Rule 1**

*The intersection of the "Area-of-Interest" attribute of all the actions (involved in an operation) should be exactly the same as that of POI or should belong to an equivalent class[30] in the respective ontology:*

$$\bigcap_{i=1}^{n} \mathrm{Act_i.AOI} \cong \mathrm{POI.AOI}$$

**SSM-Rule 2**

*The intersection of the "Purpose" attribute of all the actions should be exactly the same as that of POI or should belong to an equivalent class in the respective ontology:*

$$\bigcap_{i=1}^{n} \mathrm{Act_i.purpose} \cong \mathrm{POI.purpose}$$

**SSM-Rule 3**

*Data types of each element in the event parameters of the send-event and receive-events should be of the same class, equivalent class or should be in direct hierarchical relationship i.e., the sender's parameter data-type should belong to the direct child class of the receiver's parameter data-type (but not the inverse).*

e.g., a send-event contains a parameter of type 'second', whereas the receive-event expects a parameter of type 'time' which according to the rule it is a semantic match. **Figure 30** presents primitive data-types as an example. In real situations BOM components will have more domain specific complex data-types.

[30]In OWL two classes can be marked equivalent if they have same semantic meanings and both classes have the same individuals (instances) e.g., Healthcare and Medical are synonyms. We denote it as $\cong$

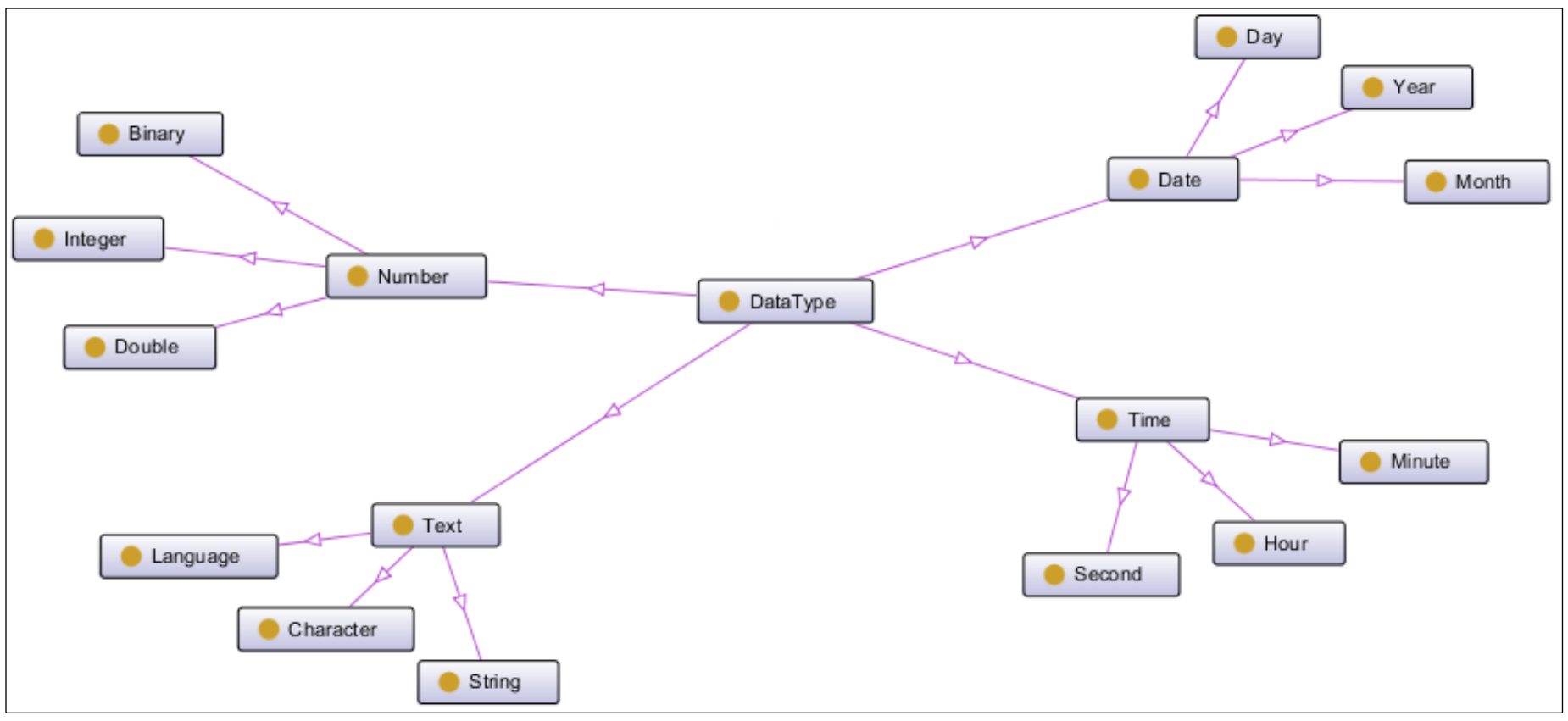

**Figure 30: Some of the sub-classes of Data Type ontololgy**

**SSM-Rule 4**

*The units of the measurements expressed in the event parameters should be same or equivalent or should belong to a direct class hierarchy such that they are convertible without (or with acceptable) loss of information.*

We assume that if two measurement units are in either of the direct relationship i.e., parent or child then their conversion loss will be acceptable e.g., a send-event has a parameter with unit m/s (meter per second) to express speed whereas the receive-event expects Km/hr (Kilometer per hour). This is a valid semantic match because the quantities are convertible without loss.

### Semantic Matching Technique

In order to match two elements we propose a semantic matching technique as shown in **Figure 31**. This technique uses OWL-API [**112**], a semantic reasoning engine (FaCT++, Pellet, or HermiT) and an OWL ontology document to process a query of any two elements A & B and outputs their semantic relationship as one of the following:

1. Exact (A = B)
2. Equivalent (A ≅ B i.e., A and B belong to equivalent classes)
3. Direct-Parent (A is a direct parent of B)
4. Direct-Child (A is a direct child of B)
5. Indirect (A and B are not in direct contact but belong to same hierarchy)
6. No relationship (A and B are not related)

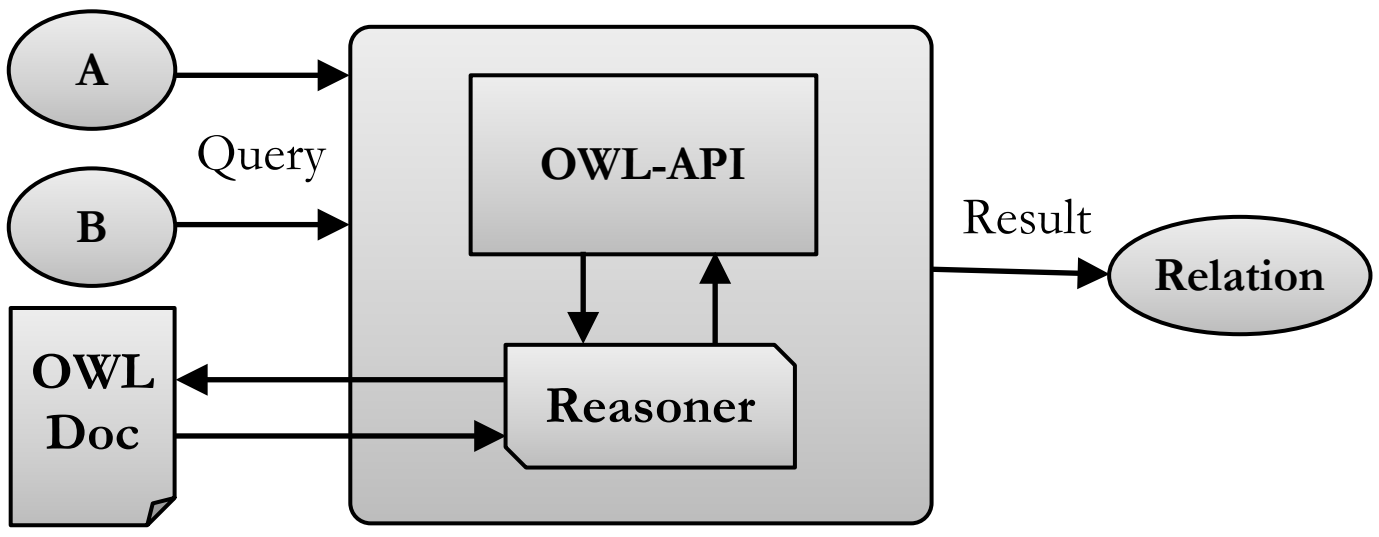

**Figure 31: Semantic Matching Technique**

This technique is used to evaluate Static-Semantic Matching Rules 1, 2, 3 & 4 using the algorithm[31] given in **Table 11**.

[31] The Pseudo-code conventions and format of the algorithms provided in this thesis, for most parts, follows the guidelines set by [**132**].

## Algorithm: Semantic Matching

**Input: {Actions}, POI, BOM-OWL**

**Output: TRUE, FALSE**

```
Owl ← Load Ontology(BOM-OWL)
{CommonAOI} ← ∩_{i=0}^{n} a_i ∈ Actions.AOI ▹ Gives a set of common area of interest of all actions
for caoi ∈ {CommonAOI} do
    SR1 ← Get-Semantic-Relation(caoi, POI.AOI, Owl) ▹ It is assumed that Get-Semantic-Relation()
    function is implemented using semantic matching technique shown in Figure 31 ◃
    if SR1 = "Exact" or "Equivalent" then ▹ Rule1 satisfy...continue
        next
    else
        Return FALSE
    end if
end for

{CommonP} ← ∩_{i=0}^{n} a_i ∈ Actions.purpose ▹ Gives a set of common purpose of all actions
for cp ∈ { CommonP } do
    SR2 ← Get-Semantic-Relation(cp, POI.purpose, Owl)
    if SR2 = "Exact" or "Equivalent" then ▹ Rule2 satisfy...continue
        next
    else
        Return FALSE
    end if
end for

{Events} ← Get-Events(Actions) ▹ gets corresponding Events of Actions
    for e ∈ Events do
        if e=Send-Event then
            f ← Get-Receive-Event(e, Events) ▹ gets corresponding Receive Event of e
            {PE} ← e.Parameters ▹ Set of parameters of send-event e
            {PF} ← f.Parameters ▹ Set of parameters of receive-event f
            ▹ No. of parameters of e and f must be same because of SM-Rule3
            for pe∈PE & pf ∈PF do
                SR3 ← Get-Semantic-Relation(pe.Type, pf.Type, Owl) ▹ Compare Parameter types
                if SR3 = "Exact" or "Equivalent" or "Direct-Child" then
                    ▹ Rule3 satisfy...continue to rule4
                    SR4 ← Get-Semantic-Relation(pe.Unit, pf.Unit, Owl) ▹ Compare Units
                    if SR3 = "Exact" or "Equivalent" or "Direct-Parent" or "Direct-Child" then
                        Return TRUE ▹ Static-Semantic Matching Successful
                    else
                        Return FALSE
                    end if
                else
                    Return FALSE
                end if
            end for
        else
            next
            ▹ Goes to the next send-event and need not to check receive-events (because SM-Rule2)
        end if
    end for
```

**Table 11: Semantic Matching Algorithm**

The semantic matching algorithm takes a set of actions (parsed from Basic BOMs which are being composed); the pattern of interplay (POI) which specifies how the actions are connected to each other and the corresponding OWL ontology document as input. The output of this algorithm is TRUE if the static-semantic matching is successful otherwise FALSE if any of the rule is violated. **Figure 32** shows steps in the verification of BOM composability at Static-Semantic level.

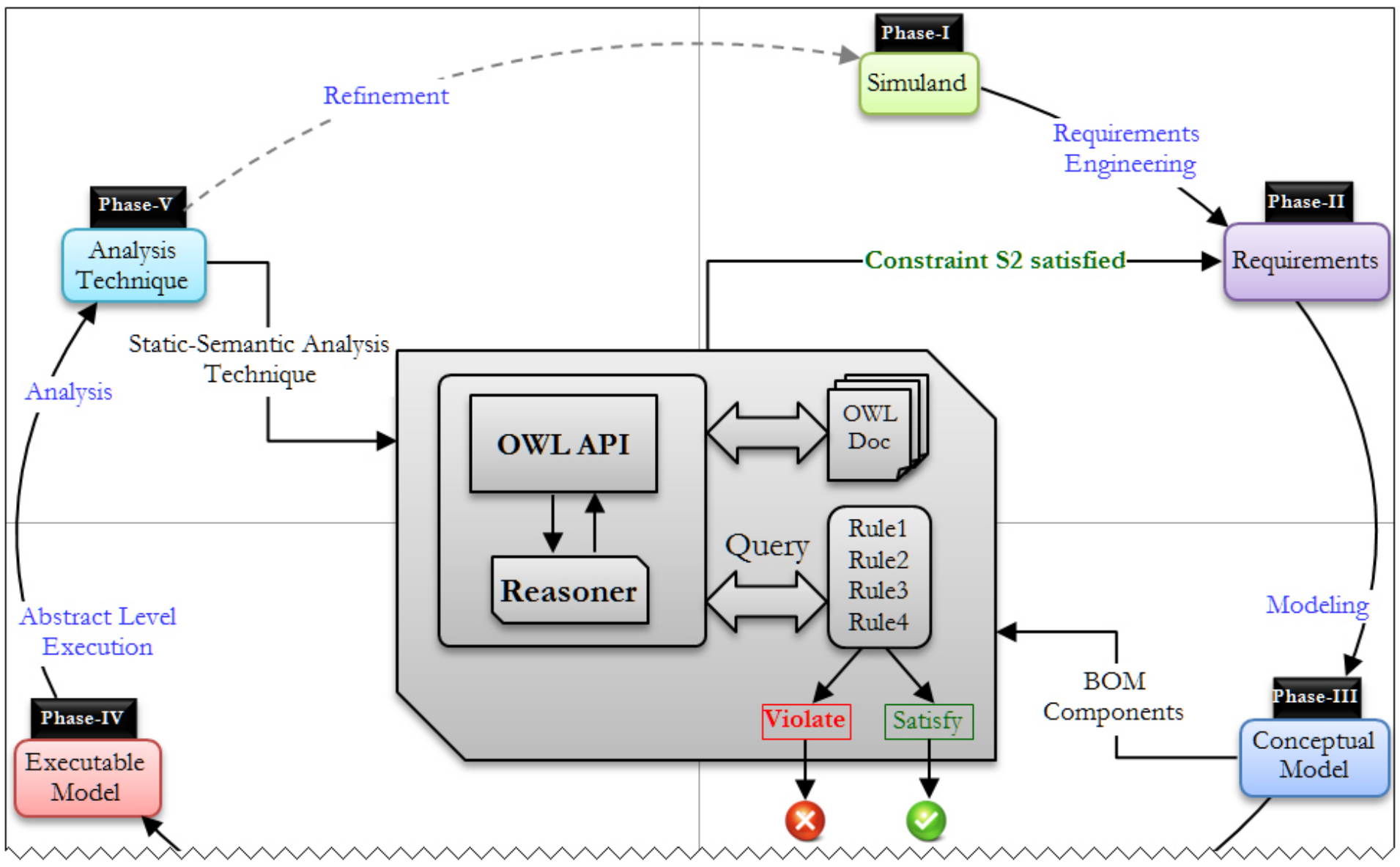

**Figure 32: Static-Semantic Matching**

If the semantic matching is successful, it will fulfill the default constraint (S2) of the requirement specification (see **Table 10**) which is a necessary condition for the overall composability verification.

## 5.6.4 Dynamic Analysis

We use Dynamic Analysis technique (see section 4.3.3) to evaluate the behavior of the conceptual model. At first the components undergo a state-machine matching process for the evaluation of the behavior consistency. When this evaluation is successful, we proceed with the in-depth verification at the dynamic-semantic composability level, choosing one of the different proposed set of dynamic analysis techniques. These analyses are called dynamic analysis because they require execution at different abstract levels as mentioned in section 5.4

### State-Machine Matching (SMM)

State-machines represent behavior of the components and are the essential dynamic part of BOM components. In the verification of BOM composability at dynamic-semantic level, it is important that the behavior of the composed components should be coherent with each other i.e., their interactions are consistent in order to make progress towards composition goals. To ensure this fact we assert (as a necessary condition) that the state-machines of the composed components should match each other. BOM state-machines are event driven in nature and make progress by exchanging events. In order to ensure that the state-machines of the composed BOM components match each other they are required to be executed at an abstract level. Therefore we proposed a technique in [**113**] which transforms each BOM state-machine to SC-XML (State-Chart XML) [**114**] format. A sample of SCXML is shown in **Figure 33**.

```
<scxml initialstate= STATE1>
    <state id = STATE1 final = True>
        <transition event=EVENT NAME
            target=STATE2/>
    </state>
    <state id = STATE2 final = false>
        <transition event=EVENT NAME
            target=STATE1/>
    </state>
</scxml>
```

**Figure 33: SCXML format**

We develop a runtime environment using SCXML API for the execution. This environment parses SCXML files (transformed BOM state-machines) and creates instances. Then it initializes all the state-machines to their initial states and simulates sending and receiving of the events to observe state-machine transitions until they reach their final state. The state-machine matching process is based on the following algorithm:

## Algorithm: State-Machine Matching

**Input: {SM} ∊ BOM State-Machines, {Actions}**

**Output: TRUE, FALSE**

```
{SCXML} ← TransformSMtoScXML(SM)
⊳ Transform all BOM-Statemachines in SCxml format ⊲

Create and Initialize EventController: EC
⊳ Event Controller controls sending and receiving of events ⊲

for scxml ∊ { SCXML } do
    SC ← Parse(scxml)  ⊳ Parse scxml document
    Create and Initialize  SCXMLExecutor(SC)
    ⊳Instantiate SCXMLExecutor thread for each state-machine ⊲

    Done ← FALSE
    while  (Done =FALSE) do
        CurrentState ← GetCurrentState() ⊳ SCXMLExecutor returns current state
        if CurrentState.IsFinal = TRUE then
            Done ← TRUE
        end if
        ⊳Get Next Action to send or receive ⊲
        {NextActions}← CurrentState.GetActions()
        for next ∊ NextActions  do
            if next.Type = "Send" then
                EC.Put(next) ⊳ Simulate sending of next action
                SCXMLExecutor.Trigger(next) ⊳Transit from the current state to next state
            else
                EC.Get(next) ⊳ Simulate recieving of next action
                SCXMLExecutor.Trigger(next) ⊳Transit from the current state to next state
            end if
        end for
    end while ⊳Due to either of the send or receive actions the state-machine will
              transit to the next state and therefore the current state will be updated.
              If the final state is reached then the state-machine matching will be
              terminated successfully⊲
end for
```

**Table 12: State-machine Matching algorithm**

**Figure 34** shows the state-machine matching process. It takes BOM state-machines as modeling objects, automatically transforms it into a SCXML executable format and perform state-machine matching using abstract level execution environment.

A successful run of this routine implies that all the state-machines match each other, which satisfies a necessary (but not sufficient) condition of BOM composability i.e. constraint S3a of the requirement specification. The fulfillment of S3a certifies consistency and completeness of the behavioral design of the composed components. Consistency is due to the fact that the components are in correct causal order and Completeness, because their inputs and outputs (send and receive-events) are complete to reach their final states. However we still cannot guarantee correctness the 3rd C of requirements, unless the composition satisfies its requirement specification i.e., all the assigned objectives and required constraints. Also the state-machine matching approach may result in reaching final-states but it does not explore all possibilities of the behavioral interaction of the composed components. So it is required to analyze the model at a greater depth using an appropriate dynamic analysis approach.

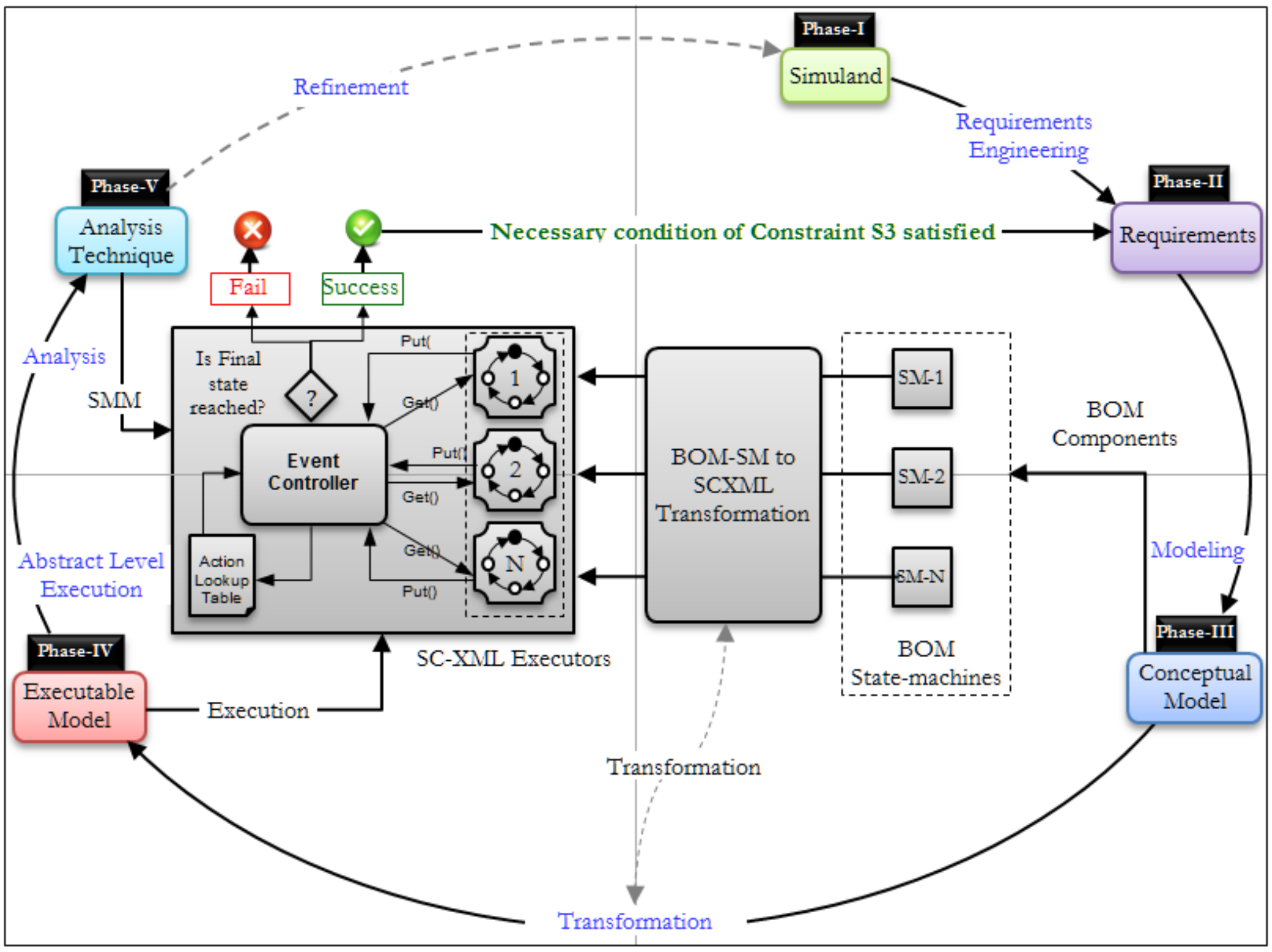

**Figure 34: State-machine Matching Process**

Therefore for deeper evaluation we propose to utilize the modeling and analytical strength of Petri Net and CSP formalism and incorporate three analysis approaches in our verification framework as introduced and discussed chapter 3. The selection of a suitable approach for the composability verification at dynamic-semantic level depends on the nature of the model. In the following subsections, each of these approaches is discussed in detail.

## 5.7 PN Algebraic Technique

The basic idea of this technique is to transform BOM into Petri Net format and verify the properties given in the requirement specifications using algebraic methods. In the verification framework, following steps are proposed to conduct algebraic analysis:

### 5.7.1 BOM to PNML Transformation

In the first step, BOM components are transformed into Petri Net Markup Language PNML format [**115**] which is an XML based form to specify Place/Transition Nets. At first BOM state-machines of all components are parsed and each **state** is transformed as a **Place** in the PN model. Similarly each **event** (send or receive event) is transformed into a **Transition** in PN with no duplication. An outgoing arc is connected from a place-P to a transition-t if the corresponding state-S (of the sender) has a corresponding event-t as its exit condition and next state S′. An incoming arc is connected from transition-t to another place-P′ which represents the next state S′. Similarly state-R (of the receiver) is transformed into place-Q and the next state R′ into Q′. The incoming and outgoing arcs are connected to t. The sender and receiver entities (of BOM) are represented as tokens in the places. **Figure 35** shows how part of a sender and receiver state-machine is transformed into a PN. The place P and Q have tokens showing the current state (or marking) of the composed model. When transition t is fired (meaning event t is sent by P and received by Q) the tokens are transported to P′ and Q′ showing the next marking of composed model.

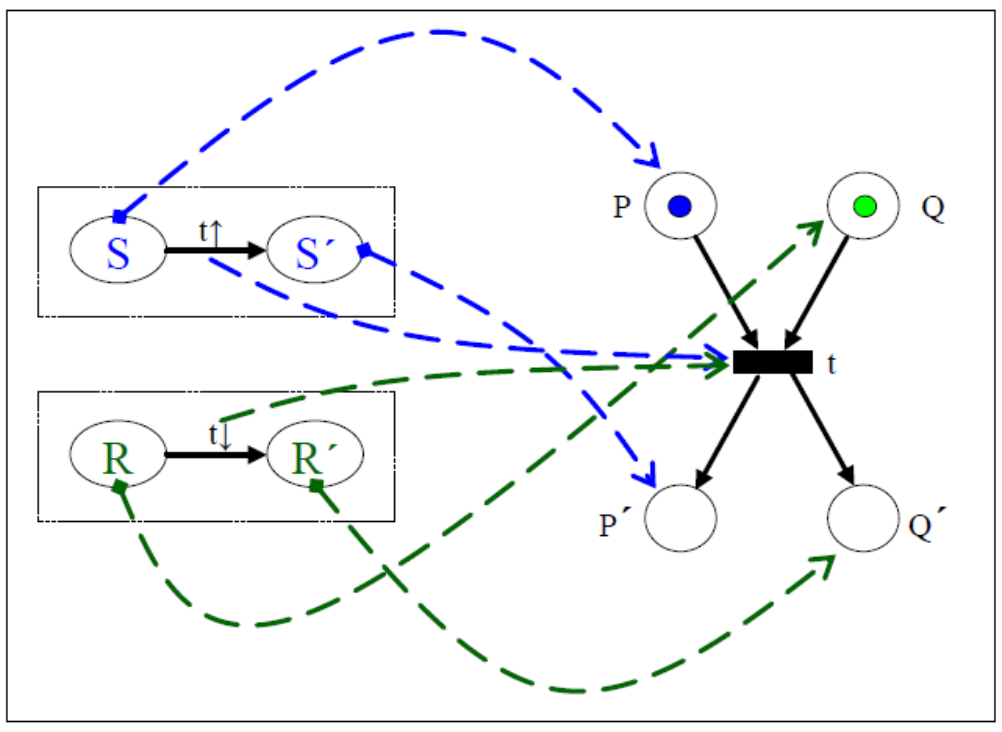

**Figure 35: BOM to PN transformation**

The transformation process is complete, when all the states and events of every state-machine in BOM are plotted in the PN model such that no element is duplicated, and each place or transition is connected so that there are no broken links.

### 5.7.2 PN Algebraic computations

In this step the PN incidence matrix and Place/Transition invariants are calculated. To perform this step we use Platform Independent Petri Net Editor (PIPE) API [**116**]. PIPE is a java based open source API for performing different Petri Net related operations. It offers API functions to automatically compute algebraic resources of a PN model such as Incidence matrix and Place/Transition invariants.

#### Incidence Matrix

An incidence matrix of a PN model is calculated by subtracting A- from A+ incidence matrices:

**Algorithm: Incidence Matrix Calculation**

**Input:** PN Model (P-places × T-transitions)

**Output:** m × n Matrix A

```
1  Initialize a Matrix A_minus of size m × n such that m=|P| and n=|T|
2  for i=0 to m do
3      for j=0 to n do
4          if pi ∈ P is connected to tj ∈ T then  ▷ i.e., p is the input place of t
5              A[i][j] ← arc weight  ▷ arc weight is always ≥ 1
6          else
7              A[i][j] ← 0
8          end if
9      end for
10 end for
11
12 Initialize a Matrix A_plus of size m × n such that m=|P| and n=|T|
13 for i=0 to m do
14     for j=0 to n do
15         if tj ∈ T is connected to pi ∈ P then ▷ i.e., p is the output place of t
16             A[i][j] ← arc weight  ▷ arc weight is always ≥ 1
17         Else
18             A[i][j] ← 0
19         end if
20     end for
21 end for
22
23 Initialize a Matrix A of size m × n
24 for i=0 to m do
25     for j=0 to n do
26         A[i][j] ← A_plus[i][j] - A_minus[i][j]
27     end for
28 end for
29 Return A
```

**Table 13: Incidence Matrix Calculation**

Lines 10 calculate the A- matrix. Lines 12-21 calculate A+ matrix and lines 23-28 calculate the final incidence matrix.

### Place and Transition Invariants

The methods for calculating P-Invariants and T-Invariants of a PN model have been extensively studied. The basic principle to compute the fundamental set of P-invariants and T-Invariants is based on Farkas Method [**117**]. The algorithm for finding P-Invariant is presented as follows. The input of the procedure is the Incidence Matrix A and an Identity matrix B, both of size m × n. The output is a matrix C whose rows are the fundamental set of P-Invariants.

**Algorithm: P-Invariant Calculation**

**Input:** Incidence Matrix A, Identity Matrix B

**Output:** Matrix C (rows of C = P-Invariants)

$C \leftarrow A \mid B$ ⊳ Augmentation of A with $m \times n$ identity matrix B<br>
**for** $i=1$ to $n$ **do** ⊳ $n = |T|$<br>
    **for** each pair of rows $c_1$, $c_2$ in $C[i-1]$ where $c_1[i]$ and $c_2[i]$ have the opposite signs **do**<br>
        $c \leftarrow |c_2[i]|.\, c_1 + |c_1[i]|.\, c_2$<br>
        $c' \leftarrow c/$g.c.d of each element of row c ⊳ g.c.d =Greatest common divisor<br>
        augment matrix $C[i-1]$ with row $c'$<br>
    **end for**<br>
    Delete all rows of $C[i-1]$ whose $i^{th}$ component is non-zero, the result is C<br>
**end for**<br>
**Return** C

**Table 14: Place-Invariants**

The same procedure is used to find T-invariants by taking the transpose of the Incidence Matrix A. Details and a discussion about the improvement of this algorithm are presented in [**118**]. These algorithms are implemented in PIPE API and can be used in form of function calls.

### 5.7.3 Property Verification Method

The outcome of algebraic analysis technique is the satisfaction or violating of a property with respect to a PN model. There are different methods to perform property verification however there is usually certain theorems behind the reasoning of necessary and sufficient conditions for the fulfillment of a property. In Petri Net literature many solutions (proofs) for the property proving theorems are contributed and can be applied to prove different properties when required. Using these theorems and the available algebraic resources a property verification method (algorithm) is developed which evaluates the conditions given in the theorem on the PN model and results in satisfaction or violation of the required property. **Figure 36** presents the mechanism of algebraic verification technique in the verification framework:

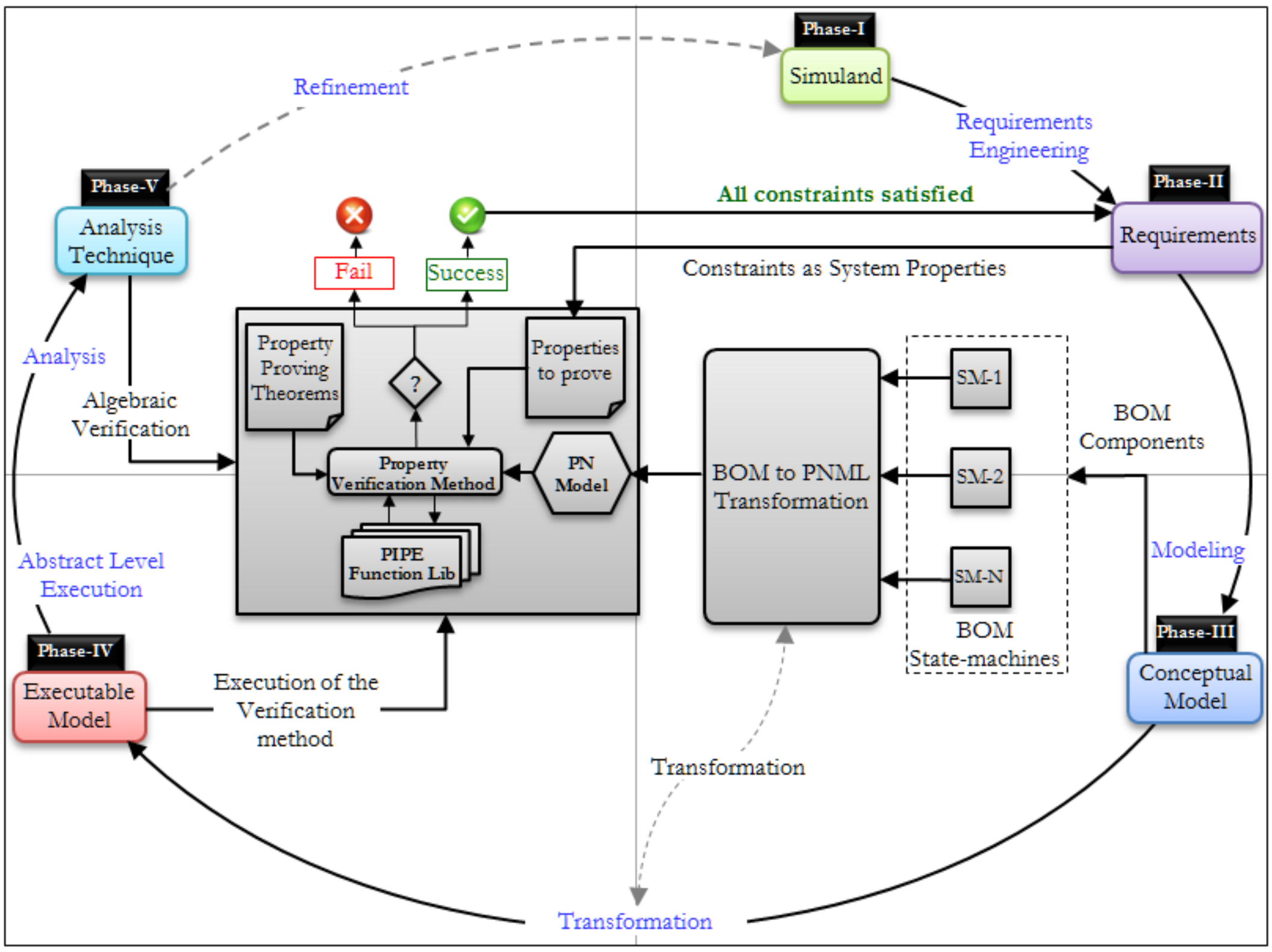

**Figure 36: PN Algebraic Technique**

To explain our approach we present the theorems and an example property verification method for the analysis of fairness property in a PN model in chapter 7.

**PNML Execution and State-space Graph**

It should be noted that PIPE library also offers an execution environment which can be used to run the transformed PNML model. If the tokens (each representing a BOM entity) eventually reaches its final state (place) then the execution is successful. This asserts that the model is correctly transformed and it correctly represents the behavior of its source i.e. the conceptual model. PIPE library also offers a function to generate and visualize state-space graph of the PNML model. This can be useful to find deadlocks and verify other system properties through graph reachability.

## 5.8 CPN based State-Space Analysis Technique

The second approach proposed for the dynamic semantic composability verification is based on Colored Petri Nets and State-space analysis technique. This approach effectively utilizes the potential of Colored Petri Net formalism, CPN modeling and programming language, its execution environment and supporting tools in order to verify a composed model at dynamic-semantic level with respect to the requirement specifications. The unique feature of this approach is its data-centric nature. As discussed in section 3.1.5 CPN supports level-3 PN modeling where tokens are structured and can represent data objects. Also the transitions cover greater details of the system behavior. Therefore the structure and the behavior of the system can be modeled with greater details. In order to exploit the data-centric nature of our approach we proposed the following stages:

### 5.8.1 BOM Extension

The current BOM standard lacks certain structural and behavioral semantics which are essential for modeling complex system behavior therefore we require specification of additional modalities that can help in capturing the structure and behavior of a system at a greater detail [**119**]. We therefore propose to extend the BOM conceptual model specification by applying the concept of *Extended Finite State-Machines* (EFSM), which is introduced and discussed with detail in [**120**]. An Extended Finite State Machine (EFSM) is defined by the tuple:

**M = (Q, I, $\Sigma_1$, $\Sigma_2$, V, Λ)** where:

Q ($\neq\emptyset$) is a finite set of states.

I $\subset$ Q is the set of initial states

$\Sigma_1$ is a finite set of (send or receive) events.

$\Sigma_2$ is a finite set of actions (Actions are the instructions to be executed and should not be confused with the BOM actions, which are used in pattern of interplay).

V is the set of state variables.

Λ is a set of transitions; each transition **λ ∈ Λ**

$$\lambda = q \xrightarrow[\{v_{in}\} \mid \{v_{out}\}]{e\,[g]\,/\,a} q'$$

Where

q and q′ ∈ Q

e ∈ $\Sigma_1$ is an event

g is a condition (or guard)

a ∈ $\Sigma_2$ is an action.

It means if the system is at a state q, an event e occurs, and the guard g is satisfied, then action a will be executed and the system will transit to the next state q′. During the firing of transition **λ ∈ Λ** the variables $\{v_{in}\}$ are used as input and the variables $\{v_{out}\}$ are used as output.

**Example:**

This example is a modified version of an extended finite state-machine of a queue discussed in [**120**] and is intended to explain the notions of EFSM. A queue component is either *empty* or *nonempty*, and in which insertions are done at the *rear* of the queue and deletions are done at the *front* of the queue. Also the queue has a maximum *size*. Two events *put* and *get* are used to update the states of the queue.

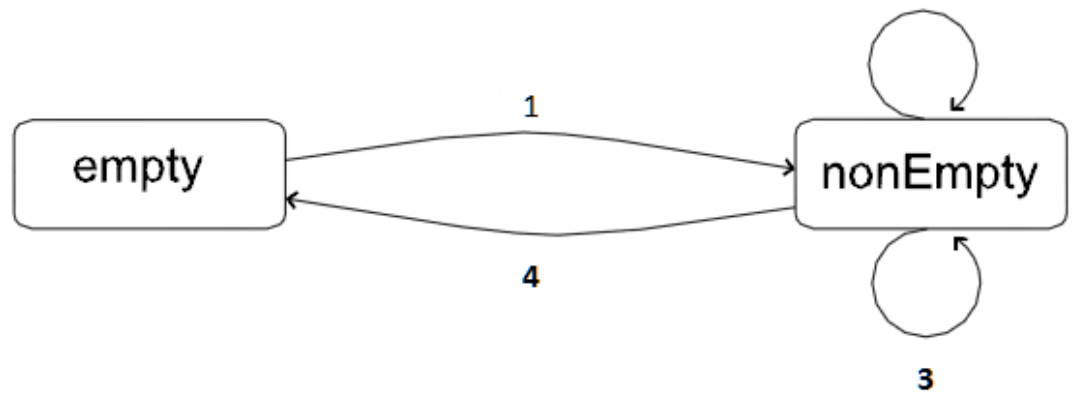

**Figure 37: Buffer Extended finite state-machine** [**120**]

The EFSM model of the buffer is: M = $(\mathbf{Q, I, \Sigma_1, \Sigma_2, V, \Lambda})$ where

Q= {empty, nonempty}

Σ = {put(string obj), get}

$q_0$: empty

V = {front, rear, M, Data}

Λ: Transition Specifications:

Put(obj) [ ] / action{ rear++; Data[rear]=obj;}
1: empty ——————→ nonempty
{rear} | {rear, Data}

Transition 1 allows Queue to transit from empty state to non-empty when put event is received. During this transition the variable rear is incremented. Also the parameter "Obj" of Put event is stored in Data at the rear location.

Put(obj) [rear<M]/ action{ rear++; Data[rear]=obj;}
2: nonempty ——————→ nonempty
{M, rear} | {rear, Data}

Transition 2 lets Queue to revisit non-empty state when put event is received if rear is less than the maximum size. During this transition the variable rear is incremented. Also the parameter "Obj" of Put event is stored in Data at the rear location.

Get(Data[front]) [rear≥front+1∧rear <M]/ action{ front++}
3: nonempty —————— nonempty
{M, front, rear, Data } | { front }

Transition 3 lets Queue to revisit non-empty state. It is fired if rear variable is greater or equal to front+1 and less than the maximum size. It will send Get event with data at the front location is sent as parameter. During this transition the variable front is incremented.

Get(Data[front]) [front+1=M]/ action{rear=0; front=0}
4: nonempty ——————→ empty
{M, front, Data} | { rear, front }

Transition 4 allows Queue to return back to empty state when if front+1 reaches the maximum size. It will send Get event with Data at front location. During this transition both front and rear variables are reset to zero.

We apply the concept of EFSM to the BOM conceptual model, so that we can introduce state-variables and extended representation for transitions (events, guards, actions), to a form, which we name: Extended BOM or E-BOM. There are several advantages in the BOM extension:

The usage of variables (or state-variables) in BOM state-machines allows to model the attributes of a component (structure) and their effects caused due to the change of states and occurrence of transitions (behavior). And values of these attributes can be used in the arithmetic or logical evaluation to assess trigger conditions of the transitions and to transfer variable values from one model to another model, thus in a composition, a value output by one component can be consumed by the other in the composition. Similarly by introducing actions we can manipulate state-variables of a component during an occurrence of a transition and thus we can model realistic and complex behavior of a system which cannot be obtained through simple labeled transitions (such as events/actions in standard BOM).

In the process of BOM to E-BOM extension some of the features are imported from the existing BOM while others are provided by the modelers. **Figure 38** provides a comparison of which parts can be imported from BOM and how they are used in E-BOM and which parts are externally provided by the modelers. A user input utility is designed for the modeler's input where the modeler can choose initial states, select state-variables from the existing BOM entity characteristic or add new variables. Assign data-types to these variables. Design transitions using events from existing BOM and adding guards (conditional statements) and actions. Currently we allow the modelers to use SML language for scripting guards and actions. This is because the extended BOM will later be used in CPN environment which uses SML for code scripts. When each BOM component is extended to the respective E-BOM we proceed to the next stage.

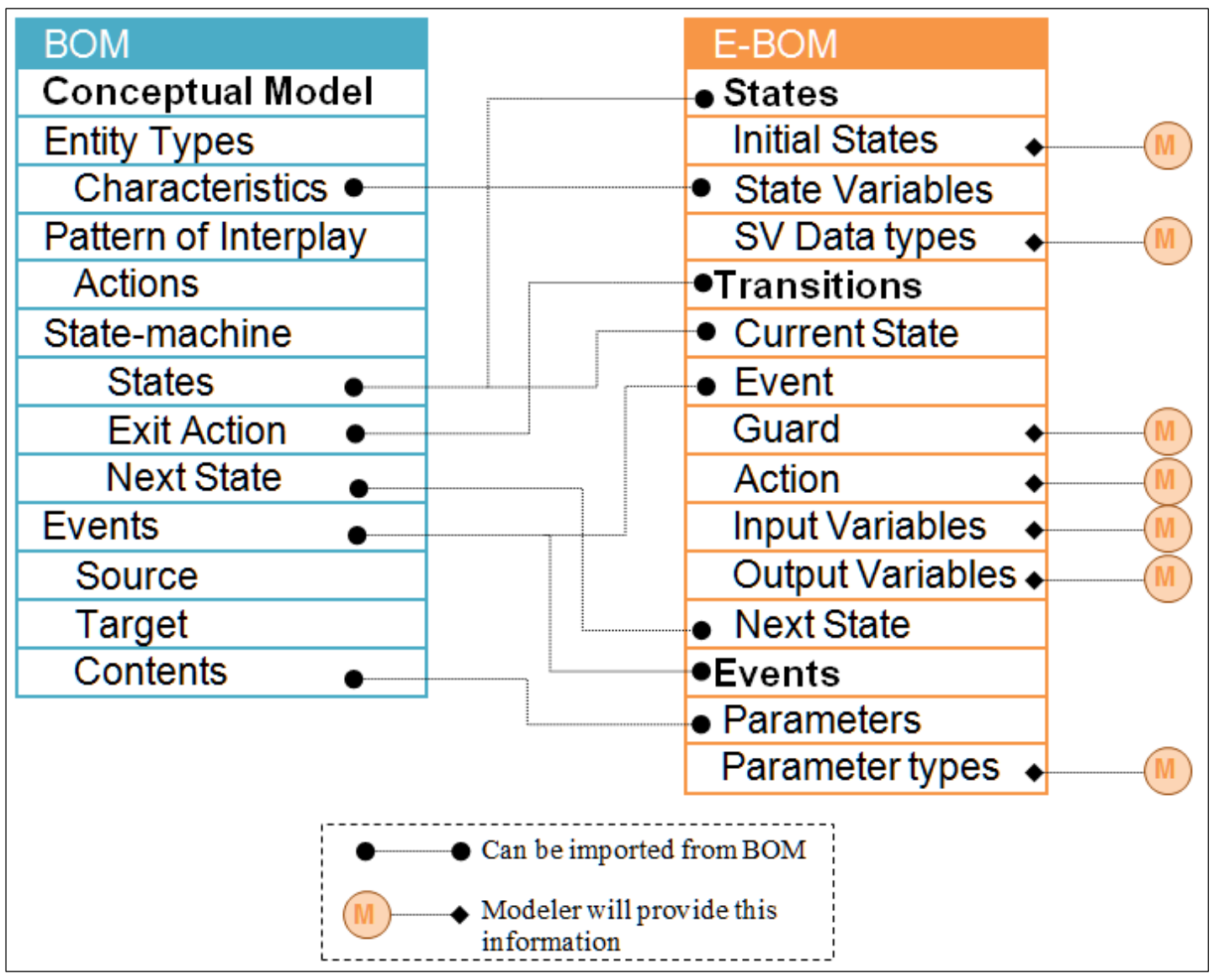

**Figure 38: BOM and E-BOM comparison**

## 5.8.2 E-BOM to CPN Component Transformation

We propose a three layered Component Model based on CPN constructs[32] and transform previously developed E-BOM into this component model using Meta-Model transformation rules (defined later in this section). Our proposed CPN component Model is defined as:

**CPN-CM = ⟨SL, BL, CL⟩ where:**

SL = Structural layer

BL = Behavioral layer

CL = Communication layer

[32] To understand this mechanism background knowledge of CPN formalism and CPN tools environment is required. We recommend to consult the online documentation [**82**]

### Structural layer

The structural layer represents physical attributes or properties of the components. The state-variables defined in E-BOM are transformed into the structural layer. In CPN-CM these State-Variables are defined in form of CPN-places with distinct data-types represented by colors. They are manipulated by the firing of connected transitions during the execution of the CPN model. The state-variables are connected to the transitions if they are defined in their input or output sets (as discussed in EFSM). The values of the state-variables are specified using CPN-tokens. The data-types of the tokens are the same as that of their respective state-variable places. The formal definition of the structural layer is:

**SL = ⟨SV, $\Sigma_{SV}$, $C_{SV}$, $V_{SV}$, $I_{SV}$⟩ where**

- SV is a set of places (that represent state-variables of the Conceptual Model)
- $\Sigma_{SV}$ represents a set of color sets that represent state-variable data types
- $C_{SV}$: SV→$\Sigma_{SV}$ is a function that assigns a color set to each state-variable
- $V_{SV}$ is a finite set of typed variables[33], assigned to each member of SV.
- $I_{SV}$: SV → EXPR is an initialization function that assigns an initial value expression of type Type [SV] to each state-variable.

### Behavioral Layer

The Behavioral Layer represents state-machine of an E-BOM component in form of CP-Net, where each state from E-BOM component becomes a CPN place of integer type, each E-BOM transition becomes a CPN transition, and the flow integer type token(s), represent the change of component's state. The state-variables of the structural layer are connected to the transitions of this layer, if they are specified as inputs or outputs in the E-BOM transition. The guards and actions from E-BOM are transformed into CPN transition inscriptions. They should satisfy CPN-ML scripting syntax otherwise the CPN-ML complier will give errors. Formally the Behavioral layer is:

**BL = ⟨S, T, $A_T$, $A_{SV}$, $A_C$, G, Act⟩** where

- S is a set of places, called *states*. Each place is assigned a type INT. This set represents the states of E-BOM state-machine
- T is a set of transitions (representing E-BOM transitions)
- **$A_T$** is the set of arcs, called *transiting-arcs*. Each transiting-arc connects a state s ∈ S to a transition t ∈ T, and t to another state s′ ∈ S, if **s → t → s′**. Each arc in **$A_T$** is assigned a variable v of type INT. When a transition t is fired a token is transferred from state-s to state-s′ to reflect a change of state in the state-machine.
- **$A_{SV}$** is another set of arcs, called "*sv-arcs*" that connects state-variables (of the structural layer) to the transitions. If a state-variable SV is assigned as an input variable in the E-BOM transition, then it is connected to the transition with an incoming arc. If a state-variable is assigned as an output variable in E-BOM transition, then the transition is connected to the state-variable with an outgoing arc. A variable v is assigned on each *sv-arc* whose type is same as that of the attached state-variable.

[33] Note that these variables are built-in CPN variables that are used to define arc bindings for transporting tokens and are different from our notions of state-variables

• **A**c is another set of arcs, called "*communication-arcs",* that connects communication ports (CP) of the communication layer to the transitions. A cp∈CP is connected to a transition t through a *communication-arc* if it corresponds to a receiving event (as specified in E-BOM). Whereas a transition t is connected to a cp∈CP through an arc if it corresponds to a sending event. A variable v is assigned to each communication-arc whose type is same as that of the attached CP.

• **G: T→EXPR** is a guard function that assigns a guard to a transition t (as specified in E-BOM) such that Type[G(t)] = **Bool**

• **Act: T→EXPR** is an action and assigned to a transition t (as specified in E-BOM). The action can be a single executable statement or a complete procedure. Usually the actions are used to read data from attached input state-variables or communication ports; to process the data and to produce an output to the attached state-variables or communication-ports. Actions can also be used to type-cast the data.

**Communication Layer**

The Communication Layer is responsible for communication with the other components. It provides interface for connecting the inputs and outputs of the components through "port places" and also provides information about the type of data exchange i.e., the tokens of complex data-types that carry message parameters contents, as described in E-BOM event types. The transitions of behavioral layer are connected to the port places of this layer. The arc direction depends on the type of the transition (send or receive). When a message (E-BOM event) is received at the In-port of the communication layer it causes the transition in the behavioral layer to fire as a result of which the state-machine (in the behavioral layer) progresses. This may also happen when a transition causes an outgoing event at the Out port which is expected by another component. Formally the Communication layer CL is:

**$\langle CP, \Sigma_{CP}, C_{CP}, V_{CP}, PT \rangle$**

• CP is a set of port-places called "*communication-ports*". A communication-port connects the inputs/outputs of the component to other components (see section 3.1.5). Each CP corresponds to an event (send or receive).

• $\Sigma_{CP}$ represents a set of color sets that represents communication ports data types (as specified by the corresponding event parameters of the E-BOM). Each color set is constructed by combining the data-types of all the parameters. If there are no parameters then NULL type is assigned, which carries a blank token.

• $C_{CP}$: CP→$\Sigma_{CP}$ is a function that assigns a color set to each port-place.

• $V_{CP}$ is a finite set of typed variables, assigned to each CP with Type[v]∈ Type[CP].

• PT: CP→ {In, Out, I/O} is a port type function that assigns a port type to each port place (as discussed in section 3.1.5).

## E-BOM to CPN-CM Transformation Rules

The rules of transformation from E-BOM to CPN-ML are summarized as follows:

| | E-BOM | → | CPN Component Model | | |
|---|---|---|---|---|---|
| | | | **Structural Layer** | **Behavioral Layer** | **Comm. Layer** |
| **1** | States | → | | State-Places of Type[int] | |
| **2** | Initial-States | → | | Add token type[int], Value = {0, 1…n} for each initial-state | |
| **3** | State-Variables (SV) | → | SV-Places of (type color-sets defined in 4), Add tokens if SV need initialization | | |
| **4** | SV Data Types | → | Define Color-Sets of SV-Places | | |
| **5** | Events | → | | | Add Communicating-Place for each Event of type color-set defined in 6 |
| **6** | Event Parameters and their types | → | | | Add a color-set of type same as event parameters. If there are multiple parameters then add color-set of type product, where each event parameter type is added in the product. (If there is no parameter, use Null type) |
| **7** | BOM-Action | → | | Transitions | |
| **8** | Exit-Condition (Action, Next-State) | → | | Connect a transiting arc from a state-place to transition and from transition to next state-place, Declare v variable of type[int] on each arc. | |
| **9** | Guards of Transitions | → | | Add guard on each transition | |
| **10** | Actions of Transitions | → | | Add action script on each transition | |
| **11** | {Input-Variables} of Transitions | → | Connect SV-arc from each SV to the transition, if it is added as input variable in that transition. Declare sv of Type SV on the arc. | | |
| **12** | {Output-Variables} of Transitions | → | Connect SV-arc to each SV from a transition if it is added as output variable in that transition. Declare sv of Type[SV] on the arc. | | |
| **13** | Receive-Events ∈ Events | → | | | Connect an arc from CP to the transition if it has a receive event. Declare cp variable of type[CP] on the arc. Mark the CP as in-port. |
| **14** | Send-Events ∈ Events | → | | | Connect an arc from transition to the CP if it has a send event. Declare cp variable of type[CP] on the arc, Mark the CP as Out port |

**Table 15: Transformation Rules**

## Example

Let us consider the E-BOM of the Queue example discussed in the previous section:

| **E-BOM: Queue Component** | | | | |
|---|---|---|---|---|
| **States** | {empty, nonempty} | | | |
| **Initial States** | {empty} | | | |
| **State Variables & their Data types** | {front : integer, rear : integer, Max : integer, data: string[ ]} | | | |
| **Events** | {Put (obj:String), Get(obj:String)} | | | |
| **Transitions** | | | | |
| | **Put** | **Put** | **Get** | **Get** |
| **Current State** | empty | nonempty | nonempty | empty |
| **Event** | Put(obj) | Put(obj) | Get(null) | Get(null) |
| **Guard** | [ ] | [rear<M ] | [ rear⩾front+1∧rear <M ] | [ front+1=M ] |
| **Action** | rear++ | rear++ | front++ | rear=0; front=0 |
| **Input Variables** | rear | rear, M | rear, front, M, data[front] | rear, front, data[front] |
| **Output Variables** | rear, data(obj) | rear, data(obj) | front | rear, front |
| **Next State** | nonempty | nonempty | nonempty | empty |

Based on the above E-BOM the transformation will result in the CPN-CM model shown in **Figure 39**. This model is completely executable in CPN environment and can be composed with other components through communication layer.

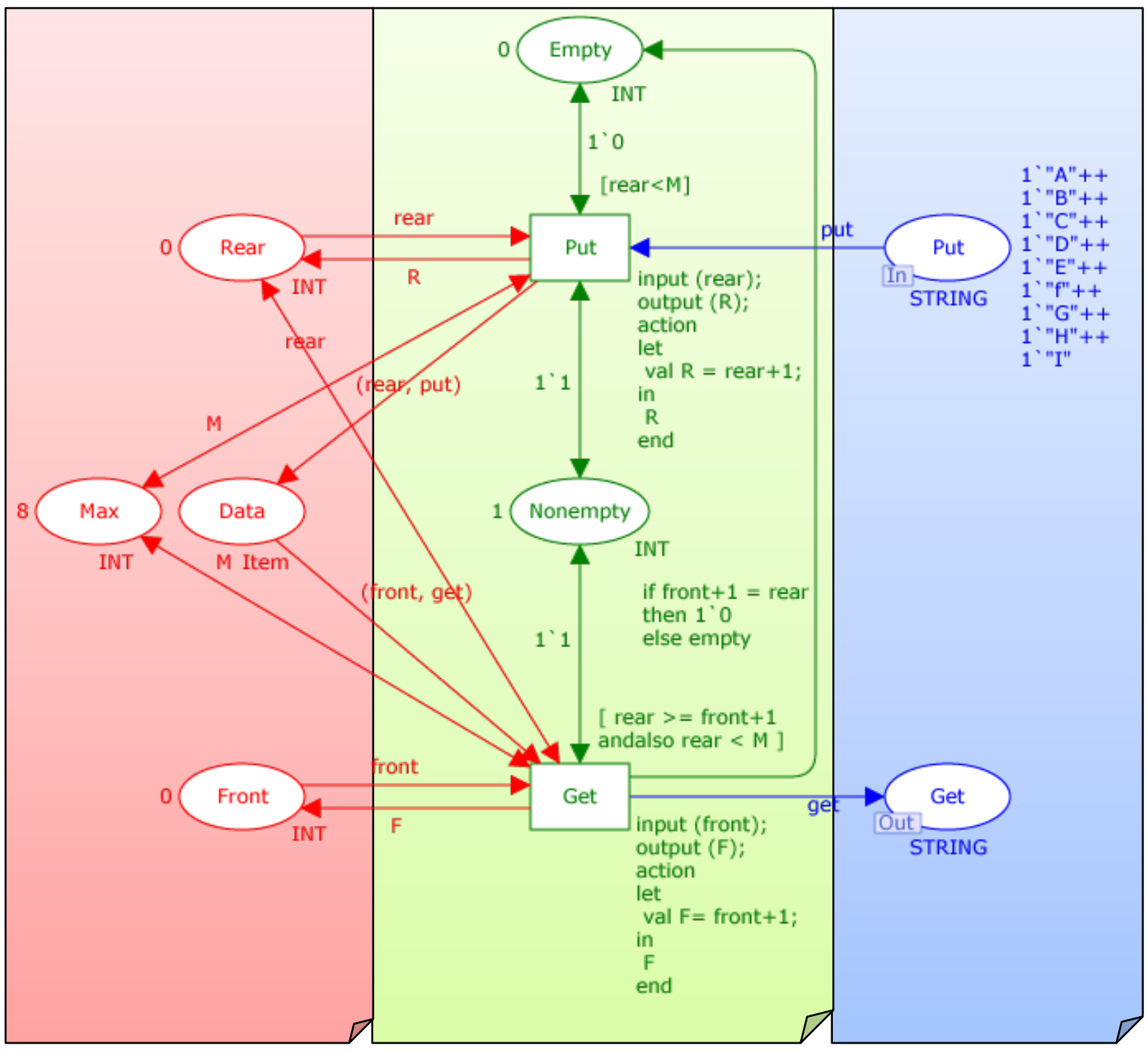

Structural Layer Behavioral Layer Comm. Layer

**Figure 39: CPN-CM represention of Queue component**

The red area represents structural layer and consists of state-variables. The green area represents behavioral layer and shows the state-machine. Blue area represents communication layer and shows the Communication Ports (CPs). When some data is received at CP-Put, the value of rear and Max variables are read as input. Guard rear<M is satisfied. A token is taken from the empty state. Action is executed that increments the rear variable then the transition Put is finally fired. After which a token is produced at the nonempty state showing the state-transition. Also rear and data variables are updated. Max variable retains the token (due to bi-directional arc). If Put is fired again it will repeat the same process. If Get is fired provided the guard is satisfied, then front, Max and data variables are read as input. The data (picked from the front of the queue) will be sent to the out-CP. When the data is emptied the token will be sent to the empty state.

### Automated Transformation Tool

In order to automate the E-BOM to CPN transformation process, we develop a transformation utility, which takes an E-BOM component as input and produces CPN- code for all three layers of CPN component model automatically. The code follows CPN-XML specifications. For each E-BOM component, a separate CPN sub-page is generated (programmatically) and the necessary CPN elements (places, transitions, arcs, color sets, variable declarations, initial markings multi-sets, guards, actions, code segments, CPN ML functions, ports, ports-tag) are generated in one CPN output file, which can be loaded in CPN tools. Once all the CPN models of the BOM composition are generated, the modeler creates a main model and "manually" combines the generated CPN-CM modules (using CPN hierarchical features). The output of this step is a composed CPN model. The modeler is also required to initialize each component with data (in form of token assignments i.e. the initial values of the tokens of state-variables and initial states of the state-machine).

### S3b Evaluation

The S3b constraint in the requirement specification requires that "If the conceptual model is transformed into an executable model, the latter should correctly represent the structure and behavior of the former" (see **Table 10**). Therefore we have to compare each CPN component (executable model) with its respective BOM (Conceptual Model) to check that its structure and behavior is preserved after the transformation. To show that S3b holds after the transformation we rely on the following assertions:

1. As BOM is extended to E-BOM hence BOM $\subset$ E-BOM. Any information added by the modeler in E-BOM cannot cause loss of structural information of BOM. Therefore E-BOM structurally preserves BOM.
2. To check that the generated CPN component contains all the Events and their parameters, States and their exit-conditions, Actions and their senders/receivers we need at least one transformation rule that is responsible to transform these elements:
    a. Rules 6 & 7 (see **Table 15**) are responsible for transforming Events and their parameters into CPN component.
    b. Rules 1 & 8 are responsible for transforming states and their exit-conditions.
    c. Rule 7 is responsible for specification of BOM-actions as transitions in CPN model. Also rule 5 defines port-places which are used to connect senders or receivers.

    Existence of Rule 1, 6, 7 & 8 confirm that the structure of corresponding BOM is preserved in the transformation.

3. CPN Tools provide a built-in compiler for the compilation of CPN models and report if there is any syntax error in the model. The absence of error confirms that the transformed model is structurally consistent and behaviorally functional.

For the behavioral bi-similarity we propose an inspection technique. At first we evaluate that all the generated components possess the same behavior as defined in the conceptual model. So we test the functional output of each CPN model by giving the needed inputs. If by giving correct inputs, the model produces desired output then its functional behavior is correct. To perform functional testing, the modeler initializes all the IN-type communicating-ports (CPs) with tokens of required parameters. (See **Figure 39** for an example where IN-CP "Put" is initialized with tokens of type String). Then the model is executed. If the model produces desired output on the corresponding Out-CPs (In **Figure 39** the desired output should be a token of type string retrieved at Get Out-CP), then the functional test is successful. The modeler performs functional test on all generated CPN components.

In the second step, when all CPN components are composed (i.e. the socket-places of the main model are connected to the Communicating-Port places of the CPN components then the modeler is required to inspect that CPN components are connect exactly according to the Pattern of Interplay of the BOM composition. Also when the composed model is executed the sequence of sending and receiving events from one component to another (which can be observed at the main model by seeing the movement of tokens) follows the pattern of interplay. If the execution is according to the pattern of interplay and the components make progress until they reach their final states, then we say that the behavior of the transformed model is bi-similar to the conceptual model. This confirms the satisfaction of S3b constraint.

The execution can be automated or interactive. In automated mode the choices between multiple progressive paths are randomly picked whereas in interactive model the modelers can pick a path of his choice. Using this option the modeler can probe paths that can lead to a successful execution scenario. During the execution CPN tool also offers *Data Collection Monitors* for recording the data values, which are very valuable for collecting statistics and results of the execution.

### 5.8.3 Verification of the composed CPN model

In the next step, the state-space analysis is performed. At first the state-space of the composed CPN model is generated using CPN state space calculation tool. As discussed in section 3.1.5 a state-space is a graph of nodes (of system-states or markings) and arcs (transitions). When the state-space is generated, different query functions can be used to explore the state space graph for various verification questions. A query function is like an algorithm that explores the state-space graph. These algorithms are based on theoretical concepts of Petri Nets state-space analysis and are used to verify PN properties. Therefore we translate a system property given in the requirement specification into a suitable PN property. There have been a lot of contributions in the PN literature in specifying PN properties and methods of reasoning of their satisfiability or violation. In CPN state-space analysis, the existing methods can be utilized in developing query functions for their respective PN properties.

CPN tools provide some built-in-functions for the common query tasks. We also propose a library of additional functions to perform queries specific to our composability verification framework and the requirement specifications. Figure 40 illustrates the state-space analysis of a composed CPN model using a query function.

We divide these query functions into two categories:

### (i) General System Properties

This category includes commonly known system properties such as *freedom of deadlock*, *live lock*, *starvation*, or *existence of boundedness*, *mutual exclusion*, *fairness*, *sequentiality*, *time-synchronization* etc. if any of these or similar system properties are included as a constraint in the requirement specification then it is translated in CPN terms and a suitable query function is selected from the Function library to perform verification using state-space of the composed CPN model.

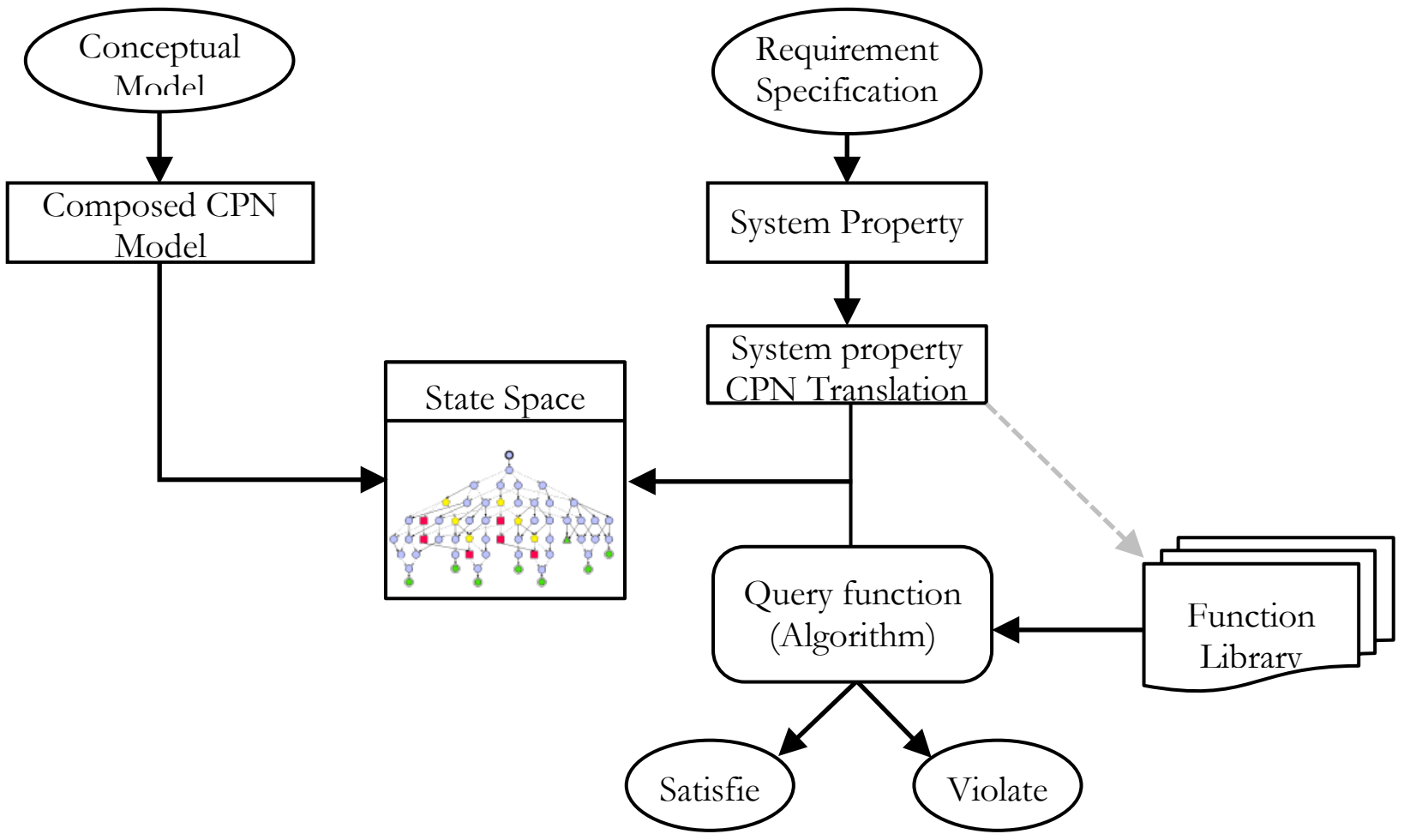

**Figure 40: CPN State-space analysis**

For instance a deadlock freedom property can be translated into CPN terms as:

"*An absence of a marking with no outgoing arcs in the entire state-space graph*"

So essentially we need to find such a node in the state-space graph that violates above condition. If no such node is found then the model is said to be deadlock free. A library function **`ListDeadMarking()`** returns a set of all those markings (if any) which have no outgoing arcs. If the result of this query is an empty list, then we assert that the model is deadlock free. Similarly there are other library functions that deal with the evaluation of other system properties.

### (ii) Scenario Specific Properties

These properties are specific to the scenario (of the real system) under which the model is built. The objectives or goals from the requirement specification are usually translated in form of scenario specific properties. In CPN terms a typical goal or objective can be translated as a certain desirable marking, where the values of state-variables in structural layer evaluate to a particular criteria or reaching of particular state(s) in behavioral layer is desired or certain data at the output port(s) of the communication layer is looked-for. A goal or objective can be expressed in a combination of all these possibilities too.

Scenario specific properties may also include certain safety or liveness assumptions, which represent certain desirable (or un-desirable) situations that must (or must not) occur in order to satisfy (or violate) the requirements. These properties are mostly the CPN translations of the constraints defined in the requirement specifications.

Unlike general system properties, verifying scenario specific properties is not a standard operation, and depends on the way they are defined. Most commonly, we make use of our proposed library functions: **IsEqual(), IsNotequal(), IsBetween(), IsUpperBound()** or **IsLowerBound()** to construct a "**predicate**", that serves as a condition evaluation criteria. Then we use **SearchNode(**predicate**)** function to find those nodes, which satisfies the predicate. If one or more nodes are found, then it is verified that the goal is reachable. In cases, where it is important to know how a sequence of the occurrence of transitions, leads to a particular situation when a property is satisfied (e.g., how an objective or goal is reached) we use **SearchArc()** function with the predicate. This tells us the path in the graph that leads to fulfillment of a property. We also develop an export function, that creates a .DOT file of the entire state-space and can be viewed in graph tools such as GraphViz or Gephi, for visualization and performing further tests on the graph such as finding certain paths/shortest paths/longest paths between two particular nodes. When, a CPN composed model satisfies all the properties in the requirement specification, we say that it is verified at dynamic semantic composability level. In chapter 8 we discuss a Field Artillery Scenario as an example of CPN state-space analysis to explain our approach.

An example of translating a scenario-specific property in CPN terms is a restaurant model where we assume that customers may leave the restaurant without paying the bill because they have been waiting for a long time for the waiter to bring bill. This act of the customers is known as "Balking" and is undesirable. Its translation in CPN can be as follows:

*"There should be no arc with the name "balk" that leads to any marking in the graph"*

Arcs are generated due to firing of the transitions. Existence of balk arc means somewhere in the model an incidence occurred when a customer balked (by firing balk transition). So essentially we need to find that such arc is absent in the state-space graph. This can be done by using **SearchArc()** function. Note that this is a simple example. There could be cases in which a sequence of transitions (called traces) or cycles are searched to verify a property. **Error! Reference source not found.** describes the overall process of state-space analysis technique.

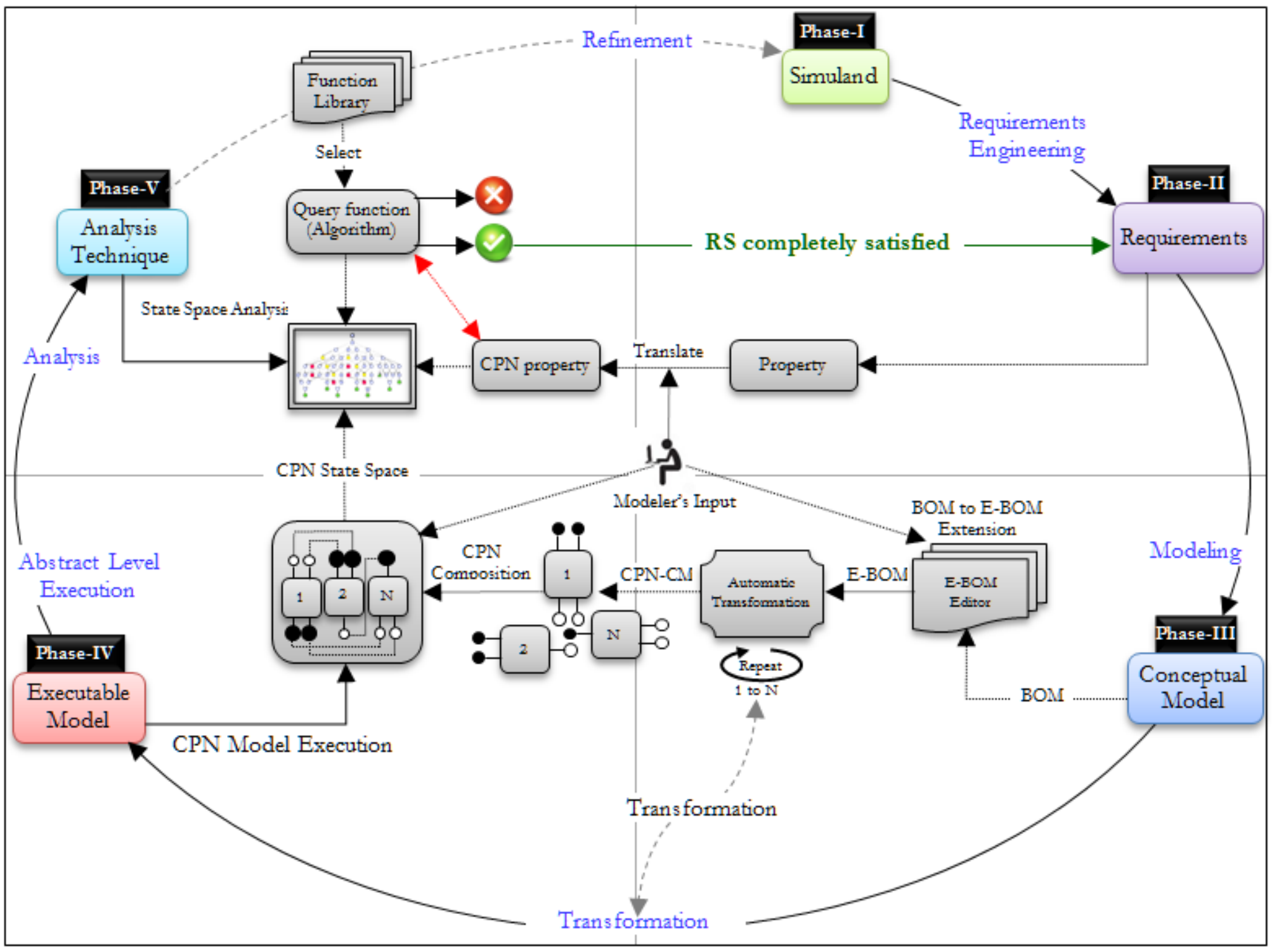

**Figure 41: State-space Analysis Technique**

## State-Space Reduction Technique

In order to alleviate the well-known problem of state-space explosion we propose a reduction technique called "*compositional state space*". The main idea of this technique is that in a hierarchical composition of CPN model, we propose to only consider the places in the main model and treat all the composed components as black boxes. The inputs and outputs of each component can be observed using the flow of Tokens and the data they carry. Therefore in the state-space graph we only keep the markings in which any token is present in the Main model (i.e., any of the place in the main model has at least one token) and delete all other nodes in the state-space graph using the algorithm presented in **Table 16**. The resultant graph will be a reduced form of the actual graph and only considers those markings that reflect a *compositional state-space*. It is called compositional state-space because it only represents a part of the actual state-space which is the result of interactions due to the composition of components. In our experience this subset state-space of the whole state-space is sufficient to evaluate whether the objectives, goals and the constraints are satisfied or not.

**Algorithm: Compositional State-Space Generation**

**Input:** Original State-Space Graph G | **Output:** Reduced State-Space Graph G

```
{Vertices} ← Get-Vertices(G) ⊳ Retrieve all the nodes of the graph in a collection
for v∈to {Vertices} do
    If False ← Is-Filtered(v) then
        G ← Remove-Vertex(G, v)
    Else
        next
    end if
end for
Return G ⊳ Reduced state-space

Procedure Remove-Vertex(Graph G, Vertex v)
{Predecessors} ← Get-Predecessors(G, v) ⊳ Retrieve all the predecessor vertices of v in G
{Successors} ← Get- Successors (G, v) ⊳ Retrieve all the successors vertices of v in G
for p∈to { Predecessors } do
        for s∈to { Successors } do
            G ← Add-Edge(p, s, "DIRECTED") ⊳ Add a directed arc from each predecessor
                                            to each successor ⊲
            G ← Delete-Vertex(v)
        end for
end for
Return G

Procedure Is-Filtered (Vertex v)
        If {places}← GetData(v) then ⊳ Each vertex is a marking, which contains data of all
        ⊳ the places of the model with their names and their token values ⊲
      for p∈to { places } do
         if p is a main place and it is not empty then
               Return  TRUE ⊳ A valid marking with a non-empty Main place is found
         else
               next
         end if
      end for
      ⊳ if the loop is complete then there is no main place which is not empty
Return FALSE
```

**Table 16: Compositional State-space generation algorithm**

Using the Compositional state-space generation algorithm we can filter unnecessary nodes and reduce the size of the graph. The current limitation of this approach is that we first need to construct the actual state-space which is a bottleneck if the model is too large. But this limitation is due to the fact that the process of CPN state-space graph generation cannot be externally modified otherwise if the principle of our reduction technique is applied to the state-space generation algorithm it will directly generate the reduced graph. In chapter 8 we will present the results of our reduction technique by applying it to the Field Artillery example model.

## 5.9 CSP based Model Checking Technique

The third approach proposed for the dynamic semantic composability verification is based on Model Checking, which is widely accepted as formal technique for software verification. In this approach we propose to use Communicating Sequential Processes formalism as a model description language and Process Analysis Toolkit (PAT) as its execution and verification environment in order to verify a composed model at dynamic-semantic level with respect to the requirement specifications. The strength of this approach is in its ability to answer a large variety of verification questions due to the fact that the verification criteria can be specified using LTL, CTL or any of the temporal logic extensions. The Model Checking technique is becoming more promising and acceptable by many software verification users since there is an abundance of improved algorithms, efficient data-structures and faster techniques which are constantly being contributed by the model checking community in order to manage large models with complex modeling requirements. We propose to integrate CSP based model checking verification approach in our composability verification framework. The following stages are proposed in order to perform composability verification using model checking approach.

### 5.9.1 BOM Extension

The E-BOM extension for CSP based Model Checking approach is also inspired from the concept of Extended Finite State-Machine as discussed in section 5.8.1. The extended BOMs for CSP can also have state-variables but since CSP# specification does not allow declaration of strings or higher data-types, the state-variable definitions are restricted to integer and Boolean[34], which in our experience are sufficient to model the behavior of BOM components using CSP (or otherwise it is required to narrow it down to a less detailed version of the component, where only the necessary behavioral details are specified). The transitions of E-BOM contain current-state, event (with parameters), guard, actions and next states. However in this case the action scripts are written in CSP# specification language instead of CPN-ML language. And instead of input and output variables, we have local variables which are accessible only to the component and global variables which are accessible to all the components of the composed model. Some additional information such as time constraints and probability factors are further proposed to be included in the BOM extension so that the behavior of complex systems such as real-time systems and probabilistic systems can be modeled and verified.

Since Timed-CSPs support a number of timed behavioral patterns to capture quantitative timing requirements, such as delay, timeout, deadline, therefore we suggest using these patterns as time functions in the BOM extension, which helps in the automatic transforming of E-BOM into Timed-CSP components. These time functions are essentially assigned to the E-BOM transitions as explained in the following table:

[34] Generally the high level or user defined data types are not permissible in most of the model checking description languages due to the economy of state size, and to avoid risk to state-space explosion. However if the use of such type is inevitable, the PAT tool do provide mechanisms of importing classes from external libraries. If this is the case then the modeler is required to program the components in PAT manually, instead of relying on our automatic transformation tool.

| Time Function | Usage and Explanation |
|---|---|
| **Wait[*duration*]** | Wait is assigned to model the delay in an activity. An enabled transition waits for the given duration before it is fired. |
| **TimeOut[*duration, next*]** | When a timeout function is assigned to a transition, it waits for an event to occur. If the event occurs before timeout, it transits to the next state described in the transition definition; otherwise it transits to the next state described in the timeout parameters. |
| **Deadline[*duration*]** | A transition is constrained to fire when the deadline is reached. The difference between **Wait[]** and **Deadline[]** function is that the former makes the system inactive, i.e., it cannot do anything but wait, whereas when the latter is used the system is active and can respond to events etc., until its deadline is reached. |

**Table 17: Time functions in E-BOM**

Using any of these time functions during the BOM extension is useful to capture the behavior of the real-time systems. In order to further capture the behavior of the complex reactive systems, we also propose to introduce probabilistic factors in the BOM extension. These probabilistic factors can either be used to model the system behavior in form of Markov Decision Processes (MDP) as discussed in section 3.2.5. Or the probability factors can be used to model random time delays, timeouts or deadline, using a particular probability distribution function. For modeling the MDP behavior probability factors can be assigned to multiple transitions of a component's state using the following notation provided by the PAT tool:

```
Pcase {
        [P1]: Transition 1
        [P2]: Transition 2
        ...
        [Pn]: Transition n
};
```

Where $\sum_{i=1}^{n} Pi = 1$

For randomizing time functions, we propose to assign the commonly used probability distribution functions as parameters in the E-BOM:

| Probability Distribution Functions | Usage and Explanation |
|---|---|
| **Normal[*mean, variance*]** | Returns a random value from a normal distribution with a given mean and variance. (Since PAT does not support higher types so we have confined these functions to use integers). |
| **Discrete[*a, b*]** | Returns a random value from a discrete uniform distribution between a and b (a and b included), such that a < b |
| **Exponential [*1/lambda*]** | Returns a random value from a an exponential distribution with parameter 1/lambda |

**Table 18: Probability Distribution Functions**

In order to implement these assignments we develop an external function library in C# which can be imported and used in PAT. A call to these functions generates a random number according to the specified probability distribution. Beside the time functions, these functions can also be used to generate random values for global or local variables, which can help in modeling different probabilistic system behaviors.

When each BOM component is extended to the respective E-BOM we proceed to the next stage.

### 5.9.2 E-BOM to CSP# Transformation

At this stage, each E-BOM component is transformed into a CSP# process component and composed into an executable system. The main idea of this transformation is based on [121], which discusses the transformation of UML state machines to CSP. We however extend this transformation with Communication channels, Time-functions and probability factors to be able to use it for E-BOM transformation. **Table 19** shows the rules used in the transformation process:

| E-BOM | | CSP# Statement and Description |
|---|---|---|
| **States** | → | State-Name()<br>Each state in E-BOM is defined as a CSP process.<br><br>Final-State() = Skip;<br>This statement defines a final state in CSP where Skip is a reserved word means the process terminates successfully. If no such statement exists in any component of the composed model then it is said to be a non-terminating model. |
| **Component** | → | Component-Name = Initial-State() or Component-Name(i) = Initial-State(i)<br>An initial state is defined and assigned to the component. If a component has multiple instances it is passed a parameter 'i' which represents the instance number. |
| **Transitions** | → | **Simple Transitions:**<br><br>State() = [guard] event !/? parameters {action} → NextState();<br><br>The transitions are defined using the above format, where State() is the current state of a component. [guard] is a conditional statement. If it is true only then the transition will be enabled. Event is sent using '!' symbol or received using '?' symbol through an event channel. For each event in an E-BOM component, we define a channel as follows:<br>channel event-name 0;<br>In CSP# "0" means the buffer size of the communication channel is zero, which further means that it is a synchronous channel. Parameters are the values that are passed during an event exchange and a separated using '.' Actions are scripts that should be executed when the transition is fired. Usually these actions are used to update local or global variables. NextState() is the new state which will be reached when the transition is fired. It must be defined within the CSP component body. |
| | | **Transitions with Time functions:**<br><br>Following statement represents a transition with timed-functions:<br><br>State() = [guard] **Wait**[d]; event !/? parameters {action} → NextState();<br><br>State()=[guard] event!/?parameters {action} → NextState() **deadline**[d];<br><br>State()=[guard]event?parameters{action}→NextState() **timeout**[d] NextState2 ();<br><br>Note that in case of timeout, the transitions should only be receiving an event. |
| | | **Markov Decision Process style Transitions:**<br>State() = pcase{<br>[**Prob1**]: [guard] event !/? parameters {actionA} → NextStateA()<br><br>[**Prob2**]: [guard] event!/? parameters {actionB} → NextStateB()<br>};<br><br>Note the postfixes A and B in action or next states of the transitions. Using this CSP# code style multiple transitions can be modeled with different probabilities for either creating a variation of the action which is fired when one of these transitions is selected in a simulation run, or the next states (or both). |
| | | **Transitions with Probability distribution functions:**<br>For using probability functions, at first it is required to import our external probability function library using:<br><br>#import "PAT.Lib.ProbabilityDistributionFunctions";<br><br>Following are some examples of how the function calls can be made:<br>var x = call(Normal, 10, 4);<br>An integer variable is defined which will randomly be assigned a value using normal distribution with mean=10 and variance=4 |
| | | Wait[call(Exponential, 1/4)];<br>Delay function with an exponential distribution, where the inter-arrival rate is ¼. |
| **State-variables** | → | var Variable-Name=Initial-Value;<br>or<br>#define Constant initial-value;<br><br>In CSP# weakly typed variables are used which means that while declaring a variable, the type is not specified. The global variables can be accessed by all components whereas the local variables can only be accessed by the component they belong. |
| **Component** | → | Component-Name = Initial-State() or Component-Name(i) = Initial-State(i)<br>An initial state is defined and assigned to the component. If a component has multiple instances it is passed a parameter 'i' which represents the instance number. |
| **Composed Model** | → | Composed-Model = Component1 \|\|\| Component2 \|\|\| ... ComponentN;<br>The composed model (name) is defined as a composition of CSP process components with an interleaving operator '\|\|\|' between each other. However if there are broadcast events (i.e., one event is sent to all components); or one to many; or many to one synchronization events are used then a parallel operator '\|\|' is used to compose CSP process components. |

**Table 19: E-BOM to CSP# transformation rules**

We develop a transform tool that takes all the E-BOMs as input, and outputs a single composed model using CSP# description language. The generated CSP# composed model can be opened in PAT tool and compiled for checking errors. If no errors are found then the transformed model is said to be structurally consistent and behaviorally functional and it is ready for simulation and verification. It can also be directly compiled, executed and verified using command line operation.

### S3b Evaluation

The S3b constraint in the requirement specification requires that "*If the conceptual model is transformed into an executable model, the later should correctly represent the structure and behavior of the former*" (see **Table 10**). In order to evaluate S3b, i.e., to check that the structure and the behavior of the generated executable model (CSP composed model) correctly represents its conceptual model (BOM composition), we propose following steps:

1. For each CSP component, manually inspect that it contains all the states that exist in its corresponding BOM component
2. Inspect that the exit condition(s) of each State in BOM correspond to a transition(s) and a next state(s) in CSP.
3. Execute the generated CSP model in PAT simulator and observer that all the components reach their final states (or in case of a non-terminating model each component re-visit its initial state iteratively).

Step 1 & 2 confirms by inspection that the structure of the generated model correctly represents its conceptual mode whereas step 3 confirms that it behavior is bi-similar to the conceptual model and therefore satisfies S3b constraint.

## 5.9.3 Verification of the composed CPN model

At this stage, the CSP composed model undergoes composability verification using PAT model checker. At first the requirement specification is translated into CSP# property (or assertion) description language. This language is based on a mix of classical Linear Temporal Logic (LTL) and its different extensions such as Real-Time LTL and Probabilistic LTL and is used to construct assertions (verification questions) of various types, such as reachability properties, safety properties, liveness properties, deadlock freeness etc. We use the syntax of assertion specification language of PAT to translate the objectives and constraints of given requirement specifications. Following are some generic examples of how to specify PAT assertions:

| | | |
|---|---|---|
| 1 | `#assert System deadlockfree;` | This assertion checks deadlock freedom in the 'System' |
| 2 | `#assert System reaches goal?0;` | This assertion checks that whether the 'System' can reach its goal (by receiving a goal event with '0' parameters) |
| 3 | `#assert System \|= <>goal?0;` | This is an equivalent LTL assertion It checks if the goal is eventually reachable. |
| 4 | `#assert System \|= []<>goal?0;` | This LTL assertion checks if the goal is always eventually reachable by the system. Note that it is different from assertion 2. |
| 5 | `#assert System \|= <>goal?0 deadline[50];` | This assertions verifies goal reachability with time constraint i.e., its checks if the goal is reachable within 50 time units or not? |
| 6 | `#assert System \|= <>goal?0 with prob;` | This assertion checks the min and max probability of the goal reachability. |
| 7 | `#define goal (Some-Variable == True);`<br>`#assert System reaches goal;` | This is another way to verify goal reachability, where the goal definition is based on some value of a variable. |

**Table 20: Some examples of PAT Assertions**

When an assertion is defined and its syntax is correct, we can verify it by running the PAT model checker and select the desired assertion from the list. The model checker will present the verification results with success, showing that the assertion is verified or it will provide a counter example if the assertion is not satisfied. In chapter 9, an example of field artillery is presented to show how a CSP composed model is verified with requirement specifications defined as PAT assertions. **Figure 42** describes the overall process of state-space analysis technique.

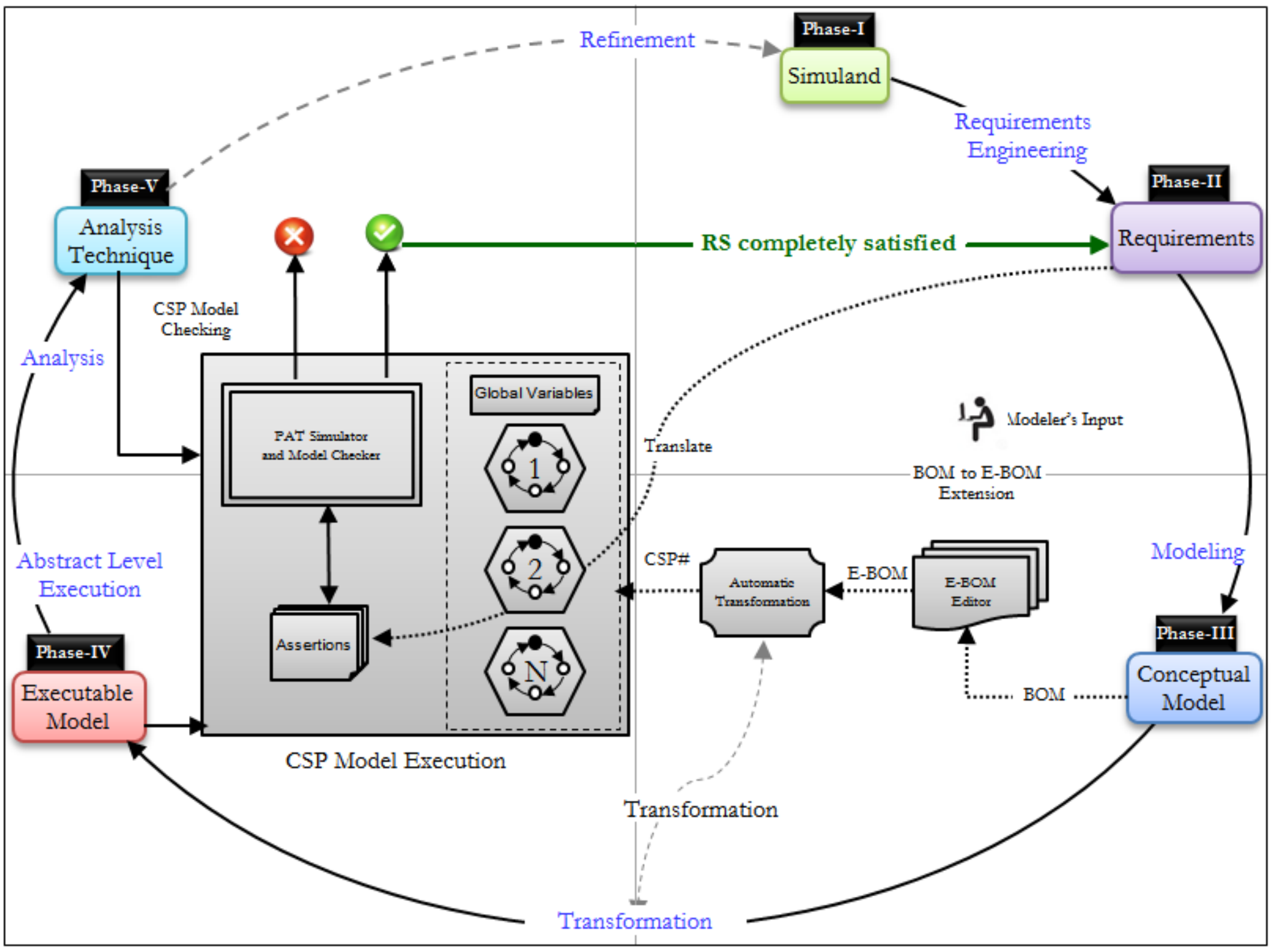

**Figure 42: CSP based Model Checking Technique**

## 5.10 Summary

In this chapter the proposed composability verification framework is discussed in details with its structural and functional specifications. Each activity, algorithm, technique and the process is explained in the perceptive of Component based M&S life-cycle. The composability verification is performed at three levels of composability called static, semantic and dynamic-semantic composability. The main objective of the proposed framework is to verify composability at these levels with respect to requirement specifications. The first two levels are suggested to be evaluated using static-analysis techniques whereas the third level is proposed to be verified using dynamic analysis techniques. At first the behavior of the composed components is evaluated using State-machine matching technique. If they pass this step, they are subjected to one of the three proposed approaches called (i) Algebraic Analysis Technique, (ii) State-space analysis technique or (iii) Model checking for dynamic-semantic composability verification. The choice of these approaches depends on the nature of the model. In chapter 10 we will present some guidelines on how to choose an appropriate approach.

When the entire composability verification process is successful, it implies that the BOM based composed model is structurally and behaviorally consistent, it is composable at syntactic, semantic and dynamic-semantic level and is correct with respect to the given requirement specifications.

# Chapter 6
# Composability Verification Process

*Chapter 5 mainly presented the specification of our proposed composability verification framework including details of different modules, their mechanics and the procedures they perform. In this chapter we present how to use our framework. It can be used as a manual of our composability verification framework. At the end of this chapter we also provide necessary recommendations for the selection of appropriate approach based on the given inputs.*

The description of shapes used in the following flow diagrams is as follows:

| | | | |
|---|---|---|---|
| Object, Data, Model, Component etc. | 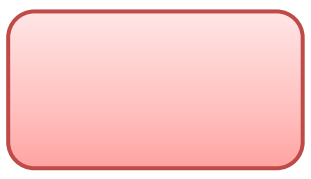 | Any shape of this color express a 3rd party tool | 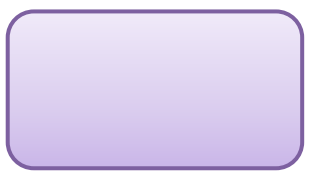 |
| List, Collection or Set of objects, Data, Model | 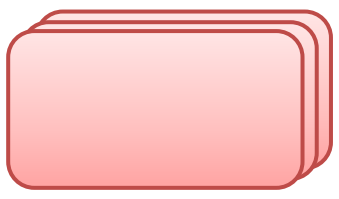 | Stop means that the process has failed. | 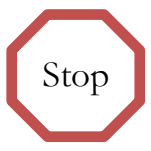 |
| Process or action | 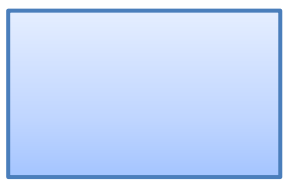 | Go means it is successful, therefore process with the implementation phase | 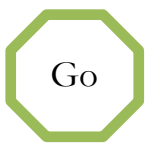 |
| Iterative process. | 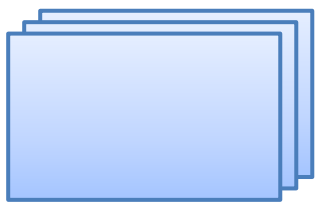 | Compare two objects. |  |
| Extension or Transformation of object | 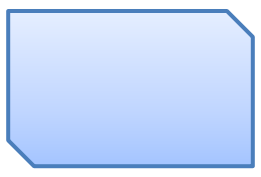 | Compare multiple objects |  |
| Extension or transformation of many objects | 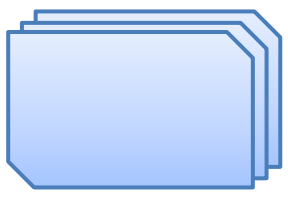 | | |
| Comments | 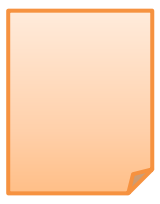 | | |
| Data, Process, Information flow | 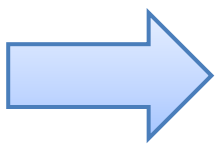 | | |
| Page connector | 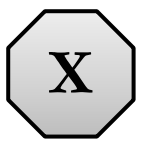 | | |
| Repeat process (Go to previous step) | 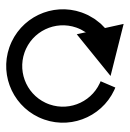 | | |

## 6.1 Composability Verification Process

**Figure 43** to **Figure 53** illustrate the composability verification process in the form of a flow chart. (The illustrated steps are explained later in this section):

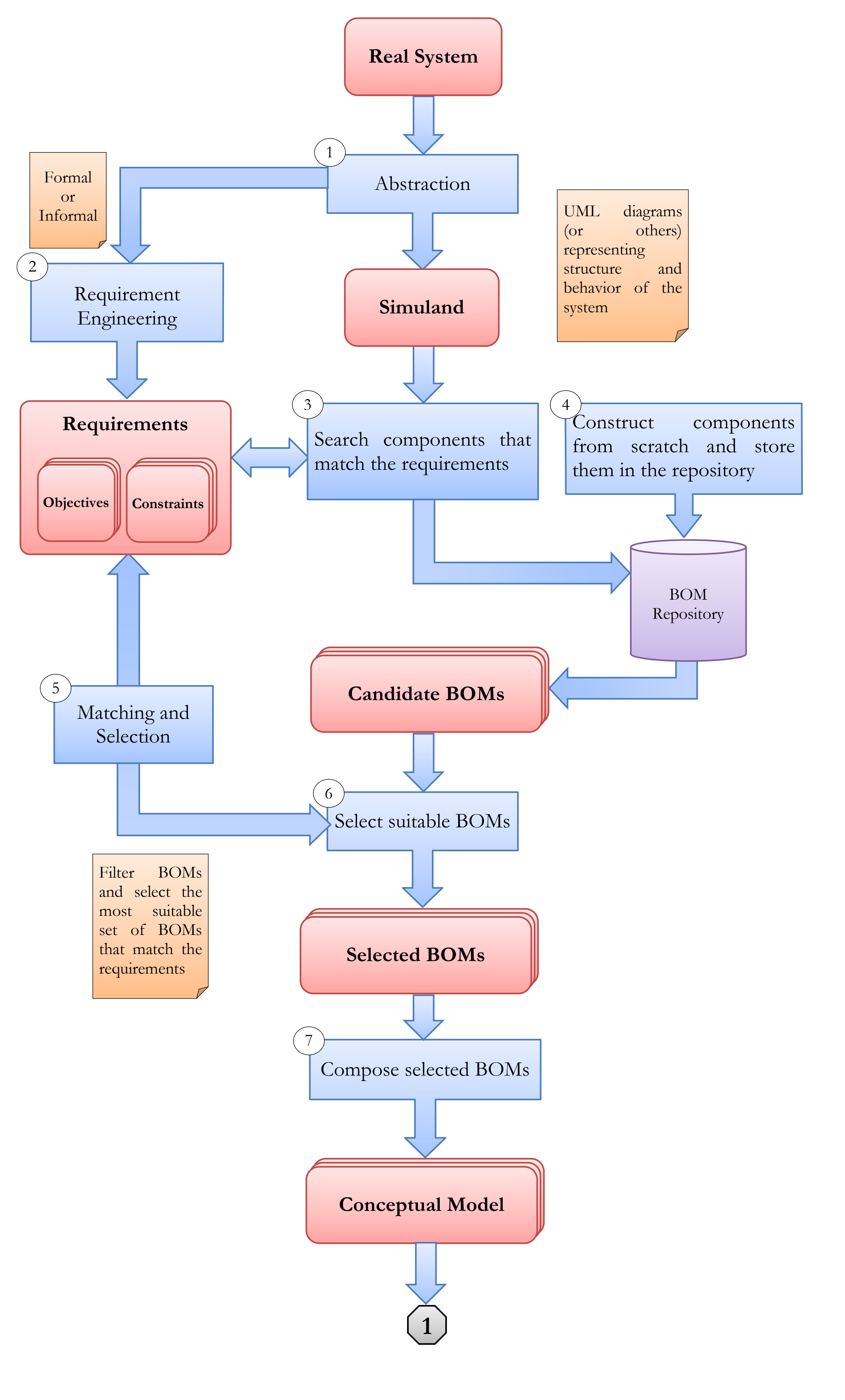

**Figure 43: Formulation of Simuland, Requirements and Conceptual Model**

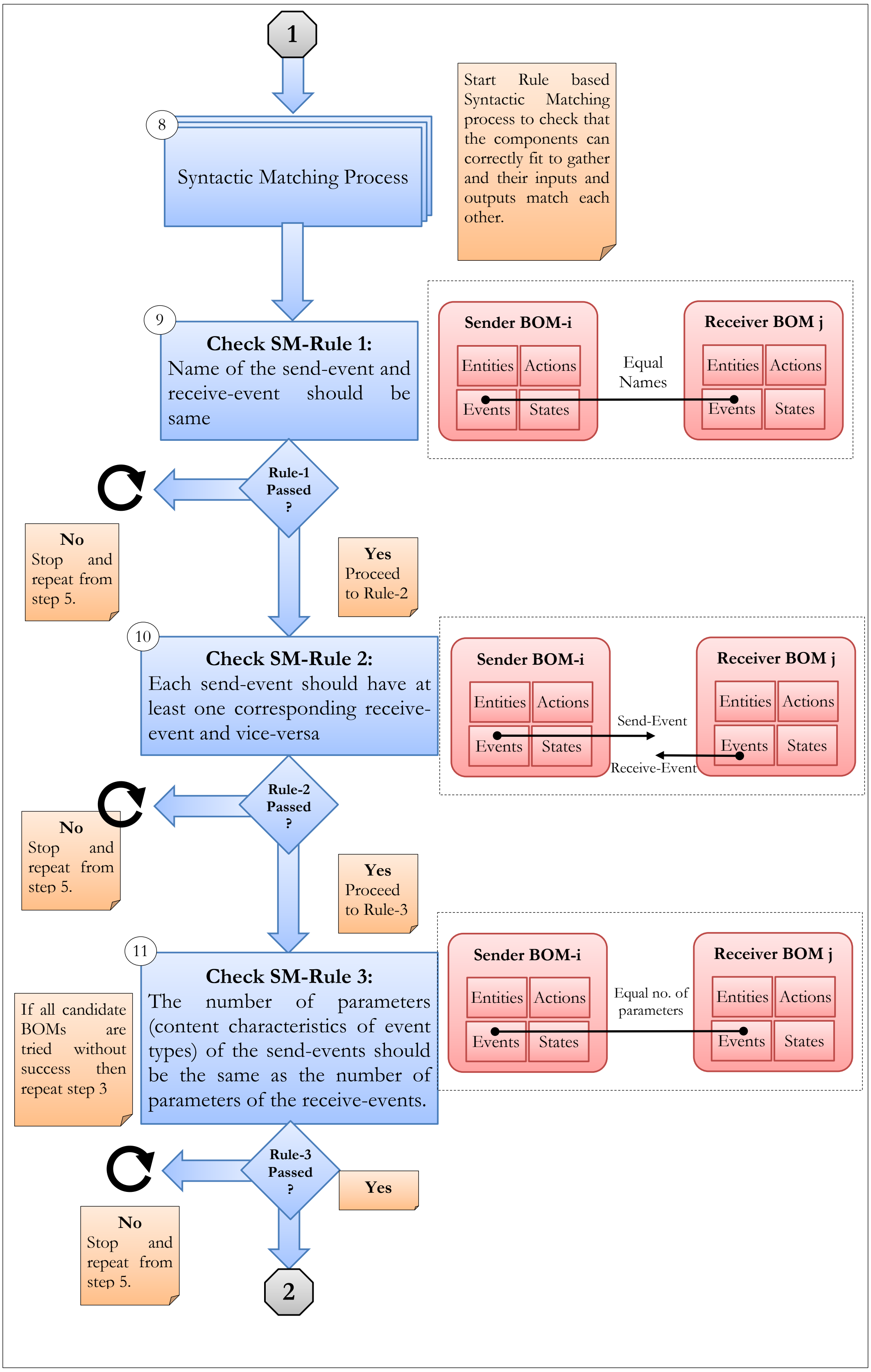

**Figure 44: Syntactic Matching Process**

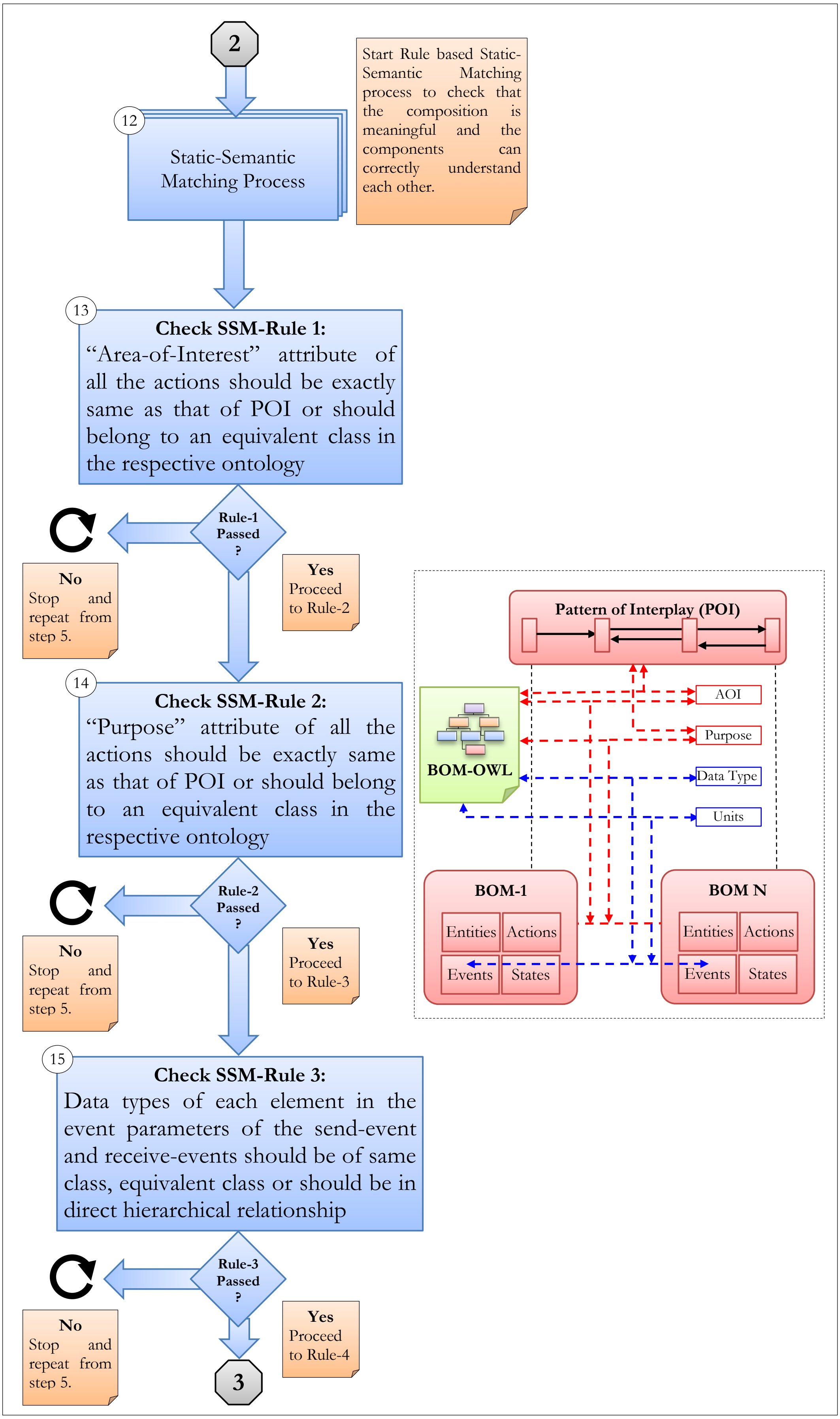

**Figure 45: Static-Semantic Matching Process**

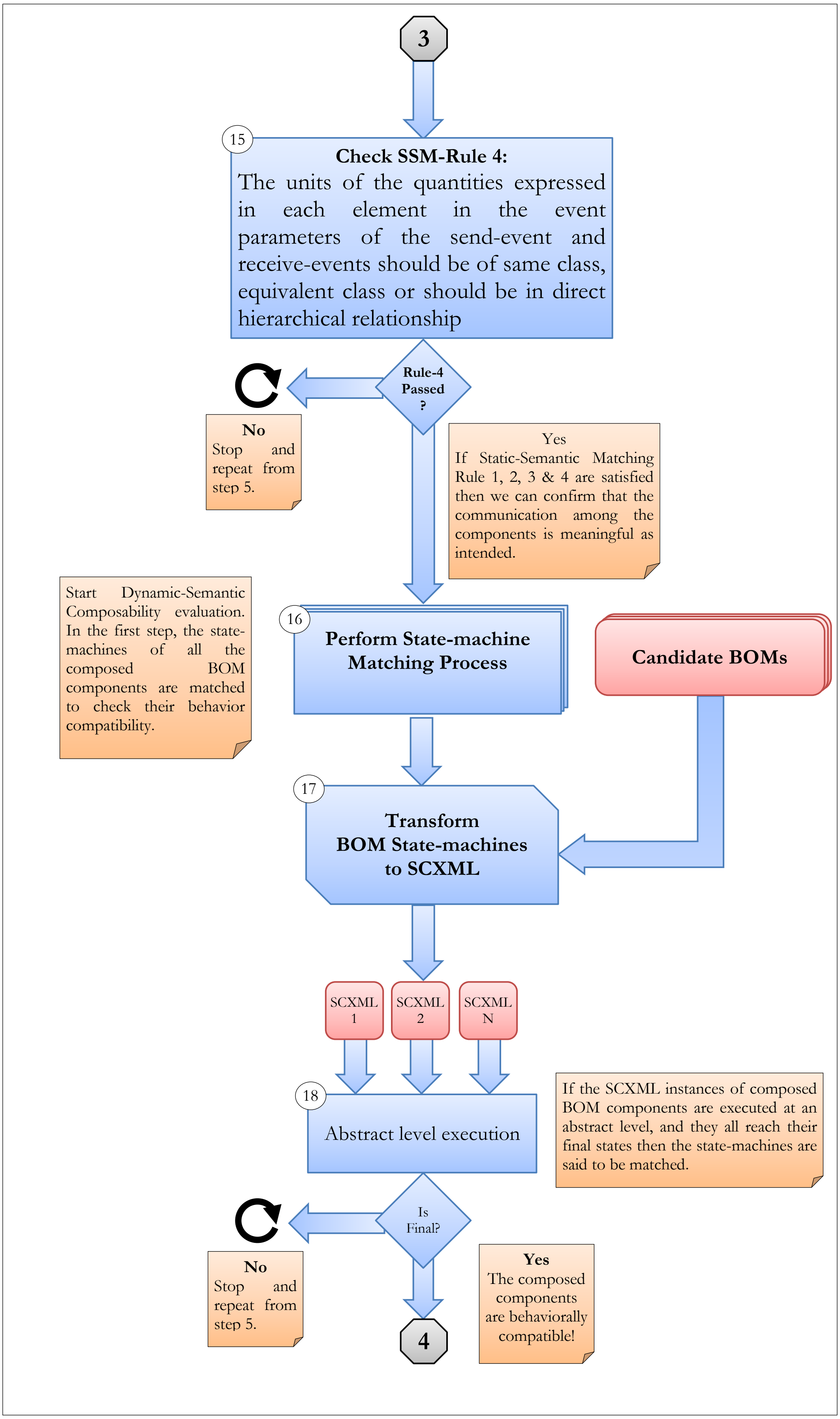

**Figure 46: State-machine Matching Process**

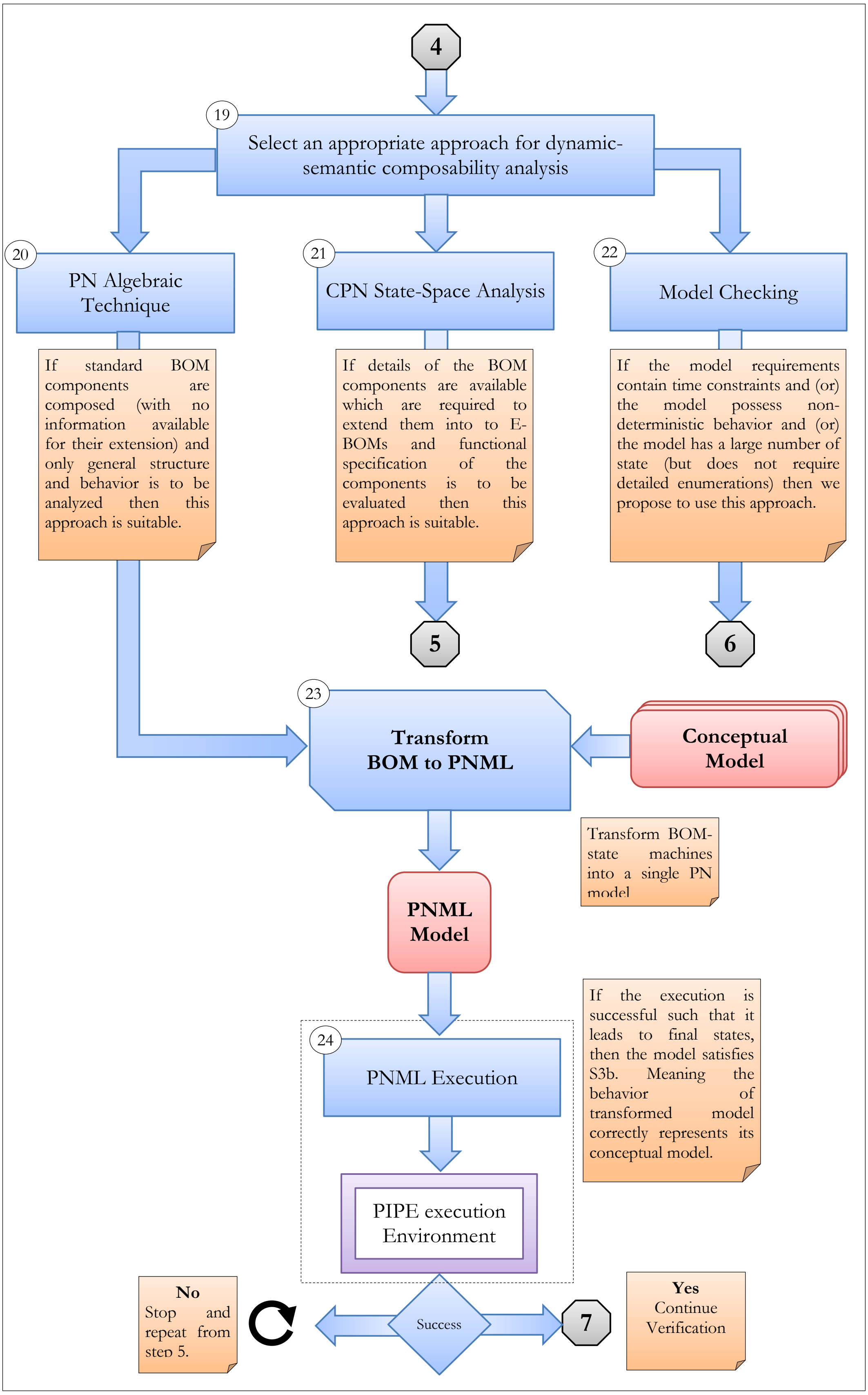

**Figure 47: Approach Selection | PN Algebraic Technique**

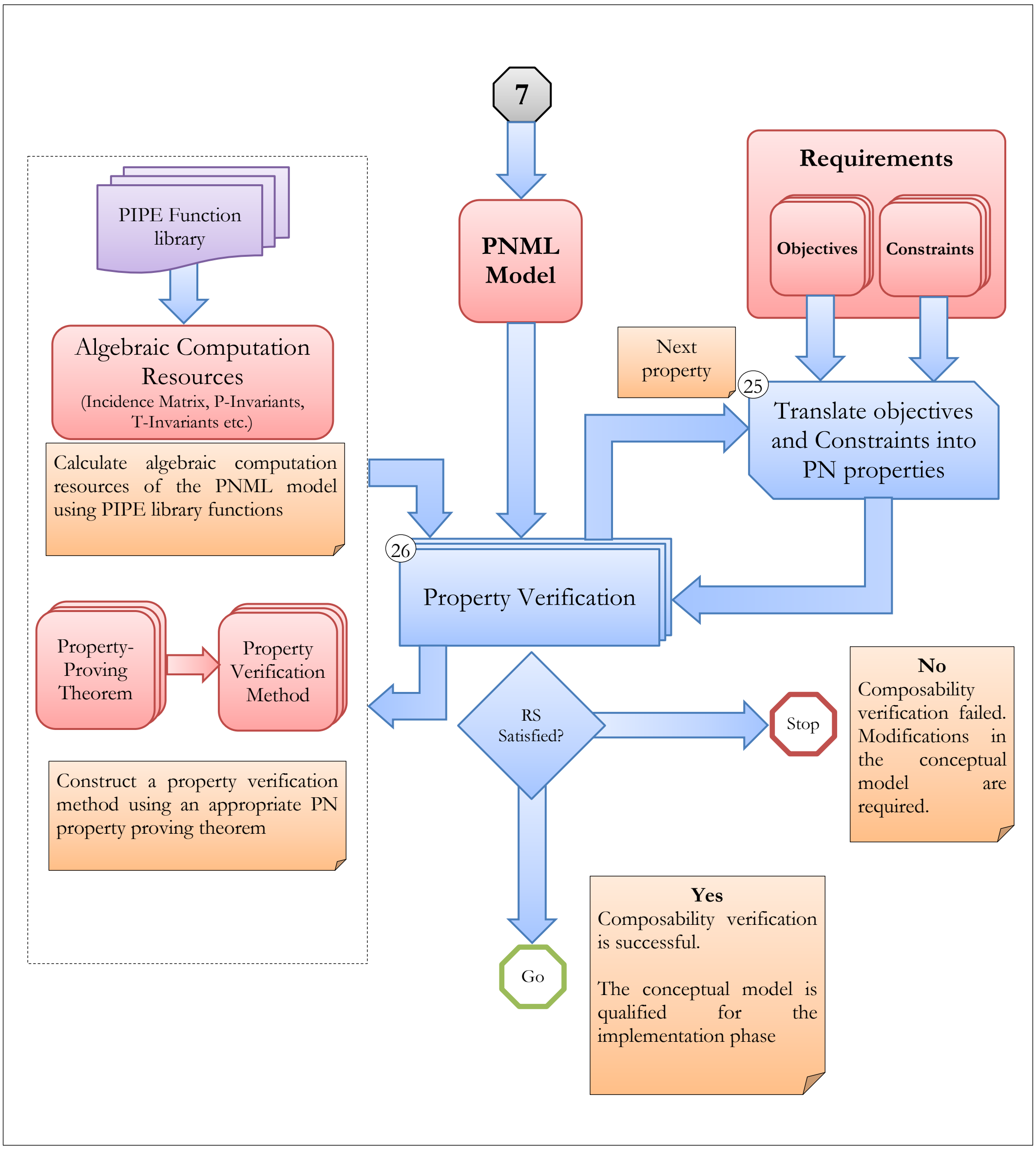

**Figure 48: PN Algebraic Technique (continued)**

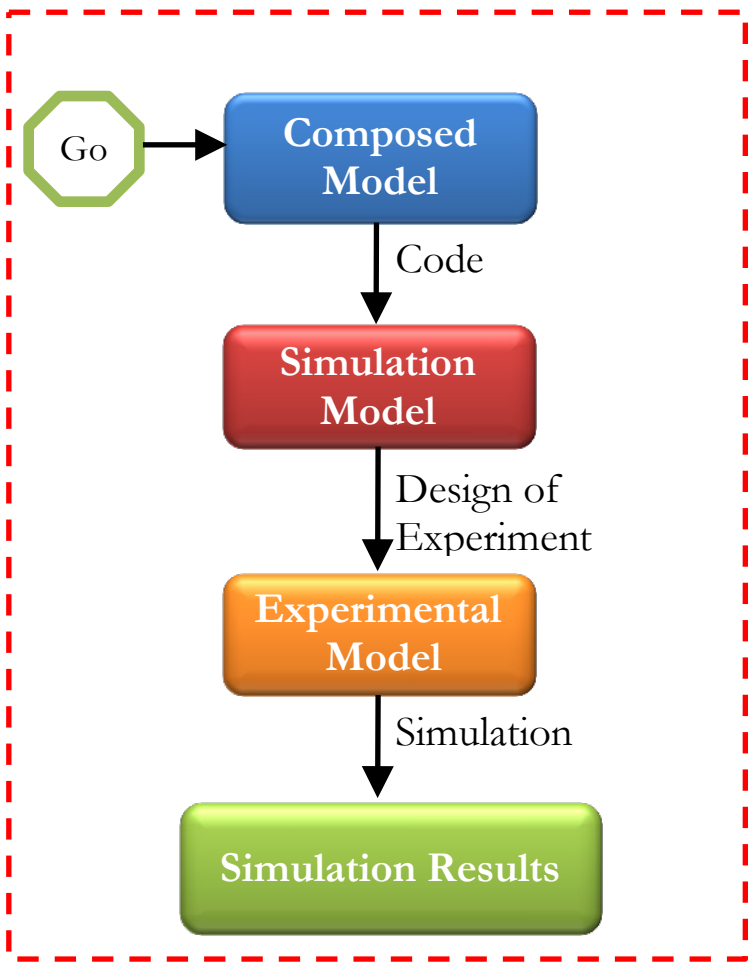

**Figure 49: Implementation**

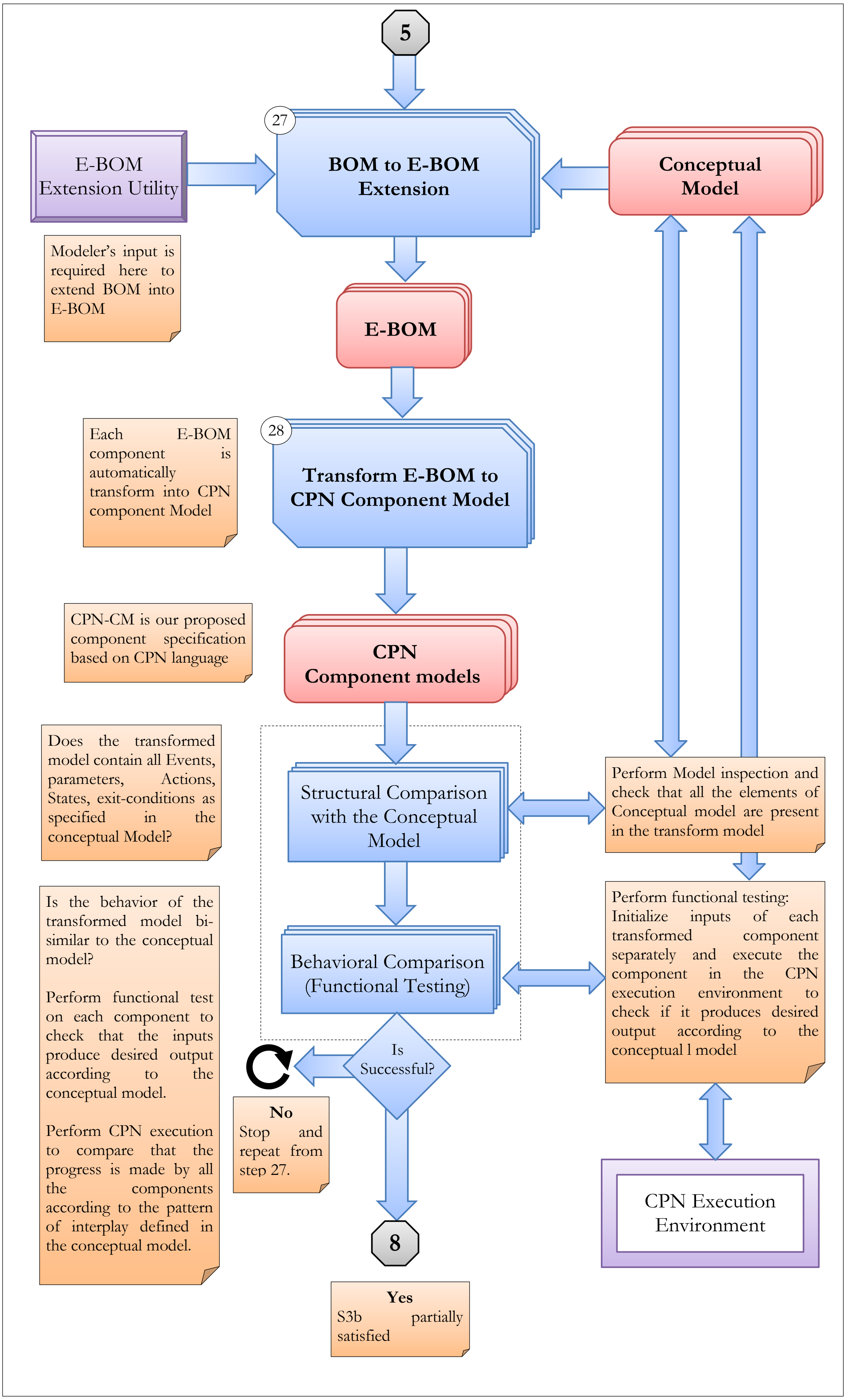

**Figure 50: State-Space Analysis Technique**

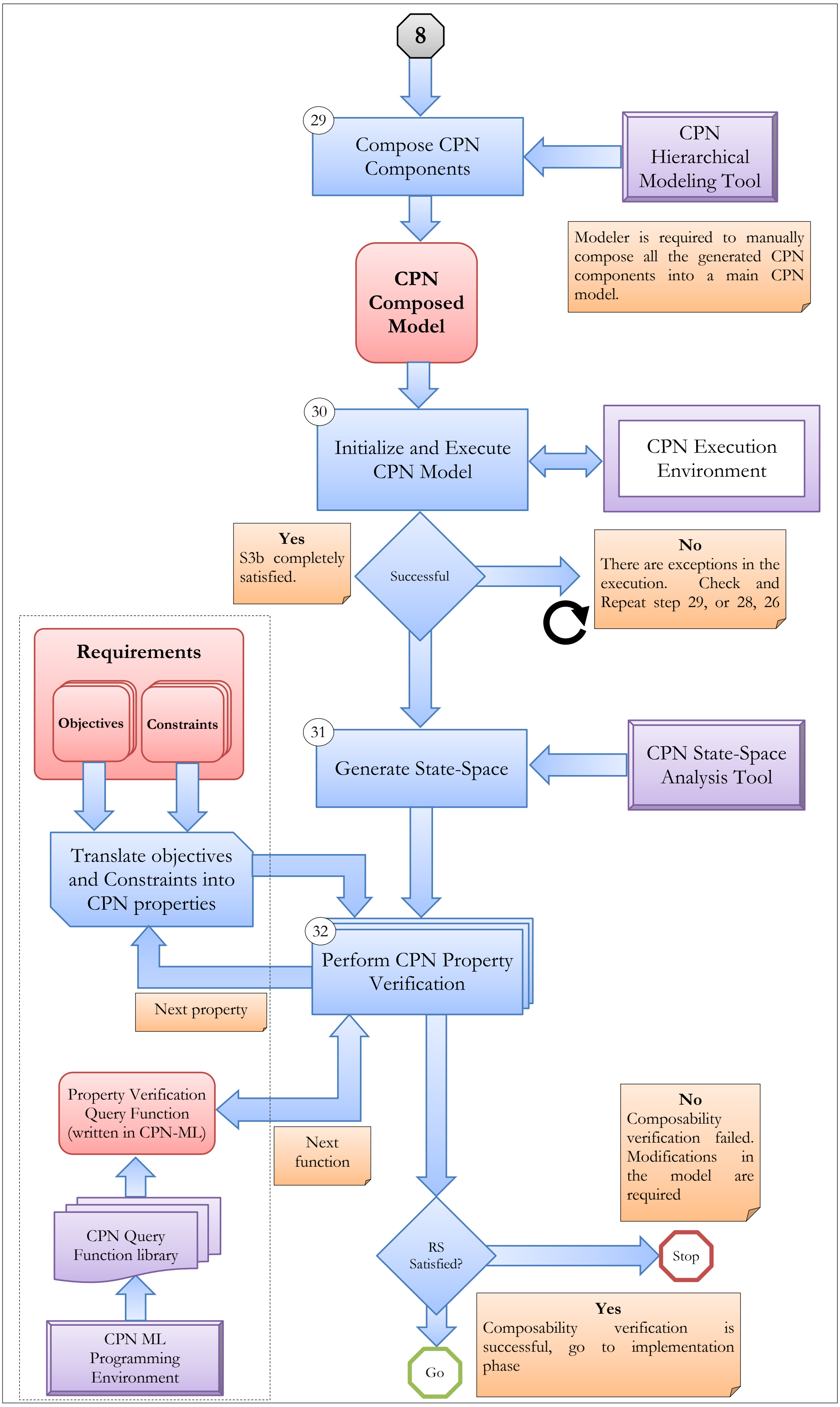

**Figure 51: State-Space Analysis Technique (continued)**

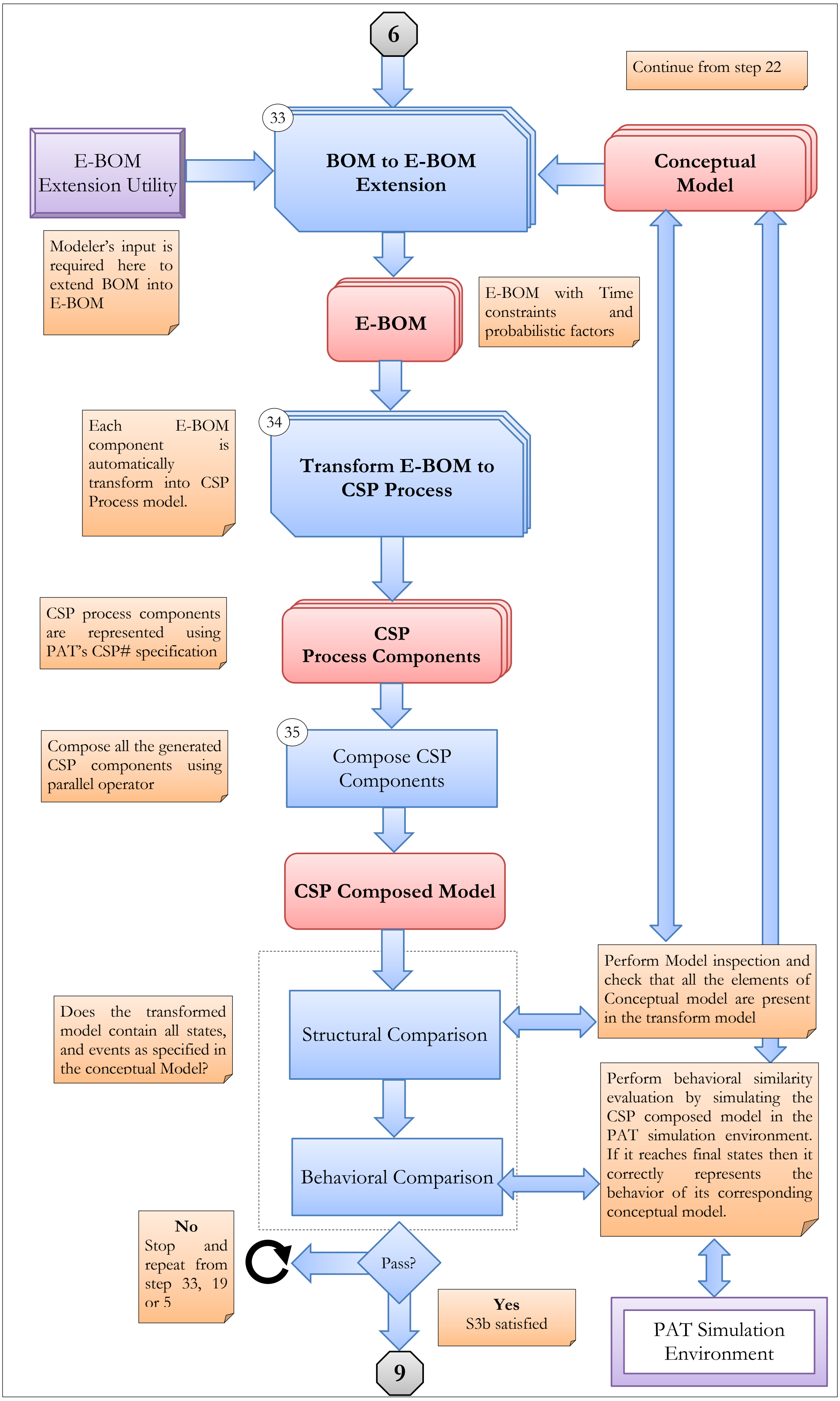

**Figure 52: Model Checking**

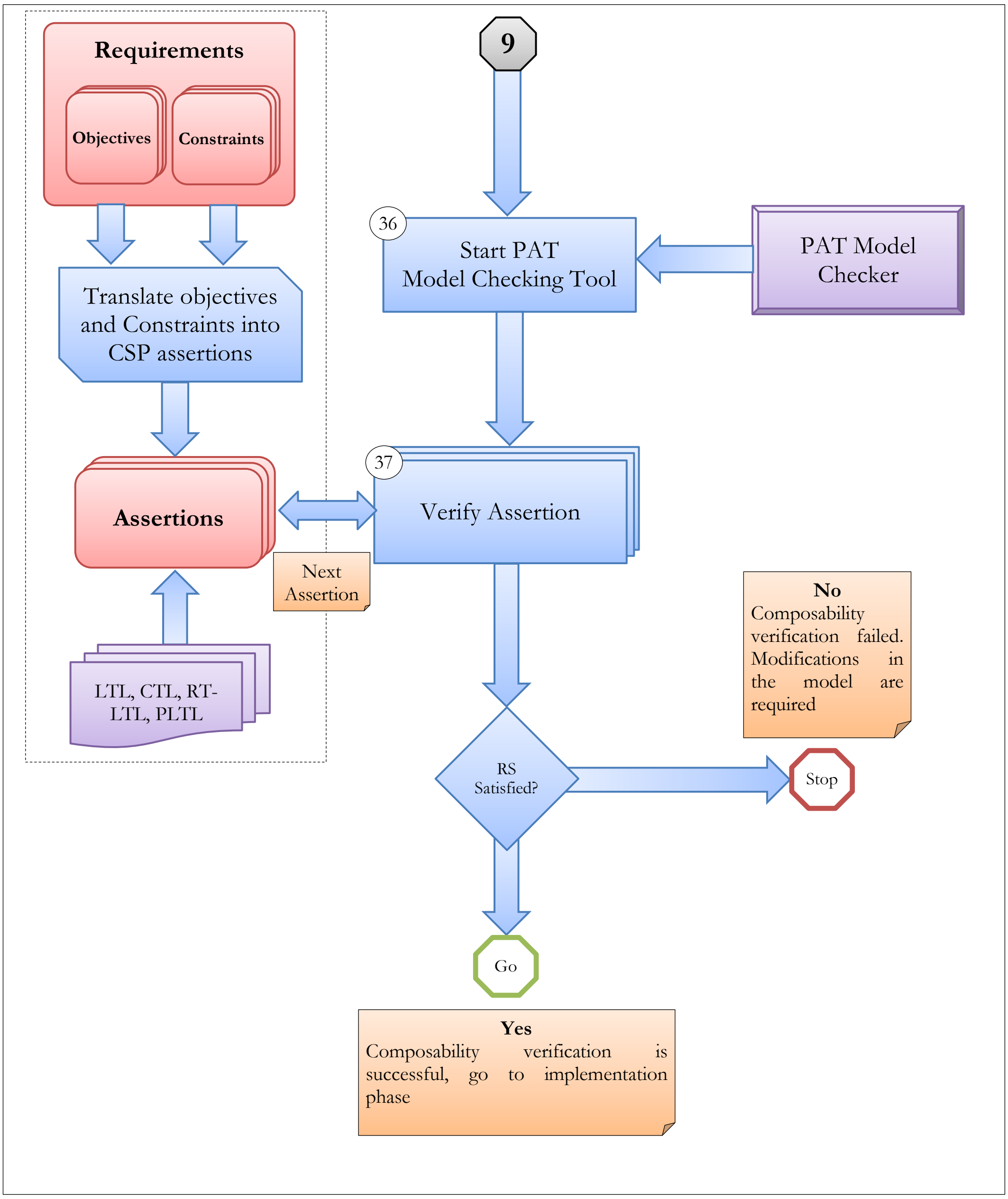

**Figure 53: Model Checking (continued)**

The Composability Verification Process is explained as follows:

### 6.1.1 Formulation of Simuland, Requirements and Conceptual Model

In step 1, the Real system is studied and a suitable simuland is formulated. It can be described formally or informally. We assume that UML diagrams are used to describe the simuland. The system is also studied to gather requirements and formulate requirements using our proposed formal requirement specification method (step 2). With this information at hand, suitable components are searched in the BOM repository, with an assumption that a composition of these components will form a conceptual model that represents the simuland (step 3). If a desired component is not found it is constructed form the scratch, added to the repository, and then used in the current context (step 4).

The discovered components are called candidate components. Among these candidates, most suitable ones i.e., those that best match the simuland and the given requirements, are selected (step 5, 6).

These BOM components are composed and a conceptual model is constructed (step 7). We recommend that the modeler also creates a formal model of the conceptual model using our proposed BOM formalism and graphical notation. This will help in documentation and understanding details of conceptual model and its composition.

### 6.1.2 Syntactic Matching Process

When the verification process starts the Composed BOM model (conceptual model) is passed through a rule-based static analyzer to verify the composability at syntactic level (step 8-11). If this level is passed then the constraint S1 (as defined in **Table 10**) is satisfied and only then the model is cleared for the next step (otherwise the verification process is stopped and another candidate selection is picked, composed and this step is revised).

### 6.1.3 Static-Semantic Matching Process

In the next step the components are analyzed at static-semantic level using the semantic analyzer (step 12-15). When this step is passed then the constraint S2 (as defined in **Table 10**) in is satisfied and the BOM composition is ready to be verified at dynamic-semantic level.

### 6.1.4 State-machine Matching Process

At this level, the first step is to perform state-machine matching of BOM components using State-machine checker (step 15 – 18). A successful state-machine matching satisfies the constraint S3a (as defined in **Table 10**).

### 6.1.5 Approach Selection for Dynamic-Semantic Composability Verification

In the next stage the verification framework offers three choices of verification technique for the analysis of dynamic-semantic composability level. The modeler can choose algebraic technique if there is no information available to extend the BOM components into E-BOM. Therefore the conceptual model will be transformed into PNML without requiring any extension. If the modeler has details and data available to transform BOM into E-BOM and the model does not represent a real-time system, then it is highly recommended that the second proposed approach (CPN state-space analysis) should be chosen. If the model represents a real-time system and it is stochastic in nature then the modeler should choose the third approach (Model checking). These are general guidelines and are not concrete rules. The ultimate choice of the approach depends on the nature of the model, nature of the requirement specification properties and the available information.

### 6.1.6 PN Algebraic Technique

When Algebraic technique is selected, at first the conceptual model is transformed into PNML model (step 23). This PNML model is executed in PIPE execution environment to evaluate S3b (step 24). If successful then the requirement specification properties are taken one by one and translated into a PN property (step 25). Thereafter a property proving theorem is selected that proves this PN property. Based on this theorem a property verification method is constructed inform of an algorithm. Running this algorithm proves or falsifies the requirements specification property (step 26). If all the properties in the requirement specification are satisfied then the model is successfully verified otherwise the process is stopped and model refinements are made.

### 6.1.7 State-Space Analysis Technique

When the CPN state-space analysis technique is selected, at first each BOM component is extended to E-BOM (step 27). This step requires modeler's input and can be delivered using the BOM-to-E-BOM extension utility. When the extension is complete, each E-BOM is transformed into our proposed CPN component model using our automatic transformation tool (step 28). The output of this step is a set of CPN components. At this step it is required to conduct structural and behavioral comparison between the generated components and the respective BOM using inspection and functional testing methods. If the comparison is successful then S3b constraint of the requirement specification is partially satisfied.

The modeler is then required to compose these generated components in a main model using CPN hierarchical tool (step 29). (Binding IN-ports and OUT-ports of each component using sockets in the main model). When the model is composed, it is executed (step 30) using CPN execution environment to test that all components correctly interact with each other and make necessary progress to reach the final states. If the execution is successful then the constraint S3b is fully satisfied, conforming that the structure and behavior of the executable model correctly represents its respective conceptual model and therefore any verification operation performed on the executable model will imply correctness of its conceptual model.

In the next step the CPN model is subjected to the state-space analysis (step 31). At first a state-space graph of the model is generated. Then for each objective and constraint in the requirement specification a verification query function is either created or selected from the function library. The execution of this function is done using CPN-ML program execution environment and the result of this function tells if the property is satisfied or violated (step 32). If all the properties are satisfied we say that the composability verification process is successful.

### 6.1.8 Model Checking

When the Model Checking technique is selected, at first each BOM component is extended to E-BOM (step 33). This step requires modeler's input and can be delivered using the BOM-to-E-BOM extension utility. It is possible to assign Time constraints and the probabilistic factors with the states or the transitions. When the extension is complete, each E-BOM is automatically transformed into CSP# process specification (step 34) and composed (step 35). The composition of each CSP process representing a BOM component be done using sequential operator ';' parallel operator '||' interleaving operator '|||' or (non-deterministic or user's) choice operator '[ ]' depending upon the nature of the composed components. We suggest composing each CSP in parallel so that each process executes in parallel and synchronizes with each other by sending or receiving events at their respective communication channels. The composed model can then be simulated using the PAT simulator. A successful simulation run with at least one path leading to the final state(s) shows that the behavior of the composed model correctly represents its conceptual model and thus satisfies constraint S3b.

In the next step, the assertions defined in CSP format (using LTL, Real-Time LTL or PLTL) are verified using PAT model checker which results in its satisfaction or violation (step 36, 37). If all the assertions are satisfied we say that the composability verification process is successful.

## 6.2 Summary

In this chapter, a flow diagram of composability verification process is presented. It indicates different steps and forms of inputs and outputs of each step in the process. This flow diagram can be used as a guideline to perform composability verification using three different approaches. Some recommendations are also presented in making a suitable choice. Once the verification process is completed successfully, the composed model can undergo implementation phase where it is programmed and simulated using a suitable simulation platform. Also the experimental model can be constructed to perform different experiments on the implemented model and simulation results are generated for study and decision making. The implementation phase is out of the scope of this thesis.

# Chapter 7
# Fairness verification using PN Algebraic Techniques

*This chapter explains how algebraic techniques can help in verifying system properties of a Composed Model, using an example of a manufacturing system in which fairness is selected to be the required system characteristic.*

## 7.1 Fairness

Fairness has been defined in section 3.1.3 in terms of a Petri Nets property. In this chapter the concept of fairness is covered in detail. Intuitively, fairness is a liveness property that means no component of the system which becomes possible (or becomes enabled) sufficiently often should be delayed indefinitely. [**122**].

On the basis of the extent of sufficiency, fairness is generally categorized in the following three types in literature:

**Unconditional Fairness**

Also called Impartial implies that every component in a system proceeds infinitely often without any condition. The term "proceed" means to make progress, (e.g., firing of a transition). Unconditional fairness is also known as non-deterministic choice and is usually present among the components that are independent of each other [**122**].

**Weak Fairness**

Also called Just, implies that every component in a system that is enabled continuously from some point onwards eventually proceeds.

**Strong fairness**

Every component in a system that is enabled infinitely often proceeds infinitely often. A noticeable difference in weak and strong fairness is that weak fairness involves persistent enabling of a component that wants to proceed, whereas strong fairness is not persistently enabled.

Some important generalizations of fairness exist in literature [**122**]:

**Equi fairness:** means to give each component an equal chance to proceed. It can be regarded as Justice. This type of fairness does not always apply in real world scenarios because of priority policies or some other reasons.

**Bounded fairness:** means to give each component an equal number of chances such that no component proceeds for more than "k-times" without letting the others to take their turn

For instance there is a check-in service at the airport that serves two types of queues: (i) Business class and (ii) Economy class at a time. It will be called fair, if it mostly serves business class passengers but not more than (say) 10 times, without serving a passenger from the economy class queue.

In Petri Nets, fairness can be viewed in two perspectives namely: Transition fairness and Marking fairness. The former corresponds to fairness of choice of transitions, and the latter deals with the fair reachability of states.

## 7.2 Fairness Verification

There are different ways to verify fairness of a model. The focus of this chapter is to discuss the technique for the verification of fairness property using PN Algebraic analysis and provide the necessary and sufficient conditions for a PN model to be fair. The evaluation of these conditions in a PN model involves theorems and linear algebraic computations; therefore it is classified as an *Algebraic technique.* Based on the theorems below, we propose an algorithm for automatic fairness verification.

In Petri Nets, fairness is mainly perceived in terms of occurrences (or firing) of transitions. Two transitions $t_1$ and $t_2$ are said to be in a fair relation if there exists a positive integer **k** such that for any reachable marking **M** and any firing sequence σ**:**

$$\#(t_1/\sigma) = 0 \Rightarrow \#(t_2/\sigma) \leq k \ \wedge \#(t_2/\sigma) = 0 \Rightarrow \#(t_1/\sigma) \leq k \quad (7.1)$$

(The symbol #(t/σ) denotes the number of times a transition t occurs in a firing sequence σ)

In words, neither of the transitions should occur more than a finite number of times (k) without letting the other to occur at least once. This is known as *bounded fairness* (or *B-Fairness*) *with upper bound = k*. If every pair of transition is in a bounded fair relation, then the entire net is said to be fair [**123**].

For the algebraic verification of fairness property in a PN model the following theorems are applied. Details and proofs of these theorems are discussed in [**123**].

**Theorem I**

*Given a PN with an incidence matrix A, if there exists a firing-count vector X, such that:*

$$A.X \geq 0 \text{ and } X \neq 0$$

*Then a necessary condition for the PN to be fair is that each entry of* ***X*** *is positive.*

**Theorem II**

*If a Petri Net* ***N*** *is* **bounded** *for any initial marking* $M_0$ *then the condition in Theorem I is necessary and sufficient for N to be fair.*

**Corollary:** *If there exists a P-Invariant Y of positive integers such that:* $A.Y=0$ *then the PN is guaranteed to be structurally bounded.*

**Theorem III**

*A fair Petri Net PN has only one reproduction vector (i.e., a minimal T-Invariant) at the most.*

Based on the above definition of the bounded fairness and theorems I, II and III a PN is said to be fair if it satisfies two conditions: (i) There must exist a **single** T-Invariant X of a given PN model whose each entry is non-zero and the product AX = 0 and (ii) There must be at least one P-Invariant, which means that the net is structurally bounded.

## 7.3 Manufacturing system

In this section, a component based composed model of a manufacturing system is presented. Using this composed model, it is shown how the proposed verification framework is used to verify the required specification, which in this example consists of fairness as an important quality constraint. Two different scenarios of this example are discussed. In the first scenario the model is shown to be unfair as verified by our verification framework. In the second scenario the model is modified. It is then verified and it satisfies the fairness constraint.

### 7.3.1 Scenario I

It is assumed that the manufacturing system model is composed of two machines M1 & M2 and a shared Robot R as shown in **Figure 54**. The robot loads raw material on the machines and operate on them for producing goods. The Robot is assigned ("loaded") to either of the machines at a time. When the Robot is loaded, it deposits raw material on the machine and process it. When the good is produced the robot is unloaded and is available for the other machine.

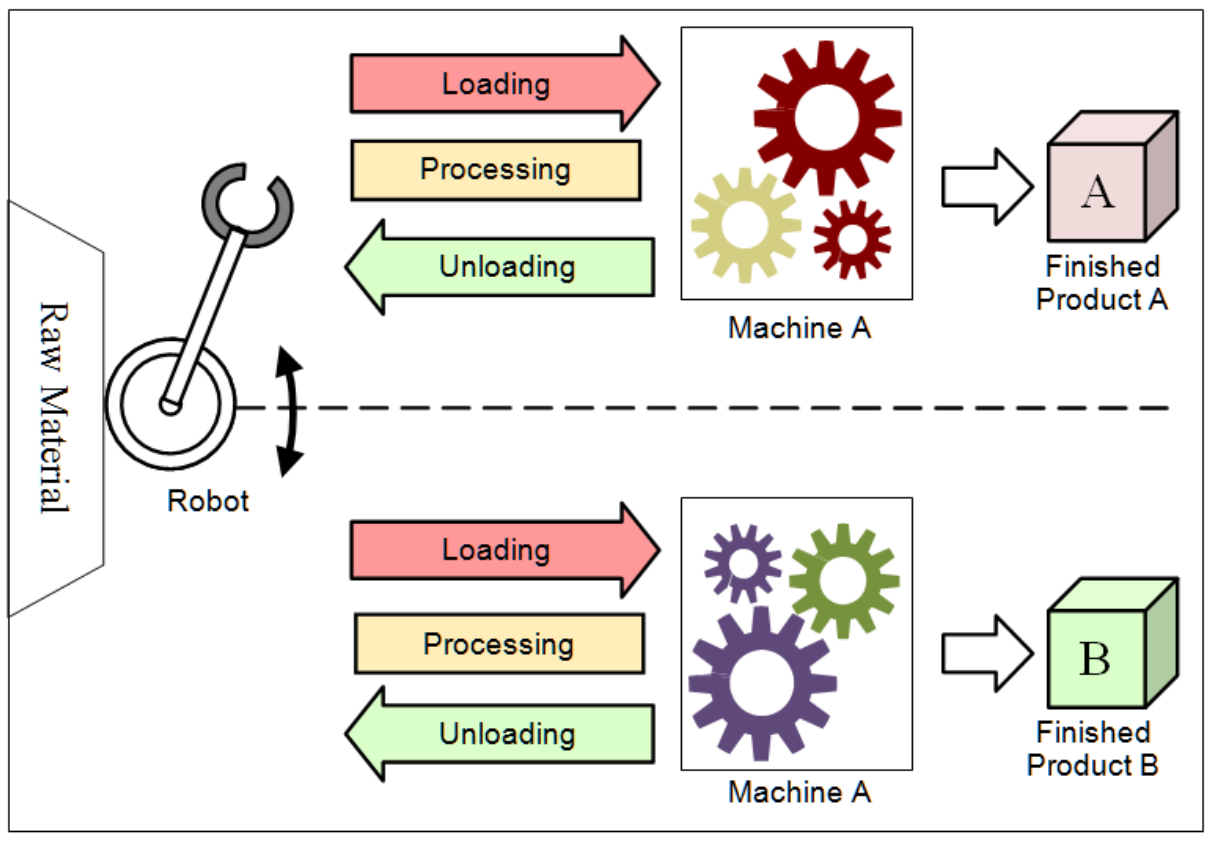

**Figure 54: Manufacturing System** (acquired from [**124**])

The process of composability verification is initiated as follows:

#### Simuland and Requirement Specification

In the first step, the entities, events and the states of the simuland are perceived according to **Figure 54**. The simuland and the requirement specifications are used to construct an appropriate conceptual model according to the steps given in the composability verification process described in Chapter 6.

We define Requirement speciation of the manufacturing system as:

$$\mathbf{RS_0} = \langle \mathbf{O}, \mathbf{S} \rangle \text{ where:}$$

| **Objectives O = {$o_1$, $o_2$, $o_3$}** |
|---|
| **$o_1$:** Machine1 should continuously produce product1 without any infinite delay |
| **$o_2$:** Machine2 should continuously produce product2 without any infinite delay |
| **$o_3$:** Both machines should produce products with a ratio of 1:1 |

**System Constraints S = {$s_1$, $s_2$, $s_3$, $s_4$}**

**$s_1$:** Machine1, Machine2 and the Robot components should be composable at syntactic level

**$s_2$:** Machine1, Machine2 and the Robot components should be composable at static-semantic level

**$s_{3a}$:** State-machine matching of the composed model should be successful. Since the models are non-terminating so there are no final states, instead the goal-states: "*Machine1 completes production*" & "*Machine2 completes production*" will be considered.

**$s_{3b}$:** The transformed executable model correctly represents the structure and behavior of the conceptual model.

**$s_4$:** The shared robot should treat both machines with fairness (i.e., k-fairness; k=1).

### Conceptual Model

The formal specification and graphical representation of each BOM model participating in the manufacturing composed model are as follows. For the ease of readability following color codes are used for different BOM elements in the formal definition:

**$BB_0$ = ⟨ EnT, EvT, S, AcT ⟩** where:

**EnT** = Machine1 {$C_0$(Id:Integer)}

**EvT** = {$E_0$(LoadingM1, Robot, Machine1, null), $E_1$(UnloadingM1, Robot, Machine1, null), $E_2$(ResetM1, Robot, Machine1, $C_0$)}

**Act** = { $A_0$(LoadingM1, Robot, Machine1, $E_0$), $A_1$(UnloadingM1, Robot, Machine1, $E_1$), $A_2$(ResetM1, Robot, Machine1, $E_2$)}

**S** = {$S_0$(M1Waiting, $A_0$, $S_1$), $S_1$(M1Processing, $A_1$, $S_2$), $S_2$(M1Completed, $A_2$, $S_0$)}

**Table 21: Formal definition of Machine1 Base-BOM**

**$BB_1$ = ⟨ EnT, EvT, AcT, S ⟩** where:

**EnT** = Machine2 {$C_1$(Id:Integer)}

**EvT** = {$E_3$(LoadingM2, Robot, Machine2, null), $E_4$(UnloadingM2, Robot, Machine2, null), $E_5$(ResetM2, Robot, Machine2, $C_1$)}

**Act** = { $A_3$(LoadingM2, Robot, Machine2, $E_3$), $A_4$(UnloadingM2, Robot, Machine2, $E_4$), $A_5$(ResetM2, Robot, Machine2, $E_5$)}

**S** = {$S_3$(M2Waiting, $A_0$, $S_4$), $S_4$(M2Processing, $A_1$, $S_5$), $S_5$(M2Completed, $A_2$, $S_3$)}

**Table 22: Formal definition of Machine2 Base-BOM**

**$BB_2$ = ⟨ EnT, EvT, AcT, S ⟩** where:

**EnT** = Robot {}

**EvT** = {$E_6$(LoadingM1, Robot, Machine1, null), $E_7$(UnloadingM1, Robot, Machine1, null), $E_8$(LoadingM2, Robot, Machine2, null), $E_9$(UnloadingM2, Robot, Machine2, null)}

**Act** = {$A_6$(LoadingM1, Robot, Machine1, $E_6$), $A_7$(UnloadingM1, Robot, Machine1, $E_7$), $A_8$(LoadingM2, Robot, Machine2, $E_8$), $A_9$(UnloadingM2, $BB_2$, Machine2, $E_9$)}

**S** = {$S_6$(Idle, {$A_6$, $S_7$},{$A_8$, $S_7$} ), $S_7$(Busy, {$A_7$, $S_6$}, {$A_9$, $S_6$})}

**Table 23: Formal definition of Robot Base-BOM**

**$CB_0$ = ⟨ $AcT_{IN}$, $AcT_{OUT}$ , POI ⟩** where:

**$AcT_{IN}$** = **$AcT_{OUT}$** = ∅

**POI** = {$POI_0$(**!$A_6$** , ?$A_0$), $POI_1$(**!$A_7$** , ?$A_1$), $POI_2$(**!$A_2$**, ?$A_2$), $POI_3$(**!$A_8$** , ?$A_3$), $POI_4$(**!$A_9$** , ?$A_4$), $POI_5$(**!$A_5$**, ?$A_5$) }

**Table 24: Formal definition of Manufacturing System composed BOM**

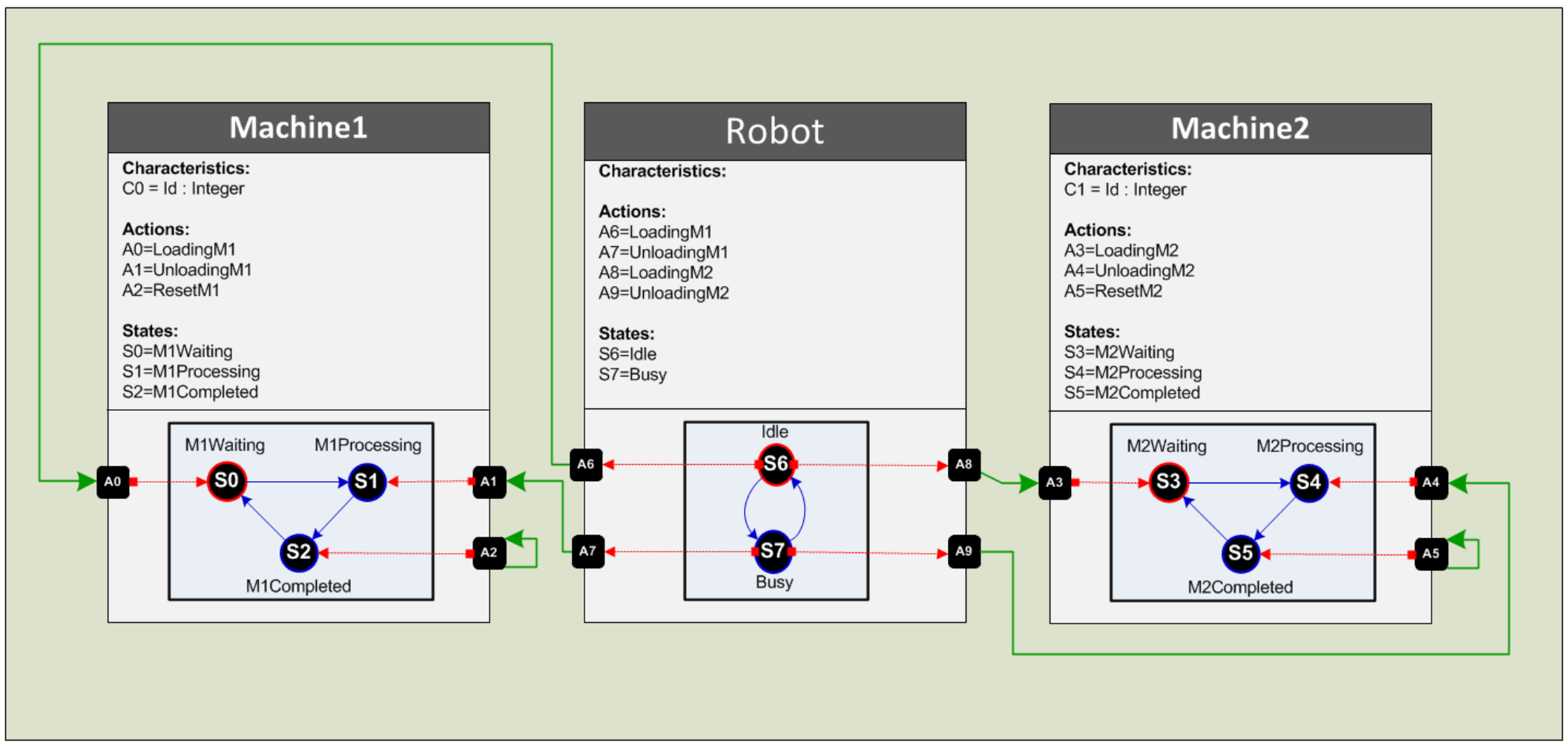

**Figure 55: Manufacturing System BOM based Composed Model**

Figure **55** represents the BOM based Conceptual Model of the manufacturing system which includes three BOMs, formally defined using our proposed graphical notation. The figure shows how the characteristics, Events, Actions and states are mapped to each other (using dotted red line). In machine 1 characteristic $C_0$ is mapped to Event $E_2$ which means event uses characteristic $C_0$ as parameter. Similarly Event $E_0$ is mapped to $A_0$, $E_1$ to $A_1$ and $E_2$ to $A_2$ respectively which means the Actions uses their mapped events. The mapping of actions to the states in the figure shows which action will cause which state to transit to the new state (shown by blue arrow). The basic BOM components are connected to each other using the formal definition shown in **Table 24**, which describes the source (!) and destination (?) of an action from one component to other. In **Figure 55** this is shown using black arrow lines with their input/output (I/O) label. This is called Pattern of Interplay (in BOM specification).

### Static Analysis

| Rules | Machine1 | Machine2 | Robot |
|---|---|---|---|
| • Name of the send-event and receive-event should be same<br>• Each send-event should have at least one corresponding receive-event and vice-versa<br>• The number of parameters (content characteristics of event types) of the send-events should be the same as the number of parameters of the receive-events. | ?LoadingM1(null) | | !LoadingM1(null) |
| | ?UnloadingM1(null) | | !UnloadingM1(null) |
| | | ?LoadingM2(null) | !LoadingM2(null) |
| | | ?UnloadingM2(null) | !UnloadingM2(null) |

**Table 25: Syntactic Matching**

It can be seen in **Table 25** that the name of the send-event and receive-events are the same. ( !=Send, ?=Receive). And they are in one-to-one relationship. Also the no. of parameters of each event is equal to 1. Based on these facts the components are said to be syntactically composable (S1 satisfied).

We assume that Machine1, Machine2 and Robot components have the semantic-attributes as shown in **Table 26** which satisfy all the static-semantic matching rules.

The attributes highlighted in red color are semantically equivalent (Exact match) therefore S2 is satisfied.

| Machine1 | Machine2 | Robot |
|---|---|---|
| AOI = {Production, Manufacturing, Production-line, Lathing} | AOI = {Production, Manufacturing, Production-line, Polishing} | AOI = {Production, Manufacturing, Conveyer, Automation} |
| Purpose = {Manufacture Product1, Manufacture Product3} | Purpose = {Manufacture Product2, Manufacture Product3} | Purpose = {Manufacture Product3} |
| Data Types of parameters= {null} | Data Types of parameters = {null} | Data Types of parameters = {null} |
| Units of Measurement = {} | Units of Measurement = {} | Units of Measurement = {} |

**Table 26: Static-Semantic Matching**

### Dynamic Analysis

The state-machine matching process is successfully conducted as both Machine1 and Machine2 reach their goal-states namely: *Mcompleted* and *M2-completed* and satisfy S3a.

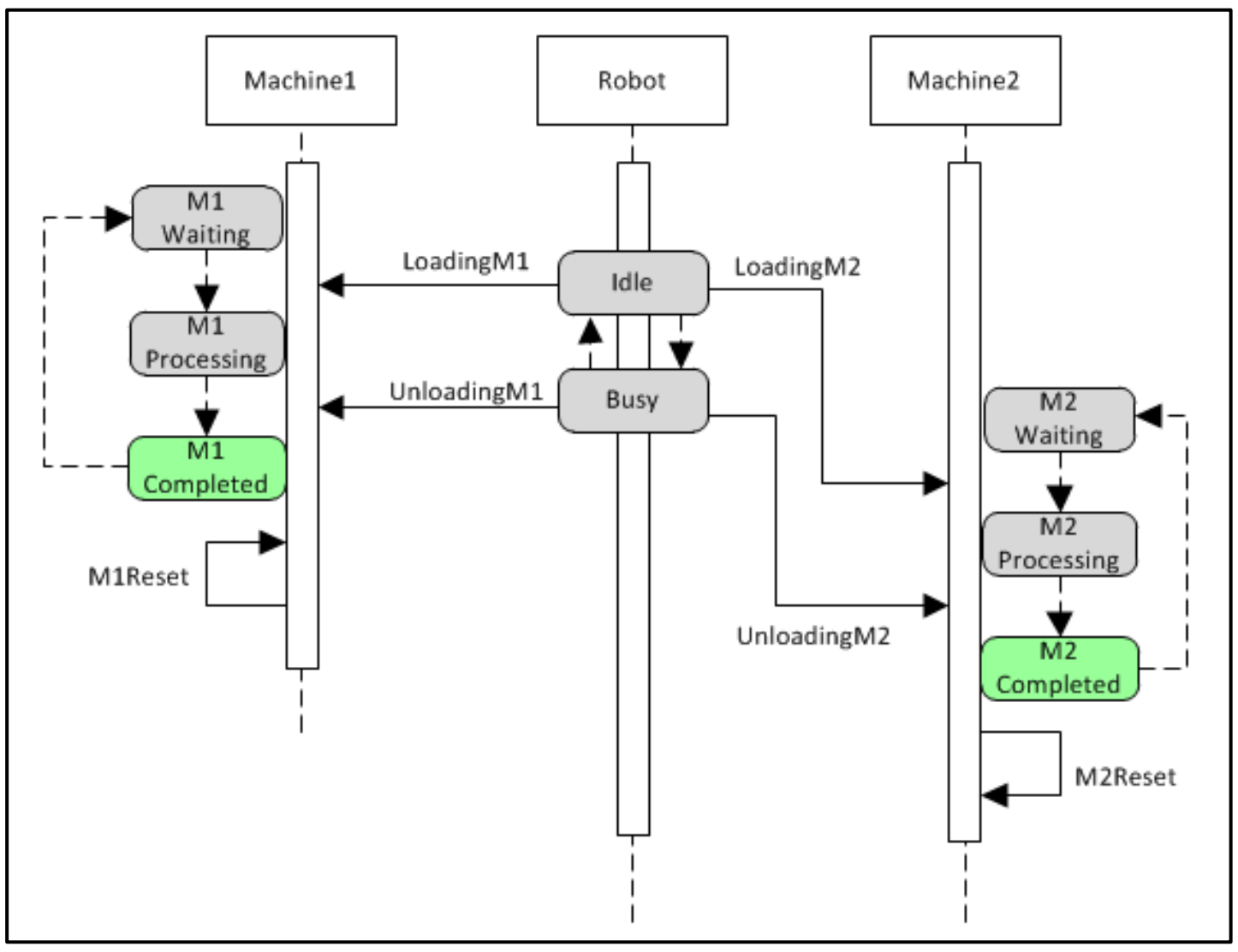

**Figure 56: State-machine matching of manufacturing system**

## BOM to PNML Transformation

In the next step the components are subjected to PNML transformation process. The output of the transformation process is a PN model shown in **Figure 57**. It can be seen from the inspection that the States and their exit conditions, Events and Actions all are present in the transformed model (as specified in the original conceptual model). Also this PN model is executed in the PIPE runtime environment. The execution is successful because the places P3 and P6 acquired tokens (showing that these goal states were reached during the execution). This satisfies S3b.

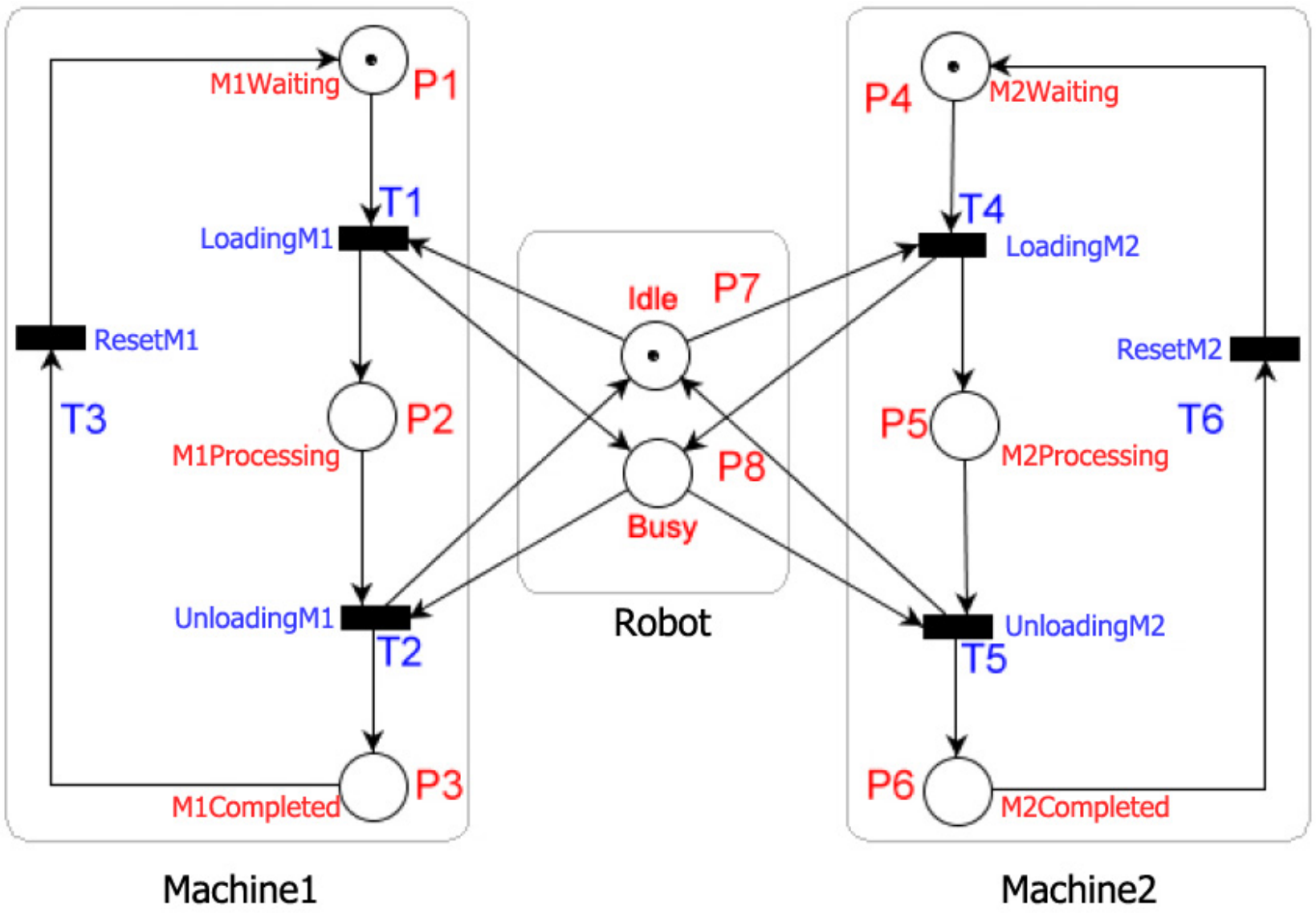

**Figure 57: PN model of the manufacturing System**

### Algebraic Resource Computation

At this step, the initial marking $\mathbf{M_0}$ and the Incidence Matrix **A** of the PN composed model shown in **Figure 57** are calculated using PIPE library functions as follows:

| $\mathbf{M_0}$ | P1 | P2 | P3 | P4 | P5 | P6 | P7 | P8 |
|---|---|---|---|---|---|---|---|---|
| | 1 | 0 | 0 | 1 | 0 | 0 | 1 | 0 |

| **A** | P1 | P2 | P3 | P4 | P5 | P6 | P7 | P8 |
|---|---|---|---|---|---|---|---|---|
| T1 | -1 | 1 | 0 | 0 | 0 | 0 | -1 | 1 |
| T2 | 0 | -1 | 1 | 0 | 0 | 0 | 1 | -1 |
| T3 | 1 | 0 | -1 | 0 | 0 | 0 | 0 | 0 |
| T4 | 0 | 0 | 0 | -1 | 1 | 0 | -1 | 1 |
| T5 | 0 | 0 | 0 | 0 | -1 | 1 | 1 | -1 |
| T6 | 0 | 0 | 0 | 1 | 0 | -1 | 0 | 0 |

**Table 27: Initial Marking and Incidence Matrix (Scenaro I)**

Note that the labels of rows and columns in **A** and elements in $\mathbf{M_0}$ correspond to places and transitions in **Figure 57**. The matrix **A** is given as input to the Invariant calculation module that calculates the following P-Invariants and T-Invariants in the PN model of the Manufacturing System:

| P1 | P2 | P3 | P4 | P5 | P6 | P7 | P8 |
|---|---|---|---|---|---|---|---|
| 1 | 1 | 1 | 0 | 0 | 0 | 0 | 0 |

| | |
|---|---|
| T1 | 1 |
| T2 | 1 |
| T3 | 1 |
| T4 | 0 |
| T5 | 0 |
| T6 | 0 |
| T1 | 0 |
| T2 | 0 |
| T3 | 0 |
| T4 | 1 |
| T5 | 1 |
| T6 | 1 |

**Table 28: P-Invariants and T-Invariants (Scenaro I)**

### Property Verification Function

In order to proceed with the verification, we have to translate the objectives and constraints of the requirement specification into PN properties:

$\mathbf{o_1}$: Machine1 should continuously produce product1 without any infinite delay

$\mathbf{o_2}$: Machine2 should continuously produce product2 without any infinite delay

$\mathbf{o_3}$: Both machines should produce products with a ratio of 1:1

$\mathbf{s_4}$: The shared robot should treat both machines with fairness (i.e., k-fairness; k=1).

It is clear from the $\{o_1, o_2, o_3$ and $s_4\}$ that if the robot serves both machines with fairness (S4) only then both of them will be able to produce their respective products continuously without indefinite delay (O1 & O2). And if the fairness is bounded such that k=1, then both machines will produce products with equal ration 1:1.

Therefore the translation of the requirement specification in PN form is as follows:

PN Property P1 = *"The model should be Bounded-Fair (with K=1) such that the robot serves both machines alternatively".*

In order to verify bounded-fairness we consider the property proving theorems I, II & III defined in section 7.2. Based on these theorems we construct a property verification function using the following algorithm:

**Algorithm: B-Fairness Verification**

**Input:** {P-Invariants}, {T-Invariants}, A; **Output:** TRUE

```
If |{T-Invariants}| = 1 then ▷ List of T-Invariants has exactly 1 invariant, meaning it is a
                                    Reproduction vector , Theorem III◁
    X_T ← T-Invariants[0]      ▷ Get the only T-Invariant from the list
    if A.X_T ≥ 0 and each element in X_T >0 then  ▷ Multiply X_T with Incidence matrix and
                                                   Check that each element of T-invariant is
                                                   positive, Theorem I◁
        if |{P-Invariants}|>0  then       ▷ Check if there is any P-Invariant, meaning PN
                                            Model is bounded, Theorem II◁
            Return TRUE
        else
            Return FALSE ▷ Theorem II violated
        end if
    else
        Return FALSE  ▷ Theorem I violated
    end if
else
    Return FALSE ▷ Theorem III violated
end if
```

**Table 29: B-Fairness Verification**

Based on this algorithm we perform property verification of the given PN model. It is evident that the T-invariants (see **Table 28**) contain zero entries which violate Theorem I. Also there is more than one T-invariant which violates Theorem III therefore the net is said to be unfair. As the PN is unfair, it is impossible to guarantee that objectives $\mathbf{o_1}$, $\mathbf{o_2}$, $\mathbf{o_3}$ will be satisfied because either of the machines may over perform by acquiring robot multiple times without letting the other to get the robot for at least once (failure of $o_1$ & $o_2$). Therefore either of the machines may face a situation in which it is unable to produce enough number of products to meet the required objectives i.e., the ratio 1:1 for producing products cannot be fulfilled (failure of $o_3$); consequently the composed model may fail to satisfy given specifications.

### 7.3.2 Scenario II

In order to understand the fairness verification process, a counterexample is presented. In this example another component is added to the composition called Controller that can supervise the robot assignments. The job of the Controller is to enforce fairness in the system. The BOM model of the controller is defined as follows:

**$BB_3$ = ⟨ EnT, EvT, AcT, S ⟩** where:

**EnT** = Controller {}

**EvT** = {$E_{10}$(LoadingM1, Controller, Robot , Machine1, null), $E_{11}$(LoadingM2, Controller, Robot, Machine2, null)}

**Act** = {$A_{10}$(LoadingM1, Controller, Robot, Machine1, $E_0$), $A_{11}$(LoadingM2, Controller, Robot, Machine2, $E_1$)}

**S** = {$S_8$(AssignM1, {$A_{10}$, $S_9$}), $S_9$(AssignM2, {$A_{11}$, $S_8$})}

**Table 30: Formal definition of Controller Base-BOM**

**$CB_0$ = ⟨ $AcT_{IN}$, $AcT_{OUT}$ , POI ⟩** where:

**$AcT_{IN}$** = **$AcT_{OUT}$** = ∅

**POI** = {$POI_0$(**!$A_{10}$** , {?$A_0$, ?$A_6$}), $POI_1$(**!$A_7$**, ?$A_1$), $POI_2$(**!$A_2$**, ?$A_2$), $POI_3$(**!$A_{11}$** , {?$A_3$, ?$A_8$}), $POI_4$(**!$A_9$** , ?$A_4$), $POI_5$(**!$A_5$**, ?$A_5$) }

**Table 31: Formal definition of Modified Manufacturing System composed BOM**

The other components have the same definition except that the sender of Event $E_0$ and $E_1$ is $BB_3$ (controller) and the receivers of event $E_0$ are $BB_0$ (Machine1) and $BB_2$ (Robot); whereas the receivers of event $E_1$ are $BB_1$ (Machine2) and $BB_2$ (Robot). **Figure 58** shows the composed BOM of modified manufacturing system.

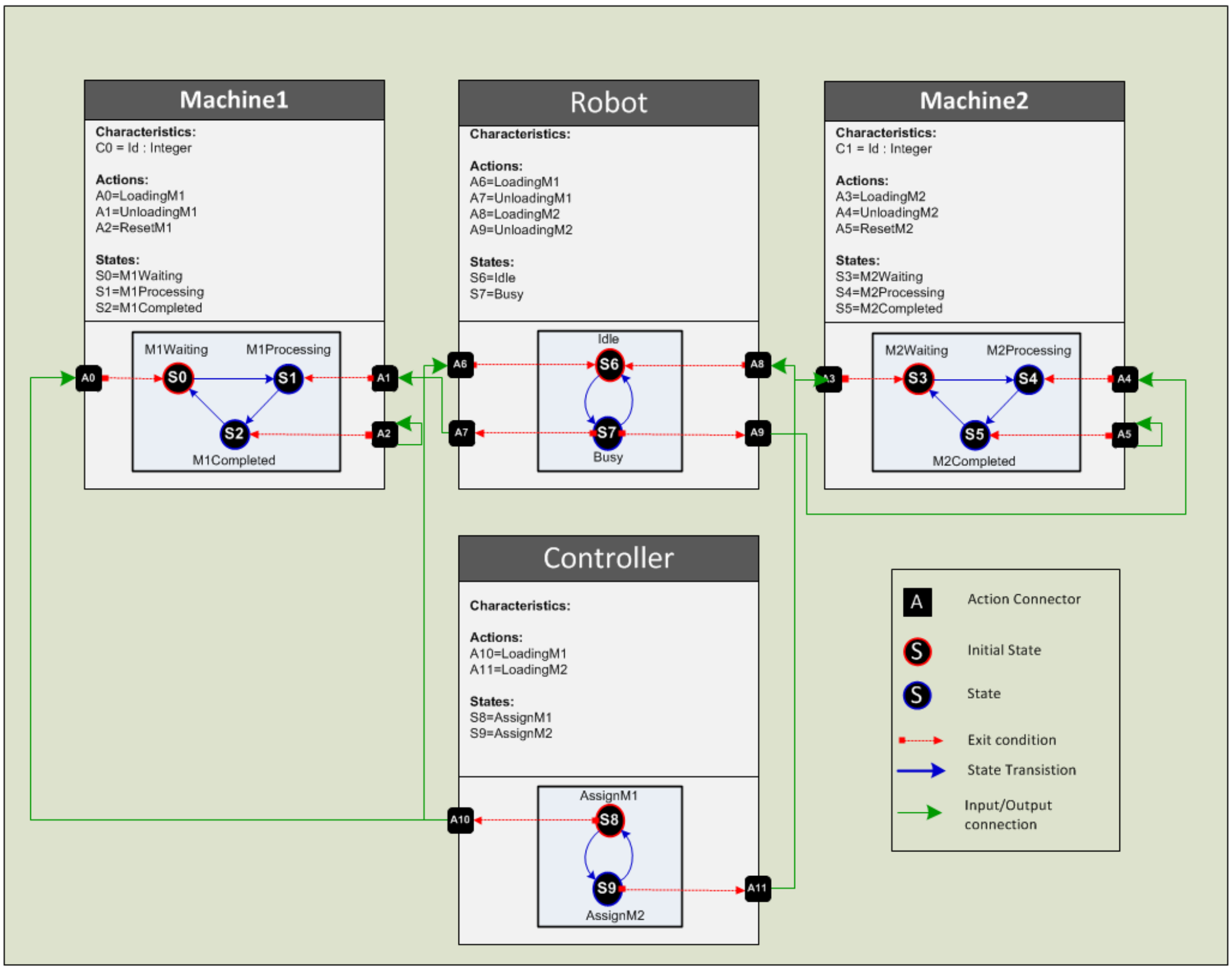

**Figure 58: Modified manufacturing system composed BOM**

When the verification process is started, the BOM components are transformed into PN model as shown in **Figure 59** where the controller component is attached to both machines and the robot and controls the machine assignment to enforce Kfairness. It is evident from the figure that when the robot is assigned to Machine1 once, it cannot be reassigned (because of the lack of token in P9), and vice-versa. If the number of tokens are increased to ‘n’, the same model can work for k=n fairness.

In the initialization phase, the initial marking $\mathbf{M_0}$ and Incidences Matrix **A** were calculated as follows:

| $M_0$ | P1 | P2 | P3 | P4 | P5 | P6 | P7 | P8 | P9 | P10 |
|---|---|---|---|---|---|---|---|---|---|---|
| | 1 | 0 | 0 | 1 | 0 | 0 | 1 | 0 | 1 | 0 |

| A | P1 | P2 | P3 | P4 | P5 | P6 | P7 | P8 | P9 | P10 |
|---|---|---|---|---|---|---|---|---|---|---|
| T1 | -1 | 1 | 0 | 0 | 0 | 0 | -1 | 1 | -1 | 1 |
| T2 | 0 | -1 | 1 | 0 | 0 | 0 | 1 | -1 | 0 | 0 |
| T3 | 1 | 0 | -1 | 0 | 0 | 0 | 0 | 0 | 0 | 0 |
| T4 | 0 | 0 | 0 | -1 | 1 | 0 | -1 | 1 | 1 | -1 |
| T5 | 0 | 0 | 0 | 0 | -1 | 1 | 1 | -1 | 0 | 0 |
| T6 | 0 | 0 | 0 | 1 | 0 | -1 | 0 | 0 | 0 | 0 |

**Table 32: Initial Marking and Incidence Matrix (Scenaro II)**

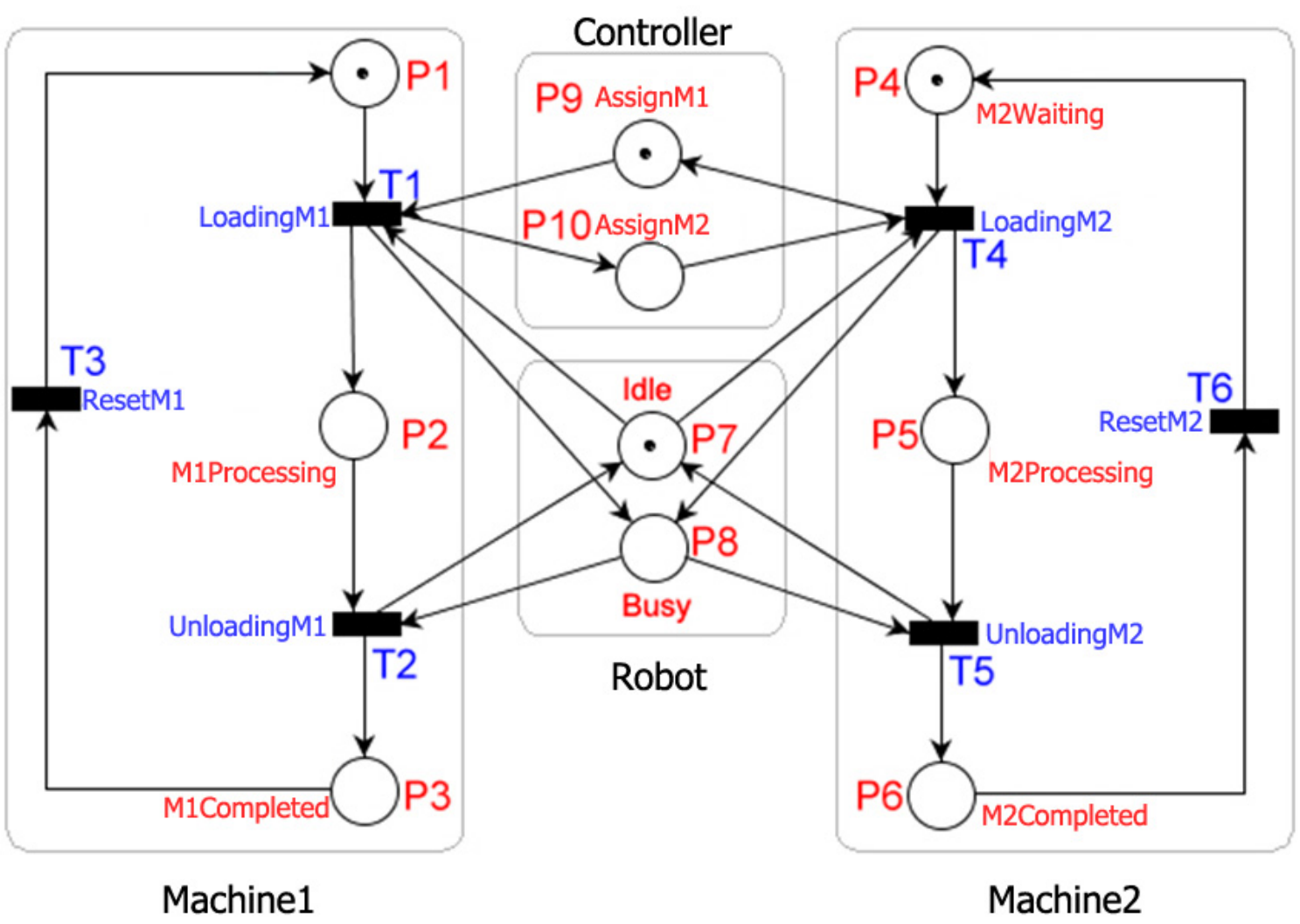

**Figure 59: Modified PN model of the manufacturing System**

When the Invariant calculation module is executed, the following T-Invariant and P-Invariant were discovered for the model shown in **Figure 59**:

| | |
|---|---|
| T1 | 1 |
| T2 | 1 |
| T3 | 1 |
| T4 | 1 |
| T5 | 1 |
| T6 | 1 |

| P1 | P2 | P3 | P4 | P5 | P6 | P7 | P8 | P9 | P10 |
|---|---|---|---|---|---|---|---|---|---|
| 1 | 1 | 1 | 0 | 0 | 0 | 0 | 0 | 0 | 0 |

**Table 33: P-Invariants and T-Invariants (Scenaro II)**

Having only one T-Invariant (and the only one) with non-zero entries and having a P-Invariant (with some non-zero entries), satisfies the conditions (of Theorem I, II & III) required for the model to be *bounded fair*.

Based on these result from PN algebraic analysis technique, we can confirm that the composed model satisfies given requirement specifications. Due to the supervised controller, the Robot is bound to operate fairly between the two machines, which results in fulfillment of the objectives $O_1$, $O_2$ and $O_3$ and also satisfied required constraint $S_1$.

## 7.4 Summary

Fairness property becomes significant in the composability verification of a composed model because it does not allow any component to dominate and excessively proceed, while other components do not proceed even for once. As illustrated by the example of the manufacturing system, fairness of Robot allocation can ensure that both machines will perform to produce a required number of products. If there is no fairness we cannot guarantee that this objective will be reached.

Using the example of Fairness verification in the manufacturing system, we explain how our Algebraic Verification Technique works. It is a notable fact that this technique does not face state-space explosion because it does not involve reachability graph construction and can work only by calculating incidence matrix and P/T invariants. There are a lot of PN properties which can be verified using these PN algebraic computation resources. On the other hand this approach can only be used to verify a limited set of PN properties (for which suitable theorems exist).

# Chapter 8
# Model Verification using State-space Analysis techniques

*Colored Petri Nets and its analysis techniques are very useful for accurate and efficient verification as it is one of the competitive formalisms in the specification of the concurrent systems. Its application in the Composability verification proves to be very constructive, especially with a focus on the dynamic semantic composability level. The analysis techniques contributed by the CPN community over a couple of decades provide a significant improvement on efficient and accurate reasoning regarding the model correctness. In this chapter a Field Artillery Model is presented as an example. It is shown how the BOM based Field Artillery Model is transformed into our proposed Colored Petri Net components and verified using state-space analysis.*

Combat Modeling is about the models that describe or represent weapon systems and combat situations. There are numerous types of combat models. These types are distinguished by their modeling objectives. Some of the fundamental objectives of combat modeling are training, war-games, weapon testing etc.

## 8.1 Combat Modeling

Combat modeling purposefully abstracts and simplifies combat entities, their behaviors, activities, and interrelations to answer defense-related research questions. There cannot be a general model that answers all questions however there is a concept of a generic situated environment and four core activities that can be found on every battlefield [**125**].

### 8.1.1 Situated Environment

Combat Modeling starts with analyzing the challenges to model the *Situated Environment*. All modeled combat entities are situated in the environment, the virtual battlefield. They perceive the environment including other entities, and map their perception to an internal representation based on the knowledge and goals. They communicate and act with other combat entities within the environment. The environment contains all objects, passive ones like obstacles, as well as active ones like enemy or friendly units [**125**].

### 8.1.2 Moving

Moving is the core activity of combat modeling that deals with the movement of individual entities. These entities could be weapon, people etc. or aggregate models that are used to model the movements of groups of entities. The models use patches and grids; they use physical models for weapon systems and reference schemas for unit movement [**125**].

thermal, and optical sensors, can contribute to perceiving the environment and the other entities as close to reality as possible. Intelligence, surveillance, and reconnaissance operations contribute to similar requirements. In order to sense special properties of an entity, each of these special properties needs to be modeled explicitly. If it is modeled explicitly, it needs to make a difference in the reconnaissance process. Furthermore, if a detail is important for the military decision process, it needs to be part of the perception, and hence needs to be observed by sensors, which requires that the respective things are modeled as properties of the entities [**125**].

### 8.1.4 Shooting

Modeling the outcomes of duels between weapon systems and battles between units is still a topic of major interest. On the weapon system level, direct and indirect fires are analyzed. Direct fire means that the target is in the line of sight of the shooter. In case of indirect fire systems such as Artillery and other ballistic weapons, they do not need to see the target and shoot at it straight. Their weapons follow a ballistic curve being described by the term indirect fire. Many models have been developed to keep up with the score. For instance a game based point systems that count how often and where a target is hit and use "hit-and-kill" probabilities (which are based on real-world data) to simulate hitting or missing a target [**125**].

### 8.1.5 Communication:

This core activity deals with the modeling of Communications, Command and Control. It ties all the earlier activities together as command and control is situated in the environment and commands the entities to shoot, move, observe, and communicate. Several models of command and control in military headquarters are discussed, as more and more simulation models have to come up with decisions based on available information where until recently human decision makers had to be involved. The better command and control is modeled the less military experts are needed to provide a realistic training environment [**125**]. Based on these principles of combat modeling an example model of Field Artillery is presented to explain the approach of composability verification using state-space analysis. **Figure 60** highlights the activities of combat modeling.

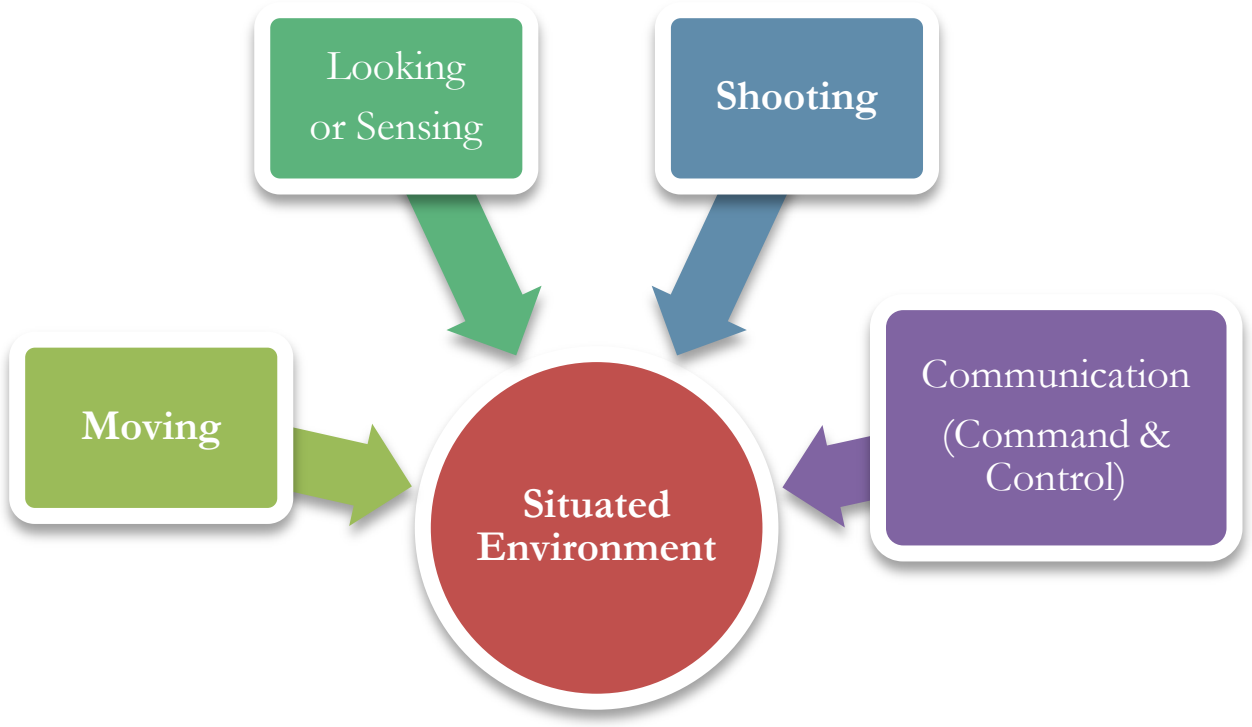

**Figure 60: Activities of Combat Modeling**

## 8.2 Field Artillery

Field Artillery (FA) is one of the indirect fire systems[35] that engage the opponent without requiring line of sight between the shooting system and the target. Infantry uses small, medium or heavy howitzers (artillery guns) that provide fire support for combat units. Similarly Navy artillery provides fire power, where missiles can be fired on land based or sea based direct or indirect targets. The general mission of FA is to destroy, neutralize or suppress the enemy by cannon, rocket, and missile fires and to help integrate all fire support assets into combined arms operations [**126**]. The field artillery system provides close support to maneuver friendly forces, counter fire and interdiction as required. These fires neutralize, canalize, or destroy enemy attack formations or defenses; obscure the enemy's vision or otherwise inhibit his ability to acquire and attack friendly targets; and destroy targets deep in the enemy rear with long-range rocket or missile fires [**127**].

FA weapons are usually located in defiladed areas in order to protect them from enemy detection. This nature of FA gunnery makes it an indirect fire problem. Observed fire (the technique that solves the indirect FA gunnery problem) is carried out by the coordinated efforts of the Forward Observers, Head Quarter (HQ), the Fire Direction Center (FDC), and firing sections of the firing unit (Batteries) [**126**]. **Figure 61** gives an overview of the essential elements of a field artillery and the situation of an indirect fire, where a forward observer spots an enemy unit and requests fire support from a nearby friendly unit. It should be noted that this scenario is only assumed and simplified for the sake of an example, whereas the today's state of the art of field artillery systems is much more modernized and technologically advanced.

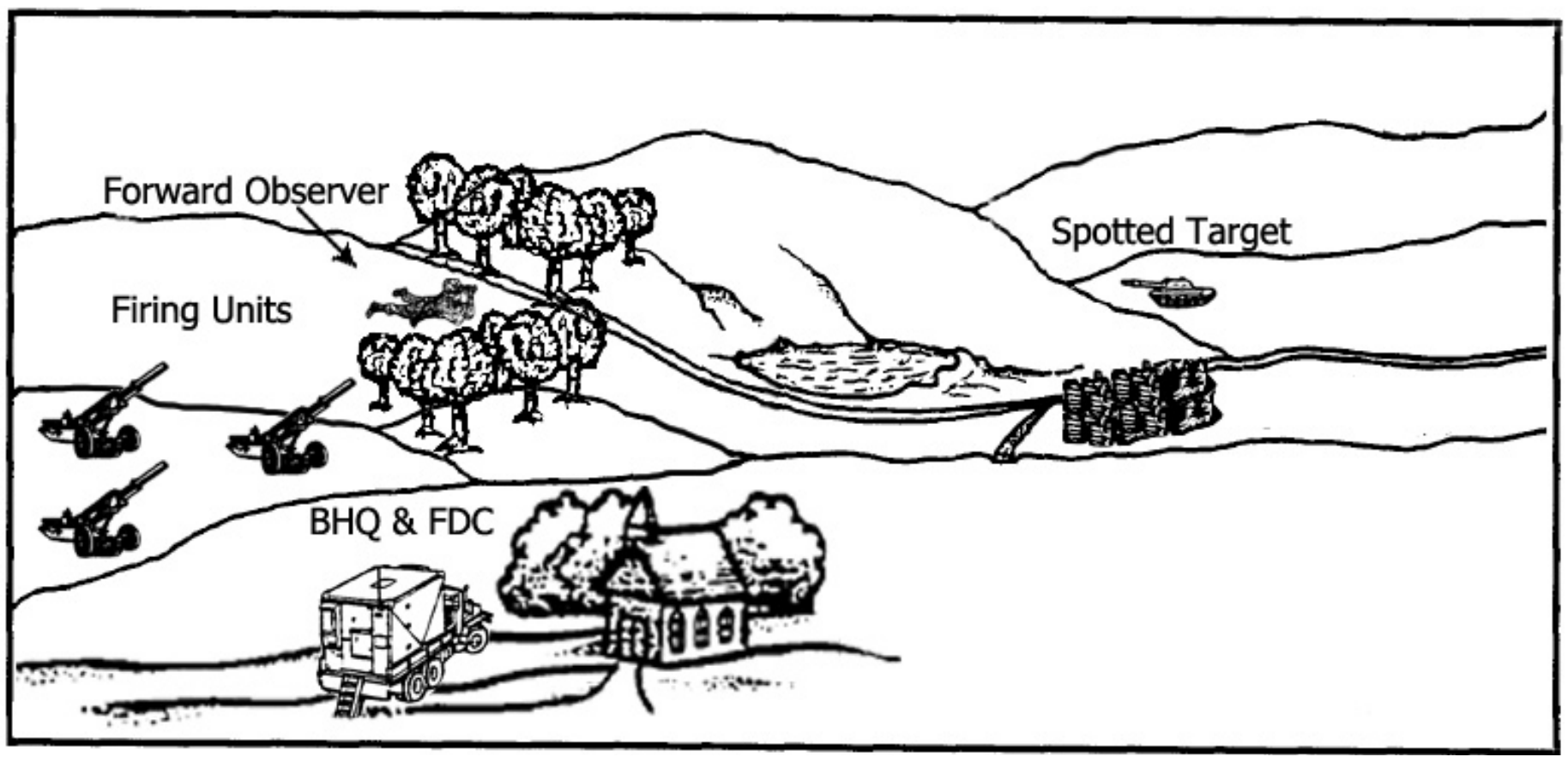

**Figure 61: Elements of Field Artliiery & Indirect Fire**

### 8.2.1 Simuland

Based on **Figure 61** an Indirect Fire Support scenario is considered. In this scenario the enemy units are not in the line of sight of the firing units. A soldier (forward observer) from the observation post observes the enemy field and detects potential targets. When a target is spotted, he calls BHQ for fire support and provides the target details. BHQ requests FDC to process the target tactically & technically. In tactical terms, the target should be of high importance to gain tactical advantage. In technical terms the target should be in the firing range of the supporting artillery. If the target is valid FDC approves the request otherwise the request is denied. If the request is approved BHQ assigns the target to the firing units (batteries). We suppose that the target can be one of three types: light (e.g., camps, troops, and trucks), medium (e.g., tanks, light guns) or heavy (e.g., artillery units, missile launchers). The target is assigned to one, two or three batteries respectively. This is because medium and heavy targets require the fire power of more than one battery for complete destruction. Based on this assumption, BHQ assigns target to the batteries. Battery components align themselves for correct orientation and elevation by computing the target's range and bearing (angle), load appropriate ammunition and fire the round. When a Field component receives fire, and if the detonation is within a destruction radius then the target is said to be destroyed otherwise it is missed, as will be observed by the observer, who provides this information to the BHQ. This process is restarted for other potential targets, until all the enemy-units are suppressed, which is the ultimate goal.

[35] Although there are some exceptions, in which Field Artillery engages in direct fire mode

### 8.2.2 Field Artillery Model

Based on the above informal description of the simuland a Field Artillery Model is constructed. There could be multiple objectives of modeling field artillery including training, exercises, weapon testing or operational optimization. The following BOM based models were discovered, selected and composed with respect to the simuland.

#### Field Artillery Conceptual Model

The BOM based conceptual model of Field Artillery is formally defined as follows:

| **Observer = ⟨ EnT, EvT, S, AcT ⟩** where: |
|---|
| **EnT** = Observer {$C_0$(Id), $C_1$(Loc), $C_2$(CurrentTarget), $C_3$(Result)} |
| **EvT** = {$E_0$(ObserveField, Observer, Field, null), $E_1$(TargetSpotted, Field, Observer, target), $E_2$(CallForFireSupport, Observer, BHQ, currtgt), $E_3$(RequestApproved, FDC, BHQ, Observer, null), $E_4$(RequestDenied, FDC, BHQ, Observer, null), $E_5$(Detonation, Field, Observer, detonation), $E_6$(TargetDestroyed, Observer, BHQ, null), $E_7$(TargetMissed, Observer, BHQ, null)} |
| **Act** = {$A_0$(ObserveField, Observer, Field, $E_0$), $A_1$(TargetSpotted, Field, Observer, $E_1$), $A_2$(CallForFireSupport, Observer, BHQ, $E_2$), $A_3$(RequestApproved, FDC, Observer, $E_3$), $A_4$(RequestDenied, FDC, Observer, $E_4$), $A_5$(Detonation, Field, Observer, $E_5$), $A_6$(TargetDestroyed, Observer, BHQ, $E_6$), $A_7$(TargetMissed, Observer, BHQ, $E_7$)} |
| **S** = {$S_0$(ObserverReady, $A_0$, $S_1$), $S_1$(ObservingField, $A_1$, $S_2$), $S_2$(RequestingFireSupport, $A_2$, $S_3$), $S_3$(WaitingForReponse, {$A_3$, $S_4$}, {$A_4$, $S_0$}) , $S_4$(WaitingForImpact, $A_5$, $S_5$) , $S_5$(EvaluateDamage, {$A_6$, $S_0$}, {$A_7$, $S_0$}} |

**Table 34: Observer Basic-BOM**

**Field = ⟨ EnT, EvT, S, AcT ⟩** where:

**EnT** = Field {$C_4$(Id), $C_5$(FD), $C_6$(Impacts)}

**EvT** = {$E_8$(ObserveField, Observer, Field, null), $E_9$(TargetSpotted, Field, Observer, target), $E_{10}$(Fire, Battery1, Field, fire), $E_{11}$(Fire, Battery2, Field, fire), $E_{12}$(Fire, Battery3, Field, fire), $E_{13}$(Detonation, Field, Observer, Impacts), $E_{14}$(UpdateField, BHQ, Field, update) }

**Act** = {$A_8$(ObserveField, Observer, Field, $E_8$), $A_9$(TargetSpotted, Field, Observer, $E_9$), $A_{10}$(Fire, Battery1, Field, $E_{10}$), $A_{11}$(Fire, Battery2, Field, $E_{11}$), $A_{12}$(Fire, Battery3, Field, $E_{12}$), $A_{13}$(Detonation, Field, Observer, $E_{13}$), $A_{14}$(UpdateField, BHQ, Field, $E_{14}$) }

**S** = {$S_6$(FieldReady, {$A_8$, $S_7$}, {$A_{10}$, $S_8$}, {$A_{11}$, $S_8$}, {$A_{12}$, $S_8$}), $S_7$(BeingObserved, $A_9$, $S_6$), $S_8$(TakingFire, $A_{13}$, $S_9$) , $S_9$(WaitingForUpdate, $A_{14}$, $S_6$)}

**Table 35: Field Basic-BOM**

**BHQ = ⟨ EnT, EvT, S, AcT ⟩** where:

**EnT** =BHQ {$C_7$(Id), $C_8$(Loc), $C_9$(CurTarget), $C_{10}$(TargetStatus) }

**EvT** = {$E_{15}$(CallForFireSupport, Observer, Field, target), $E_{16}$(ProcessRequest, BHQ, FDC, target), $E_{17}$(RequestApproved, FDC, BHQ, Observer, null), $E_{18}$(RequestDenied, FDC, BHQ, Observer, null), $E_{19}$(AssignTarget, BHQ, Battery1, Battery2, Battery3, assign_target), $E_{20}$(FiringCompleted, Battery1, BHQ, null), $E_{21}$(FiringCompleted, Battery2, BHQ, null), $E_{22}$(FiringCompleted, Battery3, BHQ, null), $E_{23}$(TargetDestroyed, Observer, BHQ, null), $E_{24}$(TargetMissed, Observer, BHQ, null), $E_{25}$(UpdateField, BHQ, Field, update) }

**Act** = {$A_{15}$(CallForFireSupport, Observer, Field, $E_{15}$), $A_{16}$(ProcessRequest, BHQ, FDC, $E_{16}$), $A_{17}$(RequestApproved, FDC, BHQ, Observer, $E_{17}$), $A_{18}$(RequestDenied, FDC, BHQ, Observer, $E_{18}$), $A_{19}$(AssignTarget, BHQ, Battery1, Battery2, Battery3, $E_{19}$), $A_{20}$(FiringCompleted, Battery1, BHQ, $E_{20}$), $A_{21}$(FiringCompleted, Battery2, BHQ, $E_{21}$), $A_{22}$(FiringCompleted, Battery3, BHQ, $E_{22}$), $A_{23}$(TargetDestroyed, Observer, BHQ, $E_{23}$), $A_{24}$(TargetMissed, Observer, BHQ, $E_{24}$), $A_{25}$(UpdateField, BHQ, Field, $E_{25}$) }

**S**={$S_{10}$(BHQReady, $A_{15}$, $S_{11}$), $S_{11}$(CallFDC, $A_{16}$, $S_{12}$), $S_{12}$(WaitingForApproval, {$A_{17}$, $S_{13}$}, {$A_{18}$, $S_{10}$}), $S_{13}$(AssigningTarget, $A_{19}$, $S_{14}$), $S_{14}$(WaitingForFire, {$A_{20}$, $S_{15}$}, {$A_{21}$, $S_{15}$}, {$A_{22}$, $S_{15}$}), $S_{15}$(WaitingForDamageReport, {$A_{23}$, $S_{16}$}, {$A_{24}$, $S_{16}$}), $S_{16}$(UpdatingField, $A_{25}$, $S_{10}$)}

**Table 36: BHQ Basic-BOM**

**FDC = ⟨ EnT, EvT, S, AcT ⟩** where:

**EnT** =FDC{$C_{11}$(Id), $C_{12}$(FD) , $C_{13}$(CurTarget), $C_{14}$(Result)}

**EvT** = {$E_{26}$(ProcessRequest, BHQ, FDC, target), $E_{27}$(RequestApproved, FDC, BHQ, Observer, null), $E_{28}$(RequestDenied, FDC, BHQ, Observer, null)}

**Act** = {$A_{26}$(ProcessRequest, BHQ, FDC, $E_{26}$), $A_{27}$(RequestApproved, FDC, BHQ, Observer, $E_{27}$), $A_{28}$(RequestDenied, FDC, BHQ, Observer, $E_{28}$)}

**S** = {$S_{17}$(FDCReady, $A_{26}$, $S_{18}$), $S_{18}$(Processing, {$A_{27}$, $S_{17}$}, {$A_{28}$, $S_{17}$})}

**Table 37: FDC Basic-BOM**

| **Battery1,2,3[36] = ⟨ EnT, EvT, S, AcT ⟩** where: |
|---|
| **EnT** = Battery1,2,3 {$C_{15}$(Id), $C_{16}$(CurTarget) } |
| **EvT** = {$E_{29}$(AssignTarget, BHQ, Battery1, Battery2, Battery3, assign_target), $E_{30}$(Fire, Battery123, Field, fire), $E_{31}$(FiringCompleted, Battery123, BHQ, null)} |
| **Act** = {$A_{29}$(AssignTarget, BHQ, Battery1, Battery2, Battery3, $E_{29}$), $A_{30}$(Fire, Battery1/2/3, Field, $E_{30}$), $A_{31}$(FiringCompleted, Battery1/2/3, BHQ, $E_{31}$)} |
| **S** = {$S_{19}$(ReadyToFire, $A_{29}$, $S_{20}$), $S_{20}$(PreparingCannon, $A_{30}$, $S_{21}$), $S_{21}$(Firing, $A_{31}$, $S_{19}$)} |

**Table 38: Battery (1,2,3) Basic-BOM**

| **FA = ⟨ $AcT_{IN}$, $AcT_{OUT}$ , POI ⟩** where: |
|---|
| **$AcT_{IN}$ = $AcT_{OUT}$ = ∅** |
| **POI** = { POI-0(!A0, ?A8), POI-1(!A9, ?A1), POI-2(!A2, ?A15), POI-3(!A16, ?A26), POI-4(!A27, {?A17, ?A3}), POI-5(!A28, {?A18, ?A4}), POI-6(!A19, ?A29), POI-7(!A30, {?A10, ?A11, ?A12}), POI-8(!A31, {?A20, ?A21, ?A22}), POI-9(!A13, ?A5), POI-10(!A6, ?A23), POI-11(!A7, ?A24), POI-12(!A25, ?A14)} |

**Table 39: Field Artillery Composed BOM**

**Figure 62** shows the formal representation of Field artillery composed model.

[36] This component has three instances.

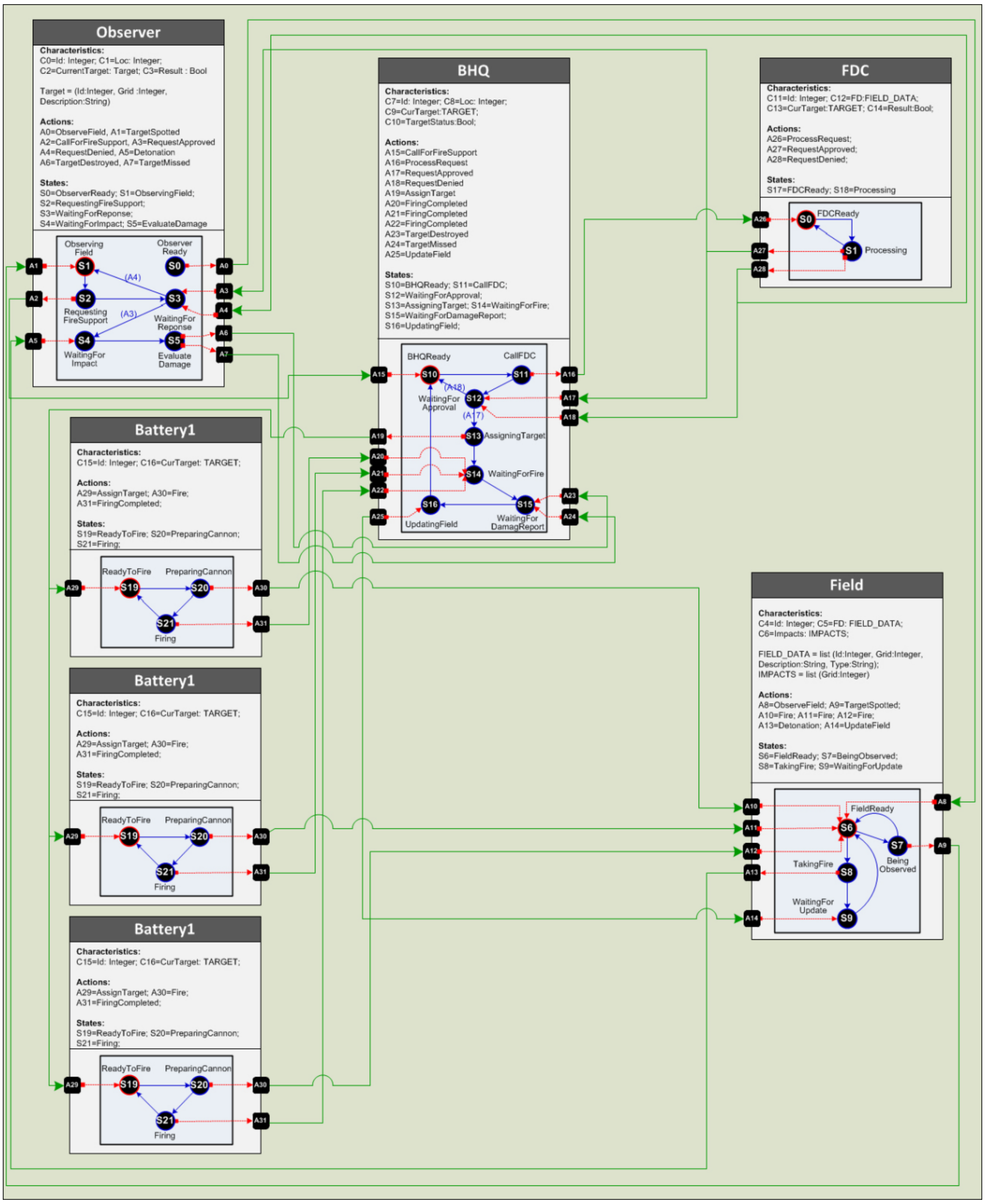

**Figure 62: Field Artillery Composed BOM**

## 8.2.3 Requirement Specification

We define Requirement speciation of the field artillery model as:

**$RS_0 = \langle O, S \rangle$** where:

**Objectives O = {$o_1$} and System Constraints S = {$s_1$, $s_2$ $s_3$, $s_4$}**

**$o_1$:** All the enemy units must be destroyed

**$s_{1,\ 2,\ \text{and}\ 3}$:** The model should be composable at syntactic and static-semantic level. The state-machines should match and the executable mode should correctly represent the conceptual model.

**$s_4$:** There should never be a friendly fire.

## 8.3 Verification of the FA model using CPN State-Space Analysis

After the BOMs are discovered, selected and composed according to **Figure 62**, the conceptual model is ready for verification. We select CPN state-space analysis technique for its verification.

### 8.3.1 Static and Dynamic Analysis

We assume that the model qualifies syntactic and static-semantic analysis. Also when it undergoes state-machine matching process it is able to make progress until the goal-states are reached. **Figure 63** shows how the components interact with each other through the exchange of events (horizontal arrows) due to which their state-machines make progress (vertical dotted arrows). Based on the fact that the constraint S1, S2 and S3a are satisfied we proceed to BOM-to-E-BOM extension which is a pre-requisite step for the transformation of conceptual model into executable model.

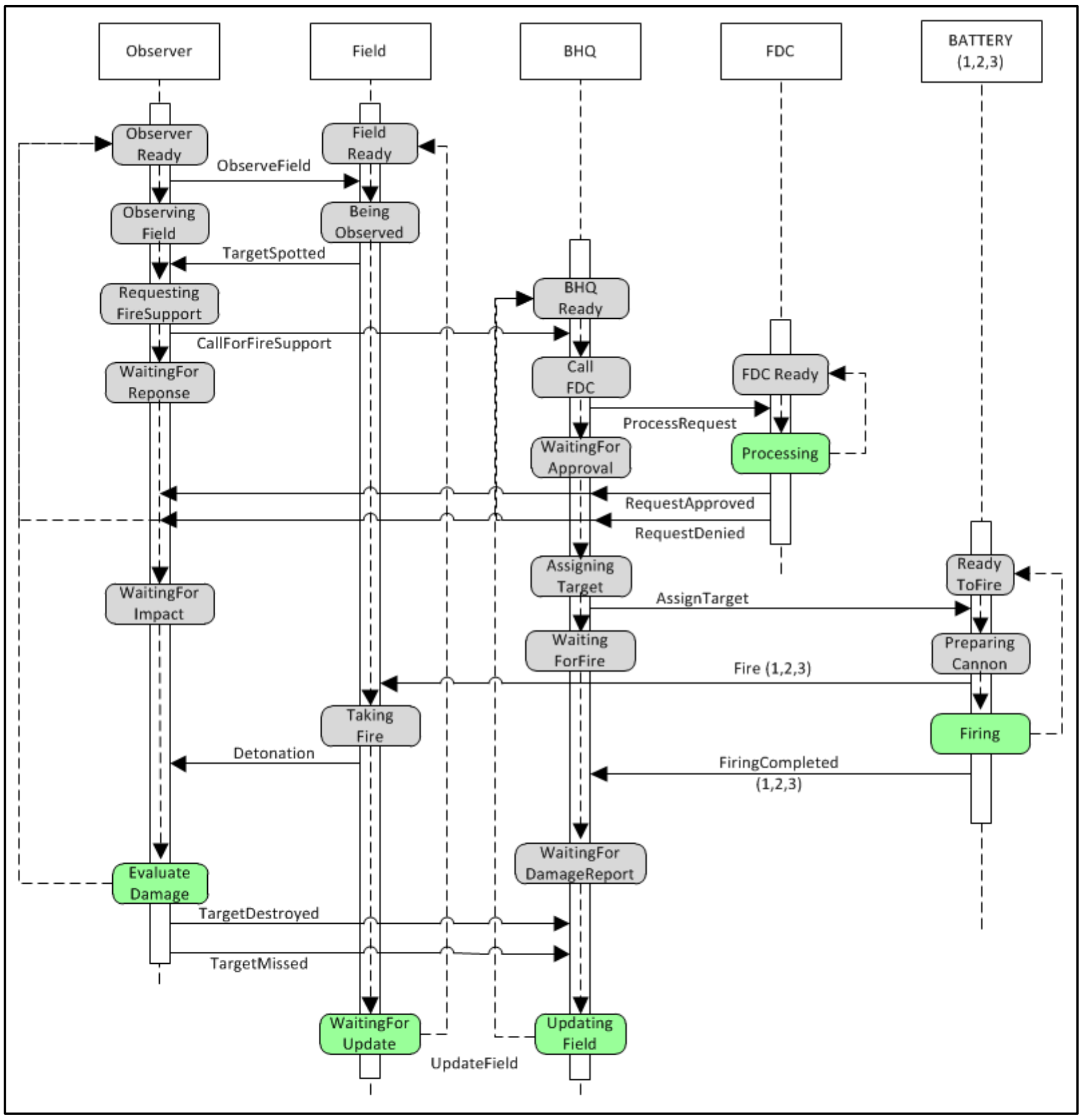

**Figure 63: State-machine Matching of Field Artillery Model**

### 8.3.2 BOM to E-BOM extension

At this stage all the BOM components are extended to our proposed E-BOM extension with the help of the modeler's input. The following tables present E-BOM extensions of BOMs in the FA model.

| **Observer E-BOM** | | | | | | |
|---|---|---|---|---|---|---|
| **SV and types** | {$C_0$(Id:Integer), $C_1$(Loc:Integer), $C_2$(CurrentTarget:TARGET), $C_3$(Result:Bool)} where TARGET = (Id:Integer, Grid[37]:Integer, Description:String) | | | | | |
| **Initial States** | {$S_0$:ObserverReady} | | | | | |
| **Transitions** | | | | | | |
| **State** | **Event** | **Guard** | **{$SV_{IN}$}** | **{$SV_{OUT}$}** | **Action** | **Next State** |
| Observer Ready | Observe Field | | | | | Observing Field |
| Observing Field | TargetSpotted | | | C2 | | Requesting FireSupport |
| Requesting FireSupport | CallForFireSupport | | C2 | | | WaitingFor Reponse |
| WaitingFor Reponse | RequestApproved | | | C2 | | WaitingFor Impact |
| WaitingFor Reponse | RequestDenied | | | | | Observer Ready |
| WaitingFor Impact | Detonation | | C2 | C3 | Action1 | Evaluate Damage |
| Evaluate Damage | Target Destroyed | [Result=true] | C3 | | | Observer Ready |
| Evaluate Damage | Target Missed | [Result=false] | C3 | | | Observer Ready |

```
Action1 {
      input (target, detonation);
      output (result);
      action
      let
       val grid= #2 target;
      in
       (Destroyed(grid, detonation))
      End
}
```

```
fun Destroyed (x, []) = false
  | Destroyed (x, h::t) = IsDestroyed(x, h)  orelse Destroyed (x, t);
```

```
fun IsDestroyed(grid, impact) =
    let
      val gridst = Int.toString(grid);
      val impactst= Int.toString(impact);
      val gridX = valOf(Int.fromString(substring (gridst, 0, 3)));
      val impactX = valOf(Int.fromString(substring (impactst, 0, 3)));
      val gridY = valOf(Int.fromString(substring (gridst, 3, 3)));
      val impactY = valOf(Int.fromString(substring (impactst, 3, 3)));
      val X = abs(gridX - impactX);
      val Y = abs(gridY - impactY);
    in
      if (abs(X)<4 andalso  abs(Y)<4)
         then true
       else false
end;
```

**Table 40: Observer E-BOM**

Action1 is a CPN-ML script that evaluates if the target is destroyed or not. We assume that the destruction radius of the rounds fired by artillery guns is 4x4 grids i.e. if the round hits the target within this radius it will be destroyed otherwise missed. Note that Action1 calls other functions which are also specified in **Table 40**.

[37] In military map, Grid reference system is used to identify a position of an object. We assume that the grid in this scenario is of 6 figures, first three integers define Easting and the other three define Northings. For details see [**131**]

| Field E-BOM | | | | | | |
|---|---|---|---|---|---|---|
| **SV:Types** | {$C_4$(Id:Integer), $C_5$(FD:FIELD_DATA), $C_6$(Impacts:IMPACTS)} FIELD_DATA = list (Id:Integer, Grid:Integer, Description:String, Type:String); IMPACTS = list (Grid:Integer) | | | | | |
| **Initial States** | {$S_6$: FieldReady} | | | | | |
| **Transitions** | | | | | | |
| **State** | **Event** | **Guard** | **{$SV_{IN}$}** | **{$SV_{OUT}$}** | **Action** | **Next State** |
| FieldReady | ObserveField | | | | | BeingObserved |
| Being Observed | TargetSpotted | [length FD>0] | C5 | C5 | Action2 | FieldReady |
| FieldReady | Fire | | C6 | | | TakingFire |
| TakingFire | Detonation | | | C6 | | WaitingFor Update |
| WaitingFor Update | UpdateField | | C5 | C5 | Action3 | FieldReady |

```
Action2 {
  input (fd);
  output (target);
  action
  let
    val indx = discrete (0,(length fd));
    val F = List.nth(fd, indx);
    in
    ((#1 F, #2 F, #3 F))
  end }
```

```
Action3 {
  input(update,fd);
  output (ufd);
  action
  let
     val status = #2 update;
     val target = #1 update;
     val U = (#1 target, #2 target, #3 target, "Enemy");
  in
     if (status=true) then
      rm U fd
     else
      fd
   end }
```

**Table 41: Field E-BOM**

Action2 randomly picks a target from a list of targets (Field Data) and sends it as parameters to the observer, simulating that the observer has spotted a target in the enemy area. Action3 is executed when *Update-Field* event is received from BHQ. This action removes an object if the target destroyed.

| BHQ E-BOM | | | | | | |
|---|---|---|---|---|---|---|
| **SV:Types** | {C7(Id:Integer),C8(Loc:Integer),C9(CurrentTarget:TARGET), C10(TargetStatus:Bool)} | | | | | |
| **Initial States** | {$S_{10}$: BHQReady } | | | | | |
| **Transitions** | | | | | | |
| **State** | **Event** | **Guard** | **{$SV_{IN}$}** | **{$SV_{OUT}$}** | **Action** | **Next State** |
| BHQReady | CallForFireSupport | | C9 | | | CallFDC |
| CallFDC | ProcessRequest | | | C9 | | WaitingFor Approval |
| WaitingFor Approval | RequestApproved | | C9 | | | Assigning Target |
| WaitingFor Reponse | RequestDenied | | | | | BHQReady |
| Assigning Target | AssignTarget | | | C9 | Action4 | WaitingForFire |
| WaitingForFire | FiringCompleted | | | | | WaitingFor DamageReport |
| WaitingFor | TargetDestroyed | | C10 | | | UpdatingField |
| DamageReport | | | | | | |
| WaitingFor DamageReport | TargetMissed | | C10 | | | UpdatingField |
| UpdatingField | UpdateField | | | C10 | | BHQReady |

```
Action4 {
    input (target);
    output (assign_target);
    action
    let
    in
        if ((#3 target) = "Artillery") then
          ((#1 target, #2 target, #3 target, [true, true, true]))
        else if ((#3 target)  = "Tank") then
          ((#1 target, #2 target, #3 target, [true, true, false]))
        else
          ((#1 target, #2 target, #3 target, [true, false, false]))
end}
```

**Table 42: BHQ E-BOM**

Action 4 is used to assign light, medium or heavy targets. If a target is heavy then all three batteries are assigned to hit the target. If the target is medium then battery 1 and 2 are assigned otherwise only battery 1 is assigned.

| **FDC E-BOM** | | | | | | |
|---|---|---|---|---|---|---|
| **SV:Types** | {$C_{11}$(Id:Integer), $C_{12}$(FD:FIELD_DATA) , $C_{13}$(CurrentTarget:TARGET), $C_{14}$(Result:Bool)} | | | | | |
| **Initial States** | {$S_{17}$: FDCReady } | | | | | |
| **Transitions** | | | | | | |
| **State** | **Event** | **Guard** | **{$SV_{IN}$}** | **{$SV_{OUT}$}** | **Action** | **Next State** |
| FDCReady | ProcessRequest | | C12, C13, C14 | | Action5 | Processing |
| Processing | RequestApproved | [Result=true] | | C13, C14 | | FDCReady |
| Processing | RequestDenied | [Result=false] | | C13, C14 | | FDCReady |

```
Action5 {
    input (target, fdcd);
    output (fdc_result);
    action
    let
       val tid = #1 target;
       val r = List.nth((listsub fdcd (GetFieldByID tid fdcd)),0);
    in
       (if (#4 r = "Enemy") then true else false)
    end
}
```

**Table 43: FDC E-BOM**

Action 5 is used to process targets. Here we only check from the internal FDC data that the target is an enemy unit. This method can be expanded to compute target priorities and process other tactical and technical fire direction rules.

| Battery E-BOM | | | | | | |
|---|---|---|---|---|---|---|
| **SV:Types** | {$C_{15}$(Id:Integer), $C_{16}$(CurrentTarget:TARGET) } | | | | | |
| **Initial States** | {$S_{19}$: ReadyToFire } | | | | | |
| **Transitions** | | | | | | |
| **State** | **Event** | **Guard** | **{$SV_{IN}$}** | **{$SV_{OUT}$}** | **Action** | **Next State** |
| ReadyToFire | AssignTarget | | C16 | | | PreparingCannon |
| PreparingCannon | Fire | | | C16 | Action6 | Firing |
| Firing | FiringCompleted | | | | | ReadyToFire |

```
Action6 {
    input (assign_target);
    output (fire);
    action
    let
        val bid = inst();
        val tid = #1 assign_target;
        val grid = #2 assign_target; val impact=grid;
    in
        (bid, tid, grid, impact)
    end
}
```

**Table 44: Battery E-BOM**

Action 6 is used to initiate the fire. It creates a token of type "Fire" to the Field component containing the information of the firing battery id, target id, grid location of the target and the location of the impact.

### 8.3.3 E-BOM to CPN Transformation

In this step, all the extended BOM models (E-BOM) are automatically transformed into our proposed CPN components in such a way that all variables from the corresponding E-BOMs are added in the *Structural Layer* (shown in Red color in the following figures) and the State-machine is transformed into the Behavioral Layer (shown in Green color). In communication layer (shown in blue color), receive-events are transformed into input ports and send-events are converted into output ports. Figure 64Figure 65Figure 66Figure 67Figure 68 represent the CPN component models of each component:

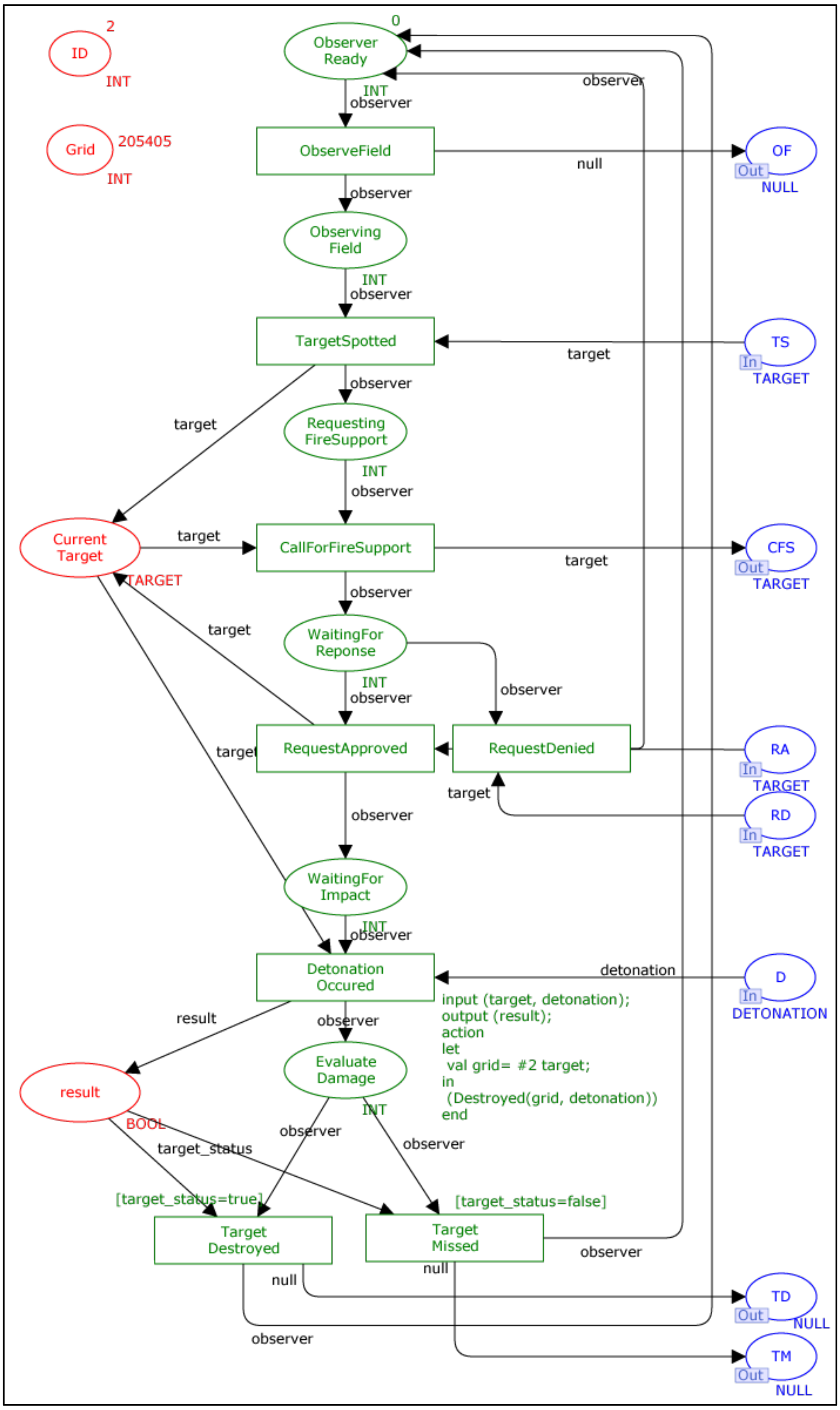

**Figure 64: Observer CPN Component**

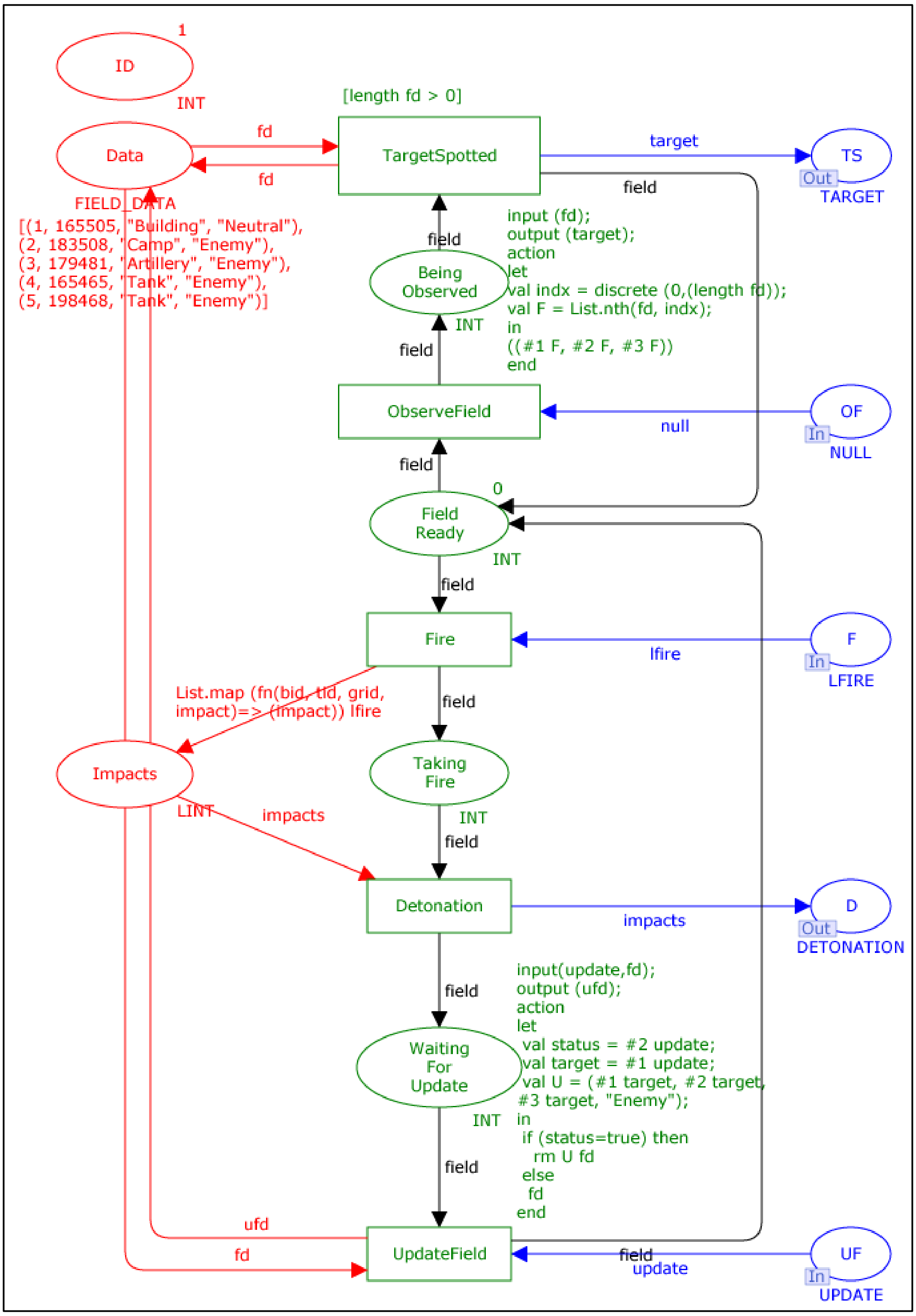

Figure 65: Field CPN Component

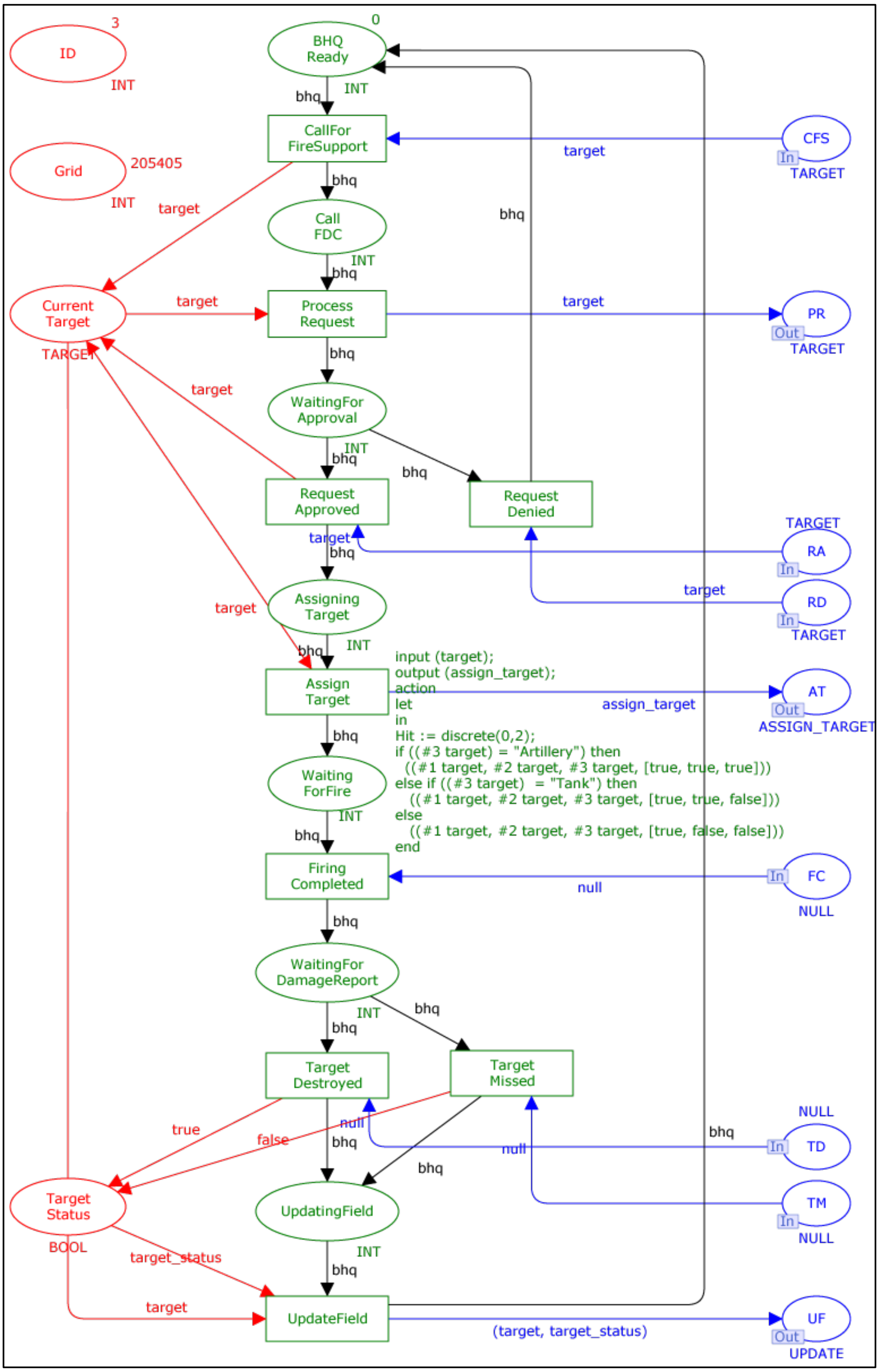

Figure 66: BHQ CPN Component

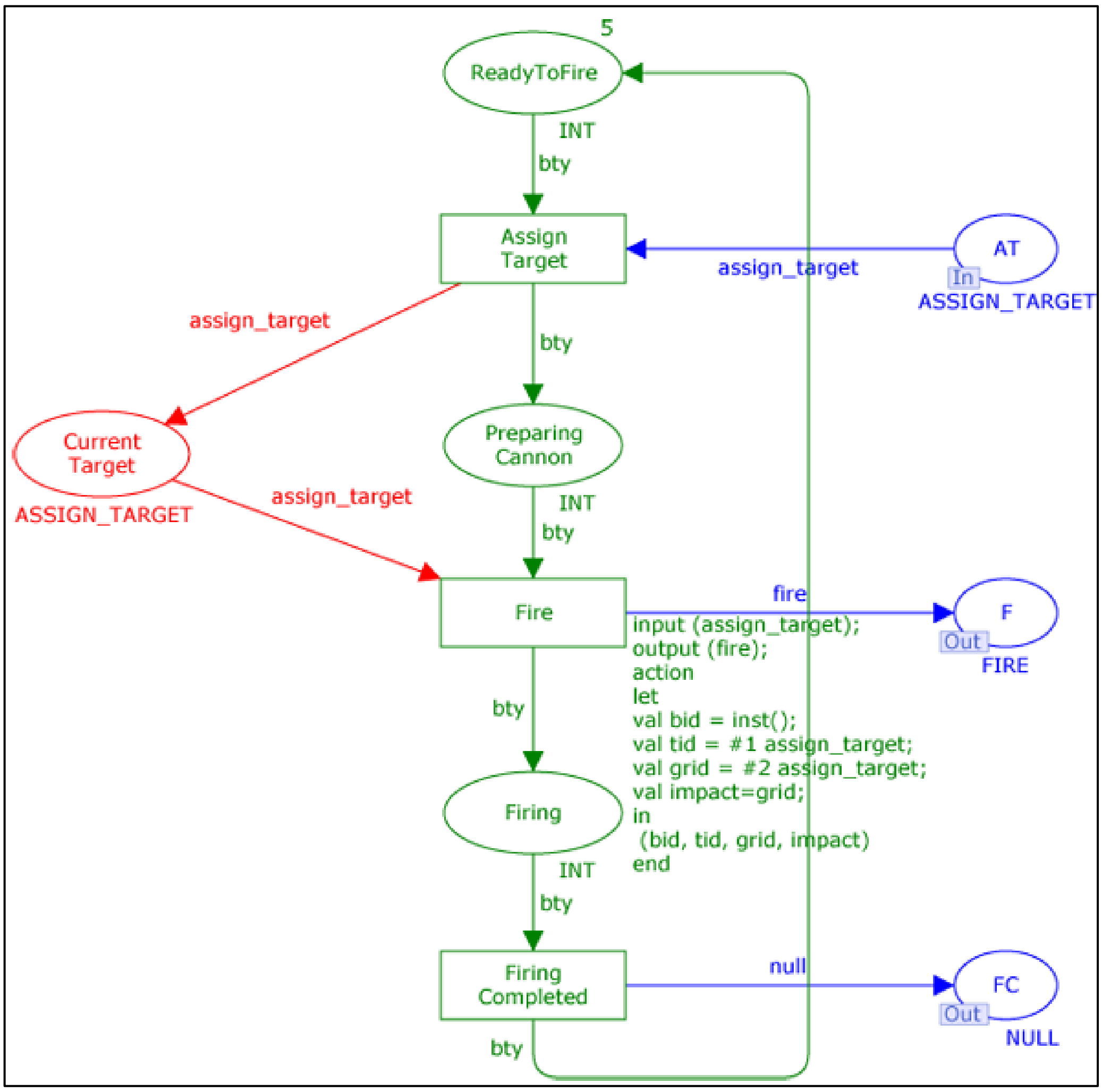

Figure 67: Battery CPN Component

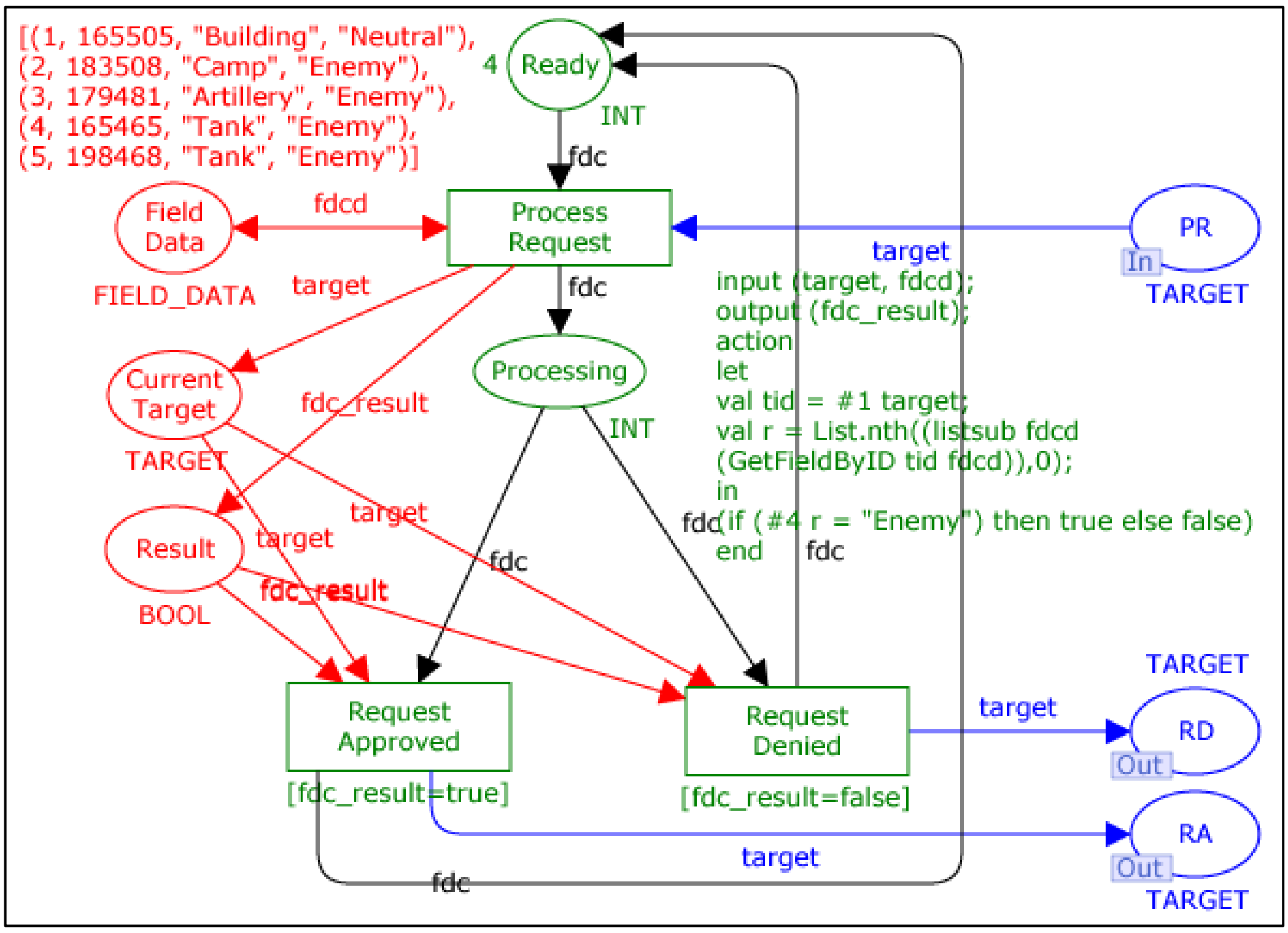

Figure 68: FDC CPN Component

We assume that each transformed CPN component has passed structural evaluation which is conducted using inspection method and behavioral evaluation conducted using Functional Testing method therefore S3b is also partially satisfied.

### 8.3.4 Composition of CPN Components

In this step all CPN modules are combined together through socket places in a CPN Composed Model as shown in Figure 69. In this composed model some general purpose auxiliary components are also introduced such as Join and Fork to facilitate the composition.

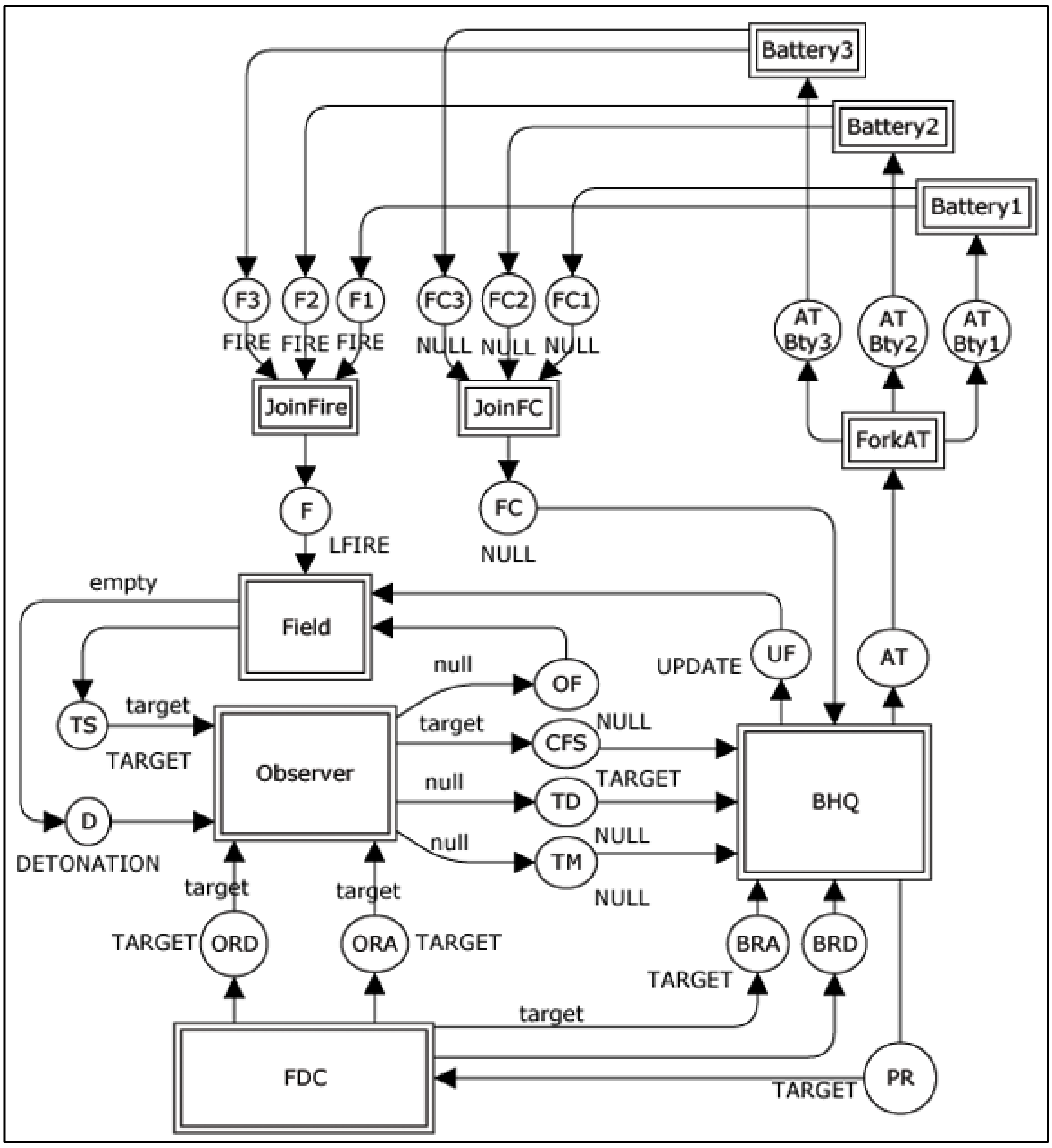

**Figure 69: Field Artillery CPN Composed Model**

When the model is composed it is executed in the CPN execution environment. The successful execution of the model (according to **Figure 63**) satisfies S3b completely.

## 8.3.5 State space Analysis

In the next step the state-space of the entire Field Artillery Model is generated using CPN state-space calculation tool, and is used to perform verification. The generated state-space graph consists of 1960 nodes and 6469 arcs as shown in **Figure 70**.

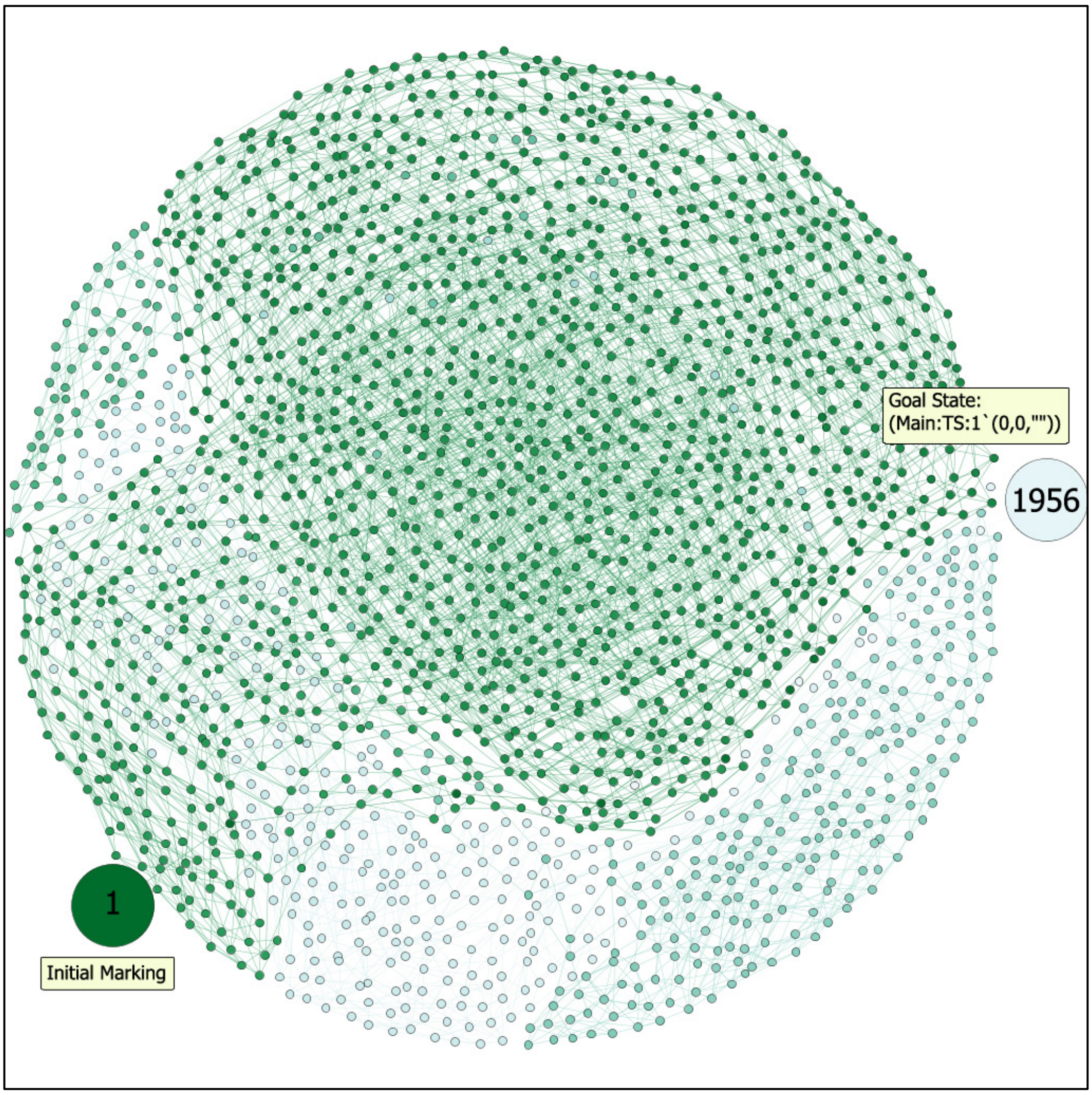

**Figure 70: State space of Field Artillery CPN Model (1960 nodes, 6469 edges)**

After the state-space is constructed in CPN tools, it is exported into a GraphML file format. It is to be noted that the layout of the state-space graph in **Figure 70** is rendered using Gephi Tool. In that "node-1" represents initial marking of the composed model whereas "node-1956" represents the goal state (explained later in this section). Shades of green color (from dark to light) represent proximity from node 1. All the nodes are connected with edges (some of which may not be visible due to light colors).

### Translation of Requirements specification into CPN Properties

To proceed with the verification we first translate Objectives and Constraints defined in the requirement specification to CPN properties. We assume that the default constraints S1, S2 and S3 are already verified.

| **Objective** | **O1:** All the enemy units must be destroyed |
|---|---|
| **CPN Translation** | As the Observer detects enemy units, therefore we say that if no more enemy units can be detected (because field-data is empty) then all the enemies should be destroyed. Therefore a marking where TS (Target-spotted) place has a null token should exist. If such marking is found then the objective is said to be reached. The following CPN function can be used to verify this property. |
| **CPN Function** | `fun AllTargetDestroyed() =`<br>`let`<br>`  val token = 1`(0,0,""); /*Create a search criteria */`<br>`  val predicate = fn n => (Mark.Main'TS 1 n) = token; /*Create a predicate function*/`<br>`  val TS = PredAllNodes (predicate); /* Built-in Node search function`<br>`in`<br>`  if (length TS > 0)`<br>`    then true`<br>`  else false`<br>`end;` |
| **Result** | When the function AllTargetDestroyed() is executed it returns True. This is also evident from **Figure 70** where the marking 1956 represents the goal-state and is reachable form the initial state 1. |

| **Constraint** | **S4:** There should never be a friendly fire. |
|---|---|
| **CPN Translation** | When "*UpdateField*" (UF) place gets a token from BHQ component (which will be taken as input by the Field component), it shows which field unit is destroyed. We can collect all such nodes in the state-space (where UF field has tokens) and compare that all field units that have been destroyed are of type "enemy". If this condition holds in the entire state-space then S4 holds. Following CPN-ML function can be used to check if friendly fire has ever happened or not. The result should be false to satisfy S4 |
| **CPN Function** | `fun CheckFriendlyFire() =`<br>`let`<br>`  val predicate = fn n => IsNotEnemy(Mark.Main'UF 1 n) = true;`<br>`  val ListOfFrieldlyUnits = PredAllNodes (predicate);`<br>`in`<br>`  if (length ListOfFrieldlyUnits > 0) /* Means there is a friendly fire */`<br>`    then true`<br>`  else false`<br>`end;`<br><br>`fun IsNotEnemy (update) =`<br>`let`<br>`  val upd:UPDATE = List.nth(update, 0);`<br>`  val target = #1 upd; /*Extract information from the token at the place: UF */`<br>`in`<br>`  if GetType(#1 target) <> "Enemy" then true /* Checks field unit type*/`<br>`else`<br>`  false`<br>`end;` |
| **Result** | The function CheckFriendlyFire() results false because no such incidence occurred with the data (initial state) provided to the model.<br>To check that this function works correctly we created a counter example in which FDC component is assumed to be erroneous (i.e. it wrongly accepts fire support requests of the friendly units), we ran the routine and found traces of the occurrence of friendly fire. |

As all the constraints and Objectives are satisfied we say that the field artillery model is composable at all levels and is verified with respect to its given specifications. Therefore the BOM based composed model is qualified for further implementation.

## 8.4 State Space Reduction

In this section the application of our proposed state-space reduction technique is presented. In order to proceed with the state-space reduction of the FA model generated by CPN tools we perform following steps.

### 1. Trimming Node Description

Each node in the CPN state-space graph has a description. This description essentially tells about the presence of tokens (or multiple tokens) in all places of the model, which is called marking. This description is very lengthy if the model has many places or sub-models. In this step we remove all the descriptions and only keep the information related to the places of the main model. To perform this step we use the library function **`NodeDescriptorOptions()`** (see manual [**73**]). When the descriptions are trimmed we will only get information of a node pertinent to the main places otherwise it will be a "Null" string. (This is an important difference for further steps).

Conceptually we hypothesize that trimming the node description does not cause loss of information because all the information other than the one in the main places is produced by the internal logic of the composed components. Since the composed components are considered as black boxes and they will eventually output important information (in form of tokens) in any of the main places. This information would be sufficient to answer any verification query related to the model under consideration.

### 2. Export to GraphML

In the next step we export the state-space graph to an external file. Since CPN state-space graph cannot be manipulated internally within the CPN environment therefore we export the graph to a standard GraphML format [**128**] along with the trimmed node descriptions and the information of the edges. To perform this step we develop a GraphML writer function in CPN-ML.

### 3. Reduction Algorithm

In the next step, we apply our reduction algorithm specified in **Table 16**. This algorithm is implemented in a Java application which uses JUNG library for graph manipulation functions. In brief, all the nodes which have "Null" descriptions are removed (because they are irrelevant). When a node is removed all its incoming and outgoing edges are removed. So we connect each predecessor of the node with each successor to preserve the structure of the graph. When all the nodes are checked the reduction is completed. The output of the reduce graph of Field Artillery mode is shown in **Figure 71**.

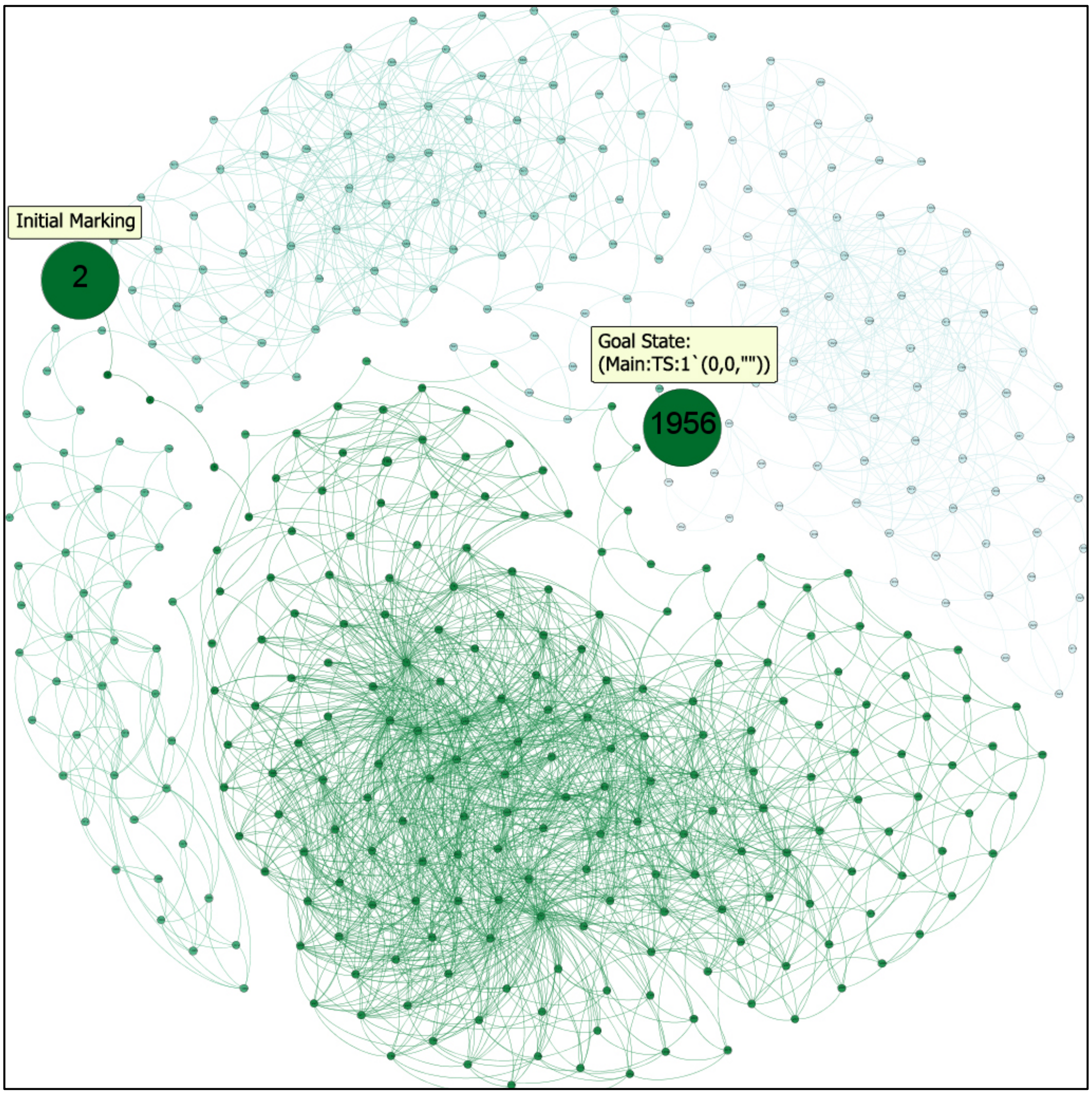

**Figure 71: Reduced State-Space graph of Field Artillery Model**

In **Figure 71** node-2 represents initial marking (note that node-1 was trimmed in the reduction process). Also Node-1956 still represents goal state. (Note that the nodes IDs remain the same in the reduction process).

| | Reduced Graph | Original Graph | Percentage |
|---|---|---|---|
| **Nodes** | 428 | 1960 | 21% |
| **Edges** | 2503 | 6469 | 38 % |

**Table 45: Reduction Statisitics**

**Table 45** shows that the nodes are reduced to 21% and the edges are reduced to 38% of the original graph.

## 8.5 Summary

In this chapter the verification of BOM based composed models is discussed using CPN based state-space analysis technique. An example model of Field Artillery is introduced and the entire verification process is applied on this model. It is shown how requirement specifications are translated into CPN properties and how they are verified using state-space analysis and Query functions.

State-space analysis is advantageous as it is exhaustive and leads the modeler to all of the possibilities that can occur during the abstract level execution of a composed model. A state-space graph helps to study all of these possibilities and to understand the dynamic behavior of the components in detail. Also a state-space query functions proposed along with the approach help in answering different verification questions and evaluate the correctness of model with respect to the requirements. This approach however is vulnerable to the state-space explosion as for a simple model of Field Artillery 2503 nodes and 6469 edges were formed. To deal with this situation we proposed an effective state-space reduction technique which not only reduces large state-spaces into reasonable size but also preserves important information for correct verification. We demonstrated how our proposed state-space reduction technique is applied to the Field Artillery model for proof of concept.

# Chapter 9
# Model Verification using CSP based Model Checking Technique

*Model Checking is becoming a standard approach for the software verification due to its numerous advantages over traditional formal methods. Communicating Sequential Processes (CSP) is an event based formal language for describing patterns of interactions in concurrent systems and very useful for concurrent behavioral specification and verification due to its theoretical foundations of process algebra (also called process calculi). The application of CSP based Model Checking technique in the Composability verification also proves to be very useful, especially with a focus on the dynamic semantic composability level. In this chapter the Field Artillery Model presented in Chapter 8 is reused and extended with information to capture the behavior of a real-time probabilistic system. It is shown how the Probabilistic-Timed Field Artillery Model is transformed into a composed CSP model and verified using PAT.*

In this chapter, the modified version of Field Artillery Model presented in chapter 8 is discussed as an example. The objective of this example is to represent a model of a system with time constraints and probabilistic behavior. To the best of our knowledge the PN algebraic approach does not support verification of the timed models or probabilistic systems at all. Also the CPN based approach has a limited support for the verification of timed system but it does not cover probabilistic systems. We therefore propose to apply Modeling Checking for real-time probabilistic systems and show how a composed model of one such system can be verified using a CSP based Model checker called PAT (see 3.2.7).

## 9.1 Field Artillery Scenario

The scenario of the Field Artillery Model is slightly different. It is assumed that a soldier observes the field and detects enemy units. When a target is spotted, he calls BHQ for fire support and provides the target details. In military practice, Time-On-Target (TOT) is a Field Artillery coordination protocol observed by multiple firing units. This technique was developed by the U.S. Army during World War II. It uses a precise pre-determination of the estimated preparation time and the time of flight of the munitions from each firing battery to the target area. When a Time on Target (TOT) is designated each battery that will join in firing on that target subtracts the time of flight from the TOT to determine the time to fire. The firing units fire their rounds so that all the munitions arrive at the target at precisely the same time. This is done in order to achieve maximum target destruction. If there is a gap between the multiple impacts the enemy soldiers get time to prone or takeover in the hideouts and mobile vehicles can escape [**129**].

BHQ assigns target to the batteries, and also schedules a certain "TOT" for the batteries to comply. Each battery needs some time to prepare for loading appropriate ammunition and setting up the correct alignment and orientation of the barrel according to the computed firing solution using range (distance) and bearing (angle) of the assigned target. It is assumed that each battery needs a random preparation delay. When each battery is ready, it will fire in its own time such that all the rounds hit the target at the given TOT. We also assume that the probability of hitting on the exact target location for each battery is '0.9'. In contrast to the previous scenario in chapter 8, we assume that there is only one target in the field component and all three batteries are taking part in the firing operation. To construct a conceptual model for this scenario, the following BOM components are composed:

**Field**: Target location (We assume there is only one target).

**Observer**: A soldier who request for the fire support from BHQ.

**BHQ**: Supervises the entire operation of fire support, responds to the calls for fire support and assigns targets to the batteries.

**Battery**: Three units of artillery batteries (cannons and crew) responsible to hit the target exactly at a given time.

(Note that FDC component is removed from the composition. Also some entity characteristics and event parameters are reduced for simplification). The modified BOM components of the Field Artillery conceptual model are formally defined as follows:

| **Observer = ⟨ EnT, EvT, S, AcT ⟩** where: |
|---|
| **EnT** = Observer { $C_0$(target)} |
| **EvT** = {$E_0$(CallForFireSupport, Observer, BHQ, target), $E_1$(Detonation, Field, Observer, detonation)} |
| **Act** = {$A_0$(CallForFireSupport, Observer, BHQ, $E_0$), $A_1$(Detonation, Field, Observer, $E_1$)} |
| **S** = {$S_0$(ObserverReady, $A_0$, $S_1$), $S_1$(WaitingForImpact, $A_1$, $S_0$) } |

**Table 46: Observer Basic-BOM**

| **Field = ⟨ EnT, EvT, S, AcT ⟩** where: |
|---|
| **EnT** = Field {$C_1$(destruction[3])} |
| **EvT** = {$E_2$(Fire, Battery1, Field, BID), $E_3$(Fire, Battery2, Field, BID), $E_4$(Fire, Battery3, Field, BID), $E_5$(Detonation, Field, Observer, destruction)} |
| **Act** = {$A_2$(Fire, Battery1, Field, $E_2$), $A_3$(Fire, Battery2, Field, $E_3$), $A_4$(Fire, Battery3, Field, $E_4$), $A_5$(Detonation, Field, Observer, $E_5$) } |
| **S** = {$S_2$(FieldReady, {$A_2$, $S_3$}+{$A_3$, $S_3$}+{$A_4$, $S_3$}) , $S_3$(TakingFire, $A_5$, $S_2$)} |

**Table 47: Field Basic-BOM**

| **BHQ = ⟨ EnT, EvT, S, AcT ⟩ where:** |
|---|
| **EnT** =BHQ {$C_2$(TOT) } |
| **EvT** = {$E_6$(CallForFireSupport, Observer, Field, target), $E_7$(AssignTarget, BHQ, Battery1, Battery2, Battery3, TOT), $E_8$(FiringCompleted, Battery1, BHQ, null), $E_9$(FiringCompleted, Battery2, BHQ, null), $E_{10}$(FiringCompleted, Battery3, BHQ, null} |
| **Act** = {$A_6$(CallForFireSupport, Observer, Field, $E_6$), $A_7$(AssignTarget, BHQ, Battery1, Battery2, Battery3, $E_7$), $A_8$(FiringCompleted, Battery1, BHQ, $E_8$), $A_9$(FiringCompleted, Battery2, BHQ, $E_9$), $A_{10}$(FiringCompleted, Battery3, BHQ, $E_{10}$)} |
| **S**={$S_4$(BHQReady, $A_6$, $S_5$), $S_5$(AssigningTarget, $A_7$, $S_6$), $S_6$(WaitingForFire, {$A_8$, $S_4$} + {$A_9$, $S_4$} + {$A_{10}$, $S_4$})} |

**Table 48: BHQ Basic-BOM**

| **Battery1,2,3 = ⟨ EnT, EvT, S, AcT ⟩ where:** |
|---|
| **EnT** = Battery1,2,3 { $C_3$(BID), $C_4$(Destroyed) } |
| **EvT** = {$E_{11}$(AssignTarget, BHQ, Battery1, Battery2, Battery3, TOT), $E_{12}$(ReadyToFire, Battery1/2/3, Battery1/2/3, null), $E_{13}$(Fire, Battery123, Field, BID, Destroyed), $E_{14}$(FiringCompleted, Battery123, BHQ, null)} |
| **Act** = {$A_{11}$(AssignTarget, BHQ, Battery1, Battery2, Battery3, $E_{11}$), $A_{12}$(ReadyToFire, Battery1/2/3, Battery1/2/3, $E_{12}$), $A_{13}$(Fire, Battery1/2/3, Field, $E_{13}$), $A_{14}$(FiringCompleted, Battery1/2/3, BHQ, $E_{14}$)} |
| **S** = {$S_7$(BatteryIdle, $A_{11}$, $S_7$), $S_8$(Preparing, $A_{12}$, $S_9$), $S_9$(ReadyToFire, $A_{13}$, $S_{10}$), $S_{10}$(Firing, $A_{14}$, $S_7$)} |

**Table 49: Battery (1,2,3) Basic-BOM**

| **FA = ⟨ $AcT_{IN}$, $AcT_{OUT}$ , POI ⟩ where:** |
|---|
| **$AcT_{IN}$** = **$AcT_{OUT}$** = ∅ |
| **POI** = { POI-0(!A0, ?A6), POI-1(!A7, ?A11), POI-2(!A12), POI-3(!A13, {?A2, ?A3, ?A4}), POI-4(!A14, {?A8, ?A9, ?A10}), POI-5(!A5,?A1} |

**Table 50: Field Artillery Composed BOM**

The composed field artillery model is shown in **Figure 72** using our proposed graphical notation.

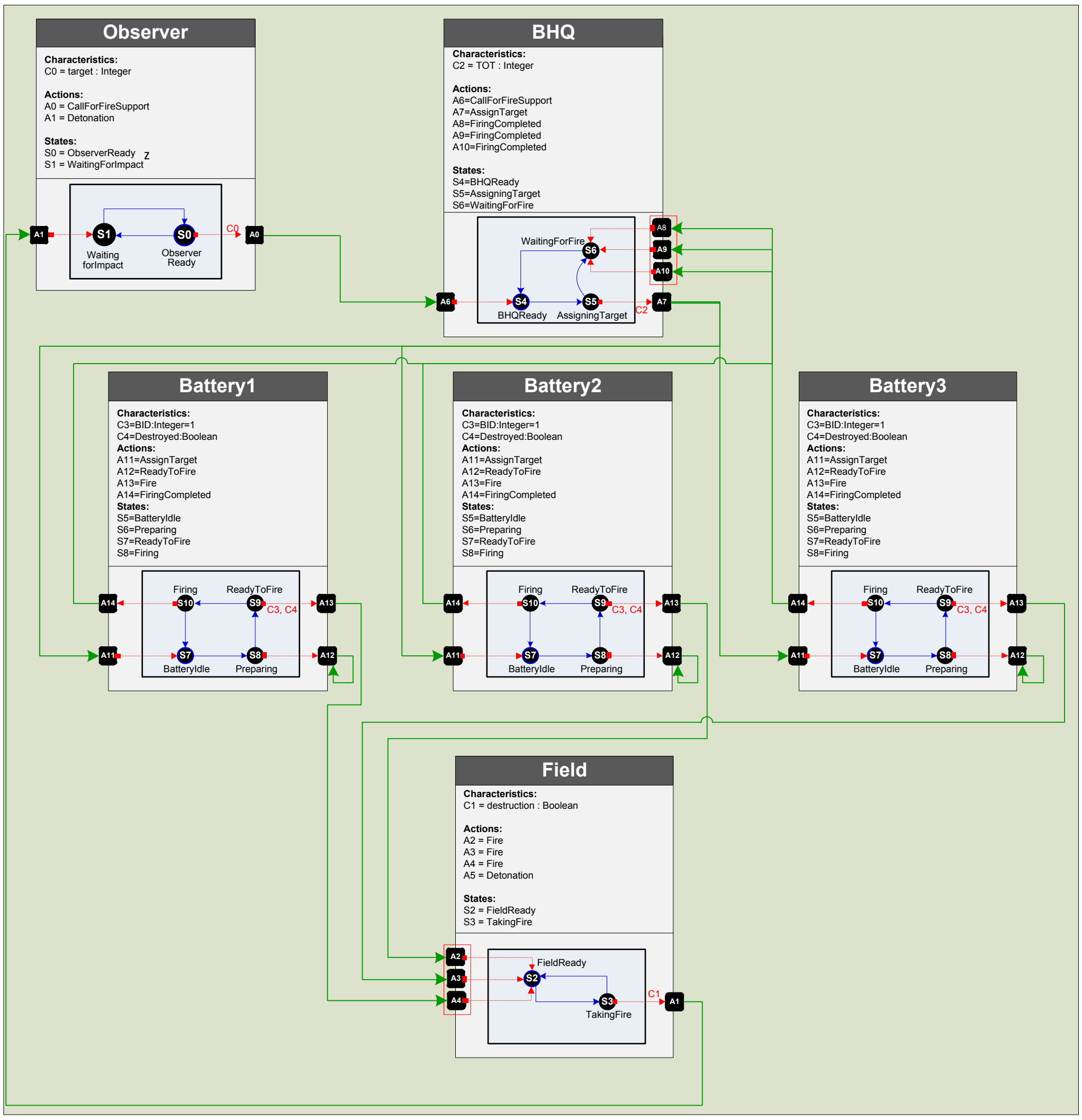

**Figure 72: Field Artillery Composed Model**

## 9.2 Requirement Specification

We define Requirement speciation of the modified field artillery model as:

**$RS_0$ = ⟨O, S⟩** where:

**Objectives O = {$o_1$, $o_2$} and System Constraints S = {$s_1$, $s_2$ $s_3$, $s_4$}**

**$o_1$:** All the firing units should fire precisely at the target location

**$o_2$:** All the firing units should fire at the target exactly at the given time (i.e., the Time on Target property should be satisfied)

**$s_{1,\ 2,\ \text{and}\ 3}$:** The model should be composable at syntactic and static-semantic level. The state-machines should match and the executable mode should correctly represent the conceptual model.

## 9.3 Verification using Model Checking

After the BOM are discovered, selected they are composed to form a conceptual model according to the simuland. This composed model is now ready for verification. At this stage we select model checking technique for its verification.

### 9.3.1 Static and Dynamic Analysis

We assume that the model qualifies syntactic and static-semantic analysis. Also when it undergoes state-machine matching process it is able to make progress until the goal-states are reached. **Figure 73** shows the interaction of the state-machine of each component during the state-machine matching process.

Based on the fact that the constraint S1, S2 and S3a are satisfied we proceed to BOM-to-E-BOM extension.

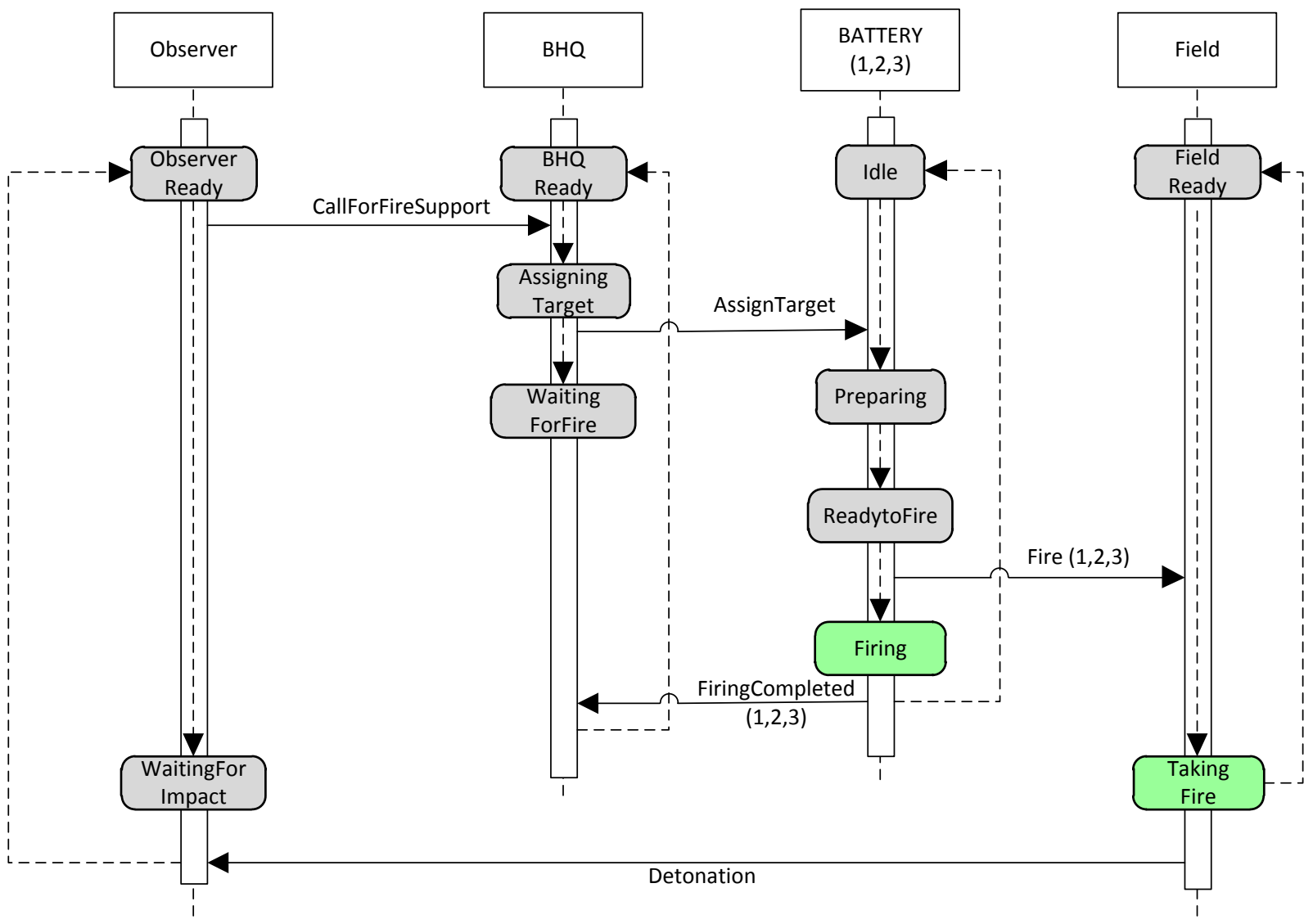

**Figure 73: State-machine Matching of Field Artillery Model**

### 9.3.2 BOM to E-BOM extension

At this stage all the BOM components are extended to our proposed E-BOM extension with the help of the modeler's input. Here additional information such as timing constraints and probabilistic factors are proposed to be included. Following tables present E-BOM extensions of BOMs in the FA model.

| **Observer E-BOM** | | | | | |
|---|---|---|---|---|---|
| **SV and types** | {$C_0$(Target:Integer} | | | | |
| **Initial States** | {$S_0$:ObserverReady} | | | | |
| **Transitions** | | | | | |
| **State** | **Event** | **Time** | **Guard** | **Action** | **Next State** |
| Observer Ready | CallForFireSupport | | | | WaitingFor Impact |
| WaitingFor Impact | Detonation | | | | Observer Ready |

**Table 51: Observer E-BOM**

| Field E-BOM | | | | | |
|---|---|---|---|---|---|
| **SV and types** | {$C_1$(Firing_Result_Of_Battery1:Boolean}, {$C_1$(Firing_Result_Of_Battery2:Boolean}, {$C_1$(Firing_Result_Of_Battery3:Boolean} | | | | |
| **Initial States** | {$S_2$:FieldReady} | | | | |
| **Transitions** | | | | | |
| **State** | **Event** | **Time** | **Guard** | **Action** | **Next State** |
| FieldReady | Fire(1) | | | Action1 | TakingFire |
| | Fire(2) | | | Action2 | |
| | Fire(3) | | | Action3 | |
| TakingFire | Detonation | | | | FieldReady |

```
/* A CSP script for defining a probabilistic action, with 95% chance that the
target will be destroyed when an event fire is received from battery1 and 5% chance
that the target will be missed */
Action1{
  pcase{
        [0.05] : fire?1 → atomic{tau{ Destruction [0]=False;} → Skip}
        default : fire?1→ atomic{tau{ Destruction [0]=True;}→ Skip}
      }
  }
/* From battery 2 */
Action2{
  pcase{
        [0.05] : fire?2 → atomic{tau{ Destruction [1]=False;} → Skip}
        default : fire?2→ atomic{tau{ Destruction [1]=True;}→ Skip}
      }
}
/* From battery 3 */
Action3{
  pcase{
        [0.05] : fire?3 → atomic{tau{ Destruction [2]=False;} → Skip}
        default : fire?3→ atomic{tau{ Destruction [2]=True;}→ Skip}
      }
}
```

**Table 52: Field E-BOM**

| BHQ E-BOM | | | | | |
|---|---|---|---|---|---|
| **SV and types** | {$C_2$(TOT:Integer} | | | | |
| **Initial States** | {$S_4$: BHQReady } | | | | |
| **Transitions** | | | | | |
| **State** | **Event** | **Time** | **Guard** | **Action** | **Next State** |
| BHQReady | CallForFireSupport | | | | AssigningTarget |
| AssigningTarget | AssignTarget | | | | WaitingForFire |
| WaitingForFire | FiringCompleted(1) | | | | BHQReady |
| | FiringCompleted(2) | | | | |
| | FiringCompleted(3) | | | | |

**Table 53: BHQ E-BOM**

| **Battery(1,2,3) E-BOM** | | | | | |
|---|---|---|---|---|---|
| **SV and types** | {$C_3$(BID:Integer} | | | | |
| **Initial States** | {$S_7$: BatteryIdle } | | | | |
| **Transitions** | | | | | |
| **State** | **Event** | **Time** | **Guard** | **Action** | **Next State** |
| BatteryIdle | AssignTarget | | | | Preparing |
| Preparing | readytofire | Wait[Prob] | | Action1 | ReadyToFire |
| ReadyToFire | Fire | | | | Firing |
| Firing | FiringCompleted | | | | BatteryIdle |

/* A CSP script for defining a probabilistic wait action, with 94% chance that each battery will launch the fire exactly at given time on target, and 6% chance that it will fire earlier or later */

```
Action1{
        pcase{
             [0.01] :  Wait[TOT-3]; readytofire → ReadyToFire(i)
             [0.01] :  Wait[TOT-2]; readytofire → ReadyToFire(i)
             [0.01] :  Wait[TOT-1]; readytofire → ReadyToFire(i)
             [0.94] :  Wait[TOT];    readytofire → ReadyToFire(i)
             [0.01] :  Wait[TOT+1]; readytofire → ReadyToFire(i)
             [0.01] :  Wait[TOT+2]; readytofire → ReadyToFire(i)
             [0.01] :  Wait[TOT+3]; readytofire → ReadyToFire(i)
             };
  }
```

**Table 54: BHQ E-BOM**

## 9.3.3 E-BOM to CSP# Transformation

In this step, all the probabilistic timed extended BOM models are automatically transformed into CSP# using our automatic BOM-to-CSP transformation tool. **Figure 74** shows the global code block which is used to define global variables and the communication channels for each BOM send-receive event pair.

```
//------------Global Block ---------------------------------
#define TOT 30; //Constant pre-defined Time on Target
enum {Hit, Miss}; //Hit or Miss flag
//Each battery has a hit/miss ratio = 95:5 %
var Firing_Result_Of_Battery1=Miss;
var Firing_Result_Of_Battery2=Miss;
var Firing_Result_Of_Battery3=Miss;

//For each event a channel is defined
channel callforfire 0;
channel detonate 0;
channel assigntarget 0;
channel firingcomplete 0;
channel fire 0;
```

**Figure 74: Global code Block of Field Artillery Model**

**Figure 75** shows the CSP code of Observer BOM component. The send-events are transformed into channels with send operator '!' and receive-events are transformed into channels with receive operator '?'.

```
//============================================================
//OBSERVER Component
//============================================================

ObserverSM = ObserverReady(); //Initial State

ObserverReady()=
 (callforfire!0 ->WaitingForImpact());

WaitingForImpact()=
 (detonate?0 -> ObserverReady());

//============================================================
```

**Figure 75: CSP representation of Observer Component**

```
//============================================================
//BHQ
//============================================================
BHQSM = BHQReady();//Initial State

BHQReady()=
(callforfire?0 ->AssigningTarget());

AssigningTarget()=
(assigntarget!> assigntarget!2 -> assigntarget!3 ->
       Waitingforfire());
//Sending assigntarget to multiple recievers

Waitingforfire()=
       firingcomplete?>firingcomplete?2->firingcomplete?3->
              BHQReady();
//Recieving firingcomplete from multiple senders

//============================================================
```

**Figure 76: CSP representation of BHQ Component**

```
//============================================================
//BATTERY i={1,2,3}
//============================================================
BatterySM(i) = BatteryIdle(i); //Initial State
BatteryIdle(i)=
(assigntarget?i->Preparing(i));
Preparing(i)= pcase{
              [0.01] :  Wait[TOT-3]; readytofire-> ReadyToFire(i)
              [0.01] :  Wait[TOT-2]; readytofire-> ReadyToFire(i)
              [0.01] :  Wait[TOT-1]; readytofire-> ReadyToFire(i)
              [0.94] :  Wait[TOT];   readytofire-> ReadyToFire(i)
              [0.01] :  Wait[TOT+1]; readytofire-> ReadyToFire(i)
              [0.01] :  Wait[TOT+2]; readytofire-> ReadyToFire(i)
              [0.01] :  Wait[TOT+3]; readytofire-> ReadyToFire(i)
              };
// TOT is a global constant
//readytofire is an internal event
ReadyToFire(i)= fire!i->Firing(i);
Firing(i)= firingcomplete!i ->BatteryIdle(i);
//============================================================
```

**Figure 77: CSP representation of Battery Component**

```
//============================================================
//FIELD Component
//============================================================
FieldSM = FieldReady(); //Initial State

FieldReady()=
       pcase{
       [0.05]: fire?1 ->
              atomic{tau{Firing_Result_Of_Battery1=Miss;} -> Skip}
        default : fire?1 ->
              atomic{tau{Firing_Result_Of_Battery1=Hit;} -> Skip}
 }     |||
/* ||| is the interleaving operator between the synchronizing events
fire(1), fire(2) and fire(3) */
       pcase{
              [0.05]: fire?2 ->
                     atomic{tau{Firing_Result_Of_Battery2=Miss;} -> Skip}
              default : fire?2 ->
                     atomic{tau{Firing_Result_Of_Battery2=Hit;} -> Skip}
 }     |||
       pcase{
              [0.05]: fire?3 ->
                     atomic{tau{Firing_Result_Of_Battery3=Miss;} -> Skip}
              default : fire?3 ->
                      atomic{tau{Firing_Result_Of_Battery3=Hit;} -> Skip}
 };
/* This code randomly sets hit or miss effect for the firing of each
battery */

Detonation();
Detonation()= detonate!0 -> FieldReady();
//============================================================
```

**Figure 78: CSP representation of Field Component**

```
// FIELD ARTILLERY COMPOSED MODEL
//=================================================================

FieldArtillery = ObserverSM || BHQSM || FieldSM || BatterySM(1)||
BatterySM(2)|| BatterySM(3)

// || is the parallel operator between all the components
// BatterySM has three instances initialized with battery id parameter.
```

**Figure 79: Field Artillery Composed Model**

**Figure 79** shows the CSP representation of how the transformed components are composed using the parallelism operator '||'. This means that all the components execute in parallel, however they perform barrier synchronization while exchanging events in their respective communication channels.

### 9.3.4 Model Checking of Field Artillery Model

The CSP based Field Artillery Model can be opened and executed in PAT tool. A successful compilation of this model shows that it has no errors. When this model is executed, and if each component reaches its final states then we say that the constraint S3b of requirement specification is satisfied i.e., the transformed executable model correctly represents the behavior of its conceptual model.

In the verification process, we define the following assertions to be verified by PAT built-in model checker. Since the nature of the input model is probabilistic and real-time, we use **Probabilistic-Real-Time module** of the PAT tool.

**Figure 80** shows how we define goal reachability assertions using PAT's Probabilistic CSP LTL specification.

```
//=========================================================
// FIELD ARTILLERY COMPOSABILITY VERIFICATION
//=========================================================

// ASSERT1: Goal state Reachability
#assert FieldArtillery |= [](callforfire.0 -> <>detonate.0);

// Goal Definition
#define goal (Firing_Result_Of_Battery1==Hit
              && Firing_Result_Of_Battery2==Hit
              && Firing_Result_Of_Battery3==Hit);

//ASSERT2: //Goal Reachability
#assert FieldArtillery |= <>goal with prob;
```

**Figure 80: Field Artillery Verificataion Assertions**

Assertion1 uses LTL construct to verify that if there is a "callforfire" then detonation at the target location will eventually occur. If assertion1 is satisfied, it shows that there exists a valid execution path, which leads to the goal state.

The result of PAT model checker is shown in **Figure 81** which shows that the goal state is reachable.

```
********Verification Result********
The Assertion (FieldArtillery() |= []( callforfire.0-><> detonate.0)) is VALID.

********Verification Statistics********
Visited States:160477
Total Transitions:475422
Time Used:8.8263759s
Estimated Memory Used:111668.696KB
```

**Figure 81: Verification Result of assertion 1**

Assertion2 uses an LTL construct to verify that the "goal" is eventually reachable where "goal" is defined as a condition that all the batteries successfully hit at the exact location of the target. Note that assertion2 uses "**with prob**" construct, which makes it a PLTL statement. **Figure 82** shows the verification result which means that the probability of reaching the goal is between 77% and 94%.

```
********Verification Result********
The Assertion (FieldArtillery() |= <> goal with prob) is Valid with Probability [0.77378, 0.94526];

********Verification Statistics********
Visited States:84019
Total Transitions:245561
MDP Iterations:63123
Time Used:5.5017483s
Estimated Memory Used:97028.192KB
```

**Figure 82: Verification result of assertion 2**

Now we check whether the goal is reachable within the time constraints defined by Time-On-Target property. To perform this evaluation we use the PAT's *deadline* operator as shown in **Figure 83.** We define three assertions: Early, Exactly and Late.

```
//===========================================================
// FIELD ARTILLERY COMPOSABILITY VERIFICATION
//===========================================================

//Goal Reachability with TOT constraint
Early = FieldArtillery deadline[TOT-3];
Exactly = FieldArtillery deadline[TOT];
Late = FieldArtillery deadline[TOT+3];


//ASSERT3: //Goal Reachability at TOT
#assert Early reaches goal with prob;
#assert Exactly reaches goal with prob;
#assert Late reaches goal with prob;
```

**Figure 83: Field Artillery Verificataion Assertions with TOT**

The verification result of assertion 3 is shown in **Figure 84** according to which the early reachability of the goal is impossible. Whereas the maximum probability of reaching the goal exactly on TOT is 86% which satisfies objectives $O_1$ and $O_2$ However the maximum probability of reaching goal at a later time is 94% which satisfies $O_1$ with a higher probability but does not satisfy $O_2$.

| |
|---|
| ********Verification Result********<br>The Assertion (**Early()** reaches goal with prob) is NOT valid.<br><br>********Verification Statistics********<br>Visited States:58414<br>Total Transitions:146548<br>MDP Iterations:2557<br>Time Used:4.46633s<br>Estimated Memory Used:69901.84KB |
| ********Verification Result********<br>The Assertion (**Exactly()** reaches goal with prob) is Valid with Probability [0, 0.86271];<br><br>********Verification Statistics********<br>Visited States:169998<br>Total Transitions:342091<br>MDP Iterations:28115<br>Time Used:9.6064619s<br>Estimated Memory Used:148703.552KB |
| ********Verification Result********<br>The Assertion (**Late()** reaches goal with prob) is Valid with Probability [0, 0.94526];<br><br>********Verification Statistics********<br>Visited States:274934<br>Total Transitions:584498<br>MDP Iterations:112797<br>Time Used:15.4390606s<br>Estimated Memory Used:232989.792KB |

**Figure 84: Verification result of assertion 3**

Based on the verification results, we can say that the field artillery model satisfies it's given requirements with a certain probability factor. Since it is a non-deterministic model, the reliability of the success depends on the threshold between how tight the Time-On-Target deadline is that BHQ can assign and how efficiently the batteries can prepare and how accurately they can fire on the target.

## 9.4 Summary

In this chapter the model checking approach is presented with an example and verified using Process Analysis Toolkit (PAT). The example of Field Artillery Model (from chapter 8) is modified to represent a Probabilistic-Timed model in order to explain how the CSP based model checking approach using PAT can be effective in the composability verification. Using the example of Field Artillery model, it is explained how the verification of time constraints is performed and how different property assertions are verified with probability using PLTL. A successful verification of this approach is a result of satisfaction of all the assertions defined in the requirement specification, with an acceptable probability factor, and hence shows that the components are composable.

# Chapter 10
# Summary and Conclusion

*This chapter makes a comparison between the composability verification approaches presented in this thesis and provides some guidelines for choosing the appropriate approach according to the nature of the composed model. This discussion is followed by a summary of the major contributions of the thesis and some suggestions for future research in this area are suggested.*

In this thesis we propose a verification framework that follows the fundamental principles of M&S domain in terms of the notions of model correctness. It integrates several methods, techniques and tools to support different tasks in the multi-tier composability verification process of a composed model. It also inherits useful technological characteristics related to model verification from other communities such as Petri Nets, Model Checking and Process-Algebra community. And utilizes the existing knowledge shared by these communities for the verification of component based simulation models. These simulation models are called component based models because they are designed in form of components and can further be composed to construct sophisticated models (called composed models or compositions). To ensure correctness, a composed simulation model is required to be verified at its different composability levels, where each level poses certain degree of difficulty in verification. The initial levels of composability require that all the components in a composition can be syntactically connected to each other through valid interfaces. And they can correctly communicate with valid semantics. Whereas a deeper level of composability is the dynamic-semantic composability which requires that all the composed components should possess suitable behavior in order to correctly interact with each other for pursuing mutual objectives. The validity of behavior in a component composition relies on two factors: (i) each component should always be at the right state while interacting with the others and (ii) the composition should satisfy required behavioral properties, as prescribed in the requirement specifications.

The proposed verification framework not only provides complete support for the verification of initial composability levels, but also the most important characteristic of this framework is its ability to verify the composed model components at the deeper level of *dynamic-semantic composability*. Composability Verification at this level is a daunting task and requires a dynamic analysis approach. The behavior of components can be studied when they are set to interplay with each other in an execution environment, where they communicate through the exchange of events and make progress by the change of their internal states. Therefore an appropriate dynamic analysis approach is required which not only provides suitable execution environment but also support built-in verification techniques to evaluate the composability behavior at the runtime.

According to our findings not a single approach completely covers all the intricacies required for proving correctness at this the dynamic-semantic composability level due to its complex nature. The effectiveness of a certain approach also varies due to the varied nature of the composed model and the modeling formalism used. Since some models have complicated structure and demand rich expressiveness in terms of data-centric details for the abstraction of a system; Whereas others have behavior of complex nature including notions of concurrency and temporal constraints. Besides the system behavior can be deterministic or stochastic. Therefore it is difficult to depend on a single approach for the challenging task of dynamic-semantic composability verification.

For this reason we investigated three different dynamic analysis approaches in our framework namely: (i) PN based Algebraic Analysis, (ii) CPN based State-Space Analysis and (iii) CSP based Model Checking Technique. These approaches inherit theories, methods, tools and techniques from their corresponding ancestry communities such as PN, CSP and Model checking. We adapt these inherited resources and integrate them in our framework. We also propose several extensions in each approach to suit the needs of dynamic-semantic composability verification. Some of these extensions are listed as follows:

- A component-based description format is proposed. This description format is used to represent the BOM based composed model in the required form in order to apply the selected approach. For instance a CPN based component model is proposed which represents the structural and behavioral aspects of a BOM component in form of a CPN model. Similarly for CSP, a Component oriented CSP process model is introduced which represents a BOM component using CSP notation.
- For each approach a rule based transformation technique is proposed which converts BOM components into the description format of the corresponding approach while keeping the structure and behavior of the model preserved. To ensure this fact methods are proposed to compare the original model (BOM) and the transformed model to assert that the latter correctly represents the former.
- For PN algebraic approach algorithms are proposed to automate the process. Also a function library is developed for the ease of conducting repeated verification tasks.
- In case of state-space analysis, a reduction technique is proposed which helps in reducing a large state-space and ease the process of verification.

The advantages and disadvantages of these three approaches are categorized as follows:

Advantage | Disadvantage | Neutral

| Category: | Kinds of properties that can be verified |
|---|---|
| **PN Algebraic Analysis** | This approach only verifies a limited number of properties because it depends on the applicability of underlying mathematical theorems which are limited in number and may not cover all types of properties |
| **CPN State-Space Analysis** | It constructs state-space of all possibilities that a system could be in. Therefore it allows to specify and verify different kinds of general system properties as well as scenario specific properties. |
| **CSP based Model Checking** | In this approach the verification depends on the specification of properties using LTL or CTL assertions, which along with their variety of extensions provide rich expressiveness to define different kinds of properties. Therefore it covers a bigger pool of verification questions both in terms of generic as well as scenario specific properties |

**Table 55: Kinds of properties that can be verified**

| Category: | Type of the models that can be verified |
|---|---|
| **PN Algebraic Analysis** | This approach supports simple event-driven PN models. It does not support models with rich data, or models of real-time or probabilistic systems. |
| **CPN State-Space Analysis** | This approach support models with rich data-centric structure and behavior since it offers flow of the tokens of complex data-types and their manipulations during the transitions. It also offers limited support for Timed systems. However it does not support model verification of probabilistic nature. |
| **CSP based Model Checking** | This approach limits size of information in the model and does not entertain models with rich data-centric expressiveness. However it offers a variety of types of systems that can be verified such as reactive systems, real-time systems, probabilistic and stochastic systems. Therefore this approach is much stronger in verifying different kinds of systems. |

**Table 56: Type of the models that can be verified**

| Category: | Scalability |
|---|---|
| **PN Algebraic Analysis** | Verification is dependent on the structure of the PN model (i.e., number of places and transitions). This factor is much less than the number of reachable markings produced by other approaches. Therefore for larger models this approach proves to be scalable |
| **CPN State-Space Analysis** | Verification is dependent on the state-space, which tends to grow large for even ordinary models and hence can easily subject to state-space explosion. Some reduction techniques (including one of our own) may minimize this risk but cannot completely omit it. |
| **CSP based Model Checking** | Model checking is also exposed to state-space explosion however it has gone through a continuous evolution of improved algorithms and compact data-structures to minimize this risk. Therefore it promises a better resolution of scalability as compared to the State-space analysis. |

**Table 57: Scalability**

| Category: | Infinite Model Verification |
|---|---|
| **PN Algebraic Analysis** | It is not affected in its reasoning if the model is finite or infinite, because in most of the cases it uses invariants for reasoning which are derived from the algebraic computations and do not depend upon the number of reachable system states |
| **CPN State-Space Analysis** | If the model is infinite it will require a construction of infinite state-space which is infeasible. |
| **CSP based Model Checking** | Infinite model verification using this approach is possible by applying bounded model checking or by abstracting an infinite system into a finite one however this may lead to results with partial correctness. |

**Table 58: Infinite Model Verification**

| Category: | Usability |
|---|---|
| **PN Algebraic Analysis** | This approach is difficult to use due to complex mathematics and requirement to underlying applicable theorems for correct reasoning. |
| **CPN State-Space Analysis** | This approach is easy to use. Most of the operations are automatic. |
| **CSP based Model Checking** | This approach requires some effort to understand the formalisms used for model input and property specifications. However its operations are easy and all run in a black box i.e., the model checker. |

**Table 59: Usability**

| Category: | Automation |
|---|---|
| **PN Algebraic Analysis** | This approach is not automatic because the definition of a property and its theorem applicability requires manual effort. When a property is defined, and a theorem is selected, the modeler has to perform mathematical computations and manually infer whether a condition is satisfied or not. |
| **CPN State-Space Analysis** | This approach is semi-automatic because defining a verification task and a suitable verification function requires modeler's input. However the execution of the function is automatic and it searches all state-space to return a result. |
| **CSP based Model Checking** | This approach is totally automatic. Once a temporal logic assertion is defined, it is executed automatically by and model checker to find out whether it is satisfied or otherwise a counter example is generated. |

**Table 60: Automation**

**Table 55** compares the proposed approaches in terms of the different types of properties that can be verified. It highlights that the PN algebraic technique is limited to verify only general properties (such as deadlock, liveness, fairness) since it depends on the underlying theorems for the proof of their satisfiability. Whereas the other two approaches are relatively more flexible to the specification and verification of properties of varied types, including general and scenario specific properties. **Table 56** presents a comparison of the proposed approaches in terms of the type of models. PN Algebraic approach only supports PN models with simple events without any parameters, guards, actions or input/output state-variables. These features are rather supported by CPN based state-space analysis approach which also provides limited support for Time based CPN models. But for models of complex real-time systems or probabilistic systems Model Checking approach is the suitable choice.

**Table 57** compares these approaches in terms of scalability of the models. In case of Algebraic technique most of the operations in the property verification require matrix computations such as Incidence matrix, P-Invariants, T-invariants. Therefore, the scalability factor is dependent on the size of the matrix i.e., the number of places × number of transitions of the composed model. Thus, the algebraic technique is relatively salable. With regard to scalability the CPN based state-space approach has serious limitations due to its rich data expressiveness and enumeration features. It is reported [**130**] that if the model is very large it generates state-space around $10^5$-$10^6$ nodes. Consequently ordinary PCs cannot easily handle such a large state-space. However there are different approaches to make it more scalable. We also believe that if our proposed state-space reduction technique is directly implemented in the CPN tools environment, this limitation can further be relaxed. Model Checking technique is relatively more scalable. Since it relies on the usage of PAT tool which can handle about $10^7$ states in a reasonable amount of time [**98**]. This should be sufficient for the verification of most industrial scale system models.

According to the **Table 58** the algebraic approach is indifferent whether the model is finite or infinite in nature. An infinite model is a non-terminating model which keeps on evolving indefinitely. Such models are difficult to be verified using State-space approach because its state-space construction is impossible. Although some techniques have been developed such as coverability graphs, to resolve this problem however they fail in some cases, such as in case of timed models. To verify infinite models using Model Checking is somewhat possible using bounded model checking or by abstracting an infinite system into a finite one. However this may lead to results with partial correctness because only a portion of the system state-space can be considered for the reasoning of property correctness. **Table 59: UsabilityTable 59** and **Table 60** compare the ease of use of these approaches in terms of their application in a verification task and the extent of automation they provide.

In short, there is no ultimate winner and making the right choice of an approach entirely depends on the kind of model under investigation and the types of verification properties in question. There are also no exact rules however some fundamental guidelines can be used to help the modeler select a suitable approach:

## 10.1 Guidelines for choosing an approach

In this section some basic guidelines are presented for the modelers in making a suitable choice

### 10.1.1 PN Algebraic Technique

This approach is most suitable when the analysis of a BOM composition is in question which with simple state/transitions and does not require any extension (i.e., it does not have state-variables, or complex notions of transitions with parameters, guards, actions, inputs and outputs etc.). Also its requirement specification includes properties which can be translated in form of PN properties (for which the solution of PN algebraic verification exist). Therefore it should be used when the requirement specifications can be defined in terms of PN properties. For instance, in chapter 7 the objectives are translated into "Fairness" which means that they can be satisfied if the model is fair so the objective of verification is to prove this assumption and can be done using PN algebraic approach. Also it is not effected by the model size, because it performs computations on matrices of the order of (No. of places × No. of Transitions) which remain static, therefore it can also be used for somewhat larger models.

It should not be used if the requirement specification contains reachability properties. Though it is possible to verify them using the PN state equation however it is rather difficult and inefficient as compared to State-Space Analysis approach. This approach cannot be used if the composed model has notions of time, colored-tokens (i.e., the BOM events have parameters) or non-determinism.

### 10.1.2 CPN based State-Space analysis Technique

This approach is best suitable when the given model has (or requires) rich data-centric structure and behavior such as state-variables, events parameters, guards and actions. In this case the BOM components are required to be extended to capture more details. If the modeler has the necessary information to extend the BOM components then he should use this approach otherwise he should choose the Algebraic technique. This approach is also suitable if the modeler wants to execute the composed model at an abstract level to study the behavior of the components before actually implementing them. Although other proposed approaches also have execution/simulation environments, but the CPN based execution is more detailed and comprehensive to study the interaction between composed components, as it provides a hierarchical interconnection between the CPN components and their execution is shown by the flow of data carrying colored tokens among inputs and outputs of each component in an interactive, step-by-step or an automatic fashion. This allows the modeler to closely inspect the composition and its dynamics in a run-time environment. Using this approach has many benefits from a component-based development point of view and the chances of its success are further elevated with our proposed state-space reduction technique called "Compositional State-Space", which reduces the risk of state-space explosion.

This approach can also be used for timed systems since CPN environment supports modeling and verifying timed systems. However few limitations exist since the state-space of timed systems is much more expensive and memory intensive, due to the fact that each state carries an overhead of timed-stamps so even for a simpler model, its state-space will be much heavier than a similar model with no time. Moreover, if the model has even one non-terminating loop, its state-space cannot be constructed as it keeps growing to infinity by incrementing the time-stamps. (i.e., the system may return back to previous states in loops and no new state is being added in the state-space but the time increases so the time-stamps keep on increasing. Therefore with different time-stamps the same states keep on adding infinitely).

This approach however completely fails when certain non-determinism is involved in the model. Even though CPN specification allows using different probability distribution functions, but when they are used, the resultant state-spaces are generated with variations, which cannot be used for verification reasoning. Therefore we do not recommend this approach if the model is stochastic in nature.

### 10.1.3 CSP based Model Checking Technique

This approach is usually favored by majority of the software verification community. It also has a greater flexibility of adopting a new technique or algorithm with specific requirements at hand, and thus can be useful in a variety of contexts. This approach allows the modeler to execute the composed model using PAT simulator (see 3.2.7) therefore it also contends with CPN based State-space analysis in terms of studying system behavior at runtime. However its main strength is revealed when it offers answers to a variety of verification questions, and to a variety of types of systems (real-time, probabilistic etc.) using model checking.

This approach however restricts model expressiveness since it limits the use of data types such as strings, products, records (unlike CPN). This requires an extra effort from the modelers to represent a model in reduced or compact forms using smaller data-types. For instance Boolean flags may be used instead of strings in the parameters such as a pair of string parameters: "*Target_Destroyed*", "*Target_Missed*" can be represented as True/False. Similarly a set of string parameters: {"Red", "Blue", "Green"} can be represented by corresponding integer values {0, 1, 2}. This kind of reduction is required for this approach to work correctly.

For example, we presented a detailed data-centric model of field artillery in chapter 8 to be verified with CPN state-space analysis approach. But when it was required to verify a specific timed property (with non-determinism) we reduced unnecessary details and presented a simpler prototype of the Field Artillery model in chapter 9, focusing only on its behavior relevant to the desired property. By doing this modification the model was useable with this approach which successfully verified the required properties that could not be verified using CPN based approach.

As a final note, each approach has its own benefits and drawback and the choice depends on the modeler's objectives, nature of the task and available information. However we also encourage using multiple approaches for a single task and comparing the results. It gives different perspectives and can better help in confirming correctness.

## 10.2 Thesis Contributions

Component based modeling and simulation is a promising approach to develop and simulate system models. It incorporates numerous benefits such as modular design, logical separation, flexible change management, reusability of existing components, cross-domain model integration and thus consequently helps in reducing cost, time and system complexity. A key characteristic in this expedient paradigm is *composability* that is the ability to add or select and assemble reusable components in order to satisfy user's requirements. In this thesis we mainly endeavored to investigate different aspects of this quality characteristic of component based model design and proposed a composability verification framework for the assessment of its correctness. Our proposed framework uses Base Object Model (BOM), a SISO standard for component based modeling, and performs composability verification of BOM based model compositions with respect to given requirement specifications. In order to prove the correctness of composability of a set of BOM components, our framework undergoes a prescribed verification process, which has different phases starting from system abstraction, requirement gathering, selection of BOM components, their composition to form a conceptual model and then verifying its different levels of composability, in an iterative top-down refinement fashion. When the entire process is completed successfully the composed model is said to be verified with respect to its specifications and can be used for implementation using an implementation architecture (such as HLA) and simulated to serve its purpose.

Following are the key contributions of this thesis:

- We developed a composability verification framework, which stands on fundamental verification principles and backed by the theoretical underpinnings of M&S, the details of which are mainly covered in Part-I. It integrates different methods, techniques, paradigms, algorithms, formalisms, templates, tools and 3rd party libraries (or APIs) to support different tasks in the multi-tier composability verification process of a composed model with respect to its requirement specification.

- We outlined a component based modeling and simulation (CBM&S) life-cycle by categorizing its different phases, and activities under each phase. A pictorial representation has been used to explain different tasks conducted under each phase. This life-cycle provides guidelines for using various features of our framework, and allows the user to conduct verification operations in a systematic fashion.
- A template to define and express requirements in a formal way is proposed. Our requirement specification template can be used to specify a set of objectives and system constraints. Objectives can be seen as ultimate goals while the constraints are necessary quality requirements that must be satisfied for achieving the objectives.
- Inspired from the Discovery, matching and composition (DMC) paradigm of model development [**19**], we propose method for rapid development of BOM based conceptual models.
- We propose a formal description of BOM components and their compositions for documentation purpose. We also propose a graphical notation[38] to describe the structure of the BOM component and to show how they are connected to each other in a compact form. This notation can be used as blue prints of different model compositions and can be shared among different teams or archived in the repository for reference.
- We propose methods for evaluating the structural consistency of the composed BOMs using rule based static analysis technique. The structural analysis involves checking that the components are correctly connected and they can communicate with each other with correct semantics. For semantic analysis, we propose an OWL based differencing approach which checks that the communication of the components is semantically consistent, meaningful and is understood as intended.
- We suggest a behavioral evaluation technique which implicates that the components can correctly interact with each other in a right causal order to reach final states or pass through the goal states. For this purpose we propose state-machine matching process, which transforms BOM state-machines of each component into an executable SCXML format and execute them to analyze their interaction. If there is no deadlock and all the state-machines make required progress then the behavior of the components is reported to be consistent.
- For the evaluation of dynamic-semantic composability level, our framework incorporates three main approaches: (a) PN Algebraic technique (b) CPN-based State-space analysis technique and (c) CSP based model checking. These three approaches are offered to be used as alternatives to each other and their selection is dependent on the nature of the model being investigated and decision of the modeler. We also present basic guidelines to help the modeler choose an appropriate approach.
- For each approach we develop automatic transformation tool that transforms a BOM based composed model into its respective executable model description formalism. This method is inspired from Model Driven Architecture, in which a platform independent model is transformed into platform specific model using some transformation rules. We also propose BOM extensions based on certain additional details that are required for correct transformation. For this purpose we develop a BOM extension editor that takes modeler's input for extending BOM components.

[38] It should be noted that different UML diagrams such as State charts and sequence diagrams are used to describe BOMs informally. Our graphical notation follows the pattern of CBSE.

- We have applied our proposed approaches in three different case studies discussed in chapter 7, 8 and 9 respectively. Each case study provides a proof of concept and validates specific characteristics of our framework. For PN based algebraic technique we presented a manufacturing system, in which fairness property is verified. For CPN-based state-space analysis approach a field artillery model is presented in which a set of scenario specific properties are verified. For model checking, the same field artillery scenario is modified into a timed non-deterministic model and a particular time property is verified with some probabilistic assumptions.
- We introduce a CPN based component model in order to describe a BOM component (or any other simulation component) in form of an executable model that can be executed using CPN execution environment. This CPN component model can also represent any other simulation component using its three layers namely (i) structural layer: which is used to define component attributes and variables; (ii) behavioral layer: which is used to describe the state-machine of a component and (iii) communication layer: which is used to describe components interfaces and how it can connect with other components and communicate. We transform all BOM components into the proposed CPN based component model and compose them to form a composed model which can be executed in CPN environment and verified using CPN based state-space analysis technique.
- We introduce a State-space reduction technique called Compositional state-space. This technique assumes that all the composed components are black-boxes and their inputs and outputs are exchanged in the main model. Therefore we can select all the nodes from the state-space which are relevant to any activity happening in the main model and filter all the other nodes, by replacing them with edges. The resultant graph will be a reduced state-space representing only those nodes which describe the interactions of components in the main model and provide sufficient information for composability verification.

## 10.3 Conclusions

The verification framework proposed in this thesis expedites the process of composability verification of BOM based composed models with respect to the requirement specifications. A verified composed model ensures consistent structure and behavior and guarantees the satisfaction of its objectives and required constraints. A rapid development of the conceptual model using Discovery, Matching and Composition paradigm, its automatic transformation into an executable form and its composability verification helps in studying its structural and behavioral correctness with respect to the given requirement specifications. This helps in rectifying any possible defects in the model design before it is actually implemented and simulated to serve its purpose, and thus saves a significant amount of time, cost and achieve robustness. Moreover this process strongly supports reusability as the entire process can easily be repeated to compose same components for different scenarios with varied configurations or with different requirement specifications (as in chapter 8 and 9).

The entire composability verification framework is acclimated by a systematic Component Based M&S life-cycle which gives an outline of different phases of component based M&S development process, where each phase has different activities. This life-cycle inherits important features and characteristic of some existing M&S development life-cycles and the Model Driven Architecture with an expansion of component based model development and guides the modelers with necessary directions to perform different tasks at different phases.

An important feature of this life-cycle is the software engineering principle of top-down refinement. According to this principle a conceptual model is refined into an executable form through a number of intermediary steps. Each step generates a relatively detailed version of the abstract model and is easier to reason about its correctness based on assumptions of its previously verified version. For instance, when the state-machine matching process is successful we can proceed to a more detailed dynamic level execution/verification with an assumption that the behavior of the composed components is consistent.

Our experience with the three different dynamic analysis approaches proves to be very constructive for composability verification. Each approach in its own way provides significant improvement on efficient and accurate reasoning regarding model correctness. We profess that the cross domain sharing of existing knowledge and valuable contributions from other communities (such as PN, CSP, model checking in our case) bridges cooperation in problem solving and helps in accomplishing quality research.

## 10.4 Future Directions

Some of the key future directions of this work include:

- We intend to deploy the composability verification framework in different application areas to evaluate its potential and to make use of its valuable features in verification. One area is the component based design for robotics applications. Many software architectures for robotic applications support component oriented design and thus can be explored for the utilization of our composability verification process, such as in studying various aspects of behavioral composability in different robotic applications.

- In the context of improvements in the verification framework following are some key future directions:
    - We intend to include verification of requirement specifications. Correctness of requirements is a necessary aspect for successful verification.
    - We also intend to produce viable solution for the validation of the composed model with respect to the real system.
    - We defined Pragmatic composability level in chapter 2 however the composability verification at this level is still under investigation. We intend to explore this direction in future.

- In general we are interested to explore the area of component based design optimization and to study the composability of component design for optimization with multiple objectives.